\documentclass[final,12pt]{clear2024} 
\def\KRR{{k\text{-RR}}}
\newcommand{\pr}{\mathbb{P}}
\newcommand{\calM}{\mathcal{M}}
\usepackage{nicefrac}
\usepackage{pbalance}
\usepackage{mathtools}
\usepackage{float}

\usepackage{microtype}
\usepackage{graphicx}
\usepackage{float}
\usepackage{float}
\usepackage{graphicx}

\definecolor{caribbeangreen}{rgb}{0.0, 0.8, 0.6}
\usepackage{xcolor}

\title{Causal Discovery Under Local Privacy}
\usepackage{times}

\clearauthor{%
\Name{Ruta Binkyte} \Email{ruta.binkyte@inria.com}\\
\Name{Carlos Pinzón} \Email{carlos.pinzon@inria.fr}\\
\Name{Szilvia Lestyán} \Email{szilvia.lestyan@inria.fr}\\
\Name{Kangsoo Jung} \Email{gangsoo.zeong@inria.fr} \\
\Name{Héber H. {Arcolezi}} \Email{heber.hwang-arcolezi@inria.fr}\\
\Name{Catuscia Palamidessi} \Email{catuscia.palamidessi@inria.fr}\\
\addr
}

\begin{document}
\sloppy

\maketitle

\begin{abstract}%
   Differential privacy is a widely adopted framework designed to safeguard the sensitive information of data providers within a data set. 
   It is based on the application of controlled noise at the interface between the server that stores and processes the data, and the data consumers. Local differential privacy is a variant that allows data providers to apply the privatization mechanism themselves on their data individually. Therefore it provides protection also in contexts in which the server, or even the data collector, cannot be trusted.
    The introduction of noise, however, inevitably affects the utility of the data, particularly by distorting the correlations between individual data components. 
    This distortion can prove detrimental to tasks such as causal discovery.
    In this paper, we consider various well-known locally differentially private mechanisms and compare the trade-off between the privacy they provide, and the accuracy of the causal structure produced by algorithms for causal learning when applied to data obfuscated by these mechanisms. 
    Our analysis yields valuable insights for selecting appropriate local differentially private protocols for causal discovery tasks. 
    We foresee that our findings will aid researchers and practitioners in conducting locally private causal discovery.
\end{abstract}


\section{Introduction}

The notion of causality is gaining popularity in machine learning (ML) because of its benefits for accuracy~\citep{richens2020improving}, robustness~\citep{tople2020alleviating, scholkopf2021toward}, explainability~\citep{madumal2020explainable} and fairness~\citep{loftus2018causal}. Many applications of causality in ML rely on knowing the ``causal structure" in the data ~\citep{binkyte2022causal, kyono2021exploiting}.
Causal discovery algorithms are computational methods that aim to infer causal relationships from observational data~\citep{nogueira2021causal, spirtes1991algorithm}. Those algorithms mostly rely on correlations between the various components (\emph{variables}) of the data. These correlations can be affected by the application of data-privatization mechanisms aiming at protecting the privacy of the data providers. However, protecting data privacy is a legal obligation in Europe and many other countries worldwide. Answering to this necessity, numerous privatization methods have been developed to maximize the trade-off between a good level of data privacy and utility.

\emph{Differential privacy} (DP)~\citep{Dwork2006} is one of the most popular data-privatization approaches. Depending on the trust model, DP can be further classified into \emph{central} and \emph{local}. Central DP, which is the original notion of DP, assumes the existence of a trusted server where the data is aggregated. Data consumers (analysts) cannot access the data set directly but only query it via the server, which is supposed to obfuscate the answer by controlled noise, before reporting it to the analysts. The DP property establishes a bound on the ratio of the probability of getting the same reported answer from two adjacent databases,  namely, two databases that differ for just one record. The bound is expressed in terms of a parameter $\epsilon$, which represents the level of privacy. DP is used nowadays in a variety of applications from programming languages \citep{dpusage1} to social networks \citep{dpusage2} and geolocation \citep{dpusage3}.

One limitation of the central DP model is that the server or the data collector cannot always be trusted: they may collude with an attacker, or just be unable to protect the data from security breaches. For this reason, local DP (LDP) has been proposed as an alternative model~\citep{first_ldp,Duchi2013}. In LDP, the individual data are obfuscated directly at the end of the data provider, before even being collected. The main advantage of LDP is that users are more willing to share their data when they don't need to rely on the trustworthiness of the data collector and the server. This model has become popular, especially thanks to the fact that has been adopted and promoted by High Tech leading companies such as Google~\citep{google}, Microsoft~\citep{microsoft} and Apple~\citep{apple}.

A variant of DP called \emph{$d$-privacy} (also known as \emph{metric privacy}),  was introduced in \citep{chatzikokolakis2013broadening}. $d$-privacy is suitable for domains provided with a notion of distance. Like in central and local DP, $d$-privacy imposes a bound on the probability that the same result is obtained from two different objects (the arguments of the mechanism). However, in contrast to DP, this bound does not depend only on the parameter $\epsilon$, but also on the distance between the objects. This means that the noise can be calibrated depending on how large the range in which we want to achieve indistinguishability is. In contrast, LDP requires indistinguishability between any pair of elements in the domain. $d$-privacy, therefore, is particularly useful in those applications in which hiding an element within a group of neighbors is a sufficient measure of privacy protection. 
$d$-privacy has been applied especially in the local model, and in particular, in the context of location privacy, where it takes the name of \emph{geo-indistinguishability}~\citep{Andrs2013}. 

In general, the addition of noise tends to reduce the utility of the information that can be extracted from the data. Many privatization approaches and denoising techniques have been optimized for the summary statistics of the individual variables in the data, such as average values. However, notions of utility also depend on the correlation between the various components of the data, especially in the case of causal discovery. Some approaches to cope with this problem have been proposed in the global DP setting when the full unobfuscated data set is available. For example, the collected data may be synthesized using generative algorithms such as GAN~\citep{jordon2019pate} or Bayesian Networks~\citep{Zhang2017}. However, little or no instances of relation-preserving local DP mechanisms are known for causal discovery. Under the local setting, the data are already obfuscated before they get to the central server, and, therefore, the methods used in global DP are not applicable.

In this work, we experimentally assess the impact of state-of-the-art LDP and $d$-privacy mechanisms on the structural accuracy of causal discovery from the data. 
More precisely, as the LDP representative, we consider the $k$-Ary Randomized Response ($k$-RR, Section~\ref{sub:KRR})~\citep{kairouz2016discrete}. 
As the local $d$-privacy representative, we considered the Geometric mechanism (Section~\ref{sec:geo}). 
We conduct extensive experiments on both real and synthetic data sets, and we evaluate their impact on 9 causal discovery algorithms, including constraint-based, score-based, and causal asymmetry-based methods.
In summary, the two main contributions of the paper are the following:

\begin{itemize}

    \item The paper systematically compares the performance of different locally differentially private mechanisms, specifically Geometric and $k$-RR, in the context of causal discovery tasks.
    With our findings, we highlight the advantages of using Geometric privatization methods over $k$-RR, shedding light on the impact of noise levels on algorithm performance.
    
    \item We introduce a unified privacy measure from an attacking perspective, allowing for the comparison of two distinct privacy notions: LDP and local $d$-privacy. 
    This measure facilitates the assessment of privacy-utility trade-offs in real-world tasks such as causal discovery.

    
\end{itemize}

\textbf{These contributions collectively enhance our understanding of locally differentially private mechanisms in the context of causal discovery and offer valuable insights into their application in real-world scenarios.}
Indeed, we hope this work can aid practitioners in collecting multidimensional user data in a privacy-preserving manner by providing insights into which locally private mechanism and causal discovery algorithms are best suited to their needs. 


\section{Related Work} \label{sec:rel_work}

Causal discovery with DP is an emerging research area that aims to combine the benefits of both identification of causal relationships among variables and privacy-preserving data analysis.
The goal is to discover causal relationships between variables while preserving the privacy of sensitive data.
One explored approach in the literature for differentially private causal discovery was to incorporate DP mechanisms directly into existing causal discovery algorithms~\citep{kusner16,Xu2017,Wang2020,Ma2022}.
These algorithms introduce controlled noise during the causal learning process to ensure privacy protection.

However, these existing differentially private causal discovery algorithms assume the centralized DP model, which requires collecting users' original data.
The approach adopted in this paper is to leverage the concept of local DP~\citep{first_ldp,Duchi2013} for causal discovery (respectively local d-privacy~\citep{chatzikokolakis2013broadening}).
In recent years, there have been several works on the local DP setting (e.g., see~\citep{tianhao2017,google,apple,microsoft,kairouz2016discrete,Duchi2013,Hadamard,Arcolezi2022,kikuchi2022castell,cormode2018privacy} and references within), and applying them to causal discovery involves sanitizing the data at the individual level. Parallel to our work, \cite{agarwal2021causal} study a class of corruptions, such as measurement error, missing values, discretization, and differential privacy in the US Census. However, their goal is to learn a causal parameter (average treatment effect) from corrupted data and they conduct experiments only in an aggregated setting. 
Similarly, \cite{ohnishi2023locally} offer causal inferential methodologies to analyze locally differentially private data. \cite{mooij2016distinguishing} experiment with causal discovery with small amount of noise added to the data. However, the noise is not produced by the privatization mechanism.
These goals differ from our work, they investigate the effect of noise in causal effect estimation of a treatment (or intervention) when randomized experiments are impossible to conduct, thus statistical theory is needed. Our work solely focuses on causal discovery, that is the inference of causal \textit{relations}, causal \textit{directions} among a set of variables (i.e., ``how the change in X influences Y?" versus ``is X the cause of Y?").
To the authors' knowledge, this is the first work that thoroughly explores and analyzes the impact of locally differentially private mechanisms on causal discovery.

\section{Preliminaries} \label{sec:preliminaries}

\subsection{Privacy Notions}\label{sec:privacy-notions}

\subsubsection{Local Differential Privacy}

One privacy model considered in this paper is LDP~\citep{first_ldp,Duchi2013}, which is formally defined as follows.

\begin{definition}[$\epsilon$-Local Differential Privacy]\label{def:ldp} Let $\epsilon>0$ be a parameter representing the level of privacy loss. A randomized mechanism ${\calM}$ satisfies $\epsilon$-local-differential-privacy ($\epsilon$-LDP) if, for any pair of input values $v_1, v_2 \in Domain(\calM)$, and any possible output $x$ of ${\calM}$, the following holds (where $\pr[e]$ represents the probability of the event $e$):

    \begin{equation*} \label{eq:ldp}
        \pr[{\calM}(v_1) = x] \leq e^\epsilon \cdot \pr[{\calM}(v_2) = x] \  \textrm{.}
    \end{equation*}
\end{definition}

In essence, LDP guarantees that it is unlikely for the data aggregator to infer the true value from the reported data.
The privacy loss $\epsilon$ controls the privacy-utility trade-off. Note that lower values of $\epsilon$ result in tighter privacy protection.
Similar to global DP, LDP also has several fundamental properties, such as robustness to post-processing and composition~\citep{dwork2014algorithmic}.

\subsubsection{Local \texorpdfstring{$d$}{d}-Privacy}
$d$-Privacy assumes that the domain of the mechanism ${\calM}$ is provided with a notion of distance $d$. 
\begin{definition}
\label{sub:d-priv}
     A mechanism ${\calM}$  satisfies   $d$-privacy, with privacy parameter $\epsilon$, iff  for all values, $v_1, v_2 \in Domain(\calM)$ and all possible outputs $x$, the following inequality holds:
    \begin{equation*} \label{eq:ldp-d-privacy}
        \pr[{\calM}(v_1) = x] \leq e^{\epsilon \, d(v_1,v_2)} \cdot \pr[{\calM}(v_2) = x] \  \textrm{.}
    \end{equation*}
\end{definition}

In essence, in the local model $d$-Privacy guarantees, like in LDP, that it is unlikely for the data aggregator or an attacker to infer the true value $v$ from the reported data. But in this case, it is because it is made indistinguishable from all the other values in the neighborhood. In other words, nearby secrets should look almost identical to any observer.

\subsection{Privacy Mechanisms}\label{sec:privacy-mechanisms}

In this section, we describe the various discrete multidimensional mechanisms used in this paper.
Visually, Figure~\ref{fig:mechanisms} in Appendix~\ref{app:privacy_mech} depicts the 4 mechanisms applied to a single point in a 4D space with shape $(2, 5, 5, 5)$, denoting the number of categories or bins per dimension.

\subsubsection{\texorpdfstring{$k$}{k}-ary Randomized Response (\texorpdfstring{$k$}{k}-RR)}
\label{sub:KRR}

Randomized Response (RR) was proposed in~\citep{Warner1965} with the aim of providing ``plausible deniability'' to individuals responding to embarrassing (binary) questions in a survey.~\citet{kairouz2016discrete} generalized RR to domains of arbitrary size $k$ (with $k \geq 2$), and proposed the so-called $k$-RR mechanism, which is one classical technique for achieving LDP on categorical/discrete data. Given a data domain $V$, and the privacy parameter $\epsilon$, let $k=|V|$ and $p\coloneqq\frac{e^{\epsilon}}{k-1+ e^{\epsilon}}\in (0,1)$. For each $v\in V$, let $\eta_{\neq v}\in V$ be a uniform  random variable  (i.e., exogenous noise with uniform distribution) over $V\setminus\{v\}$. We let $\KRR{}: V \to V$ be the random variable given by:

\begin{equation*}
    \KRR{(v;\epsilon)} \; \coloneqq \;\begin{cases}
        v ,            & \textrm{with probability } p                \\
        \eta_{\neq v}, & \textrm{with probability } 1-p \ \textrm{.}
    \end{cases}
\end{equation*}

This mechanism satisfies $\epsilon$-LDP ~\citep{kairouz2016discrete}, because $\frac{p}{q}=e^{\epsilon}$, where $q\coloneqq \nicefrac{(1-p)}{(k-1)}$.
When collecting data in practice, one is often interested in multiple attributes of a population, i.e., \textit{multidimensional data}. We assume there are $d$ attributes with domains $A_1,A_2,\ldots,A_d$, where each  $A_i$ is a discrete set of finite size $k_i=|A_i|$. Each data provider $u_j$ for $j \in \{1,2,...,n\}$ contributes to the data set  with a tuple (record) $\textbf{v}^{(j)}=(v^{(j)}_{1},v^{(j)}_{2},...,v^{(j)}_{d})$, where $v^{(j)}_{i}$ represents the value of the attribute $A_i$. We now describe the two main known methods for applying $k$-RR on multidimensional data~\citep{Arcolezi2022,kikuchi2022castell,Domingo_Ferrer2020}. 

\begin{description}
    \item[$k$-RR Component-wise]  ($k$-RR C-wise). This is a naive approach that applies $k$-RR independently on each attribute. More precisely, $k$-RR C-wise splits the privacy budget $\epsilon$ among the $d$ attributes uniformly or proportionally to their size, and reports each attribute in $A_i$ using $k_i$-RR parameterized with $\epsilon_i$-LDP, for $\sum_{i=1}^{d} \epsilon_i=\epsilon$. In this paper, we set $\epsilon_i=\epsilon\cdot\frac{k_i}{k_1+k_2+\ldots+k_d}$. 

    \item[$k$-RR Combined] ($k$-RR Comb). This mechanism considers the Cartesian product $A_1 \times A_2 \times \ldots \times A_d$ as a single attribute and sanitizes it using $k$-RR parameterized with $\epsilon$-LDP, where
        $k=k_1\cdot k_2\cdot\ldots\cdot k_d$. 
\end{description}

\subsubsection{Bounded geometric mechanism}
\label{sec:geo}

The geometric mechanism is the discrete analogous of the Laplace mechanism.
The output $Y$ is related to the input $X$ by the formula:

\begin{equation}\label{eq:geo}
    \pr[Y=y|X=x] = p_{\max} \exp(-\epsilon |y-x|)
\end{equation}

\noindent for some parameters $\epsilon$ that represents the level of privacy.  $p_{\max}$ is a normalization factor, i.e., it is  chosen so that $\sum_y \pr[Y=y|X=x]=1$.
This formula is valid in 1D, in which $|\cdot|$ denotes the absolute value, as well as in multidimensional Euclidean space, in which $x$ and $y$ are discrete vectors and $|\cdot|$ denotes the Euclidean norm, or any other $p$-norm chosen in advance (see Figure~\ref{fig:geometric-p-norm} for a comparison).
From the definition of the geometric mechanism, it is immediate that it satisfies local $d$-privacy with privacy parameter $\epsilon$, where the metric $d$ is the chosen $p$-norm based distance.

In this paper, we are interested in bounding the geometric mechanism so that the output domain equals the input domain, as in $k$-RR. 
There are three natural ways to do it, namely (1) clipping, (2) replacing samples that are out of the box with uniform noise, and (3) resampling whenever a sample is out of the box.
Let us review them in more detail.

The method (1), clipping, consists of replacing all the output values that lie outside the box with the closest values that lie inside the box, i.e., with the maximum or minimum values of the domain in the 1D case.
In this case, the two extremes of the box may increase their probabilities excessively, and the property that the output $y$ with maximum probability is always $y=x$ might be lost, especially when the input $x$ is close to the border.
In method (2), whenever the output $y$ is outside the box, it is replaced with a uniform sample from the box.
In terms of the probability distribution of the mechanism, this method crops it from the background (two tails in the 1D case), and rescales the cropped distribution by adding a constant.
This addition results in combinations of exponential terms with additive constants, which adds complexity to the formulas unnecessarily and distorts the exponential shape and its decay properties.
Instead, in method (3), which corresponds to sampling as many times as necessary until the output is inside the box, the cropped distribution is simply multiplied by a constant.
This preserves the main shape of the distribution while also keeping the formulas relatively simple.
For this reason, we prefer method (3) over the other two.

Notice that bounding is not symmetric, except for the input in the center of the box.
This means, that we should have different values of $p_{\max}$ or $\epsilon$ for different values of $x$ so that the bounded summation is $1$ on all $x$.
As it will be justified in Section~\ref{sec:tuning-priv}, we opt for fixing $p_{\max}$, so the formula that characterizes the bounded geometric mechanism becomes:
\[
    \pr[Y=y|X=x] = p_{\max} \exp(-\epsilon_x |y-x|)
\]
where both $x$ and $y$ are constrained to a fixed bounded discrete set, and $\epsilon_x$ are chosen so that \(\sum_y \pr[Y=y|X=x] = 1\).
These values always exist (assuming $p_{\max} \geq 1/k$), and we provide an algorithm for finding them.

The computation of $\epsilon_x$ for every $x$ is not possible symbolically through a formula.
    It is required that \(\sum_y \pr[Y=y|X=x] = 1\), or equivalently, \(\sum_y \exp(-\epsilon_x |y-x|) = \frac{1}{p_{\max}}\), where both $x$ and $y$ are constrained to a fixed bounded discrete set.
In the 1D case, the domain is a set of $k$ contiguous integers and for the smallest value of $x$, only one tail of the geometric distribution intersects the domain, which allows us to write $\frac{1}{p_{\max}} = \sum_y \exp(-\epsilon_x |y-x|) = \sum_{\delta=0}^{k-1} \exp(-\epsilon_x k) = \frac{1-\exp(-k \epsilon_x)}{1-\exp(\epsilon_x)}$.
However, there is no analytical solution for $\epsilon_x$ from this formula.
Moreover, for the remaining values of $x$, the expression becomes more complex, as an additional term is added for the second tail, and even more for the multidimensional case.

Nevertheless, the computation of each $\epsilon_x$ can be carried out numerically by exploiting the fact that $\sum_y \exp(-\epsilon_x |y-x|)$ is decreasing on $\epsilon_x$.
At one extreme, if $\epsilon_x \to 0$, the sum approaches $k$, and at the other, if $\epsilon_x\to\infty$, the sum approaches $1$.
This implies, first, that there is a unique point $\epsilon_x$ for which this function crosses the threshold $\frac{1}{p_{\max}}$, and more importantly, that we can use a binary search to compute $\epsilon_x$.
In the multivariate domain, the summations still satisfy the monotonicity property.
Therefore, this method can be used to implement the multidimensional geometric distribution.
Similar to $k$-RR, we compare two versions of the Geometric mechanisms, i.e., Geo Comb and Geo C-Wise.

\subsection{Causality Notions}\label{sec:causality-notions}
\subsubsection{Causal Graph}

A directed acyclic graph (DAG) $\mathcal{G} = (\mathbf{V},\mathcal{E})$ is composed of a set of variables/vertices $\mathbf{V}$ and a set of (directed) edges $\mathcal{E}$ between them such that no cycle is formed.
Let $\pr$ be the probability distribution over the same set of variables $\mathbf{V}$.
$\mathcal{G}$ and $\pr$ satisfy the Markov condition if every variable is conditionally independent of its non-descendants given its parents.
Assuming the Markov condition, the joint distribution of variables $V_1,V_2,\ldots \in \mathbf{V}$ can be factorized as:
\begin{equation}
    \pr[V_1,V_2,\ldots, V_d] = \prod_{i}\pr[V_i | Pa(V_i)] \label{eq:markov}\ .
\end{equation}
where $Pa(V_i)$ denotes the set of parents of $V_i$.
A partially directed acyclic graph (PDAG) is a special type of DAG that contains directed and undirected edges.

\subsubsection{Causal Discovery Algorithms}

Causal discovery is concerned with the identification of causal relations from the data. More precisely, it aims to learn the fully directed DAG or partly directed PDAG that best describes the given data set. 
Several causal discovery algorithms exist for a wide range of different assumptions, for a survey see \citep{glymour2019review}.
\vspace*{-3mm}
\section{Tuning the Level of Privacy} \label{sec:tuning-priv}

The parameter $\epsilon$ in LDP does not have the same meaning as the $\epsilon$  in $d$-privacy, i.e., they represent different levels of privacy.
In order to compare the mechanisms of these two families, we need to tune the respective $\epsilon$'s so as to represent the same level of privacy.
To avoid confusion for the readers that know the standard notion of DP, and are not so familiar with LDP, it  is important to remind that the standard notion for privacy in the local framework is not the same as in the central one: In central DP, the challenge for an attacker is to distinguish between two adjacent data sets, i.e., data sets that differ for presence or absence of one record. In other words, the attacker wants to infer whether or not a certain record is in the data set or not. In LDP, on the contrary, the aim of the attacker is to infer the true value of the individual data provider.

To measure the level of privacy, therefore, we consider the probability that an attacker has to infer the true value from the reported value. Naturally, the attacker
will put her bet on the  value that has the maximum posterior probability, given the obfuscated value~\citep{Arcolezi2023,Chatzikokolakis2023}.
We note that this measure of privacy is directly related to the notion of \emph{advantage of an attacker} in security, and to the notion used to assess the vulnerability of the training set in ML. 

In both $k$-RR and $d$-privacy, the value that has the highest probability to be reported is the true value itself, hence the  level of privacy provided by these mechanisms (assuming a uniform prior) is the probability to report the true value.
Specifically, the level of privacy
provided by $k$-RR with parameter $\epsilon$ is:
\begin{equation*}
    \mathit{Priv}_{k\textit{-RR}}(\epsilon)\; \coloneqq \;\frac{e^\epsilon}{k-1+ e^\epsilon}\ .
\end{equation*}
whereas, for a Geometric with parameter $\epsilon'$, the level of privacy is:
\begin{equation*}
    \mathit{Priv}_{Geo}(\epsilon') \; \coloneqq \; \pr_{\max}\cdot e^{\epsilon'\cdot 0} \; = \; \pr_{\max}\ .
\end{equation*}
where $p_{\max}$ is the normalization factor used in the definition of the geometric mechanism 
(Equation~\eqref{eq:geo}). 
Tuning the parameters of $k$-RR and $L$ to provide the same level of privacy means adjusting the above $\epsilon$ and $\epsilon'$ so that $\mathit{Priv}_{k\textit{-RR}}(\epsilon)$ and $\mathit{Priv}_{Geo}(\epsilon')$ give the same result.

\vspace*{-3mm}
\section{Experimental Results} \label{sec:results}

In this section, we empirically assess how locally private mechanisms impact causal discovery.
We evaluate the performance of 9 causal discovery algorithms in multidimensional, two-dimensional, real and synthetic data sets obfuscated using the various mechanisms described in Section~\ref{sec:privacy-mechanisms}. 
We start by applying the causal discovery algorithms to discretized non-privatized data. Then we select the algorithms that performed best at a particular data set and apply them to the privatized versions of this data set. We measure the effect of each privatization method on the algorithms by comparing the Structural Hamming Distance (SHD) score and the F1 score or Accuracy on the non-privatized and privatized data. 
We use the Benchpress causal discovery benchmarking framework~\citep{rios2021benchpress} to generate synthetic data and run causal discovery algorithms for multidimensional experiments. 
As (L)DP mechanisms are randomized, we report average results over 5 runs.
Due to space constraints, we have included all of our additional experiments in Appendix~\ref{app:all_exp}.
\vspace*{-3mm}
\subsection{Data Sets}\label{sec:datasets}

We use real benchmark and synthetic data sets for the experiments. The details can be found in Table~\ref{tb:datasets} and in Appendix~\ref{app:data}.

 \begin{table}[H]
     \centering

     \begin{tabular}{|l|c|c|c|c|c|}
   \hline
         Name & Type & Nodes& Bins & Size & Origin\\
           \hline
         Sachs & real & 11 &10& 902 &  \citep{sachs2005causal}  \\
         Human Stature & real&3 &10 & 898 & \citep{han2015galton} \\
         Synth10 & synthetic &10&10 & 5000  & random DAG, IID, Linear, Gaussian\\
         Synth5 & synthetic &5& 5& 50000  & random DAG, IID, Linear Gaussian\\
         CEP & real & 2 & 2-100 & 94-16382
         
 & \citep{mooij2016distinguishing}\\
 \hline
     \end{tabular}
     \caption{Data sets used for causal discovery. For CEP the number of bins was determined by $min(u, 100, u*0.1 )$, where $u$ denotes the number of distinct values.}
     \label{tb:datasets}
 \end{table}

\vspace*{-5mm}

\subsection{Causal Discovery Algorithms}\label{sec:aglorithms}

We apply constraint-based and score-based causal discovery algorithms for multidimensional data. We select several well-known algorithms that can run on discretized data. For pairwise data sets, we apply algorithms that are capable of identifying the causal direction for two variables. We test the performance of the discrete and continuous data-specific versions of the algorithms, as well as various parameter values.
The details can be found in Table~\ref{tb:algorithms} and in the Appendix~\ref{app:algs}.

We have used two libraries for implementation: we used the Benchpress \citep{rios2021benchpress} package for the PC, FCI, FGES, Iterative MCMC and MMHC causal discovery algorithms and metrics. For the RECI, IGCI, CDS and ANM methods we used the Causal Discovery Toolbox \citep{kalainathan2020causal}.

  \begin{table}[H]
  \scriptsize
      \centering
      \begin{tabular}{|l|c|c|c|c|}
        \hline
          Algorithm  & CI Test/Score &  Parameter\\
          \hline
          PC \citep{spirtes1991algorithm} & Gaussian, Chi-square  &Alpha (0.001,0.05, 0.1 )\\
          FCI \citep{entner2010causal} &  Fisher-Z, Chi-square  &Alpha (0.01,0.05,0.1) \\
        FGES \citep{ramsey2017million}&  BIC & Penalty discount (0.75,0.8,1,1.5) \\
        Iterative MCMC (\cite{kuipers2022efficient})& BGe & Alpha (0.001,0.01,0.1) \\ 
        MMHC \citep{tsamardinos2006max}& BDe & Alpha (0.01,0.05, 0.1)  \\
        \hline
        RECI \citep{blobaum2018cause} & MSE & \\
        IGCI \citep{daniusis2012inferring} & sp1 & \\
        CDS \citep{fonollosa2019conditional} & std. dev. & Forced Decision\\
        ANM \citep{hoyer2008nonlinear} & HSIC & \\
        \hline
      \end{tabular}
 
      \caption{The structure learning algorithms.}
     \label{tb:algorithms}
  \end{table}

\vspace*{-3mm}
\subsection{Discretization}

In order to apply the discrete mechanisms of interest to our data set, it was necessary to discretize the original continuous data.
Discretization is a critical step in the process, as it plays a pivotal role in the subsequent data analysis.
There are several approaches to discretizing data, each with varying effects on the quality of the results.
Some of these methods yield higher average precision, up to the highest possible~\citep{pinzon2020approach}, but rely on knowledge of properties about the underlying data distribution, such as quantiles or an estimation of the density function.
However, in situations where the underlying data is sensitive and private, revealing such properties can risk privacy breaches, so it is safer to assume that they are unknown.
To address this challenge, we opted for the simplest method of discretization, namely uniform bins within a fixed range.
In practice, this fixed range corresponds to estimations of the minimum and maximum values of the population.

The only parameter we can freely choose in this process is, therefore, the number of bins and it should be chosen taking into account that more bins imply more accurate information being revealed.
Moreover, the number of dimensions of the data, which corresponds to the number of columns in the data set, also plays a role in the choice of the number of bins, as it increases exponentially the total number of bins.
We chose between 5 and 10 bins for data sets with 3 or more dimensions, and for the CEP data set, which has two dimensions but contains several different data sets, we applied a dynamic number of bins (see Table \ref{tb:datasets}). Some of these data sets were already discretized (e.g., had only 2 distinct values), and some had continuous data. We determined the number of bins by $min(u, 100, u*0.1 )$, where $u$ denotes the number of distinct values in a given data set. 

\vspace*{-3mm}

\subsection{Evaluation Metrics}
For the data sets with more than two-dimensions we used structural hamming distance (SHD) to measure the difference between the ground truth adjacency matrix and the output of the causal discovery algorithm. It assigns a distance of 1 for every missing, redundant or reversed edge in the graph. Intuitively, SHD provides a number of edges that are need to be added, removed and re-directed to make the two graphs identical.
We have also calculated the F1 score, that combines the precision and recall of a model, and  is used to evaluate the recovery of the skeleton of the DAG.\\
In case of the CEP data set, we have applied the same method as in \citep{mooij2016distinguishing}. Forced-decision: given a sample of a pair $(X, Y)$ the methods \textit{must} decide on a causal direction. Then, we evaluate the weighted\footnote{Not all pairs can be considered as independent. Weights' list was acquired from the authors' website.} accuracy of the decisions. We also calculate the confidence intervals assuming a binomial distribution using the method by \citet{clopper1934use}.

\vspace*{-3mm}
\subsection{Results on Multidimensional Data}

\begin{figure}[h]
    \centering
    \includegraphics[width=0.8\linewidth]{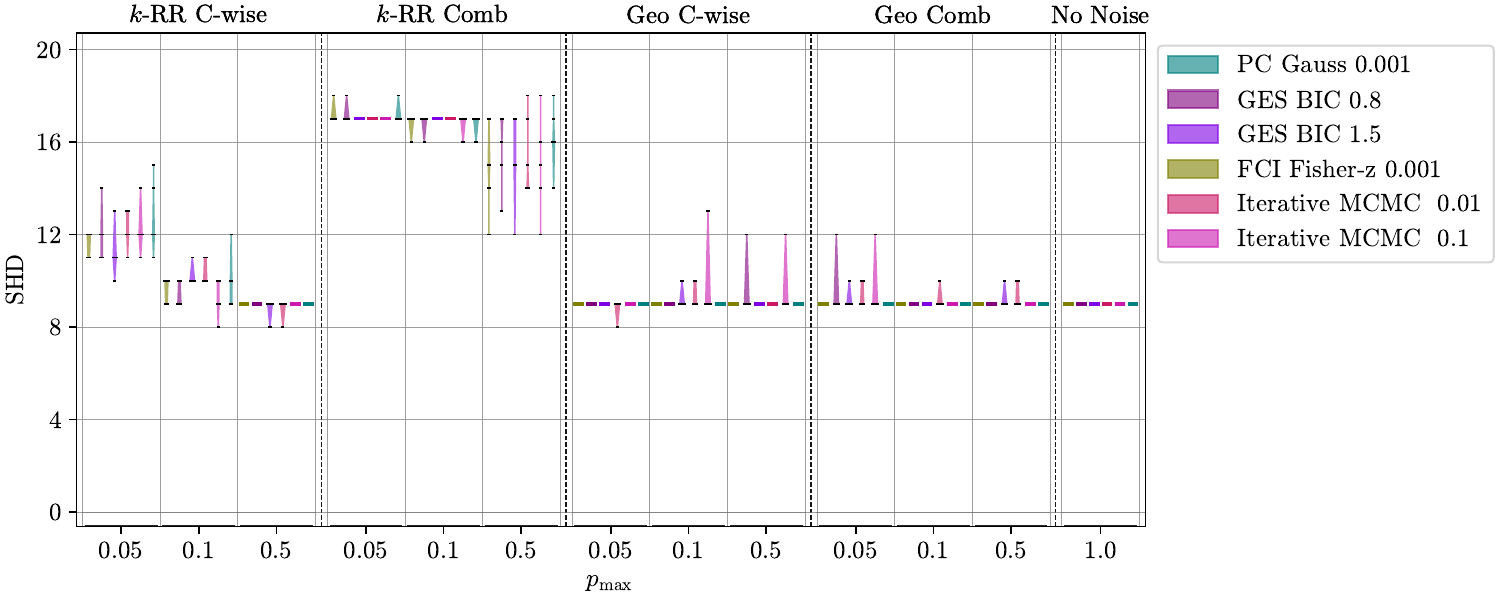}
    \vspace*{-6mm}
    \caption{Sachs data, SHD. The results for PC algorithm with Gaussian CI test and alpha value 0.001; GES algorithm with BIC score and penalty discount values 0.8 and 1.5; FCI algorithm with Fisher-z CI test and alpha values 0.001; Iterative MCMC algorithm with BGe score and alpha values 0.01 and 0.1. The width of each bar varies for different values on the y-axis proportionally to the number of samples attaining that value.}
    \label{fig:sachs_shd}
\end{figure}

\begin{figure}[h]
    \centering
    \includegraphics[width=0.8\linewidth]{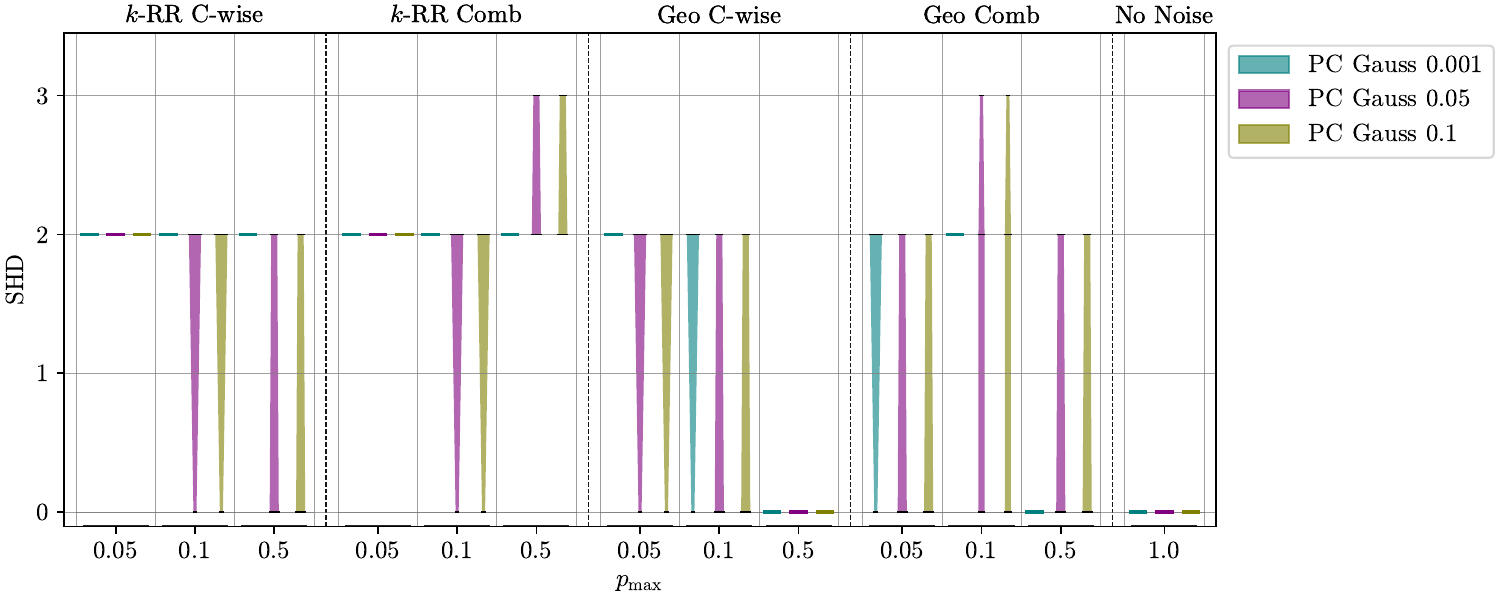}
    \vspace*{-6mm}
    \caption{Human Stature data, SHD. Results for PC algorithm with Gaussian CI test and alpha values 0.001, 0.05 and 0.1.The width of each bar varies for different values on the y-axis proportionally to the number of samples attaining that value.}
    \label{fig:HS_shd}
\end{figure}

\begin{figure}[h]
    \centering
    \includegraphics[width=0.8\linewidth]{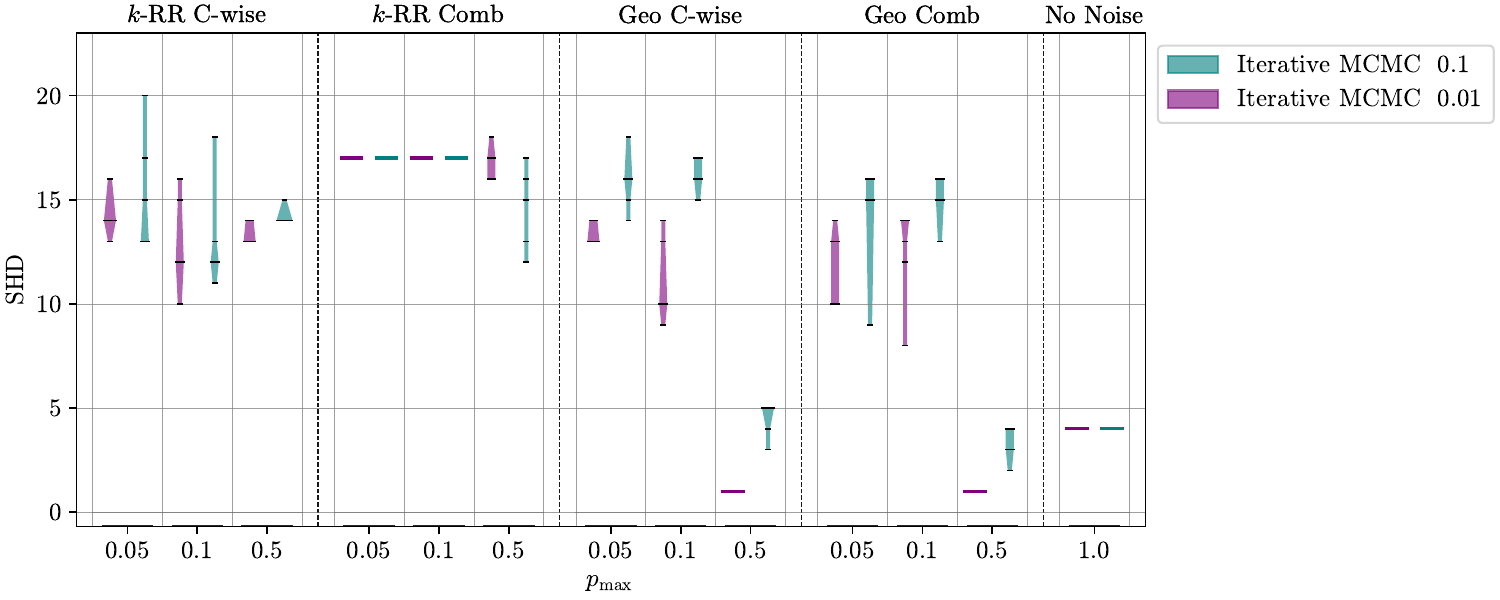}
     \vspace*{-6mm}
    \caption{Synthetic data, 10 nodes, SHD. The results for Iterative MCMC algorithm with BGe score and alpha values 0.01 and 0.1. The width of each bar varies for different values on the y-axis proportionally to the number of samples attaining that value. }
    \label{fig:SYNTH10_shd}
\end{figure}

\begin{figure}[h]
    \centering
    \includegraphics[width=0.8\linewidth]{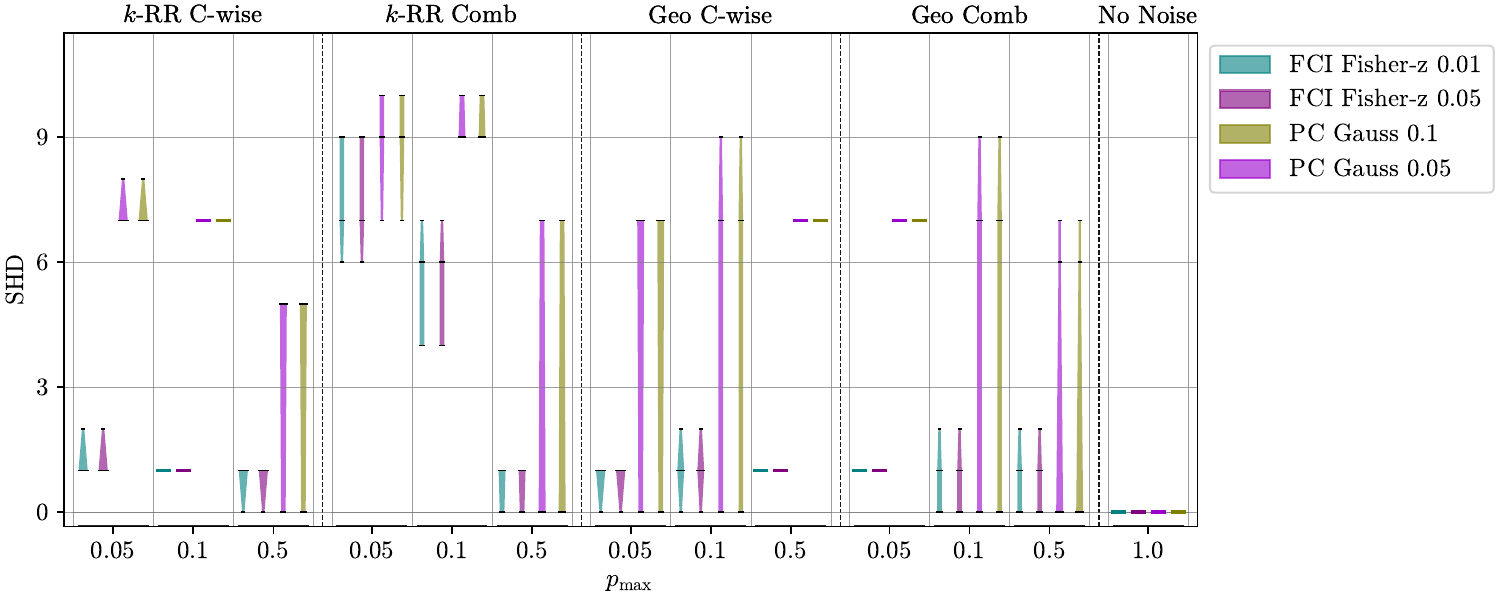}
    \vspace*{-6mm}
    \caption{Synthetic data, 5 nodes, SHD. The results for FCI algorithm with Fisher-z CI test, alpha values 0.01 and 0.05; PC algorithm with Gaussian CI test, alpha values 0.1 and 0.05. The width of each bar varies for different values on the y-axis proportionally to the number of samples attaining that value.}
    \label{fig:SYNTH5_shd}
\end{figure}

We report the results for the algorithms that performed the best on the discretized, but not privatized data. PC algorithm performed the best on most of the data sets. The iterative MCMC algorithm was performing better on the data sets with 10 or more nodes. Both data sets with 10 or more nodes show that causal discovery algorithms in general perform better under geometric privatization methods rather than $k$-RR. For the Sachs data set (Figure~\ref{fig:sachs_shd}), PC and GES algorithms perform almost the same on Geo C-wise and Geo Comb. The performance on data privatized with the geometric mechanism is very close to the performance on the original data without the noise. For the Synth10 data set (Figure~\ref{fig:SYNTH10_shd}), the performance on data privatized with geometric mechanisms with $p_{\max}=0.5$ outperforms the results on the original data. However, this result can also be accidental. For the Synth10 data set, we also observe a slightly better performance when Geo Comb is applied as compared to Geo C-wise. Performance is better with $k$-RR C-wise privatization than with $k$-RR Comb privatization on the Sachs and Synth10 data sets. 
For smaller multidimensional data sets (Figures~\ref{fig:HS_shd} and~\ref{fig:SYNTH5_shd}) the variation of the performance is too large to draw reliable conclusions. This is probably due to the high influence of chance on recovering the data structure when the true graph is small. However, we still observe a slight advantage in applying geometric mechanisms to Synth5 and Human Stature data sets. We can also observe slightly better SHD results with $K$-RR C-wise privatization than with $K$-RR Comb privatization on Synth5 and Human Stature data sets.

In our additional experiments in Appendix~\ref{app:f1_score}, we observe similar results when measuring the F1 score for the causal discovery of an undirected graphs (Figures \ref{fig:sachs_f1}, \ref{fig:HS_f1}, \ref{fig:SYNTH5_f1}, \ref{fig:SYNTH10_f1}). 
\vspace*{-3mm}
\subsection{Results on Two-dimensional Data}
\begin{figure}[h]
    \centering
    \includegraphics[width=0.8\linewidth]{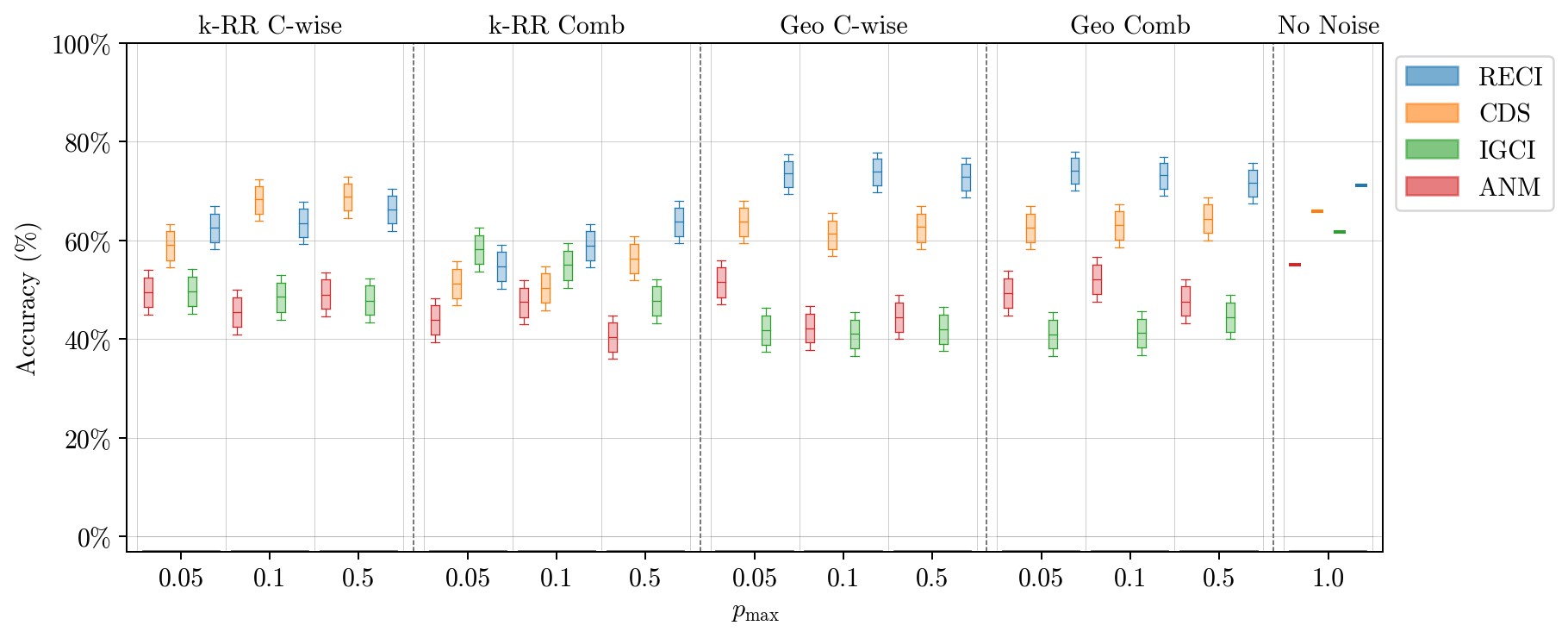}
    \vspace*{-6mm}
    \caption{CEP data set with 2 nodes, weighted accuracy. Box whiskers are at 95\%, body is at 80\% confidence.}
    \vspace*{-6mm}
    \label{fig:CEP}
\end{figure}

We report the results of all causal discovery algorithms applied for the CEP data set. In Figure \ref{fig:CEP}, we show the results before (``No Noise") and after privatization. It is evident that, similar to previous experiments, the Geometric mechanism consistently outperforms $k$-RR, with notable improvements, especially in the case of RECI, where the accuracy surpasses the baseline. We hypothesize that this phenomenon could be attributed to the potential data augmentation properties of noise addition, although further research is required to confirm this.
The CDS algorithm performs similarly after privatization, except when applying the $k$-RR Comb mechanism. But $k$-RR Comb generally has the poorest performance (also with Sachs and HS data sets), and we think this is due to the available small sample size, and because the mechanism is affected by the curse of dimensionality. 
ANM exhibited unsatisfactory performance even before noise introduction, and its performance deteriorated further (sometimes falling below chance levels) after privatization.

\vspace*{-5mm}
\section{Discussion} \label{sec:discussion}

Our results consistently demonstrate that \textbf{geometric privatization methods (both component-wise and combined) exhibit higher accuracy in terms of SHD compared to $k$-RR methods (both component-wise and combined).}
In the case of geometric noise, the algorithms do not seem to perform much worse as the noise increases. This can be expected because this privatization method is not disruptive of the correlations in the data. It would be an interesting extension to also evaluate its effect on the model parameters.
On the other hand, $k$-RR noise deteriorates the data structure and more noise results in worse performance of the causal discovery algorithms. We observe similar results when measuring the performance of causal discovery algorithms with the F1 score. The reason the geometric noise has less negative impact on causal discovery algorithms is that geometric noise in general tends to substitute a data point with one that is “similar” (in the sense of being not too distant, numerically). More precisely, the closer the point, the more likely it is chosen for replacement. In contrast, k-RR substitutes (with a certain probability) the original data point with any other point in the domain, chosen with uniform probability, regardless of the distance. Hence, the causal relation is preserved better by the geometric noise, especially when the relation is preserved by proximity, in the sense that if two data points are related, also their immediate neighbours are.
We observe some dependence between the higher parameter alpha (PC) or penalty discount (GES) parameters and better F1 scores on the noisy data in the experiments on multidimensional data. Higher parameter values result in sparser graphs and help avoid spurious edges in the graphs. We also observe that algorithms that are less accurate on the original data are also less sensitive to data privatization. 
More precisely, when applied to privatized data, their performance drops less compared to the baseline on the original data (the detailed results can be found in Appendix~\ref{app:all_exp}). 
However, the algorithms which are best on the original data still provide best overall results under geometric noise (despite being more sensitive to $k$-RR noise).

Although this paper focuses on empirical studies, we would like to extend the discussion with some theoretical considerations. For this aim, we considered two viewpoints: (1) how the LDP noise affects the independence tests, and (2) how the noise affects the causal discovery algorithms that are not based on independence tests (e.g., IGCI and RECI).
\vspace*{-5mm}
\subsection{Independence Test}

In the main body of our paper, we used the Fisher Z-test because we observed that it performs better than the $\chi$-square test (see Appendix~\ref{app:all_exp}). This is in line with the results of \cite{gaboardi2018local}, which show that locally adding Laplacian noise (the continuous version of the geometric mechanism) changes the $\chi$-square statistic so that it is no longer a $\chi$-square random variable. Indeed, our results shown in Figures \ref{fig:sachs_shd}, \ref{fig:SYNTH5_shd}, and \ref{fig:sachs_f1} show that the output of causal discovery algorithms that use the Fisher Z-test had little to no change in 3 out of 4 settings (the exception is the $k$-RR Comb method that gives consistently bad results that is due to the curse of dimensionality). We did not find any previous work on the effect of LDP on the Fisher Z-test, but this could be an interesting line of work whether the Fisher Z test is generally robust to LDP.
In \citep{gaboardi2018local}, the authors design new hypothesis testing algorithms to compensate for the noise and to get a more precise estimation of the $\chi$-square. We think that it would also be an promising research direction to explore how these modifications could be applied to perform causal discovery with $\chi$-square tests on the locally privatized data. The paper by \cite{gopi2020locally} proves the existence of an algorithm to test independence on the data resulting from the application of the $k$-RR mechanism, which coincides $2/3$ of the time with the test on the distribution of the original data. This has not been used for causal discovery on data sanitized with $k$-RR yet. The paper of~\cite{kap2021effect} explores the effect of “natural” noise (e.g., measurement error) on the performance of causal-discovery on ANMs that use independence scores, like HSIC in our paper. Their results indicate that, although ANMs are based on noise analysis, certain types of noise can hinder the detection of the causal direction. A study of whether LDP noise can give rise to the same effect could be an interesting  direction for future research.
\vspace*{-3mm}
\subsection{Other Tests}

 \textit{RECI} defines causality using polynomial regression. We have not found any results in the local privacy setting about how the LDP noise influences polynomial regression. This could be one interesting research direction. \textit{IGCI} uses a 1-spacing entropy estimation \citep{mooij2016distinguishing}, in which, after adding noise, the values are sorted and the average distance of neighboring values is computed. We corroborated in Figure \ref{fig:CEP} that adding noise degrades the performance of the IGCI algorithm, as shown by \cite{mooij2016distinguishing}. \textit{CDS} uses the variance of the conditional probability after discretizing the input into 13 equally spaced bins based on the standard deviation of the distribution. After adding noise, if the bins remained unchanged, the noise should affect the conditional probabilities, however, since the standard deviation increases, the bins change as well, and compensation may explain why the output of the CDS were not strongly affected by the LDP noise.

\vspace*{-3mm}
\section{Conclusion and Future work} \label{sec:conclusion}
In this work, we investigated the challenging problem of preserving causal structure when learning over locally differentially private data.
To allow the comparison between two distinct privacy notions, namely LDP and local $d$-privacy, we introduced a unified privacy measure based on an attacking perspective.
We performed extensive experiments on both synthetic and real-world data sets comparing the privacy-utility trade-off of 9 causal discovery algorithms when applied to locally private data.
Overall, our results demonstrate that locally $d$-private mechanisms offer a more promising approach for tackling this problem by preserving the causal structure of multidimensional data at an equivalent level of privacy.
Based on the findings of this paper, there are several areas that could be explored for future work. 
Some potential avenues for further research include investigating the same problem on continuous data and quantifying the effect of the sample size on the variability of the output metrics. Another possible extension is exploring the effect of privatization on the parameters of the causal model. Finally, we identify the need for designing a locally private mechanism specifically for causal discovery tasks.


\vspace*{-2mm}
\acks{This work was supported by the European Research Council (ERC) project HYPATIA under the European Union’s Horizon 2020 research and innovation programme. Grant agreement n. 835294. And this work was also supported by the ELSA – European Lighthouse on Secure and Safe AI funded by the European Union. Grant agreement n. 101070617. H.H. Arcolezi has been partially supported by MIAI @ Grenoble Alpes (ANR-19-P3IA-0003).}

\bibliography{clear2024}

\begin{thebibliography}{61}
\providecommand{\natexlab}[1]{#1}
\providecommand{\url}[1]{\texttt{#1}}
\expandafter\ifx\csname urlstyle\endcsname\relax
  \providecommand{\doi}[1]{doi: #1}\else
  \providecommand{\doi}{doi: \begingroup \urlstyle{rm}\Url}\fi

\bibitem[Acharya et~al.(2019)Acharya, Sun, and Zhang]{Hadamard}
Jayadev Acharya, Ziteng Sun, and Huanyu Zhang.
\newblock Hadamard response: Estimating distributions privately, efficiently, and with little communication.
\newblock In Kamalika Chaudhuri and Masashi Sugiyama, editors, \emph{Proceedings of the Twenty-Second International Conference on Artificial Intelligence and Statistics}, volume~89 of \emph{Proceedings of Machine Learning Research}, pages 1120--1129. PMLR, 16--18 Apr 2019.

\bibitem[Agarwal and Singh(2021)]{agarwal2021causal}
Anish Agarwal and Rahul Singh.
\newblock Causal inference with corrupted data: Measurement error, missing values, discretization, and differential privacy.
\newblock \emph{arXiv preprint arXiv:2107.02780}, 2021.

\bibitem[Andr{\'{e}}s et~al.(2013)Andr{\'{e}}s, Bordenabe, Chatzikokolakis, and Palamidessi]{Andrs2013}
Miguel~E. Andr{\'{e}}s, Nicol{\'{a}}s~E. Bordenabe, Konstantinos Chatzikokolakis, and Catuscia Palamidessi.
\newblock Geo-indistinguishability: Differential privacy for location-based systems.
\newblock In \emph{Proceedings of the 2013 {ACM} {SIGSAC} conference on Computer {\&} communications security - {CCS} {\textquotesingle}13}. {ACM} Press, 2013.
\newblock \doi{10.1145/2508859.2516735}.

\bibitem[Arcolezi et~al.(2022)Arcolezi, Couchot, Bouna, and Xiao]{Arcolezi2022}
H{\'{e}}ber~H. Arcolezi, Jean-Fran{\c{c}}ois Couchot, Bechara~Al Bouna, and Xiaokui Xiao.
\newblock Improving the utility of locally differentially private protocols for longitudinal and multidimensional frequency estimates.
\newblock \emph{Digital Communications and Networks}, 2022.
\newblock \doi{10.1016/j.dcan.2022.07.003}.

\bibitem[Arcolezi et~al.(2023)Arcolezi, Gambs, Couchot, and Palamidessi]{Arcolezi2023}
H\'{e}ber~H. Arcolezi, S\'{e}bastien Gambs, Jean-Fran\c{c}ois Couchot, and Catuscia Palamidessi.
\newblock On the risks of collecting multidimensional data under local differential privacy.
\newblock \emph{Proc. VLDB Endow.}, 16\penalty0 (5):\penalty0 1126–1139, jan 2023.
\newblock ISSN 2150-8097.
\newblock \doi{10.14778/3579075.3579086}.

\bibitem[Binkyt{\.e}-Sadauskien{\.e} et~al.(2022)Binkyt{\.e}-Sadauskien{\.e}, Makhlouf, Pinz{\'o}n, Zhioua, and Palamidessi]{binkyte2022causal}
R{\=u}ta Binkyt{\.e}-Sadauskien{\.e}, Karima Makhlouf, Carlos Pinz{\'o}n, Sami Zhioua, and Catuscia Palamidessi.
\newblock Causal discovery for fairness.
\newblock \emph{arXiv preprint arXiv:2206.06685}, 2022.

\bibitem[Bl{\"o}baum et~al.(2018)Bl{\"o}baum, Janzing, Washio, Shimizu, and Sch{\"o}lkopf]{blobaum2018cause}
Patrick Bl{\"o}baum, Dominik Janzing, Takashi Washio, Shohei Shimizu, and Bernhard Sch{\"o}lkopf.
\newblock Cause-effect inference by comparing regression errors.
\newblock In \emph{International Conference on Artificial Intelligence and Statistics}, pages 900--909. PMLR, 2018.

\bibitem[Chatzikokolakis et~al.(2023)Chatzikokolakis, Cherubin, Palamidessi, and Troncoso]{Chatzikokolakis2023}
K.~Chatzikokolakis, G.~Cherubin, C.~Palamidessi, and C.~Troncoso.
\newblock Bayes security: A not so average metric.
\newblock In \emph{2023 2023 IEEE 36th Computer Security Foundations Symposium (CSF) (CSF)}, pages 159--177, Los Alamitos, CA, USA, jul 2023. IEEE Computer Society.
\newblock \doi{10.1109/CSF57540.2023.00011}.

\bibitem[Chatzikokolakis et~al.(2013)Chatzikokolakis, Andr{\'e}s, Bordenabe, and Palamidessi]{chatzikokolakis2013broadening}
Konstantinos Chatzikokolakis, Miguel~E Andr{\'e}s, Nicol{\'a}s~Emilio Bordenabe, and Catuscia Palamidessi.
\newblock Broadening the scope of differential privacy using metrics.
\newblock In \emph{Privacy Enhancing Technologies: 13th International Symposium, PETS 2013, Bloomington, IN, USA, July 10-12, 2013. Proceedings 13}, pages 82--102. Springer, 2013.

\bibitem[Chickering(2002)]{chickering2002optimal}
David~Maxwell Chickering.
\newblock Optimal structure identification with greedy search.
\newblock \emph{Journal of machine learning research}, 3\penalty0 (Nov):\penalty0 507--554, 2002.

\bibitem[Clopper and Pearson(1934)]{clopper1934use}
Charles~J Clopper and Egon~S Pearson.
\newblock The use of confidence or fiducial limits illustrated in the case of the binomial.
\newblock \emph{Biometrika}, 26\penalty0 (4):\penalty0 404--413, 1934.

\bibitem[Cormode et~al.(2018)Cormode, Jha, Kulkarni, Li, Srivastava, and Wang]{cormode2018privacy}
Graham Cormode, Somesh Jha, Tejas Kulkarni, Ninghui Li, Divesh Srivastava, and Tianhao Wang.
\newblock Privacy at scale: Local differential privacy in practice.
\newblock In \emph{Proceedings of the 2018 International Conference on Management of Data}, pages 1655--1658, 2018.

\bibitem[Daniusis et~al.(2012)Daniusis, Janzing, Mooij, Zscheischler, Steudel, Zhang, and Sch{\"o}lkopf]{daniusis2012inferring}
Povilas Daniusis, Dominik Janzing, Joris Mooij, Jakob Zscheischler, Bastian Steudel, Kun Zhang, and Bernhard Sch{\"o}lkopf.
\newblock Inferring deterministic causal relations.
\newblock \emph{arXiv preprint arXiv:1203.3475}, 2012.

\bibitem[Differential Privacy~Team(2017)]{apple}
Apple Differential Privacy~Team.
\newblock Learning with privacy at scale.
\newblock In \emph{Apple Machine Learning Journal}, volume~1. Apple, 2017.

\bibitem[Ding et~al.(2017)Ding, Kulkarni, and Yekhanin]{microsoft}
Bolin Ding, Janardhan Kulkarni, and Sergey Yekhanin.
\newblock Collecting telemetry data privately.
\newblock In \emph{Proceedings of the 31st International Conference on Neural Information Processing Systems}, NIPS'17, pages 3574--3583, Red Hook, NY, USA, 2017. Curran Associates Inc.
\newblock ISBN 9781510860964.

\bibitem[Domingo-Ferrer and Soria-Comas(2022)]{Domingo_Ferrer2020}
Josep Domingo-Ferrer and Jordi Soria-Comas.
\newblock Multi-dimensional randomized response.
\newblock \emph{IEEE Transactions on Knowledge and Data Engineering}, 34\penalty0 (10):\penalty0 4933--4946, 2022.
\newblock \doi{10.1109/TKDE.2020.3045759}.

\bibitem[Duchi et~al.(2013)Duchi, Jordan, and Wainwright]{Duchi2013}
John~C. Duchi, Michael~I. Jordan, and Martin~J. Wainwright.
\newblock Local privacy and statistical minimax rates.
\newblock In \emph{2013 {IEEE} 54th Annual Symposium on Foundations of Computer Science}, pages 429--438. {IEEE}, October 2013.
\newblock \doi{10.1109/focs.2013.53}.

\bibitem[Dwork et~al.(2006)Dwork, McSherry, Nissim, and Smith]{Dwork2006}
Cynthia Dwork, Frank McSherry, Kobbi Nissim, and Adam Smith.
\newblock Calibrating noise to sensitivity in private data analysis.
\newblock In \emph{Theory of Cryptography}, pages 265--284. Springer Berlin Heidelberg, 2006.
\newblock \doi{10.1007/11681878\_14}.

\bibitem[Dwork et~al.(2014)Dwork, Roth, et~al.]{dwork2014algorithmic}
Cynthia Dwork, Aaron Roth, et~al.
\newblock The algorithmic foundations of differential privacy.
\newblock \emph{Foundations and Trends{\textregistered} in Theoretical Computer Science}, 9\penalty0 (3--4):\penalty0 211--407, 2014.

\bibitem[Entner and Hoyer(2010)]{entner2010causal}
Doris Entner and Patrik~O Hoyer.
\newblock On causal discovery from time series data using fci.
\newblock \emph{Probabilistic graphical models}, pages 121--128, 2010.

\bibitem[Erlingsson et~al.(2014)Erlingsson, Pihur, and Korolova]{google}
\'Ulfar Erlingsson, Vasyl Pihur, and Aleksandra Korolova.
\newblock {RAPPOR}: Randomized aggregatable privacy-preserving ordinal response.
\newblock In \emph{Proceedings of the 2014 ACM SIGSAC Conference on Computer and Communications Security}, pages 1054--1067, New York, NY, USA, 2014. ACM.
\newblock \doi{10.1145/2660267.2660348}.

\bibitem[Fonollosa(2019)]{fonollosa2019conditional}
Jos{\'e}~AR Fonollosa.
\newblock Conditional distribution variability measures for causality detection.
\newblock \emph{Cause Effect Pairs in Machine Learning}, pages 339--347, 2019.

\bibitem[Gaboardi and Rogers(2018)]{gaboardi2018local}
Marco Gaboardi and Ryan Rogers.
\newblock Local private hypothesis testing: Chi-square tests.
\newblock In \emph{International Conference on Machine Learning}, pages 1626--1635. PMLR, 2018.

\bibitem[Glymour et~al.(2019)Glymour, Zhang, and Spirtes]{glymour2019review}
Clark Glymour, Kun Zhang, and Peter Spirtes.
\newblock Review of causal discovery methods based on graphical models.
\newblock \emph{Frontiers in genetics}, 10:\penalty0 524, 2019.

\bibitem[Gopi et~al.(2020)Gopi, Kamath, Kulkarni, Nikolov, Wu, and Zhang]{gopi2020locally}
Sivakanth Gopi, Gautam Kamath, Janardhan Kulkarni, Aleksandar Nikolov, Zhiwei~Steven Wu, and Huanyu Zhang.
\newblock Locally private hypothesis selection.
\newblock In \emph{Conference on Learning Theory}, pages 1785--1816. PMLR, 2020.

\bibitem[Han et~al.(2015)Han, Ma, and Zhu]{han2015galton}
Hao Han, Yeming Ma, and Wei Zhu.
\newblock Galton's family heights data revisited.
\newblock \emph{arXiv preprint arXiv:1508.02942}, 2015.

\bibitem[Hoyer et~al.(2008)Hoyer, Janzing, Mooij, Peters, and Sch{\"o}lkopf]{hoyer2008nonlinear}
Patrik Hoyer, Dominik Janzing, Joris~M Mooij, Jonas Peters, and Bernhard Sch{\"o}lkopf.
\newblock Nonlinear causal discovery with additive noise models.
\newblock \emph{Advances in neural information processing systems}, 21, 2008.

\bibitem[Johnson et~al.(1985)Johnson, McClearn, Yuen, Nagoshi, Ahern, and Cole]{johnson1985galton}
Ronald~C Johnson, Gerald~E McClearn, Sylvia Yuen, Craig~T Nagoshi, Frank~M Ahern, and Robert~E Cole.
\newblock Galton's data a century later.
\newblock \emph{American Psychologist}, 40\penalty0 (8):\penalty0 875, 1985.

\bibitem[Jordon et~al.(2019)Jordon, Yoon, and Van Der~Schaar]{jordon2019pate}
James Jordon, Jinsung Yoon, and Mihaela Van Der~Schaar.
\newblock Pate-gan: Generating synthetic data with differential privacy guarantees.
\newblock In \emph{International conference on learning representations}, 2019.

\bibitem[Kairouz et~al.(2016)Kairouz, Bonawitz, and Ramage]{kairouz2016discrete}
Peter Kairouz, Keith Bonawitz, and Daniel Ramage.
\newblock Discrete distribution estimation under local privacy.
\newblock In \emph{Int. Conf. on Machine Learning}, pages 2436--2444. PMLR, 2016.

\bibitem[Kalainathan et~al.(2020)Kalainathan, Goudet, and Dutta]{kalainathan2020causal}
Diviyan Kalainathan, Olivier Goudet, and Ritik Dutta.
\newblock Causal discovery toolbox: Uncovering causal relationships in python.
\newblock \emph{The Journal of Machine Learning Research}, 21\penalty0 (1):\penalty0 1406--1410, 2020.

\bibitem[Kap et~al.(2021)Kap, Aleksandrova, and Engel]{kap2021effect}
Benjamin Kap, Marharyta Aleksandrova, and Thomas Engel.
\newblock The effect of noise level on the accuracy of causal discovery methods with additive noise models.
\newblock In \emph{Benelux Conference on Artificial Intelligence}, pages 120--140. Springer, 2021.

\bibitem[Kasiviswanathan et~al.(2008)Kasiviswanathan, Lee, Nissim, Raskhodnikova, and Smith]{first_ldp}
Shiva~Prasad Kasiviswanathan, Homin~K. Lee, Kobbi Nissim, Sofya Raskhodnikova, and Adam Smith.
\newblock What can we learn privately?
\newblock In \emph{2008 49th Annual {IEEE} Symposium on Foundations of Computer Science}, pages 531--540. {IEEE}, October 2008.
\newblock \doi{10.1109/FOCS.2008.27}.

\bibitem[Kikuchi(2022)]{kikuchi2022castell}
Hiroaki Kikuchi.
\newblock Castell: Scalable joint probability estimation of multi-dimensional data randomized with local differential privacy.
\newblock \emph{arXiv preprint arXiv:2212.01627}, 2022.

\bibitem[Kuipers et~al.(2022)Kuipers, Suter, and Moffa]{kuipers2022efficient}
Jack Kuipers, Polina Suter, and Giusi Moffa.
\newblock Efficient sampling and structure learning of bayesian networks.
\newblock \emph{Journal of Computational and Graphical Statistics}, 31\penalty0 (3):\penalty0 639--650, 2022.

\bibitem[Kusner et~al.(2016)Kusner, Sun, Sridharan, and Weinberger]{kusner16}
Matt~J. Kusner, Yu~Sun, Karthik Sridharan, and Kilian~Q. Weinberger.
\newblock Private causal inference.
\newblock In Arthur Gretton and Christian~C. Robert, editors, \emph{Proceedings of the 19th International Conference on Artificial Intelligence and Statistics}, volume~51 of \emph{Proceedings of Machine Learning Research}, pages 1308--1317, Cadiz, Spain, 09--11 May 2016. PMLR.

\bibitem[Kyono and Van~der Schaar(2021)]{kyono2021exploiting}
Trent Kyono and Mihaela Van~der Schaar.
\newblock Exploiting causal structure for robust model selection in unsupervised domain adaptation.
\newblock \emph{IEEE Transactions on Artificial Intelligence}, 2\penalty0 (6):\penalty0 494--507, 2021.

\bibitem[Loftus et~al.(2018)Loftus, Russell, Kusner, and Silva]{loftus2018causal}
Joshua~R Loftus, Chris Russell, Matt~J Kusner, and Ricardo Silva.
\newblock Causal reasoning for algorithmic fairness.
\newblock \emph{arXiv preprint arXiv:1805.05859}, 2018.

\bibitem[Ma et~al.(2022)Ma, Ji, Pang, and Wang]{Ma2022}
Pingchuan Ma, Zhenlan Ji, Qi~Pang, and Shuai Wang.
\newblock Noleaks: Differentially private causal discovery under functional causal model.
\newblock \emph{IEEE Transactions on Information Forensics and Security}, 17:\penalty0 2324--2338, 2022.
\newblock \doi{10.1109/TIFS.2022.3184263}.

\bibitem[Machanavajjhala et~al.(2008)Machanavajjhala, Kifer, Abowd, Gehrke, and Vilhuber]{dpusage3}
Ashwin Machanavajjhala, Daniel Kifer, John Abowd, Johannes Gehrke, and Lars Vilhuber.
\newblock Privacy: Theory meets practice on the map.
\newblock In \emph{Proceedings of the IEEE 24th International Conference on Data Engineering (ICDE)}, pages 277--286, 04 2008.
\newblock \doi{10.1109/ICDE.2008.4497436}.

\bibitem[Madumal et~al.(2020)Madumal, Miller, Sonenberg, and Vetere]{madumal2020explainable}
Prashan Madumal, Tim Miller, Liz Sonenberg, and Frank Vetere.
\newblock Explainable reinforcement learning through a causal lens.
\newblock In \emph{Proceedings of the AAAI conference on artificial intelligence}, volume~34, pages 2493--2500, 2020.

\bibitem[Meek(1997)]{meek1997graphical}
Christopher Meek.
\newblock \emph{Graphical Models: Selecting causal and statistical models}.
\newblock PhD thesis, Carnegie Mellon University, 1997.

\bibitem[Mooij et~al.(2016)Mooij, Peters, Janzing, Zscheischler, and Sch{\"o}lkopf]{mooij2016distinguishing}
Joris~M Mooij, Jonas Peters, Dominik Janzing, Jakob Zscheischler, and Bernhard Sch{\"o}lkopf.
\newblock Distinguishing cause from effect using observational data: methods and benchmarks.
\newblock \emph{The Journal of Machine Learning Research}, 17\penalty0 (1):\penalty0 1103--1204, 2016.

\bibitem[Narayanan and Shmatikov(2009)]{dpusage2}
Arvind Narayanan and Vitaly Shmatikov.
\newblock De-anonymizing social networks.
\newblock \emph{Proceedings - IEEE Symposium on Security and Privacy}, 04 2009.
\newblock \doi{10.1109/SP.2009.22}.

\bibitem[Nogueira et~al.(2021)Nogueira, Gama, and Ferreira]{nogueira2021causal}
Ana~Rita Nogueira, Jo{\~a}o Gama, and Carlos~Abreu Ferreira.
\newblock Causal discovery in machine learning: Theories and applications.
\newblock \emph{Journal of Dynamics \& Games}, 8\penalty0 (3):\penalty0 203, 2021.

\bibitem[Ohnishi and Awan(2023)]{ohnishi2023locally}
Yuki Ohnishi and Jordan Awan.
\newblock Locally private causal inference.
\newblock \emph{arXiv preprint arXiv:2301.01616}, 2023.

\bibitem[Pinzón et~al.(2020)Pinzón, Rocha, and Finke]{pinzon2020approach}
Carlos Pinzón, Camilo Rocha, and Jorge Finke.
\newblock An approach to optimal discretization of continuous real random variables with application to machine learning, 2020.

\bibitem[Ramsey et~al.(2017)Ramsey, Glymour, Sanchez-Romero, and Glymour]{ramsey2017million}
Joseph Ramsey, Madelyn Glymour, Ruben Sanchez-Romero, and Clark Glymour.
\newblock A million variables and more: the fast greedy equivalence search algorithm for learning high-dimensional graphical causal models, with an application to functional magnetic resonance images.
\newblock \emph{International journal of data science and analytics}, 3:\penalty0 121--129, 2017.

\bibitem[Reed and Pierce(2010)]{dpusage1}
Jason Reed and Benjamin Pierce.
\newblock Distance makes the types grow stronger a calculus for differential privacy.
\newblock \emph{Sigplan Notices - SIGPLAN}, 45:\penalty0 157--168, 09 2010.
\newblock \doi{10.1145/1932681.1863568}.

\bibitem[Richens et~al.(2020)Richens, Lee, and Johri]{richens2020improving}
Jonathan~G Richens, Ciar{\'a}n~M Lee, and Saurabh Johri.
\newblock Improving the accuracy of medical diagnosis with causal machine learning.
\newblock \emph{Nature communications}, 11\penalty0 (1):\penalty0 3923, 2020.

\bibitem[Rios et~al.(2021)Rios, Moffa, and Kuipers]{rios2021benchpress}
Felix~L. Rios, Giusi Moffa, and Jack Kuipers.
\newblock Benchpress: a scalable and versatile workflow for benchmarking structure learning algorithms for graphical models, 2021.

\bibitem[Sachs et~al.(2005)Sachs, Perez, Pe'er, Lauffenburger, and Nolan]{sachs2005causal}
Karen Sachs, Omar Perez, Dana Pe'er, Douglas~A Lauffenburger, and Garry~P Nolan.
\newblock Causal protein-signaling networks derived from multiparameter single-cell data.
\newblock \emph{Science}, 308\penalty0 (5721):\penalty0 523--529, 2005.

\bibitem[Sch{\"o}lkopf et~al.(2021)Sch{\"o}lkopf, Locatello, Bauer, Ke, Kalchbrenner, Goyal, and Bengio]{scholkopf2021toward}
Bernhard Sch{\"o}lkopf, Francesco Locatello, Stefan Bauer, Nan~Rosemary Ke, Nal Kalchbrenner, Anirudh Goyal, and Yoshua Bengio.
\newblock Toward causal representation learning.
\newblock \emph{Proceedings of the IEEE}, 109\penalty0 (5):\penalty0 612--634, 2021.

\bibitem[Spirtes and Glymour(1991)]{spirtes1991algorithm}
Peter Spirtes and Clark Glymour.
\newblock An algorithm for fast recovery of sparse causal graphs.
\newblock \emph{Social science computer review}, 9\penalty0 (1):\penalty0 62--72, 1991.

\bibitem[Tople et~al.(2020)Tople, Sharma, and Nori]{tople2020alleviating}
Shruti Tople, Amit Sharma, and Aditya Nori.
\newblock Alleviating privacy attacks via causal learning.
\newblock In \emph{International Conference on Machine Learning}, pages 9537--9547. PMLR, 2020.

\bibitem[Tsamardinos et~al.(2006)Tsamardinos, Brown, and Aliferis]{tsamardinos2006max}
Ioannis Tsamardinos, Laura~E Brown, and Constantin~F Aliferis.
\newblock The max-min hill-climbing bayesian network structure learning algorithm.
\newblock \emph{Machine learning}, 65\penalty0 (1):\penalty0 31--78, 2006.

\bibitem[Wang et~al.(2020)Wang, Pang, and Song]{Wang2020}
Lun Wang, Qi~Pang, and Dawn Song.
\newblock Towards practical differentially private causal graph discovery.
\newblock In H.~Larochelle, M.~Ranzato, R.~Hadsell, M.F. Balcan, and H.~Lin, editors, \emph{Advances in Neural Information Processing Systems}, volume~33, pages 5516--5526. Curran Associates, Inc., 2020.

\bibitem[Wang et~al.(2017)Wang, Blocki, Li, and Jha]{tianhao2017}
Tianhao Wang, Jeremiah Blocki, Ninghui Li, and Somesh Jha.
\newblock Locally differentially private protocols for frequency estimation.
\newblock In \emph{26th {USENIX} Security Symposium ({USENIX} Security 17)}, pages 729--745, Vancouver, BC, August 2017. {USENIX} Association.
\newblock ISBN 978-1-931971-40-9.

\bibitem[Warner(1965)]{Warner1965}
Stanley~L. Warner.
\newblock Randomized response: A survey technique for eliminating evasive answer bias.
\newblock \emph{Journal of the American Statistical Association}, 60\penalty0 (309):\penalty0 63--69, March 1965.
\newblock \doi{10.1080/01621459.1965.10480775}.

\bibitem[Xu et~al.(2017)Xu, Yuan, and Wu]{Xu2017}
Depeng Xu, Shuhan Yuan, and Xintao Wu.
\newblock Differential privacy preserving causal graph discovery.
\newblock In \emph{2017 IEEE Symposium on Privacy-Aware Computing (PAC)}, pages 60--71, 2017.
\newblock \doi{10.1109/PAC.2017.24}.

\bibitem[Zhang et~al.(2017)Zhang, Cormode, Procopiuc, Srivastava, and Xiao]{Zhang2017}
Jun Zhang, Graham Cormode, Cecilia~M. Procopiuc, Divesh Srivastava, and Xiaokui Xiao.
\newblock Privbayes: Private data release via bayesian networks.
\newblock \emph{ACM Trans. Database Syst.}, 42\penalty0 (4), oct 2017.
\newblock ISSN 0362-5915.
\newblock \doi{10.1145/3134428}.

\end{thebibliography}

\appendix

\section{Privacy Mechanisms} \label{app:privacy_mech}

 \begin{figure}[H]
    \noindent
    \centering
    \includegraphics[width=0.49\linewidth]{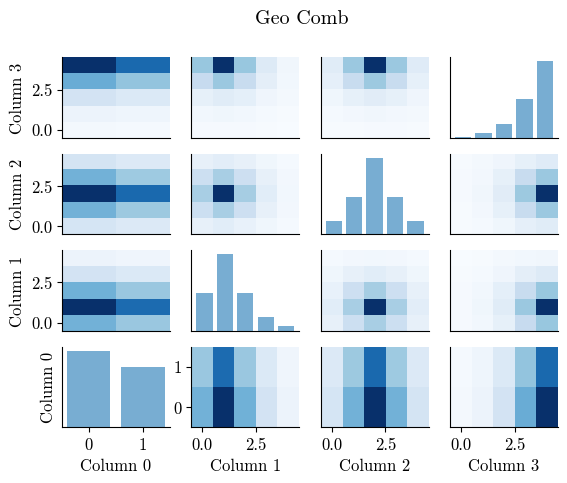}
    \includegraphics[width=0.49\linewidth]{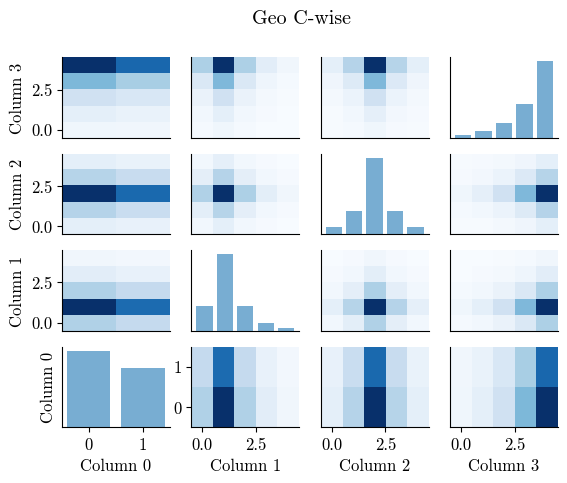}
    \includegraphics[width=0.49\linewidth]{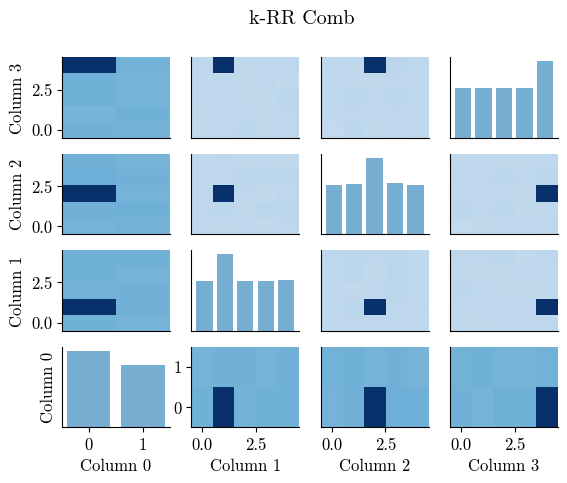}
    \includegraphics[width=0.49\linewidth]{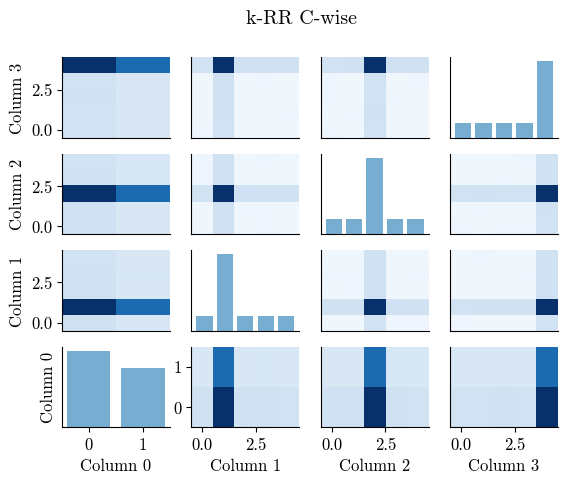}
    \caption{
        Illustration of 4 multidimensional mechanisms discussed in this paper: 4D bounded Geometric, 4x1D bounded Geometric, 4D $k$-RR and 4x1D $k$-RR.
    }
    \label{fig:mechanisms}
\end{figure}

\begin{figure}[H]
    \noindent
    \centering
    \includegraphics[width=0.49\linewidth]{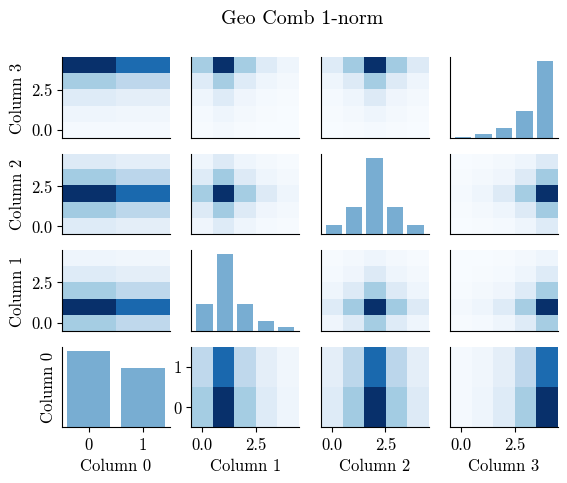}
    \includegraphics[width=0.49\linewidth]{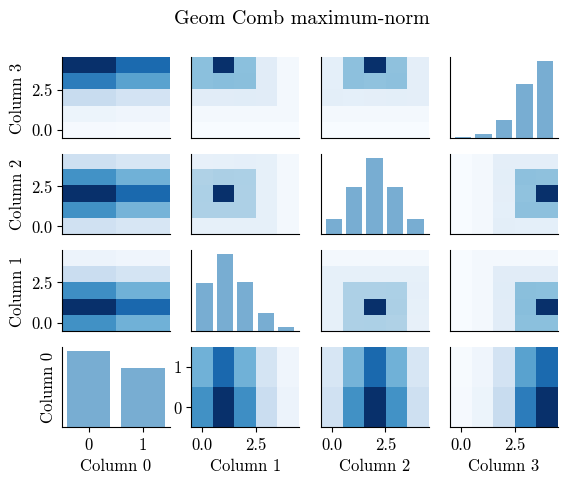}
    \caption{
        Comparison between Manhattan ($p=1$) and Chebyshev ($p=\infty$) distances for bounded geometric mechanisms.
        Refer to Figure~\ref{fig:mechanisms} for euclidean ($p=2$).
    }
    \label{fig:geometric-p-norm}
\end{figure}

\section{The Data Sets}\label{app:data}

The Sachs data set measures the expression levels of various proteins and phospholipids within human cells. It was originally generated by~\cite{sachs2005causal}. The data set consists of 11 variables and 902 samples. Sachs is a popular benchmarking data set in causal discovery, because of availability of the ground truth causal structure. 

Human Stature data is a classic historical data set collected by the statistician Francis Galton and first used for regression analysis~\citep{johnson1985galton}. Later it has been re-used as one of the benchmark data sets for causal discovery. The data set consists of four variables: father height, mother height, gender and child height, and has 898 samples. We remove the binary gender variable for our experiments. We do it, because when applied to binary data, geometric noise becomes equivalent to $k$-RR method. 

Synth10 and Synth5 are synthetic data sets with 10 and 5 nodes respectively. The background structure DAG is generated randomly using the benchpress framework~\citep{rios2021benchpress}. We specify the number of nodes and the maximum number of parents for each node. The data is generated to using a generation process compatible with the underlying structure DAG.

\section{The Algorithms}\label{app:algs}

The Peter and Clark (\textbf{PC})~\citep{spirtes1991algorithm} algorithm is a constraint-based method with two primary stages. The initial stage, known as ``adjacency search", involves identifying the undirected skeleton of the Directed Acyclic Graph (DAG). The second stage focuses on estimating a completed partially directed acyclic graph (CPDAG).
PC can be applied to linear, Gaussian data (the Fisher Z test), discrete multinomial data (the Chi Square test) and mixed multinomial/Gaussian data (the Conditional Gaussian test).
PC uses an alpha parameter which is a cutoff, which signifies the threshold at which test results are considered indicative of dependence in a statistical test of independence, typically defaults to 0.05. When using a higher alpha value, PC leads to a sparser graph. In other words, a higher alpha makes the test more stringent, and it requires stronger evidence to conclude that variables are dependent, resulting in fewer edges in the graphical model.

The \textbf{FCI} (Fast Causal Inference)~\citep{entner2010causal} algorithm is a constraint-based method designed to work with sample data, and it can also consider optional background knowledge. In the large sample limit, FCI provides an equivalence class of Conditional Bayesian Networks (CBNs) that encompass the set of conditional independence relations believed to be valid in the population, even when there are hidden confounding variables. However, FCI has limitations and is most suitable for data sets with several thousand variables. When applied to realistic sample sizes, it can be inaccurate in determining both adjacencies and orientations. FCI consists of two phases: the adjacency phase and the orientation phase. During the adjacency phase, the algorithm begins with a complete undirected graph and then conducts a series of conditional independence tests. These tests lead to the removal of edges between pairs of variables that are determined to be independent, given some subset of the observed variables. The conditioning sets that result in the removal of an edge are stored. By the end of the adjacency phase, the undirected graph correctly represents the set of adjacencies among variables, but all edges remain unoriented. FCI then proceeds to the orientation phase, where it uses the stored conditioning sets to orient as many edges as possible, adding directionality to the graph.

\textbf{FGES}~\citep{ramsey2017million} is an enhanced and parallelized variant of the Greedy Equivalence Search (GES) algorithm, initially developed by~\cite{meek1997graphical} and later studied by~\cite{chickering2002optimal}. GES is a Bayesian algorithm that uses a heuristic approach to explore the space of Conditional Bayesian Networks (CBNs) and identify the model with the highest Bayesian score. Specifically, GES commences its search with an empty graph and proceeds with a forward stepping search, where it adds edges between nodes to maximize the Bayesian score. This process continues until no further single edge addition improves the score. Subsequently, it performs a backward stepping search, eliminating edges until no single edge removal can enhance the score.  These algorithms are capable of handling both continuous data, utilizing the Structural Equation Modeling Bayesian Information Criterion (SEM BIC) score, and discrete data, making use of the Bayesian Dirichlet equivalent uniform (BDeu) score. FGES takes the penalty discount parameter. Higher penalty discount yield sparser graphs.

\textbf{Iterative MCMC}~\citep{kuipers2022efficient} is a hybrid optimization technique based on Markov chain Monte Carlo (MCMC) methods. The algorithm's initial step involves generating a skeleton, obtained through the Greedy Equivalence Search (GES) algorithm. Subsequently, it conducts a score-based search within the space defined by this initial skeleton, exploring various Directed Acyclic Graphs (DAGs).

The Max-min hill-climbing (\textbf{MMHC})~\citep{tsamardinos2006max} method is a hybrid approach that follows a two-step process. Firstly, it estimates the skeleton of a Directed Acyclic Graph (DAG) using an algorithm known as Max-Min Parents and Children. Then, it applies a greedy hill-climbing search to determine the orientation of edges within the graph based on Bayesian scoring.  MMHC is particularly suitable for domains with a high number of dimensions.

\textbf{RECI} (Regression Error based Causal Inference)~\citep{blobaum2018cause} addresses non-deterministic and nonlinear relations and allows dependency between cause and noise. The algorithm's key idea is to fit regression models in both possible directions and compare the MSE. No independence tests are used, but the assumptions on the model depend on the regressor used for the model. In our experiments we used a polynomial regressor with degree 3 after rescaling to $[0,1]$.

\textbf{IGCI} (Information Geometric Causal Inference)~\citep{daniusis2012inferring} is a pairwise causal discovery model that able to determine the causal relationship in a deterministic setting $Y = f(X)$ (where $f$ is invertible), under the ‘independence assumption’ $Cov[\log f', p_X]= 0$. In our experiments we have used a Gaussian reference measure\footnote{Our experiments with the uniform reference measure produced almost identical results, thus we exclude it from this paper.} and the sp1 or ``1-spacing" method for entropy estimation used in \cite{mooij2016distinguishing}.

\textbf{CDS} (Conditional Distribution Similarity Statistic)~\citep{fonollosa2019conditional} first normalizes the conditional distribution $P(Y|X=x)$ (for all x) to have zero mean and unit variance, then quantizes it. In our experiments as conditional distribution variability measure we used standard deviation of the preprocessed conditional distributions. The lower the standard deviation, the more likely the pair to be $X\to Y$.

\textbf{ANM}~\citep{hoyer2008nonlinear} assumes that $Y = f(X) + E$, where $f$ is nonlinear. The causal inference bases itself on the independence between $X$ and $E$. We used a Gaussian process regression for the prediction and normalized HSIC for the evaluation of the causal direction.

\section{Additional Experiments} \label{app:all_exp}

We perform experiments using real and synthetic data. Data sets are distinguished into two main groups. The first category is pairwise data, which have two variables $A$ and $B$ where $A$ causes $B$ or $B$ causes $A$. The task is to determine the causal direction between the two variables. The second category is the data that has more than two variables. The task here is to determine the causal structure (the skeleton) and the causal direction between the pairs within this structure.

\subsection{F1 Score results Sachs data set} \label{app:f1_score}
 
\begin{figure}[t]
    \centering
    \includegraphics[width=1\linewidth]{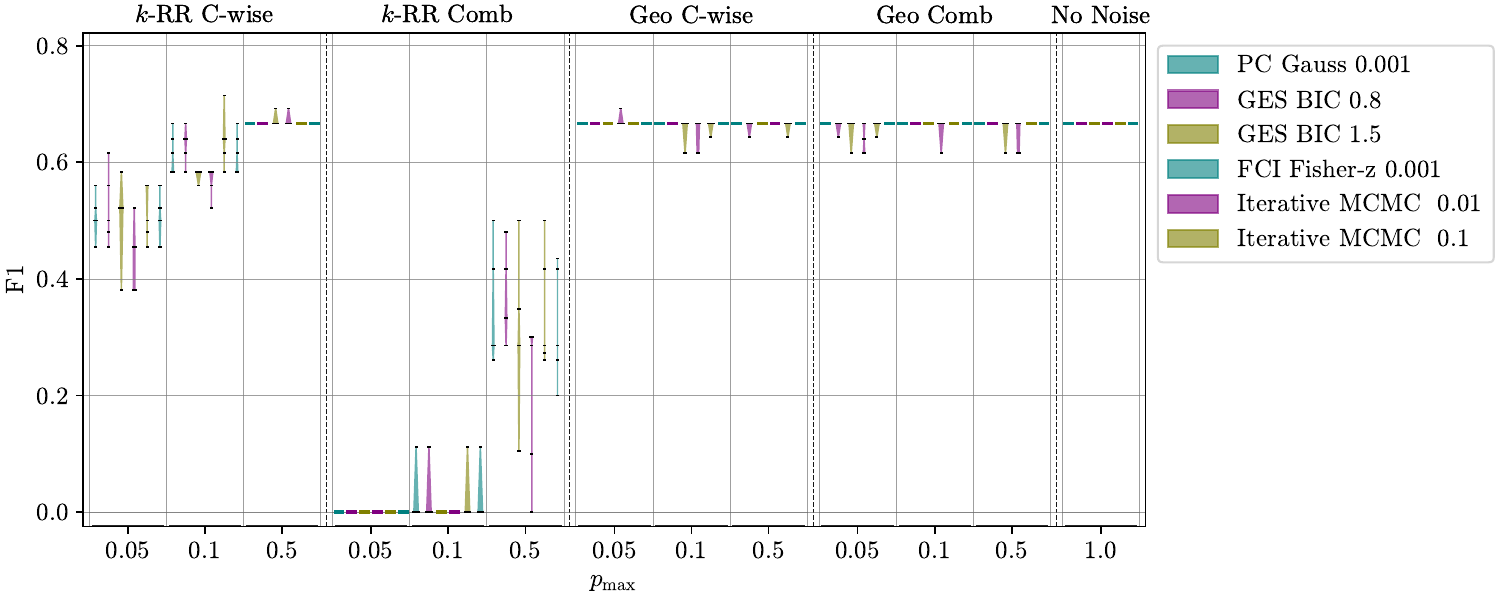}
    \vspace*{-10mm}
    \caption{Sachs data, F1.}
    \label{fig:sachs_f1}
\end{figure}

  \begin{figure}[H]
    \noindent
    \centering
    \includegraphics[width=0.45\linewidth]{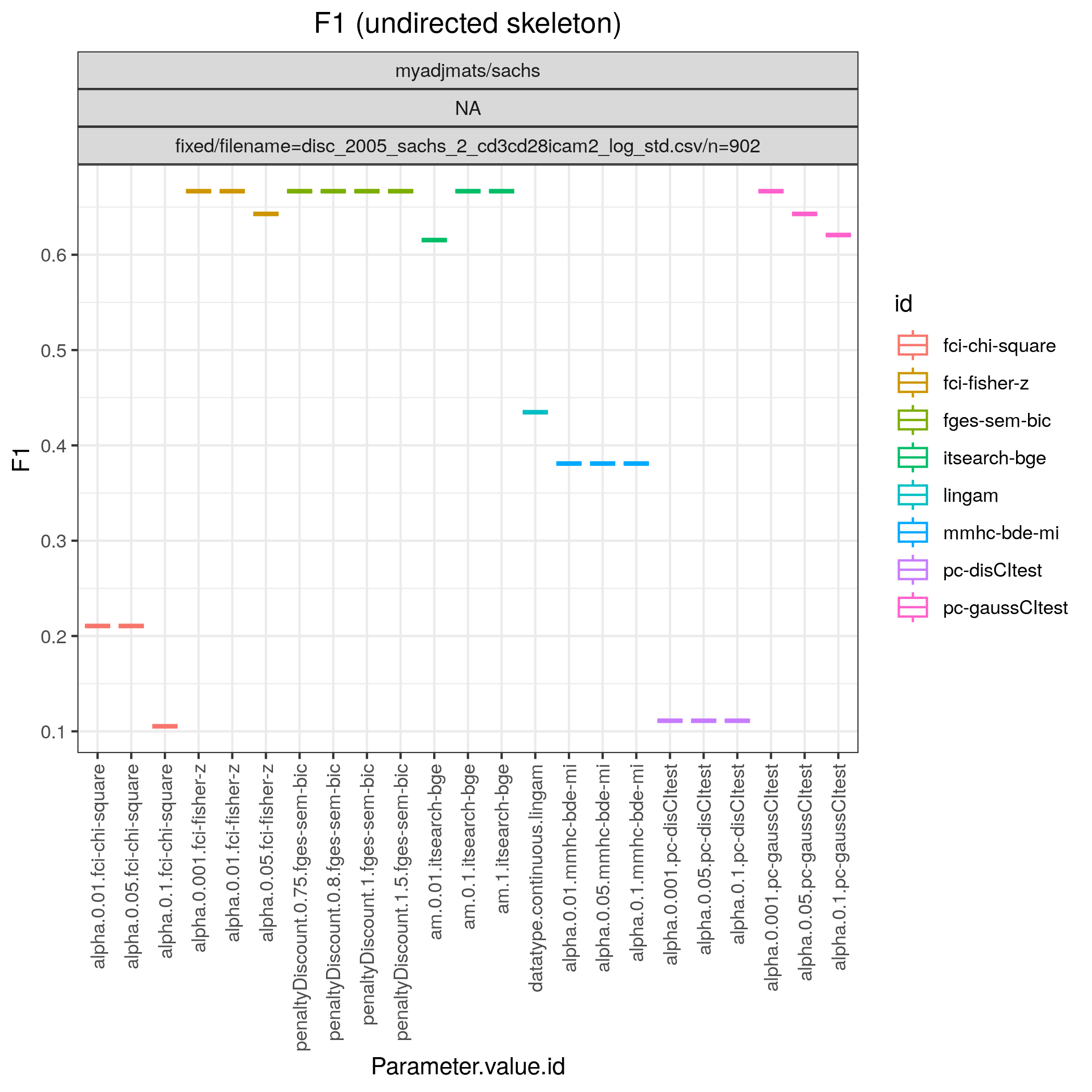}
    \caption{
        F1 Scores on the Sachs data set. Discretized, no noise.
    }
    \label{fig:synth_3}
\end{figure}

\noindent
\begin{figure}[H]
\begin{minipage}{0.31\linewidth}
\centering
		\includegraphics[scale=0.36]{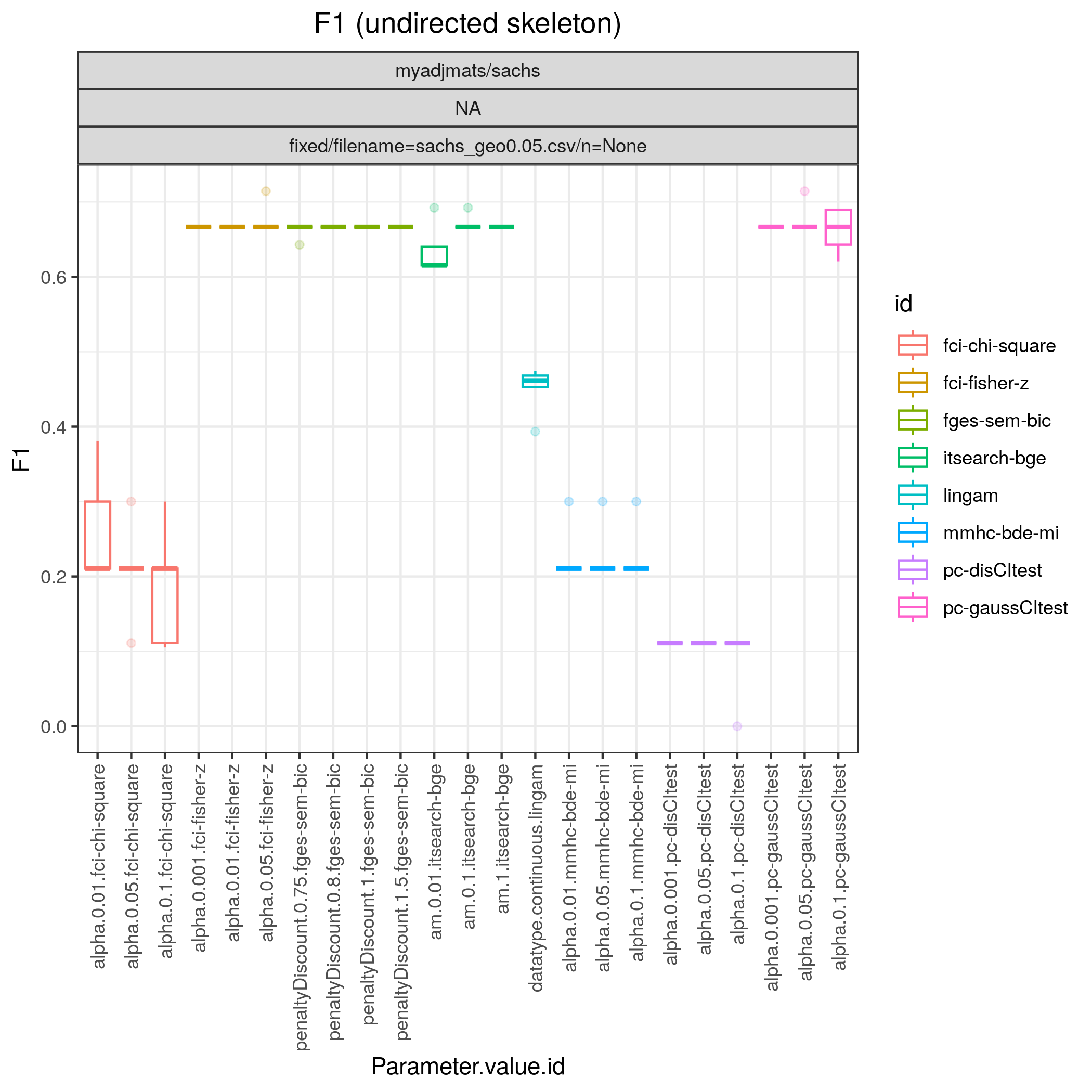}
	\caption{Sachs data, Geo C-wise mechanism, max probability 0.05.}
\end{minipage}
\begin{minipage}{0.31\linewidth}
\centering
  \includegraphics[scale=0.36]{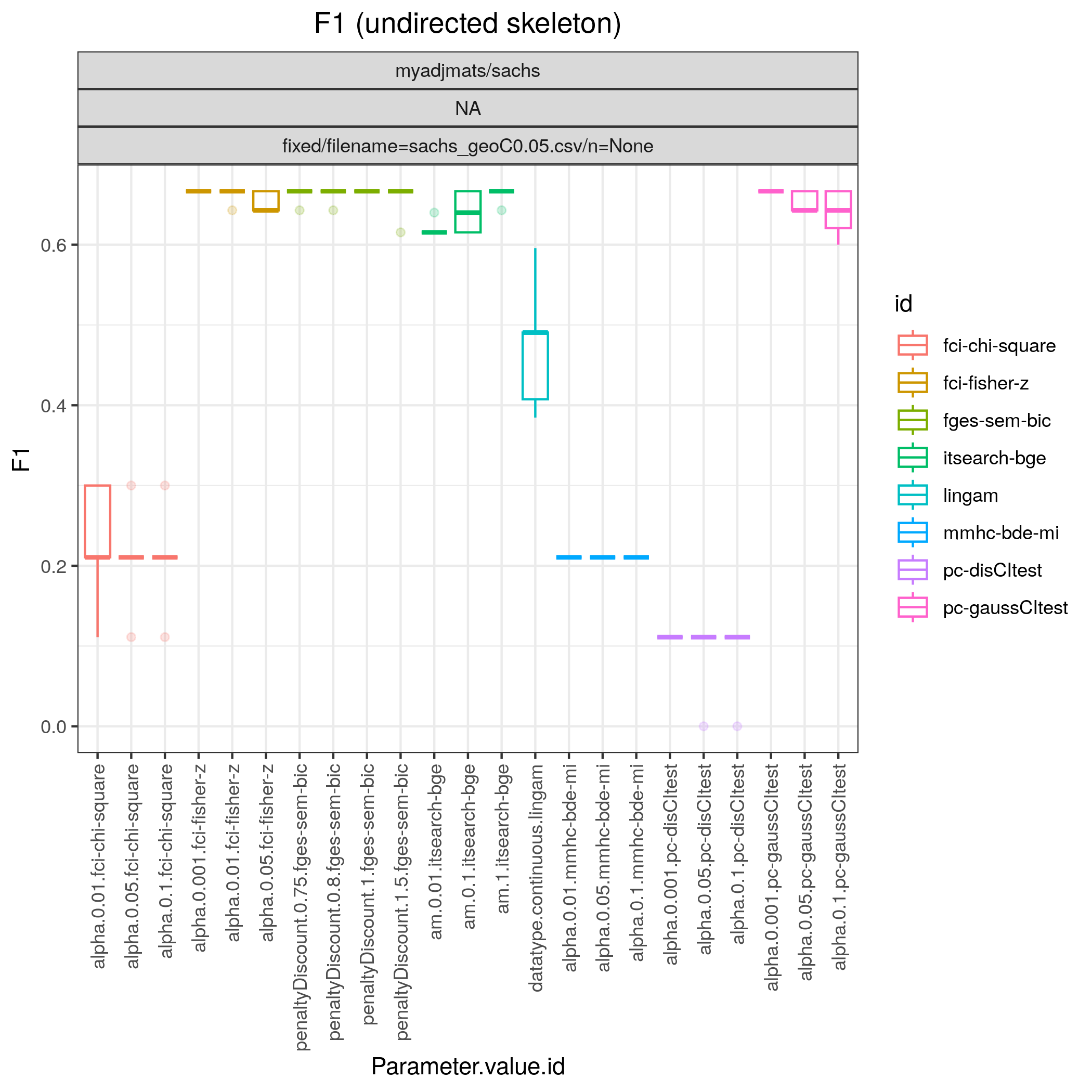}
	\caption{Sachs data, Geo Comb mechanism, max probability 0.05.}
 \end{minipage}
\begin{minipage}{0.31\linewidth}
\centering
  \includegraphics[scale=0.36]{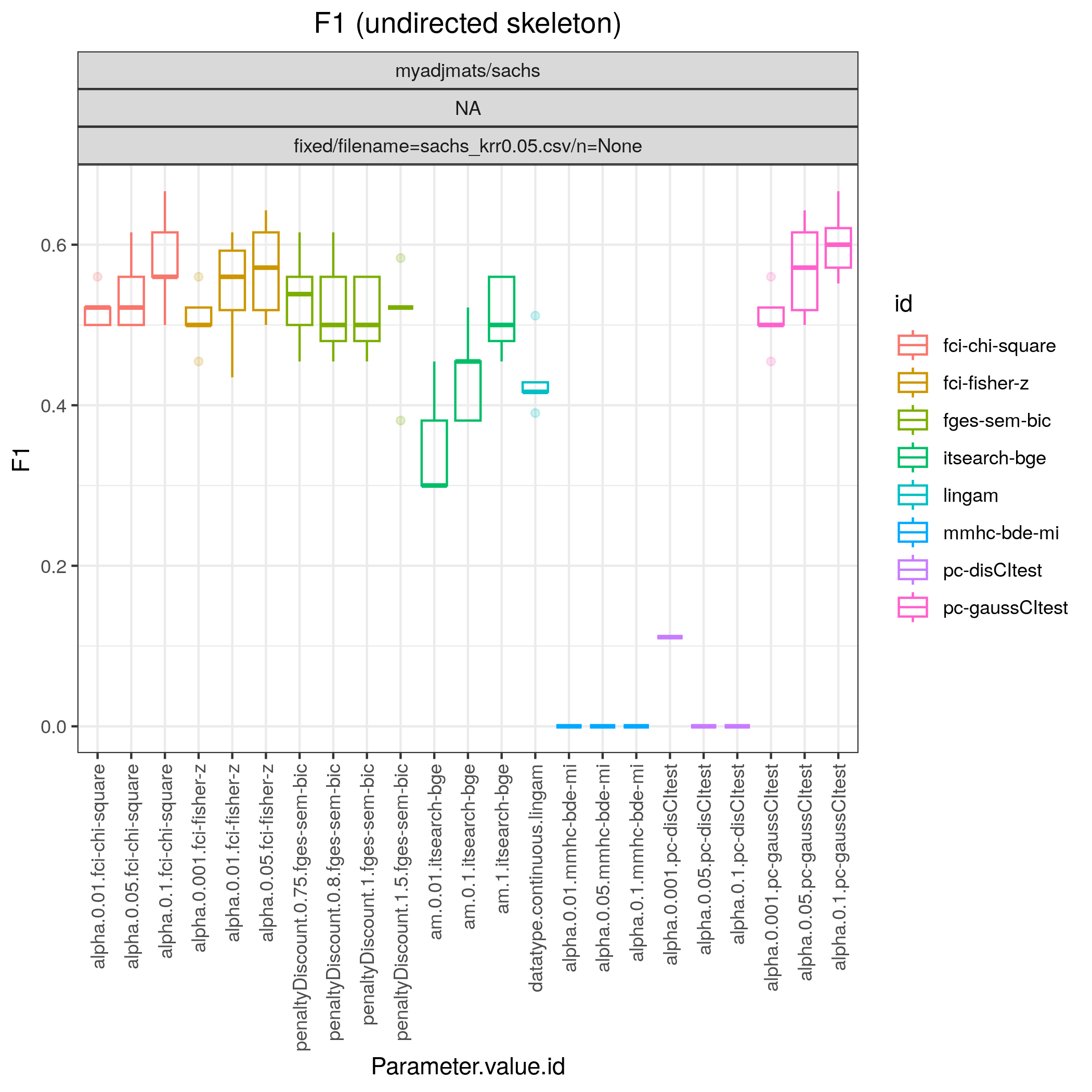}
	\caption{Sachs data, $k$-RR C-wise mechanism, max probability 0.05.}
\end{minipage}
\begin{minipage}{0.31\linewidth}
\centering
  \includegraphics[scale=0.34]{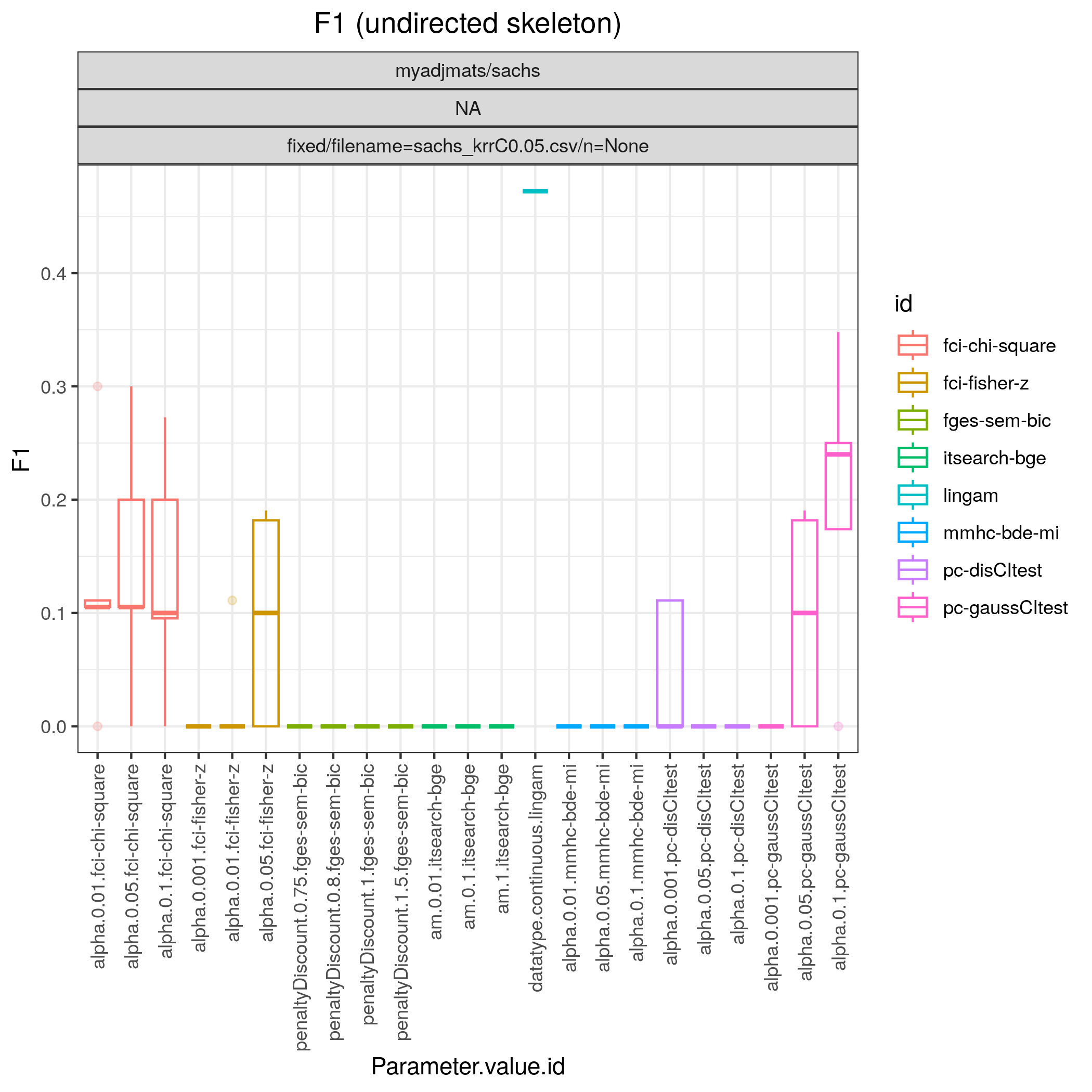}
	\caption{Sachs data, $k$-RR Comb mechanism, max probability 0.05.}
\end{minipage}
\end{figure}

\noindent
\begin{figure}[H]
\begin{minipage}{0.31\linewidth}
\centering
		\includegraphics[scale=0.36]{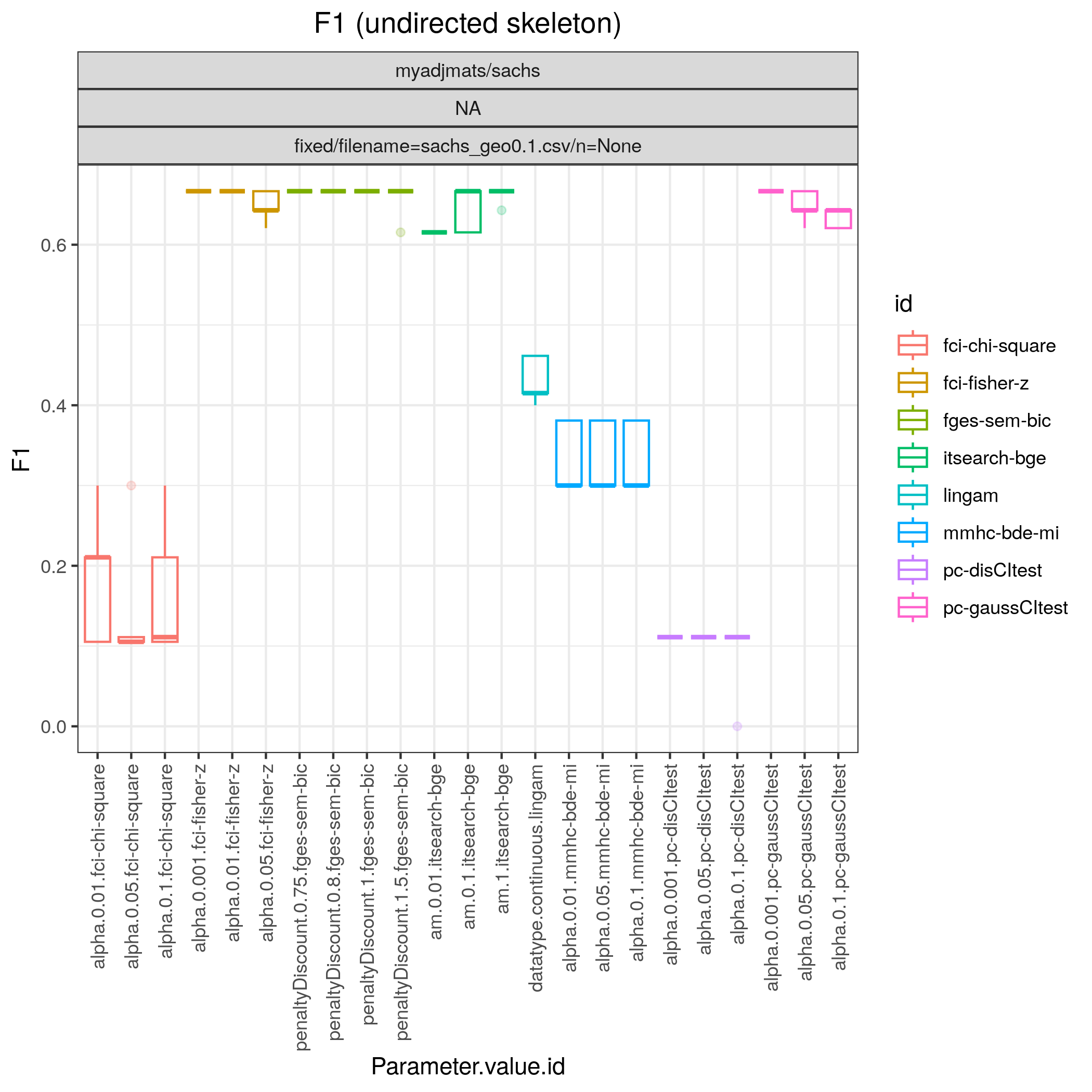}
	\caption{Sachs data, Geo C-wise mechanism, max probability 0.1.}
\end{minipage}
\begin{minipage}{0.31\linewidth}
\centering
  \includegraphics[scale=0.36]{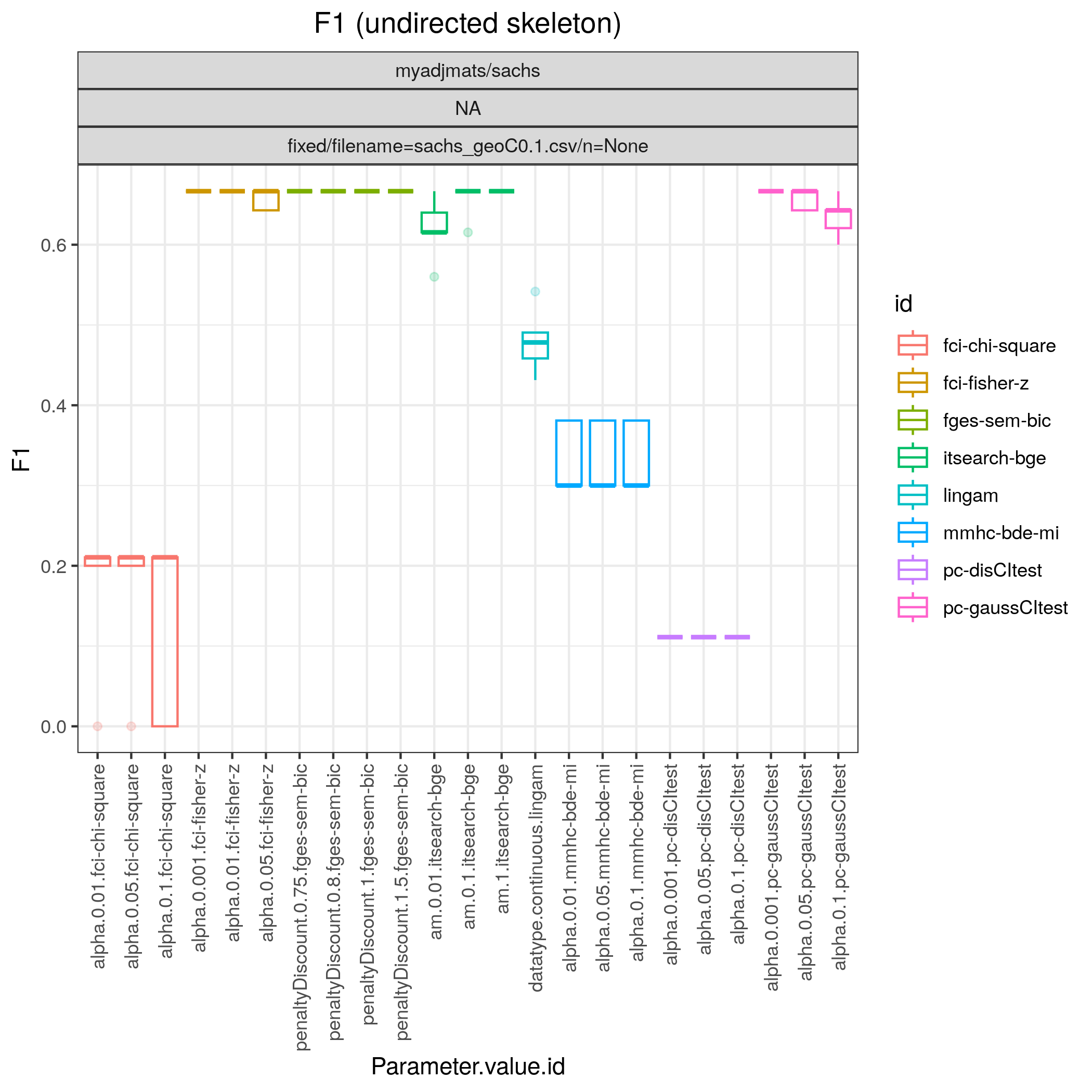}
	\caption{Sachs data, Geo Comb mechanism, max probability 0.1.}
 \end{minipage}
\begin{minipage}{0.31\linewidth}
\centering
  \includegraphics[scale=0.36]{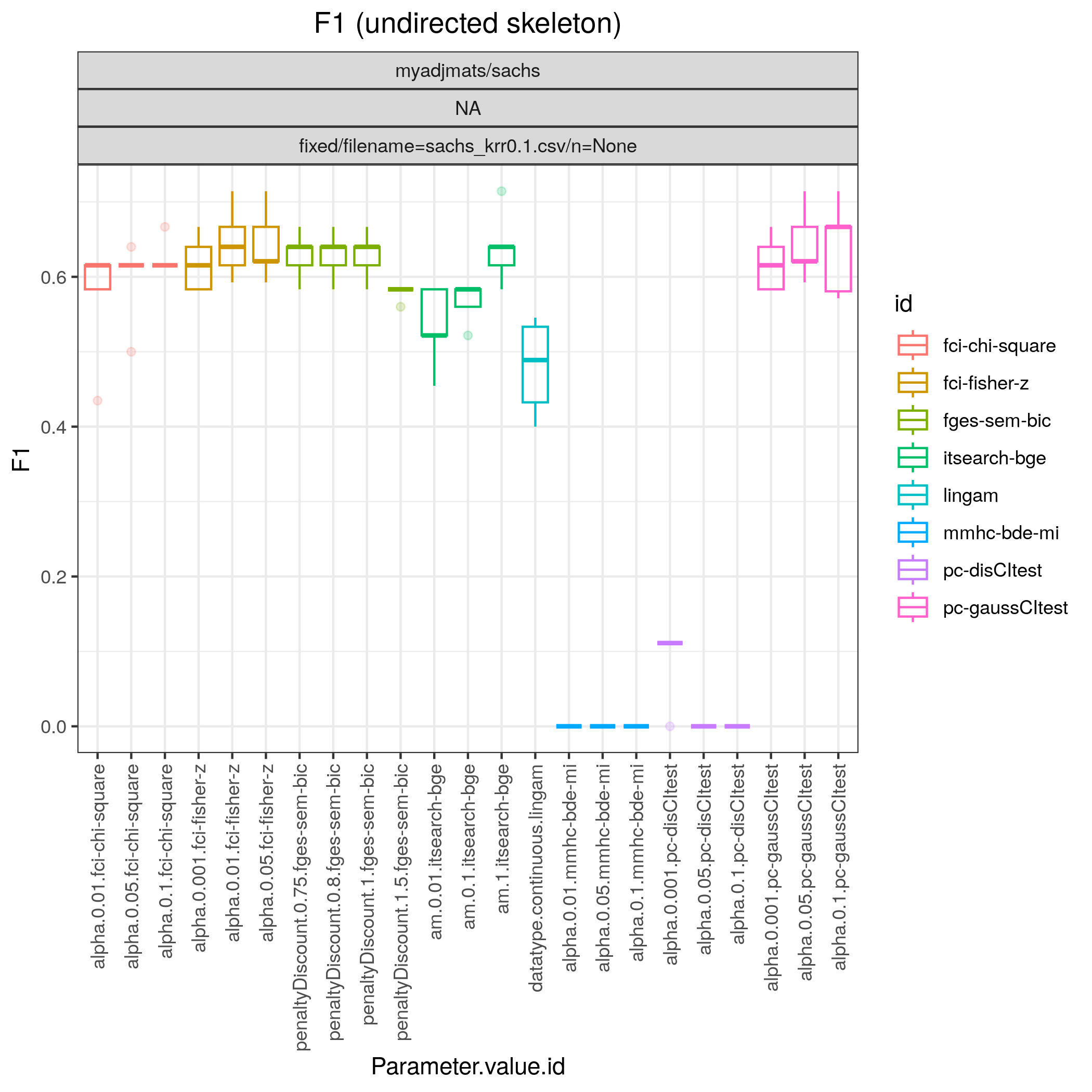}
	\caption{Sachs data, $k$-RR C-wise mechanism, max probability 0.1.}
\end{minipage}
\begin{minipage}{0.31\linewidth}
\centering
  \includegraphics[scale=0.34]{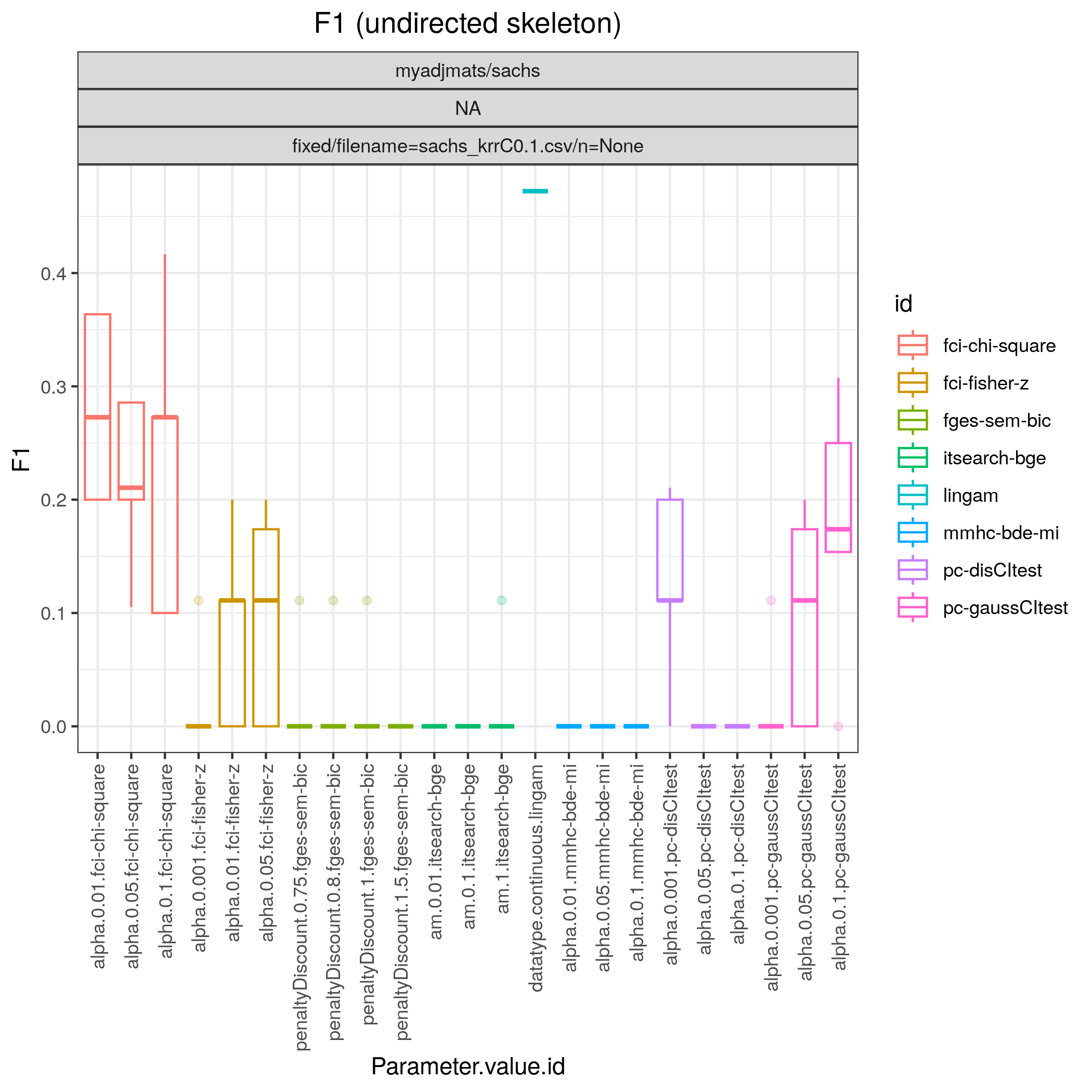}
	\caption{Sachs data, $k$-RR Comb mechanism, max probability 0.1.}
\end{minipage}
\end{figure}

\noindent
\begin{figure}[H]
\begin{minipage}{0.31\linewidth}
\centering
		\includegraphics[scale=0.36]{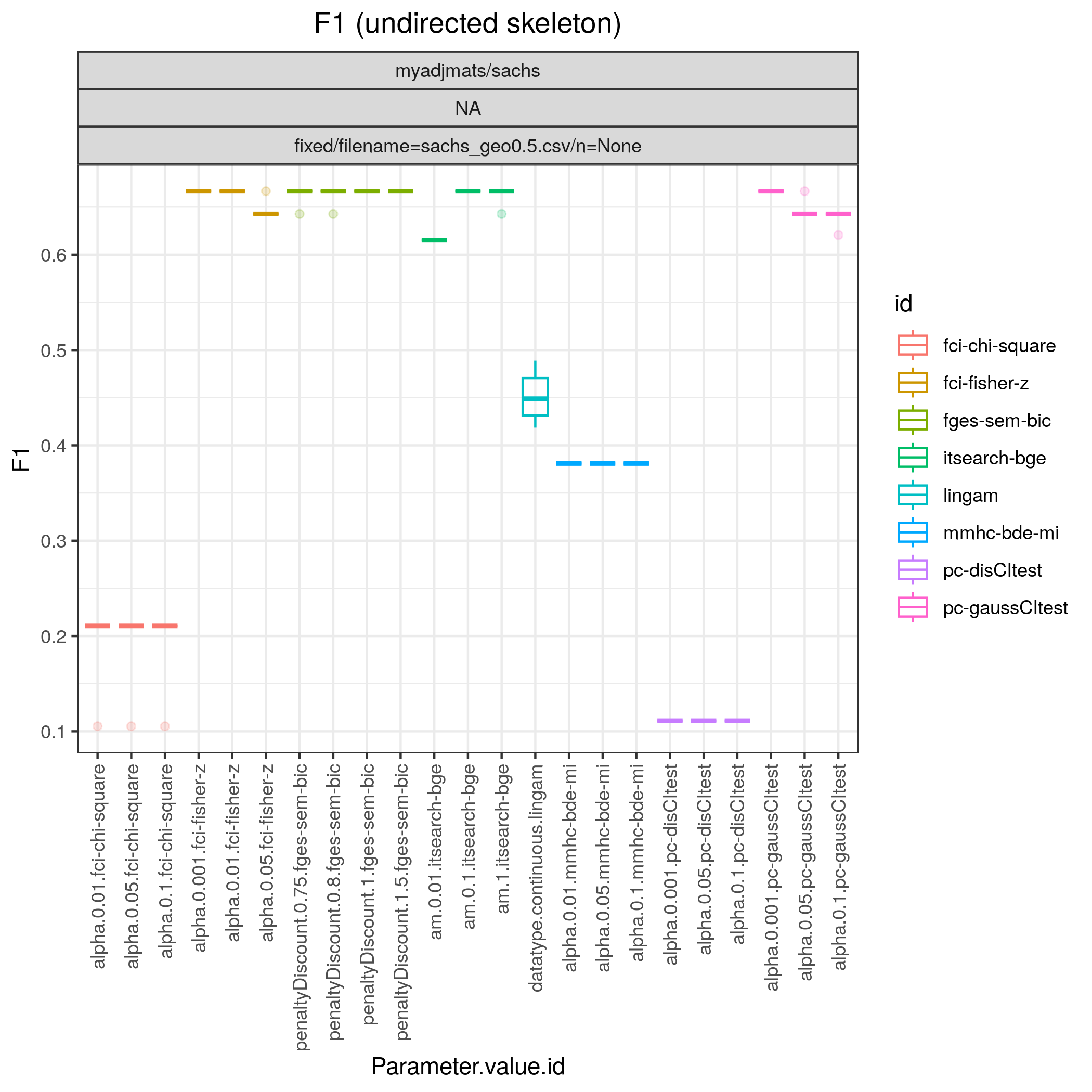}
	\caption{Sachs data, Geo C-wise mechanism, max probability 0.5.}
\end{minipage}
\begin{minipage}{0.31\linewidth}
\centering
  \includegraphics[scale=0.36]{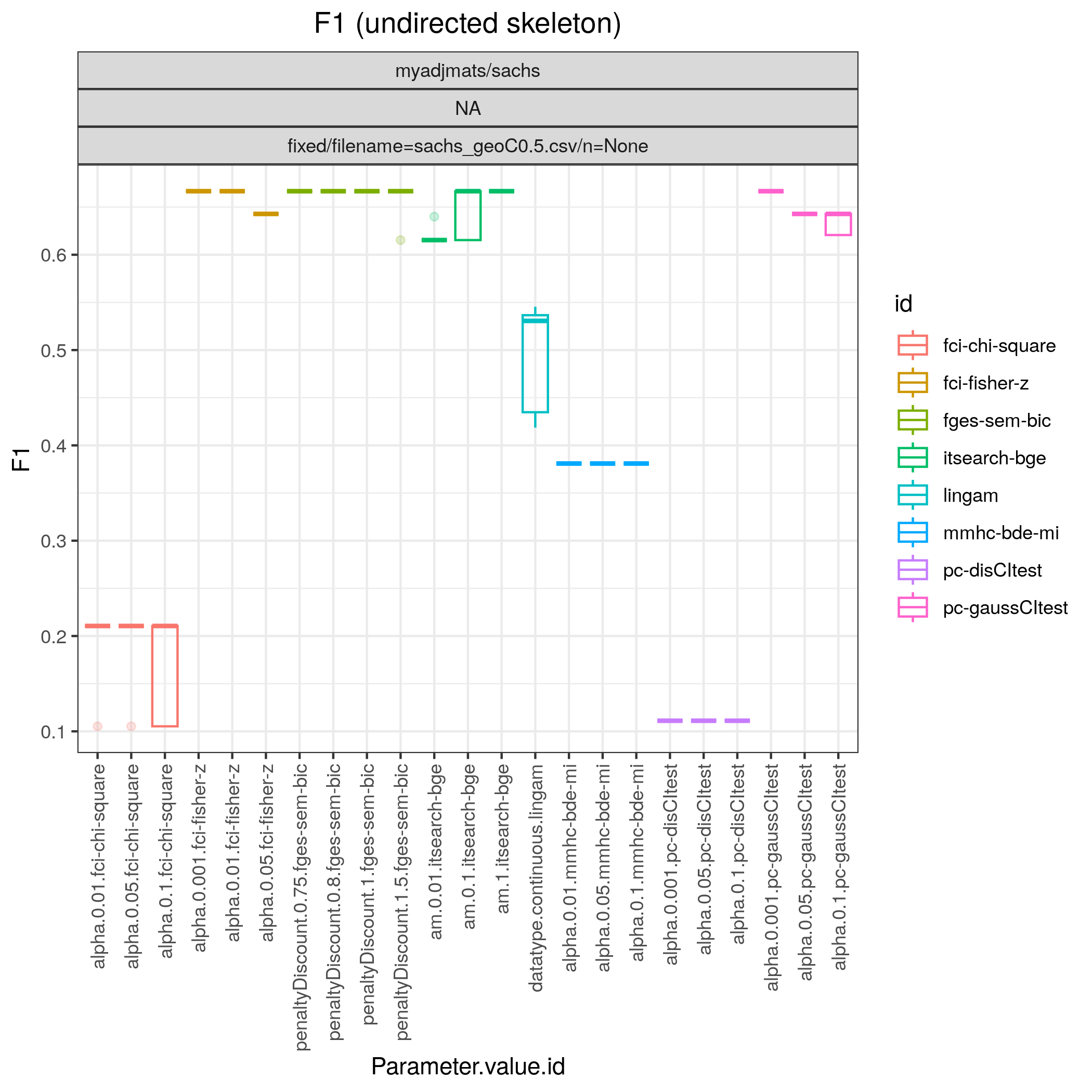}
	\caption{Sachs data, Geo Comb mechanism, max probability 0.5.}
 \end{minipage}
\begin{minipage}{0.31\linewidth}
\centering
  \includegraphics[scale=0.36]{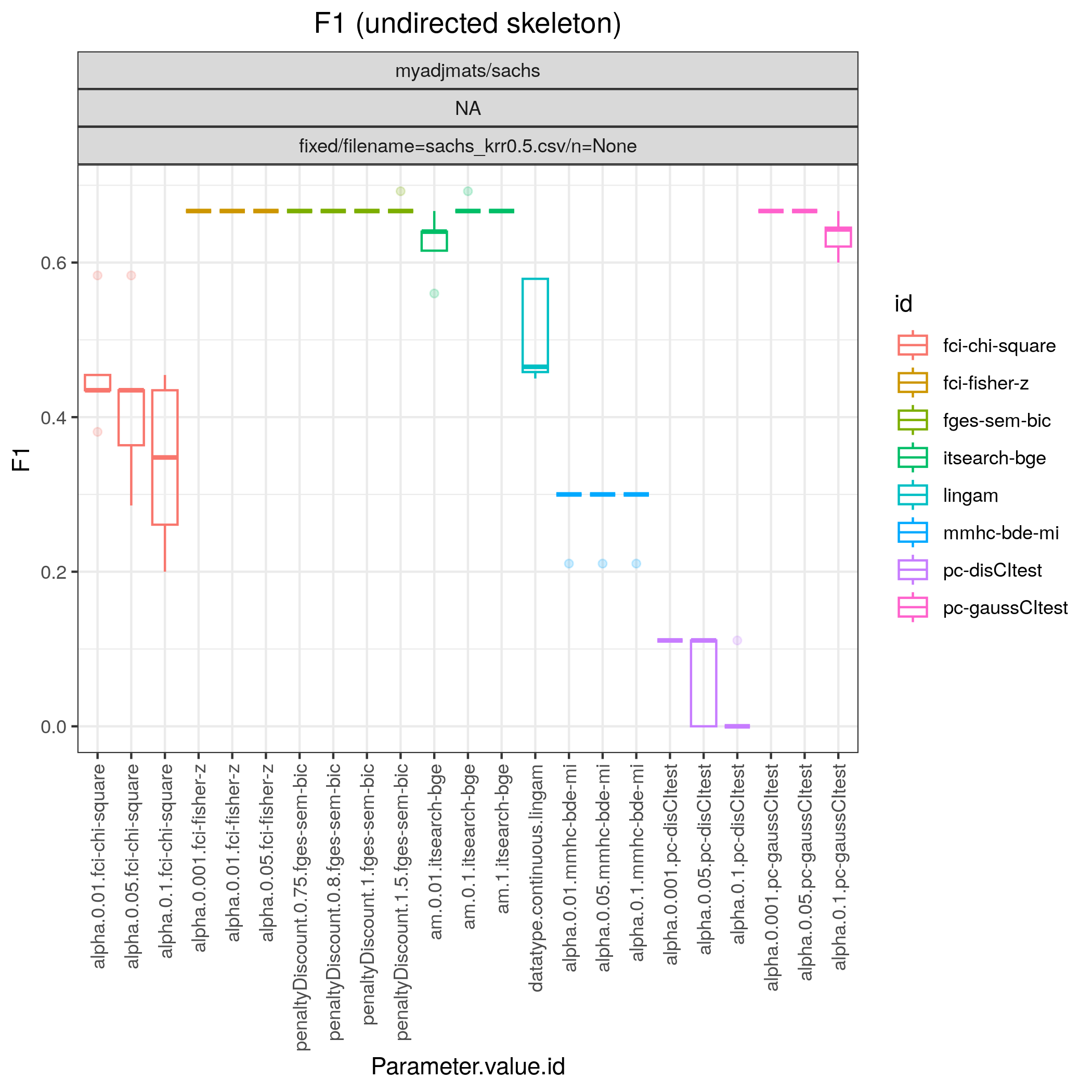}
	\caption{Sachs data, $k$-RR C-wise mechanism, max probability 0.5.}
\end{minipage}
\begin{minipage}{0.31\linewidth}
\centering
  \includegraphics[scale=0.34]{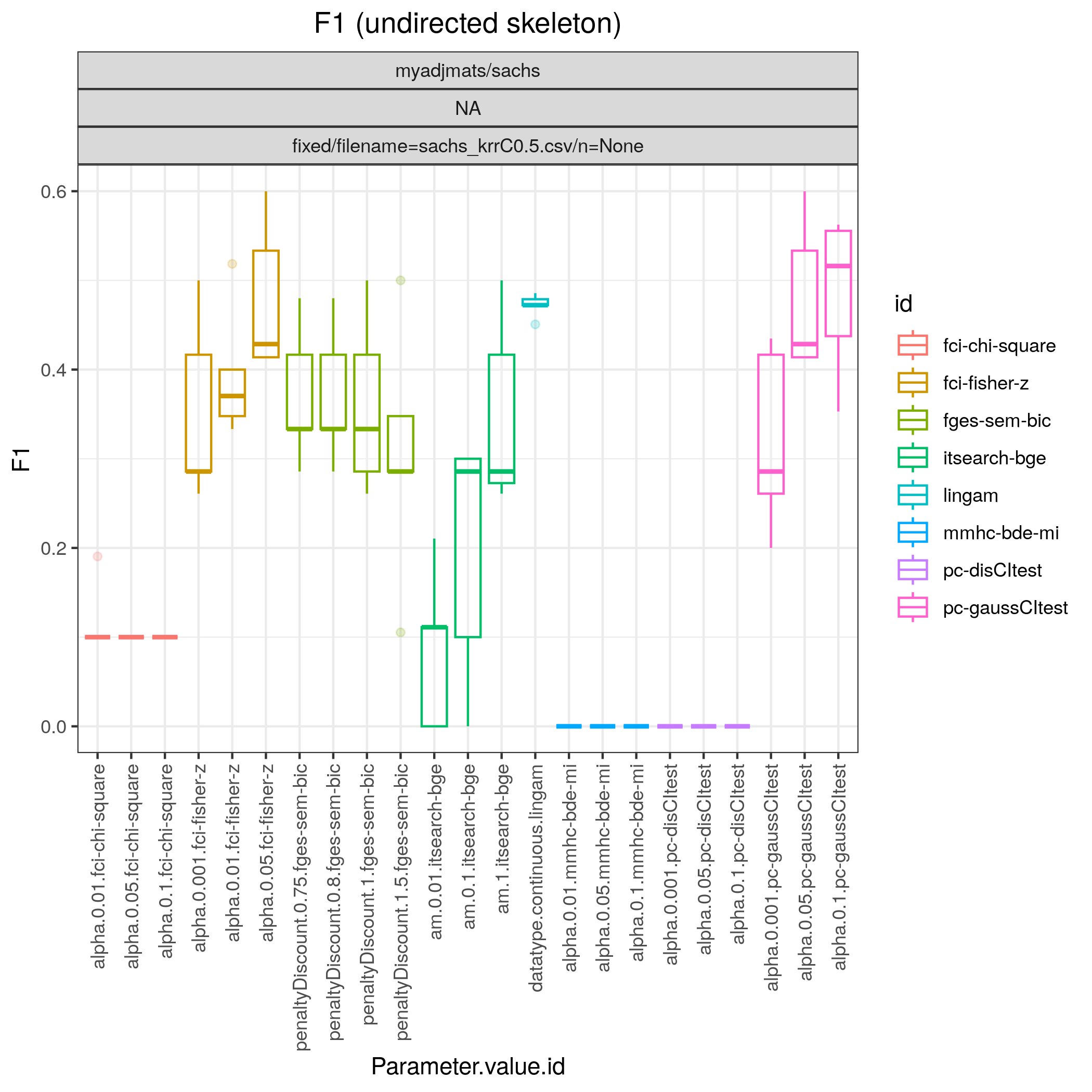}
	\caption{Sachs data, $k$-RR Comb mechanism, max probability 0.5.}
\end{minipage}
\end{figure}

 \subsection{SHD Score results Sachs data set}

 \begin{figure}[H]
    \noindent
    \centering
    \includegraphics[width=0.45\linewidth]{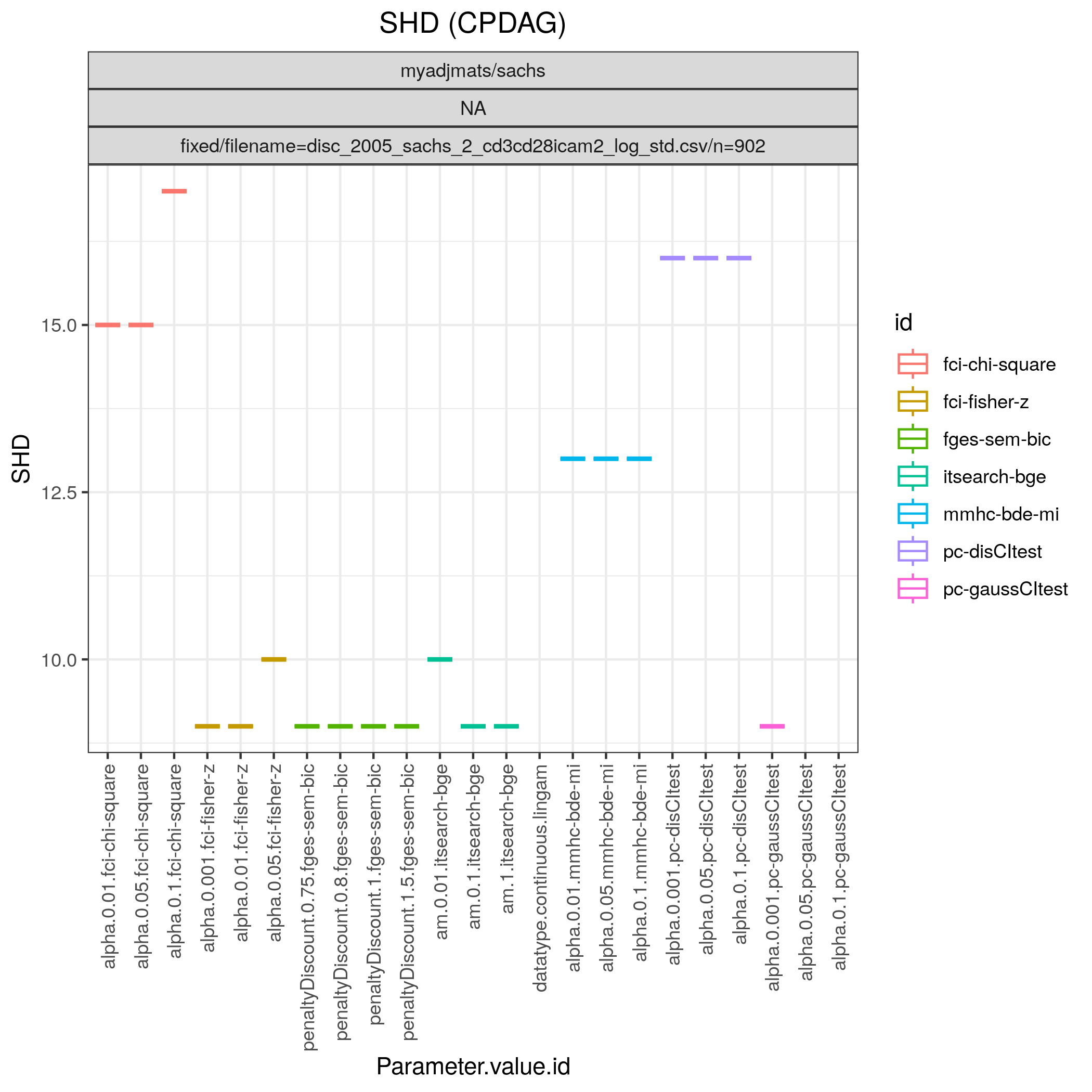}
    \caption{
        SHD Scores on the Sachs data set. Discretized, no noise.
    }
\end{figure}

\noindent
\begin{figure}[H]
\begin{minipage}{0.31\linewidth}
\centering
		\includegraphics[scale=0.36]{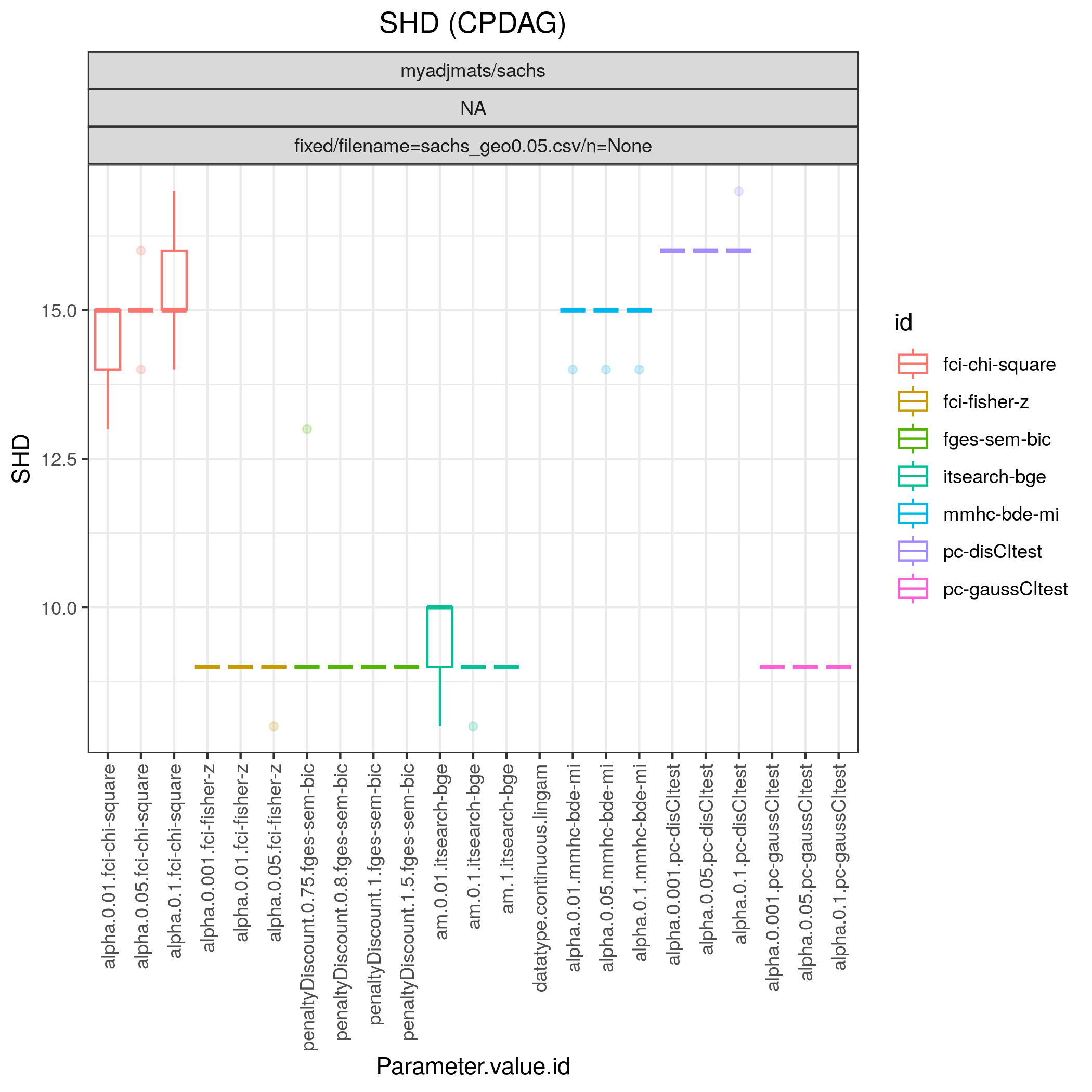}
	\caption{Sachs data, Geo C-wise mechanism, max probability 0.05.}
\end{minipage}
\begin{minipage}{0.31\linewidth}
\centering
  \includegraphics[scale=0.36]{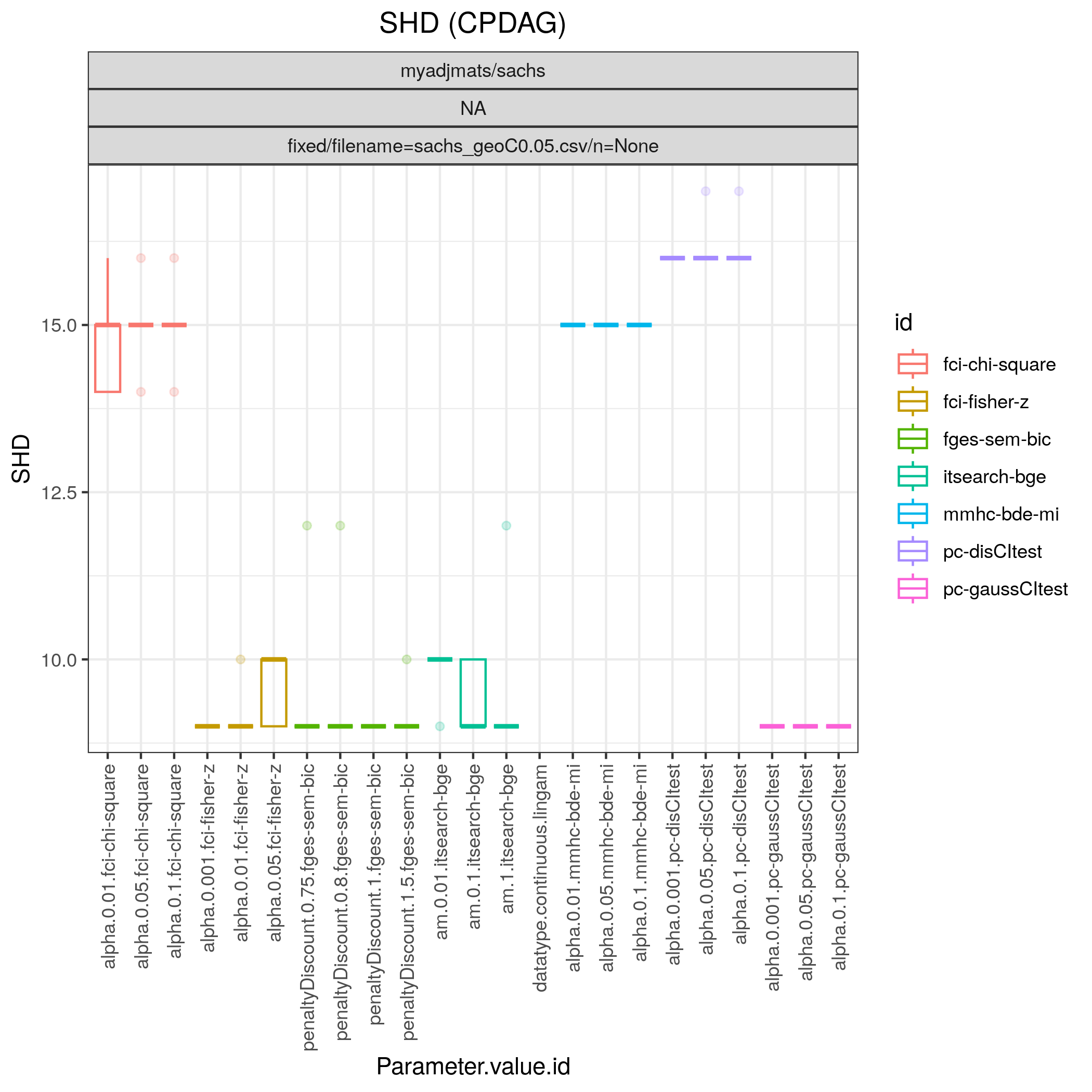}
	\caption{Sachs data, Geo Comb mechanism, max probability 0.05.}
 \end{minipage}
\begin{minipage}{0.31\linewidth}
\centering
  \includegraphics[scale=0.36]{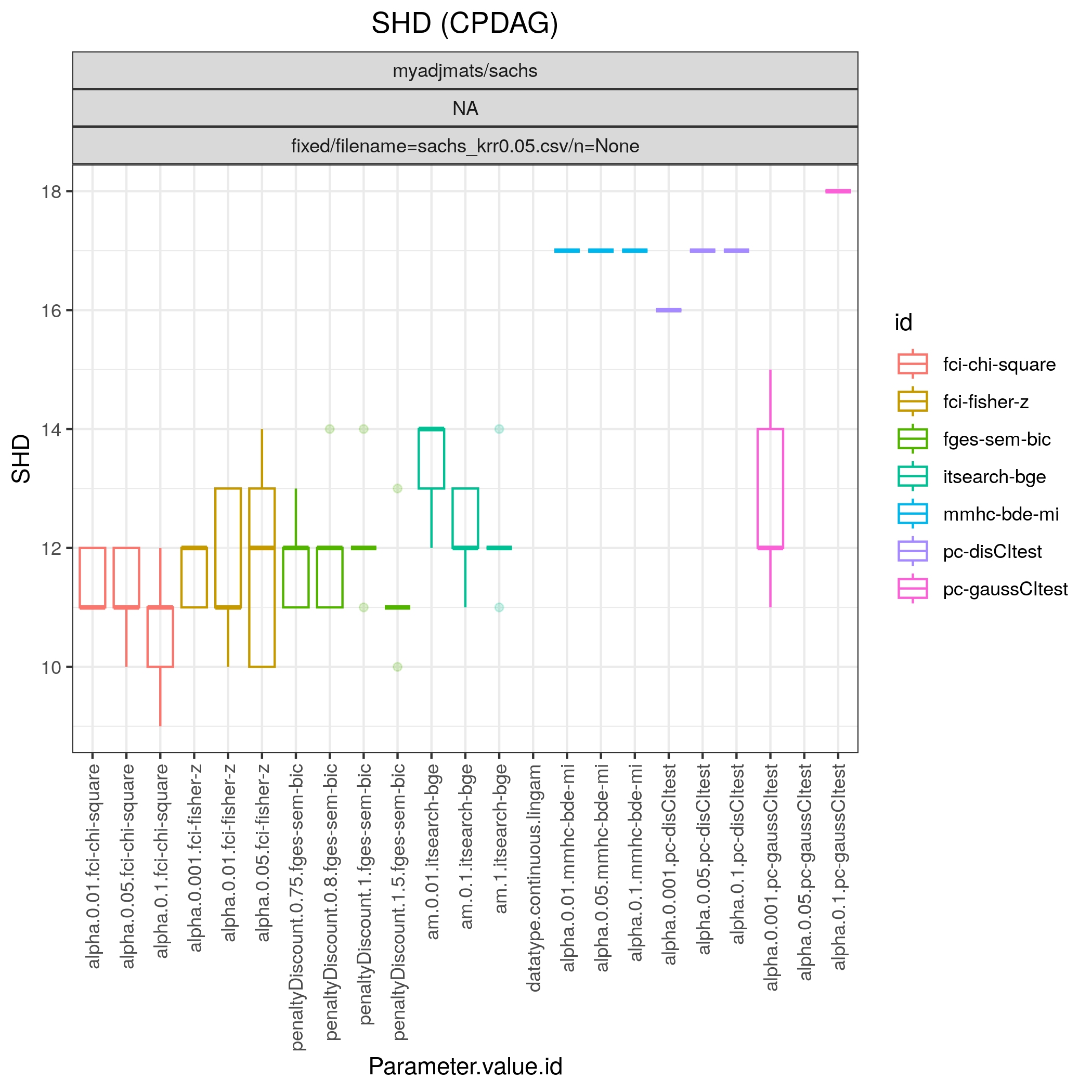}
	\caption{Sachs data, $k$-RR C-wise mechanism, max probability 0.05.}
\end{minipage}
\begin{minipage}{0.31\linewidth}
\centering
  \includegraphics[scale=0.34]{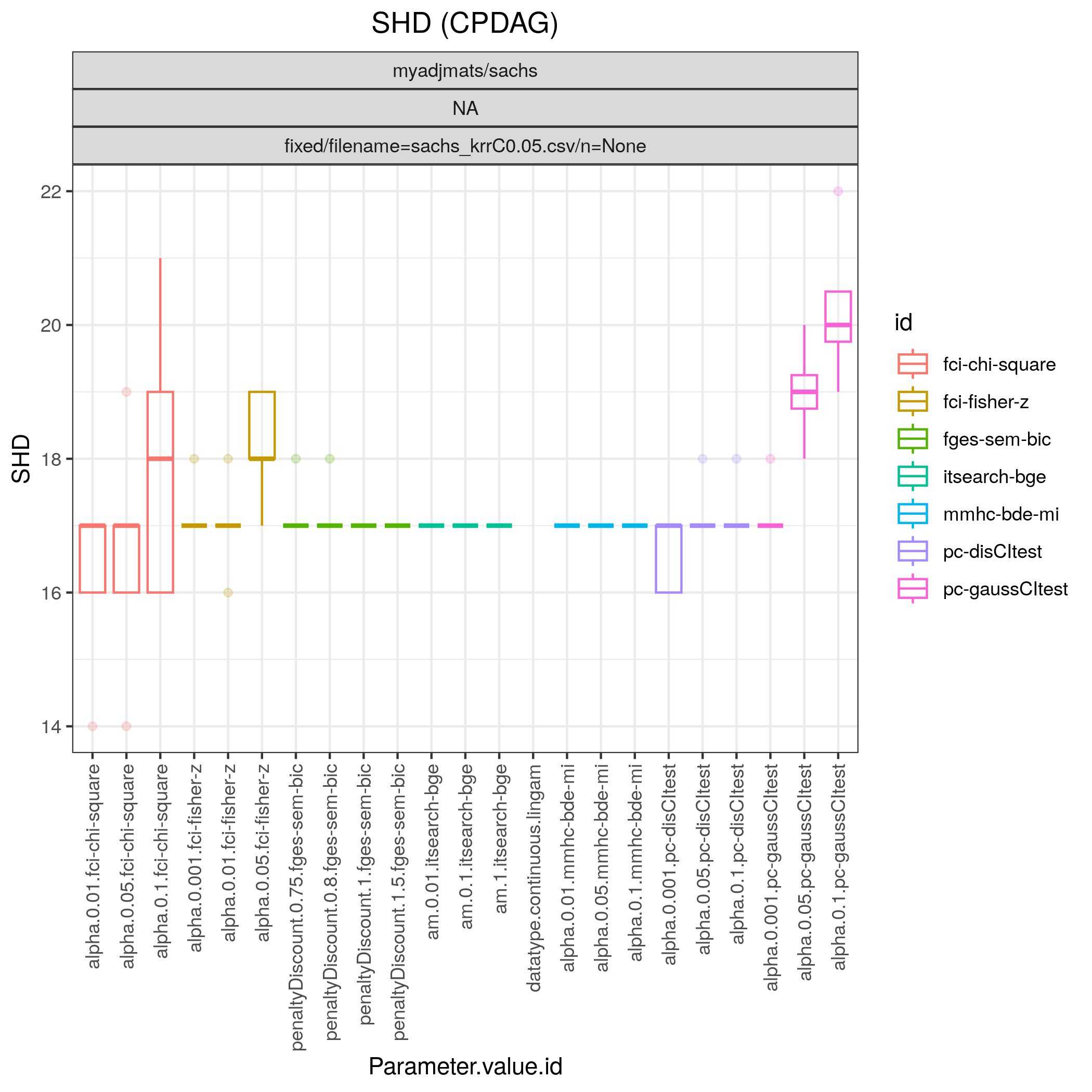}
	\caption{Sachs data, $k$-RR Comb mechanism, max probability 0.05.}
\end{minipage}
\end{figure}
\noindent
\begin{figure}[H]
\begin{minipage}{0.31\linewidth}
\centering
		\includegraphics[scale=0.36]{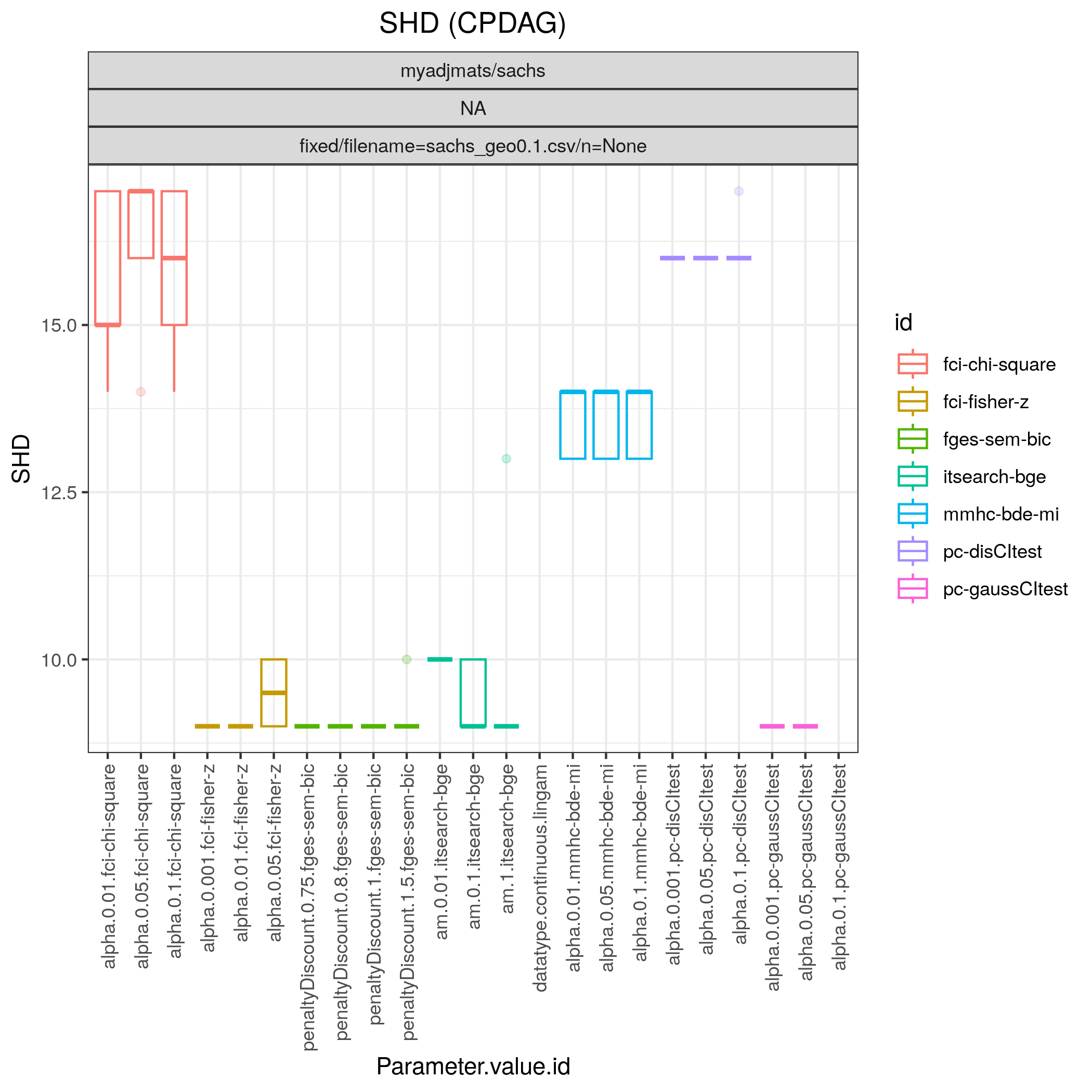}
	\caption{Sachs data, Geo C-wise mechanism, max probability 0.1.}
\end{minipage}
\begin{minipage}{0.31\linewidth}
\centering
  \includegraphics[scale=0.36]{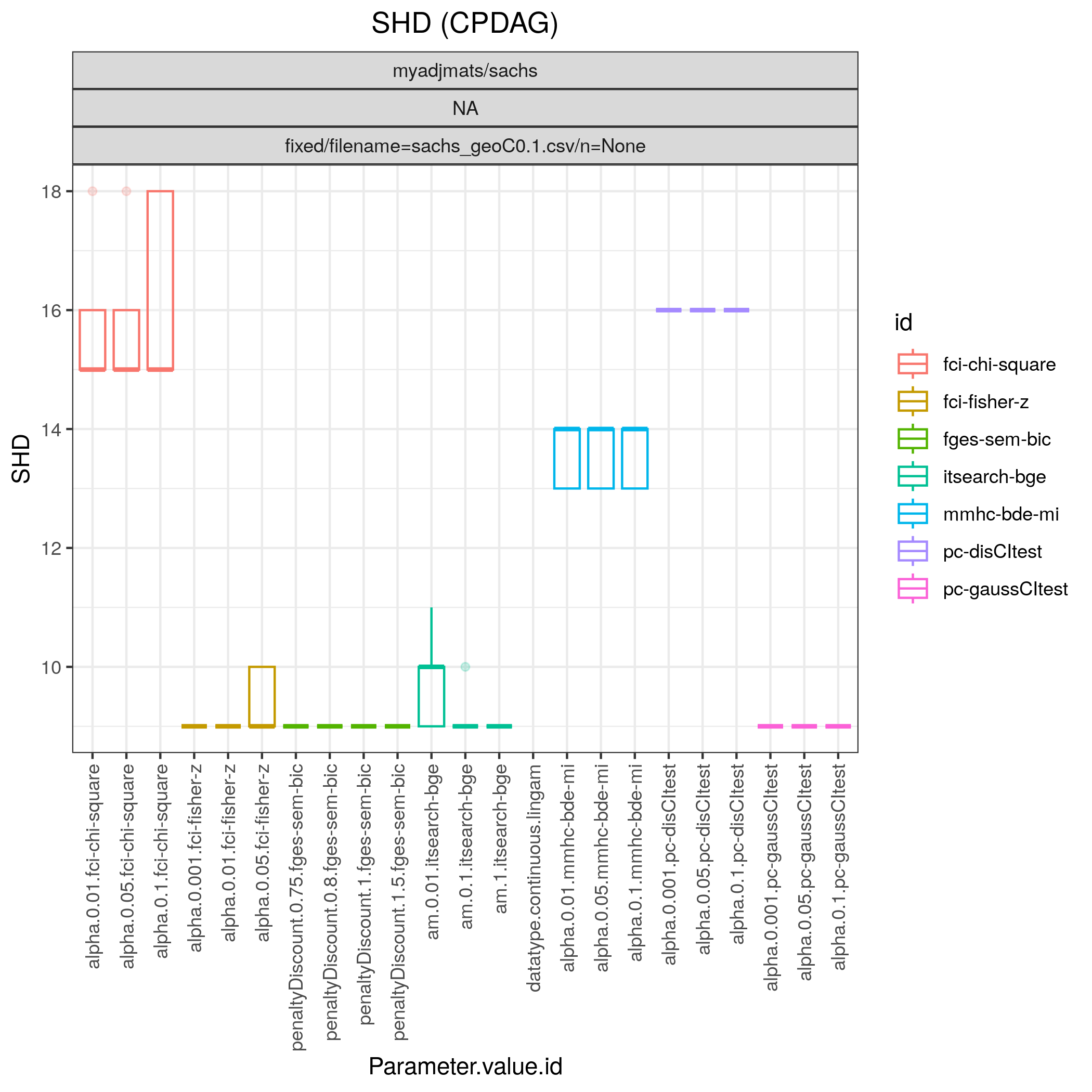}
	\caption{Sachs data, Geo Comb mechanism, max probability 0.1.}
 \end{minipage}
\begin{minipage}{0.31\linewidth}
\centering
  \includegraphics[scale=0.36]{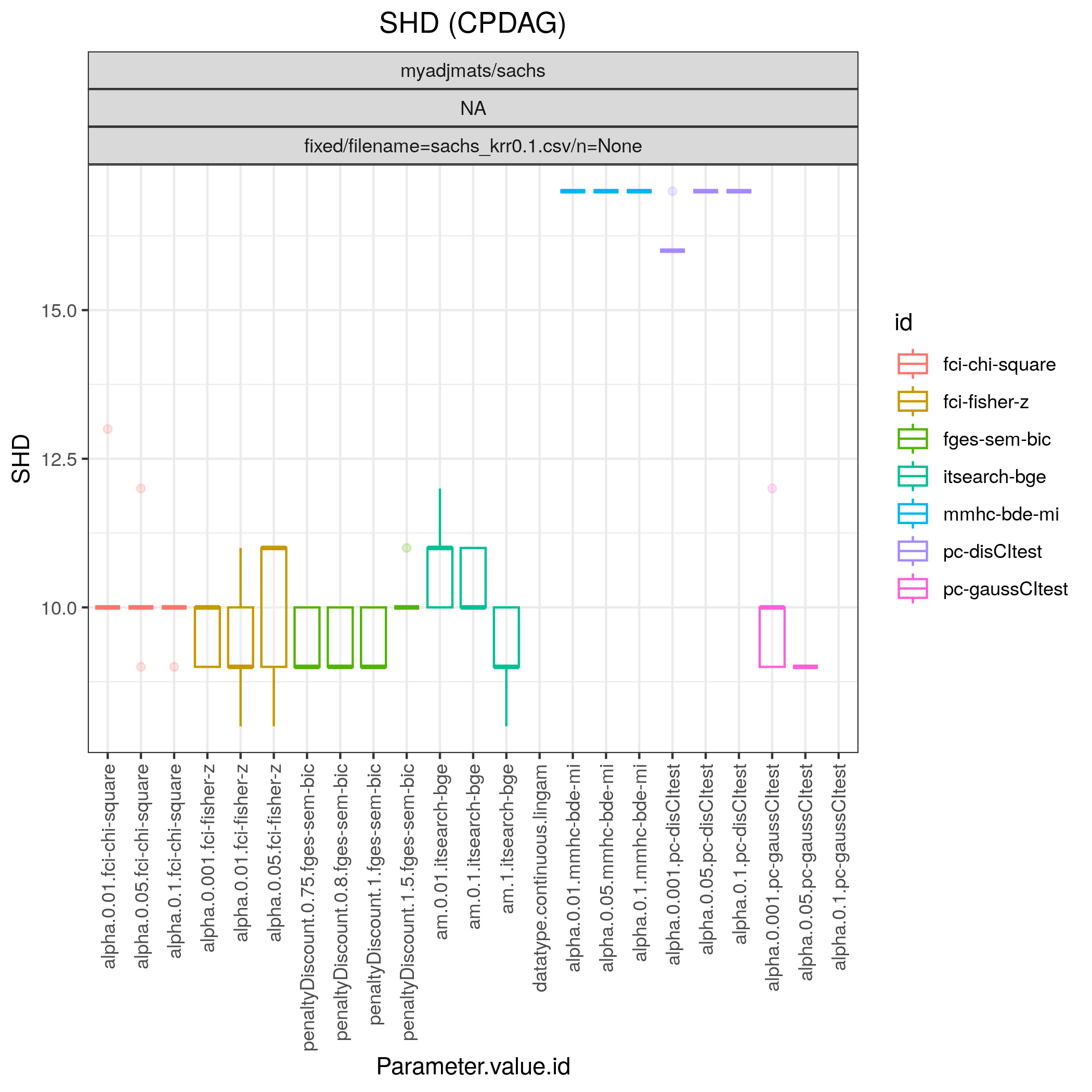}
	\caption{Sachs data, $k$-RR C-wise mechanism, max probability 0.1.}
\end{minipage}
\begin{minipage}{0.31\linewidth}
\centering
  \includegraphics[scale=0.34]{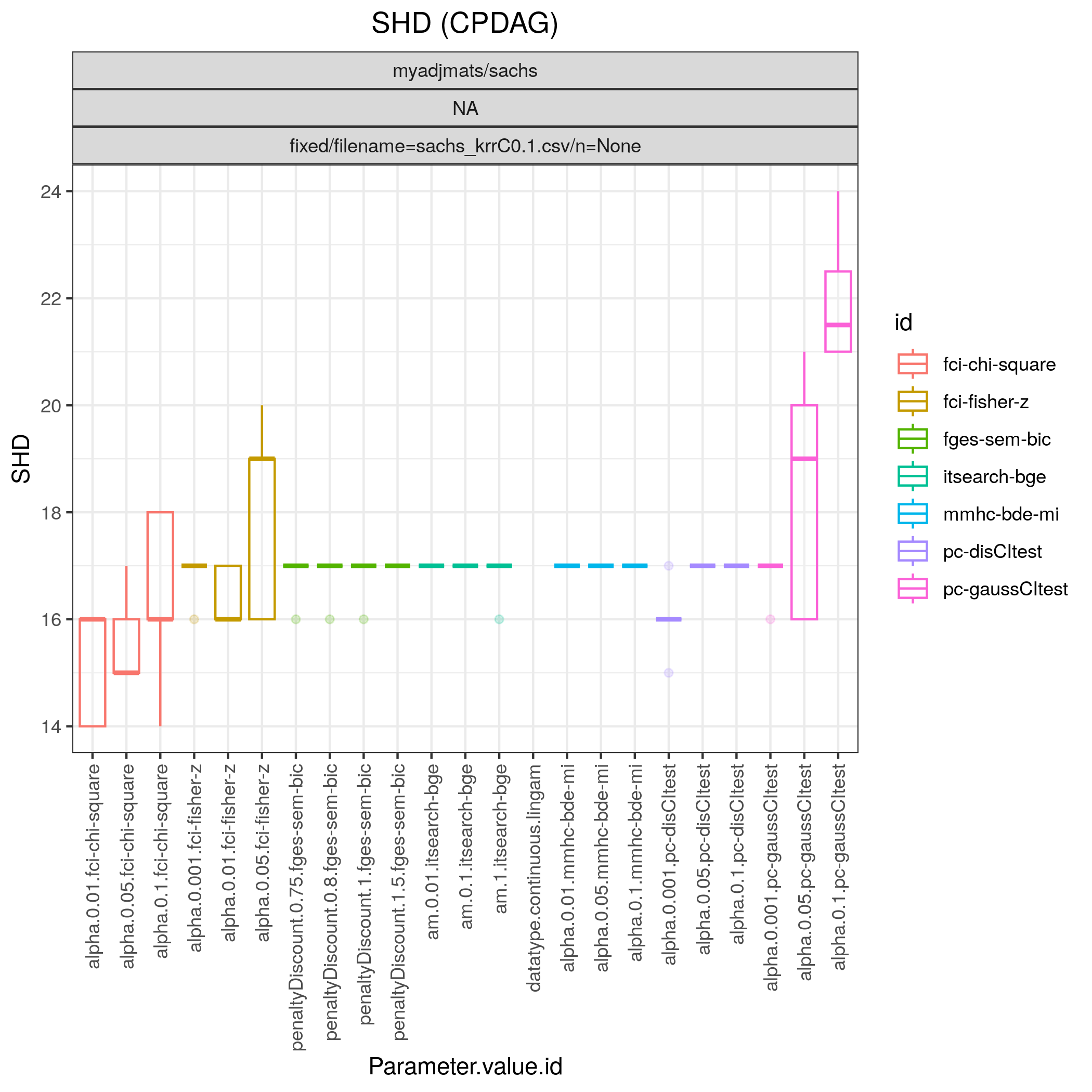}
	\caption{Sachs data, $k$-RR Comb mechanism, max probability 0.1.}
\end{minipage}
\end{figure}

\noindent
\begin{figure}[H]
\begin{minipage}{0.31\linewidth}
\centering
		\includegraphics[scale=0.36]{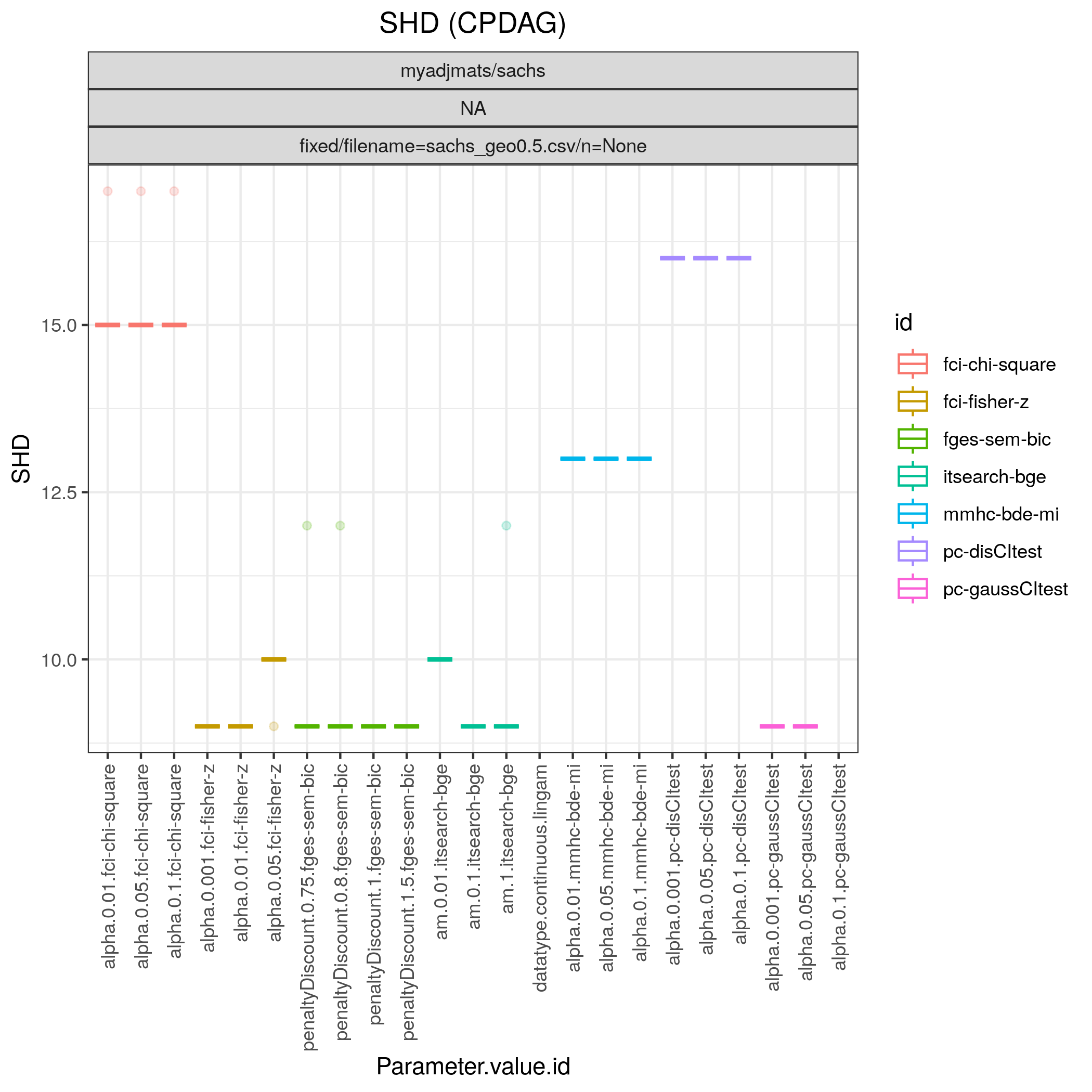}
	\caption{Sachs data, Geo C-wise mechanism, max probability 0.5.}
\end{minipage}
\begin{minipage}{0.31\linewidth}
\centering
  \includegraphics[scale=0.36]{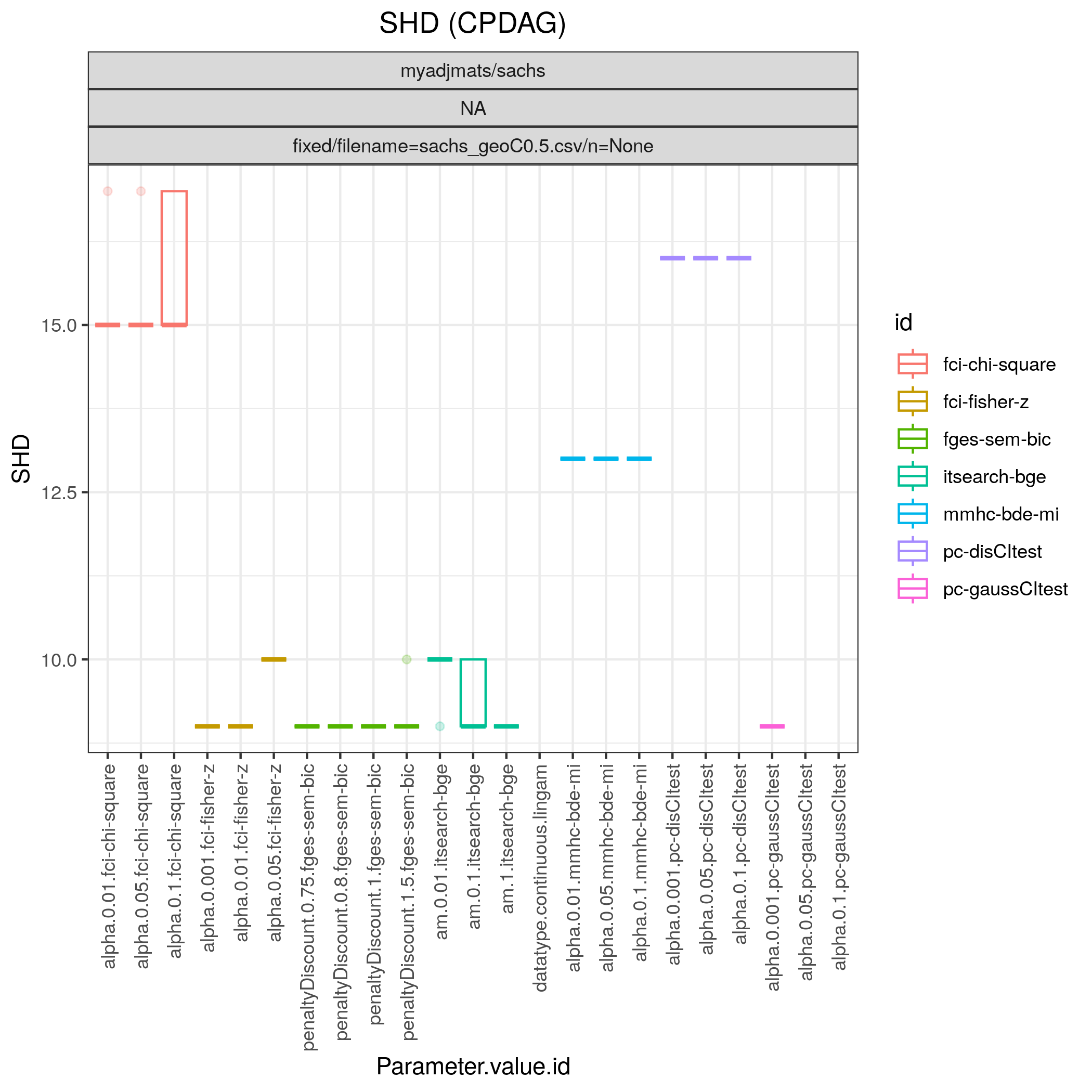}
	\caption{Sachs data, Geo Comb mechanism, max probability 0.5.}
 \end{minipage}
\begin{minipage}{0.31\linewidth}
\centering
  \includegraphics[scale=0.36]{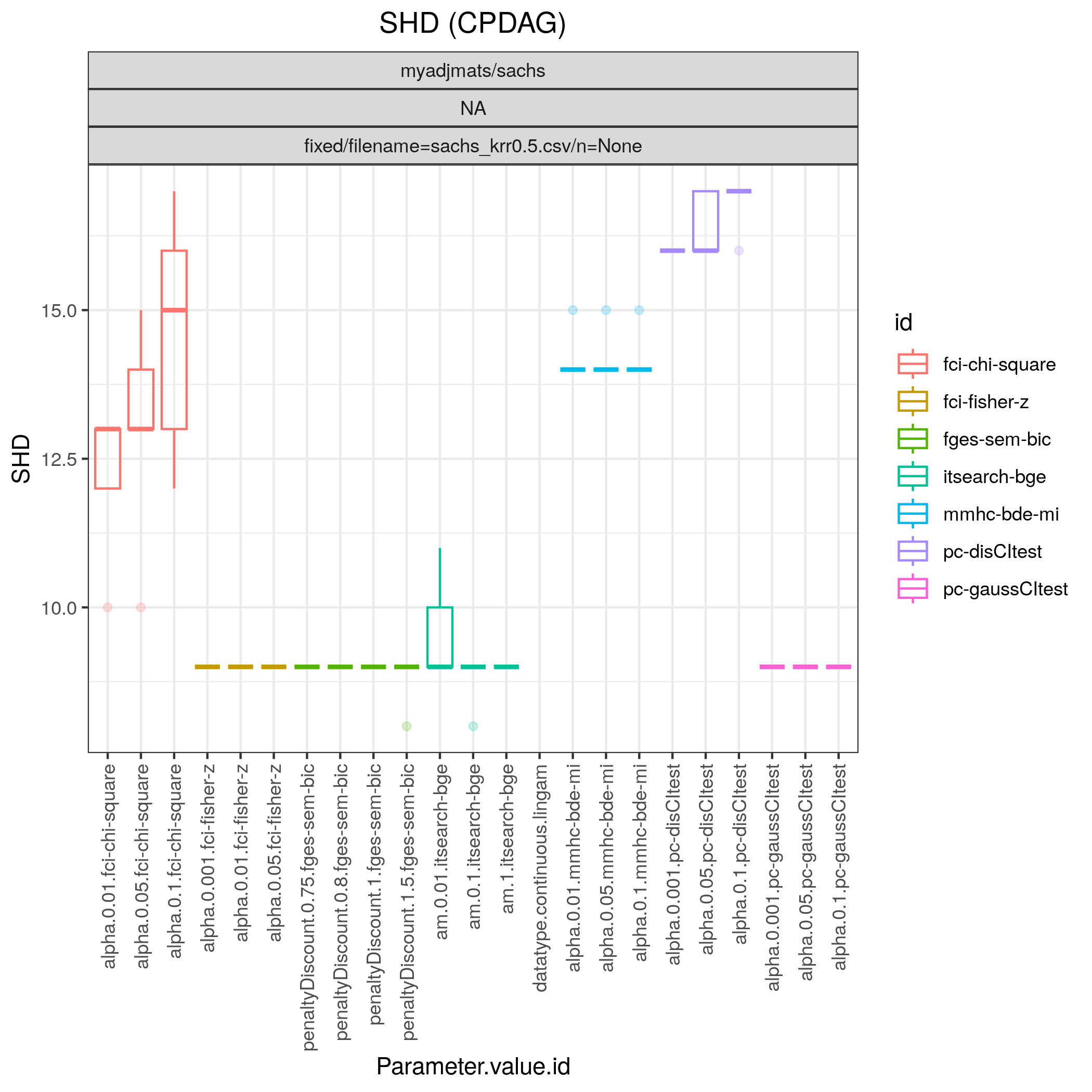}
	\caption{Sachs data, $k$-RR C-wise mechanism, max probability 0.5.}
\end{minipage}
\begin{minipage}{0.31\linewidth}
\centering
  \includegraphics[scale=0.34]{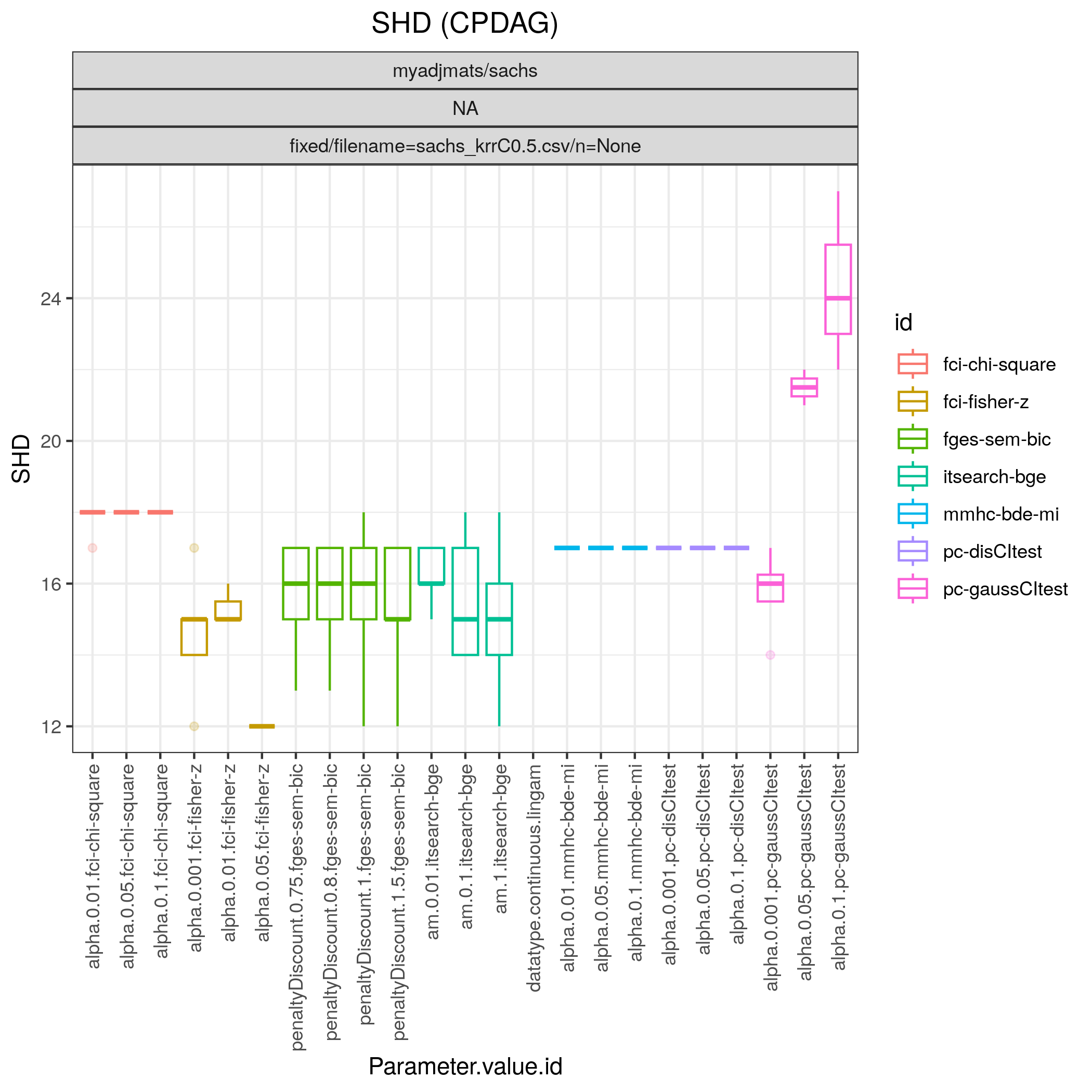}
	\caption{Sachs data, $k$-RR Comb mechanism, max probability 0.5.}
\end{minipage}
\end{figure}

 \subsection{F1 Score results Human Stature data set}
 
\begin{figure}[t]
    \centering
    \includegraphics[width=1\linewidth]{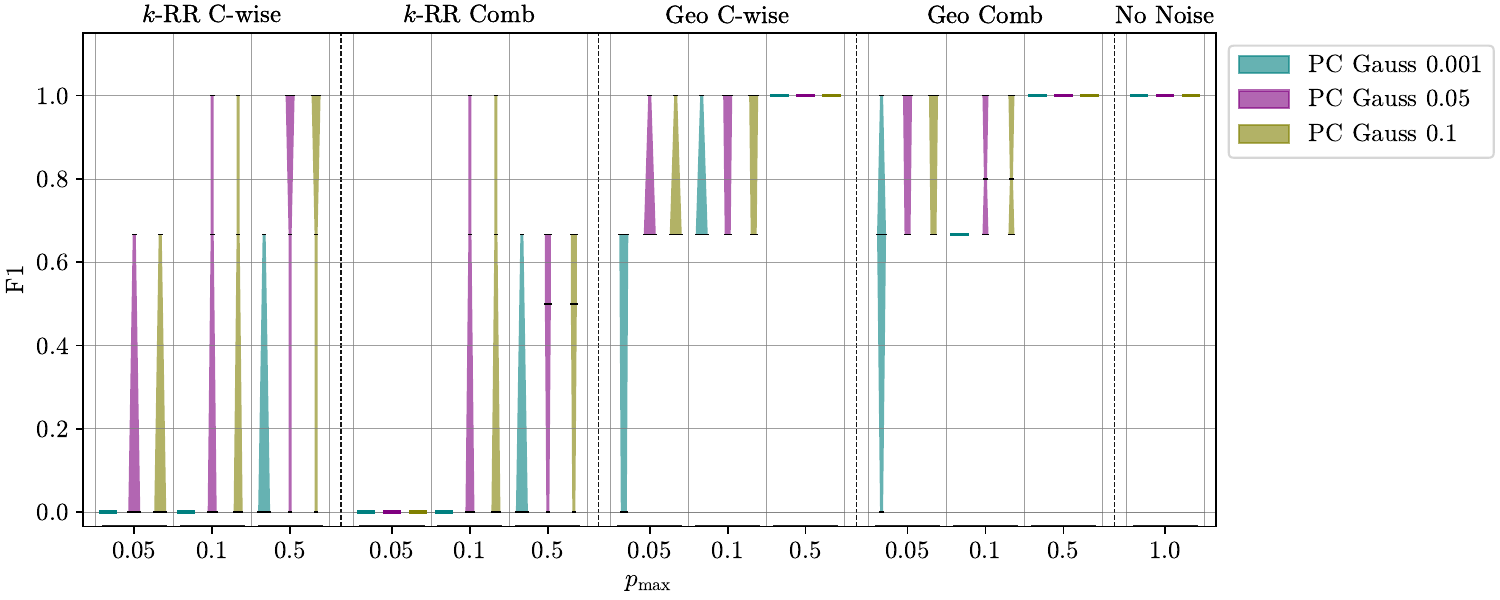}
    \vspace*{-10mm}
    \caption{Human Stature data, F1.}
    \label{fig:HS_f1}
\end{figure}

\noindent
\begin{figure}[H]
\begin{minipage}{0.31\linewidth}
\centering
		\includegraphics[scale=0.34]{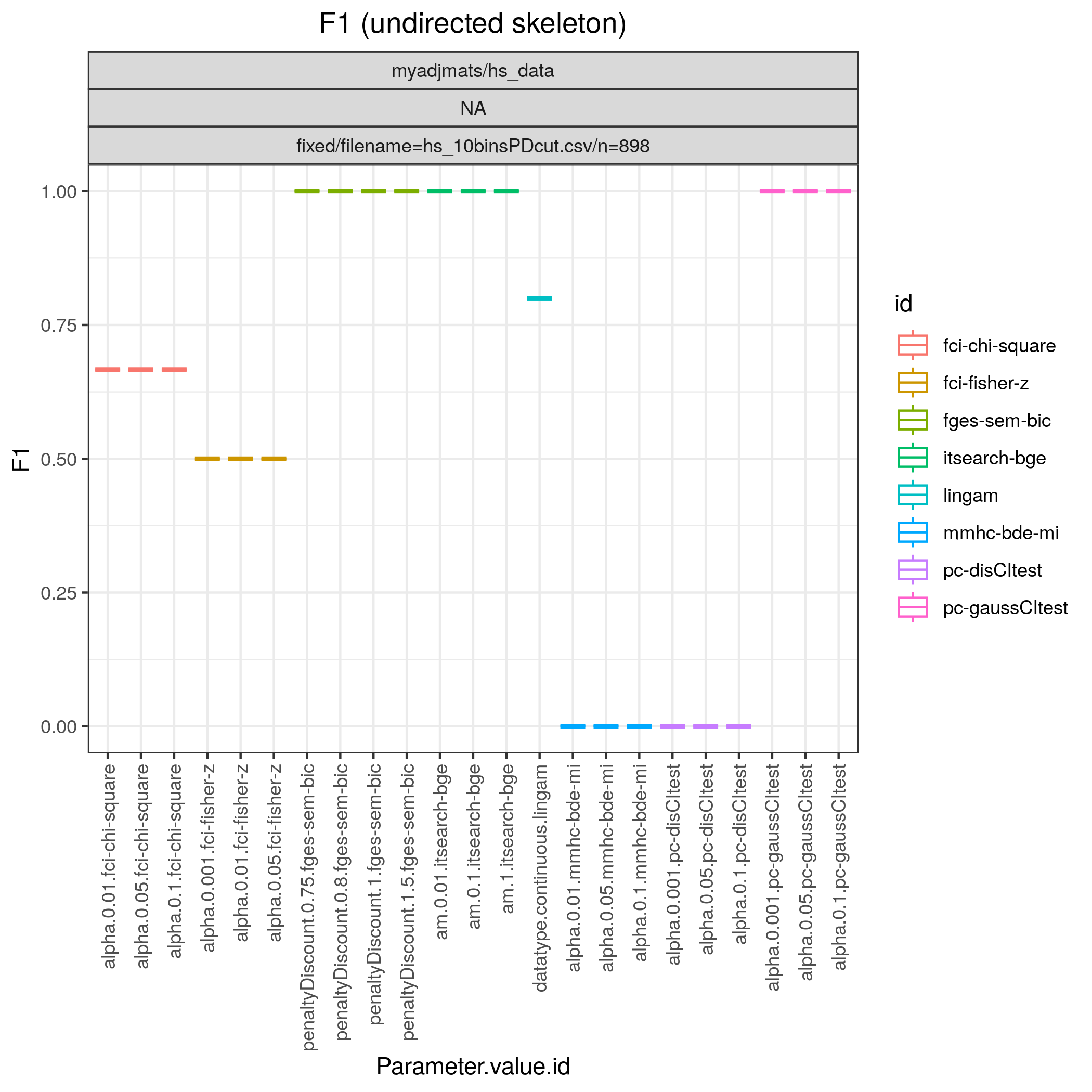}
	\caption{F1 Scores on the Human Stature data set. Discretized, no noise.}
\end{minipage}
\begin{minipage}{0.31\linewidth}
\centering
		\includegraphics[scale=0.34]{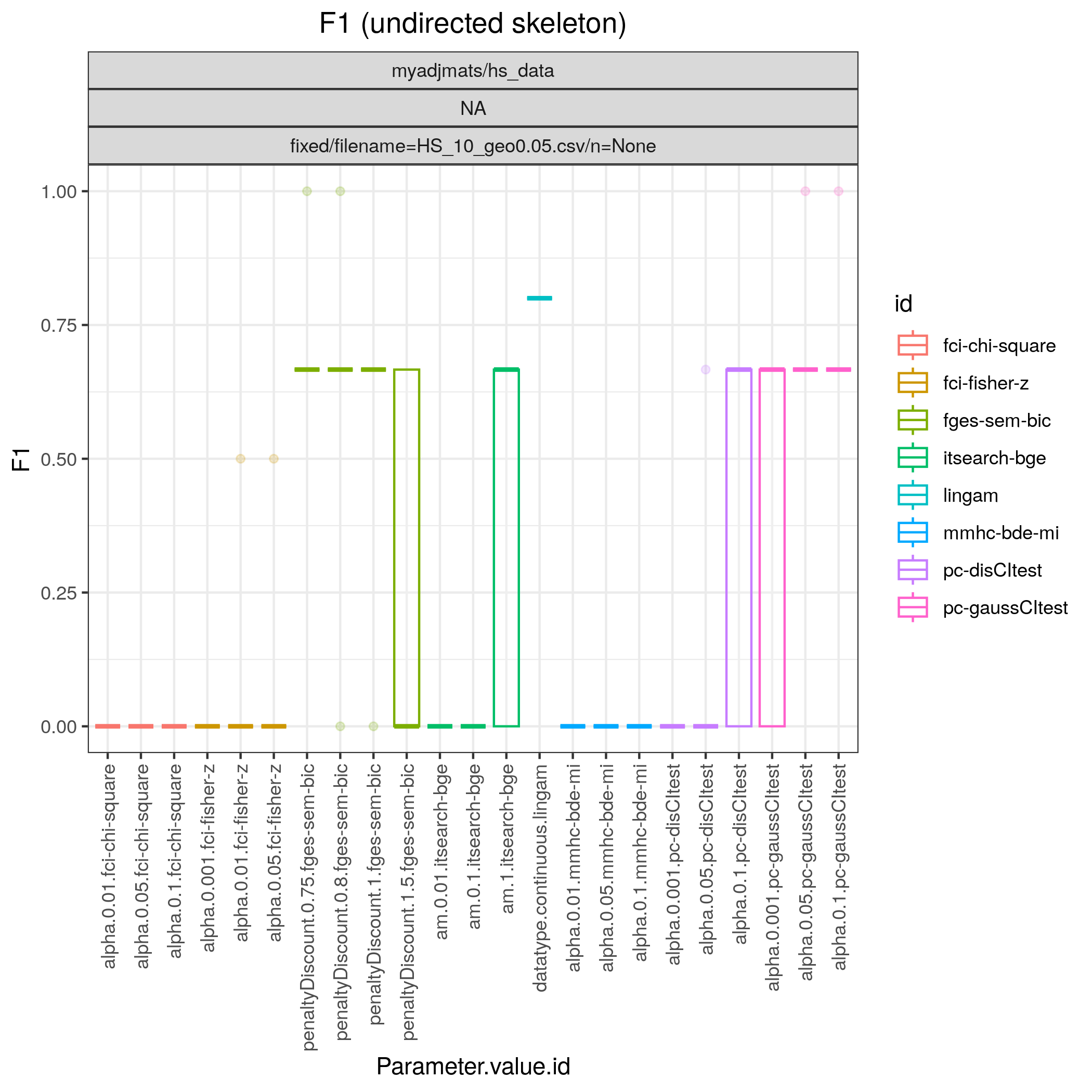}
	\caption{Human Stature data, Geo C-wise mechanism, max probability 0.05.}
\end{minipage}
\begin{minipage}{0.31\linewidth}
\centering
  \includegraphics[scale=0.34]{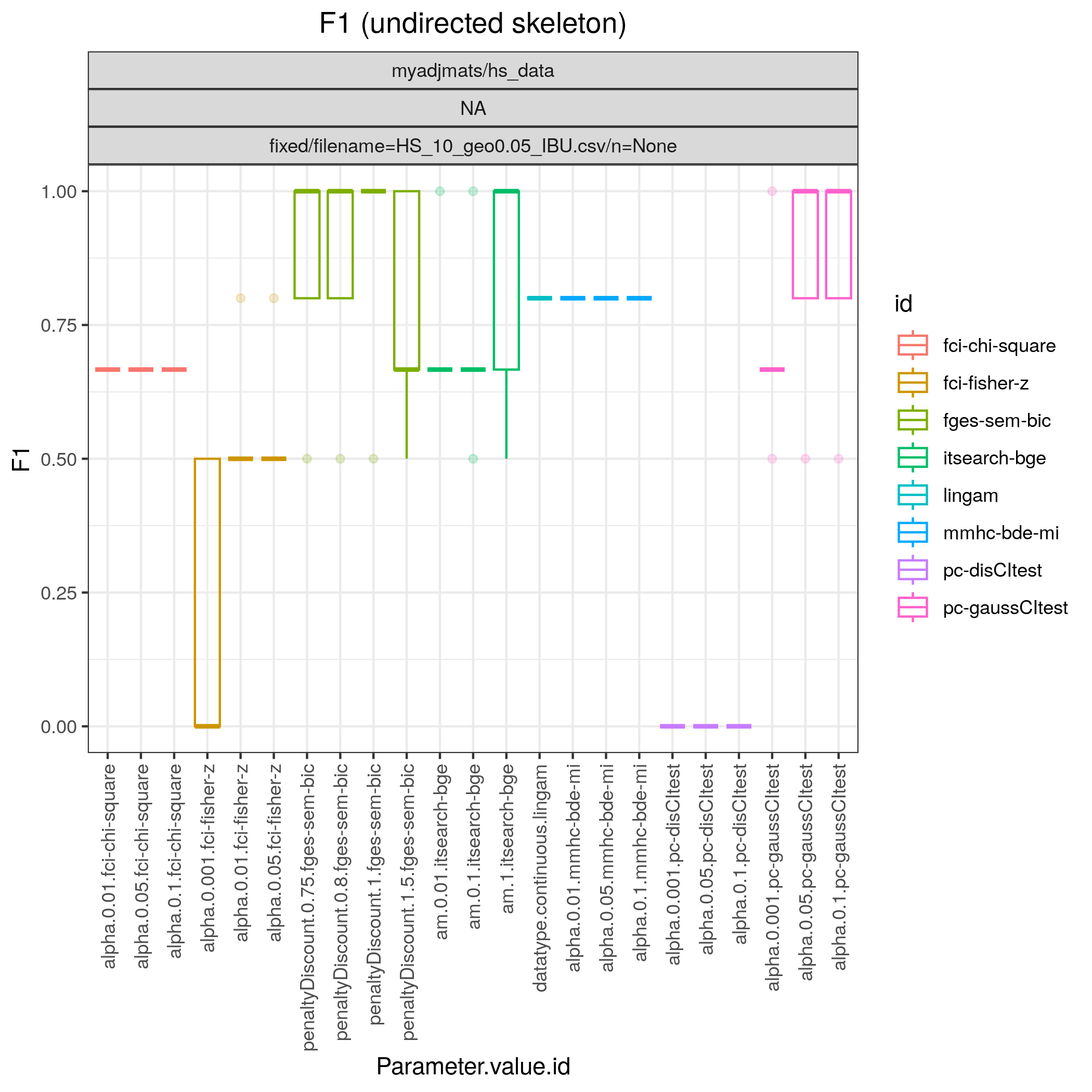}
	\caption{Human Stature data, Geo C-wise IBU mechanism, max probability 0.05.}
 \end{minipage}
\begin{minipage}{0.31\linewidth}
\centering
  \includegraphics[scale=0.34]{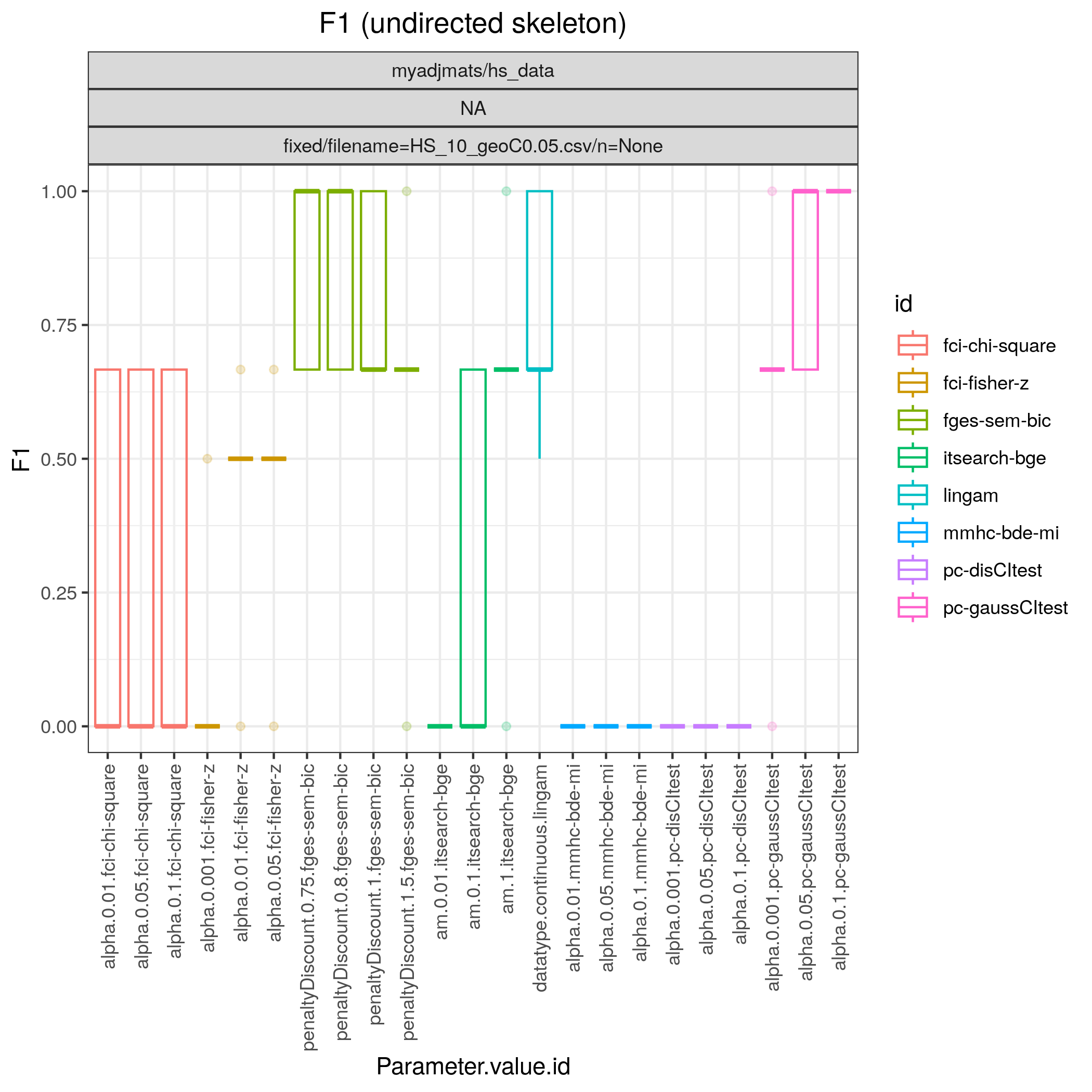}
	\caption{Human Stature data, Geo Comb mechanism, max probability 0.05.}
\end{minipage}
\begin{minipage}{0.31\linewidth}
\centering
  \includegraphics[scale=0.34]{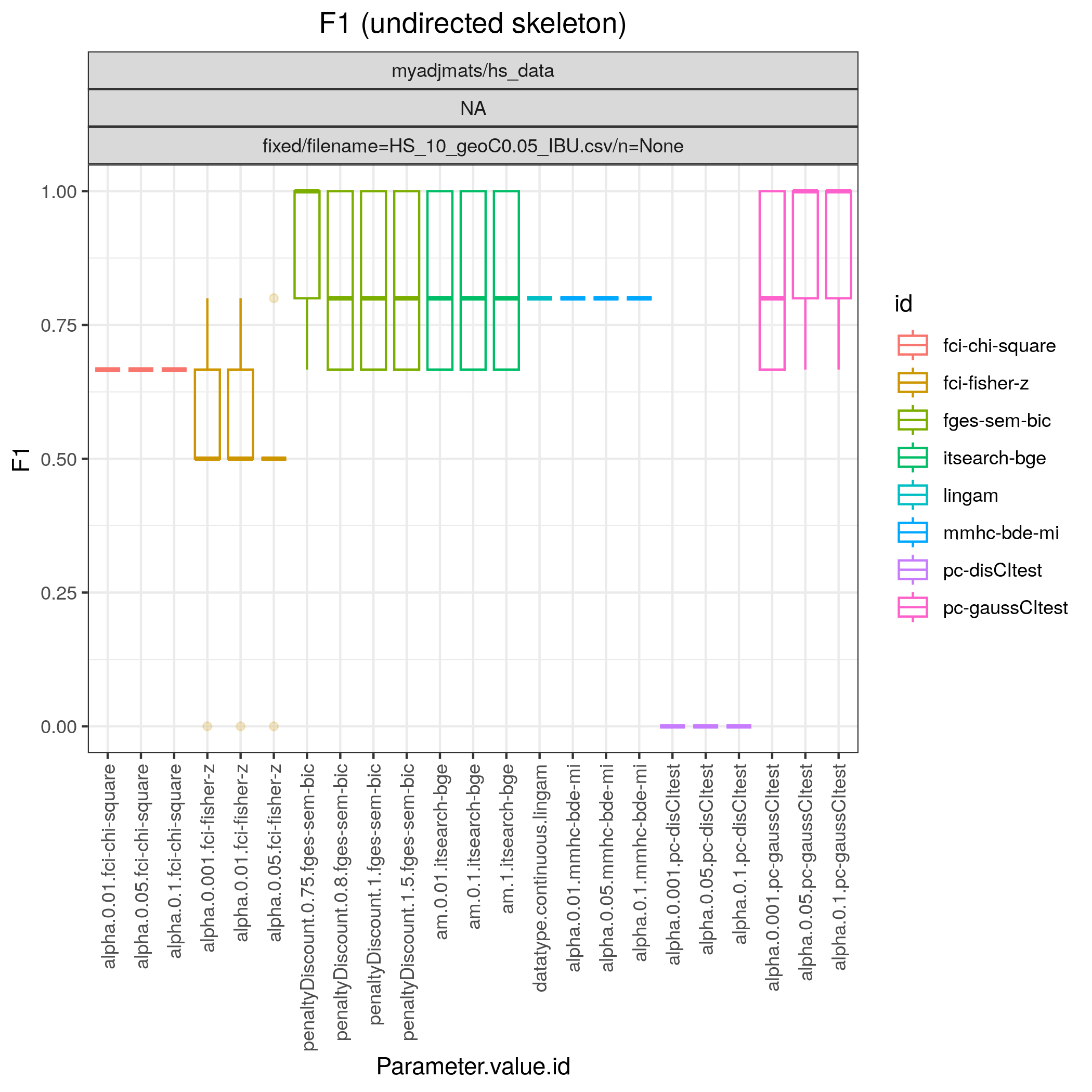}
	\caption{Human Stature data, Geo Comb IBU mechanism, max probability 0.05.}
\end{minipage}
\begin{minipage}{0.31\linewidth}
\centering
  \includegraphics[scale=0.34]{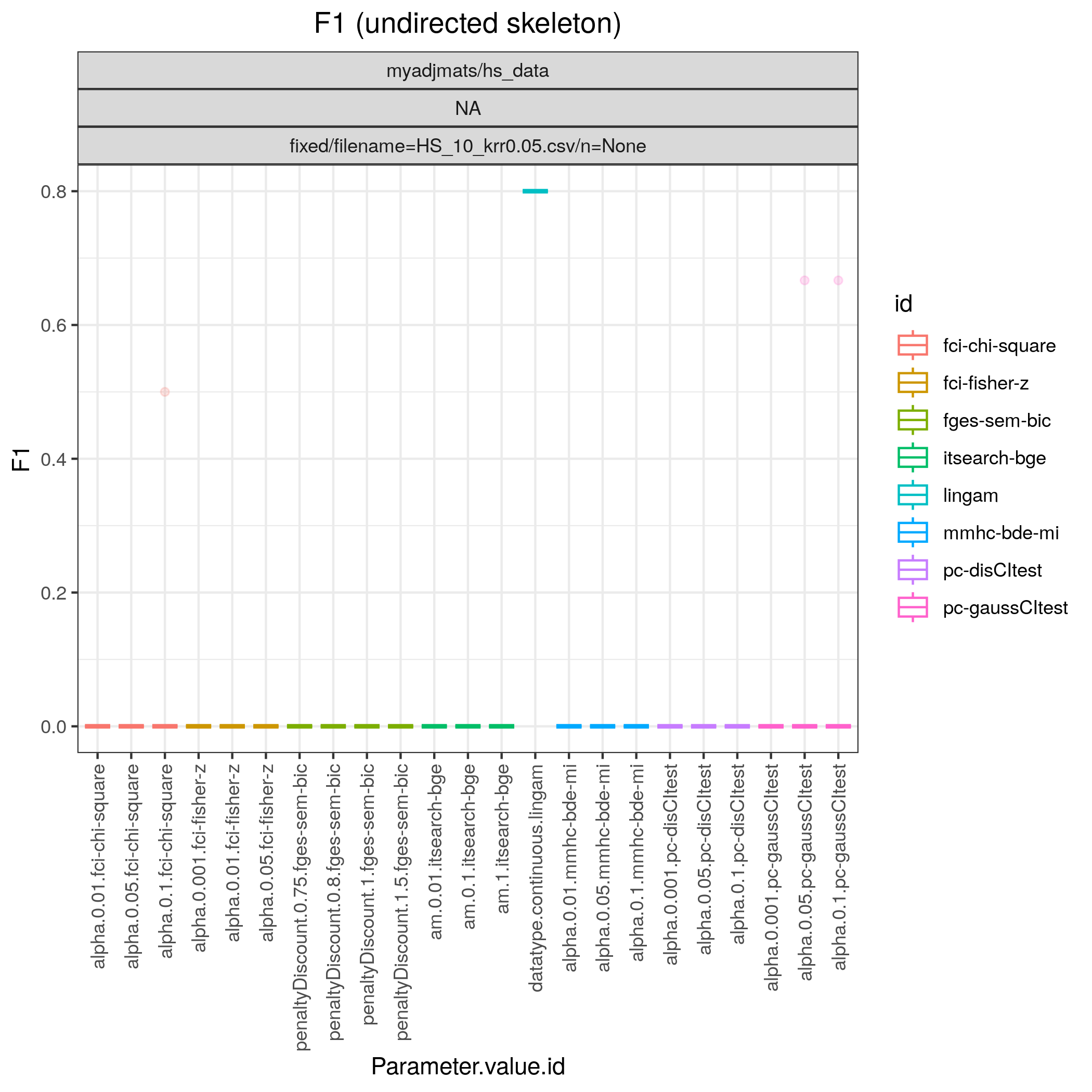}
	\caption{Human Stature data, $k$-RR C-wise mechanism, max probability 0.05.}
\end{minipage}

\end{figure}

\begin{figure}[H]
    \centering
   \begin{minipage}{0.31\linewidth}
\centering
  \includegraphics[scale=0.34]{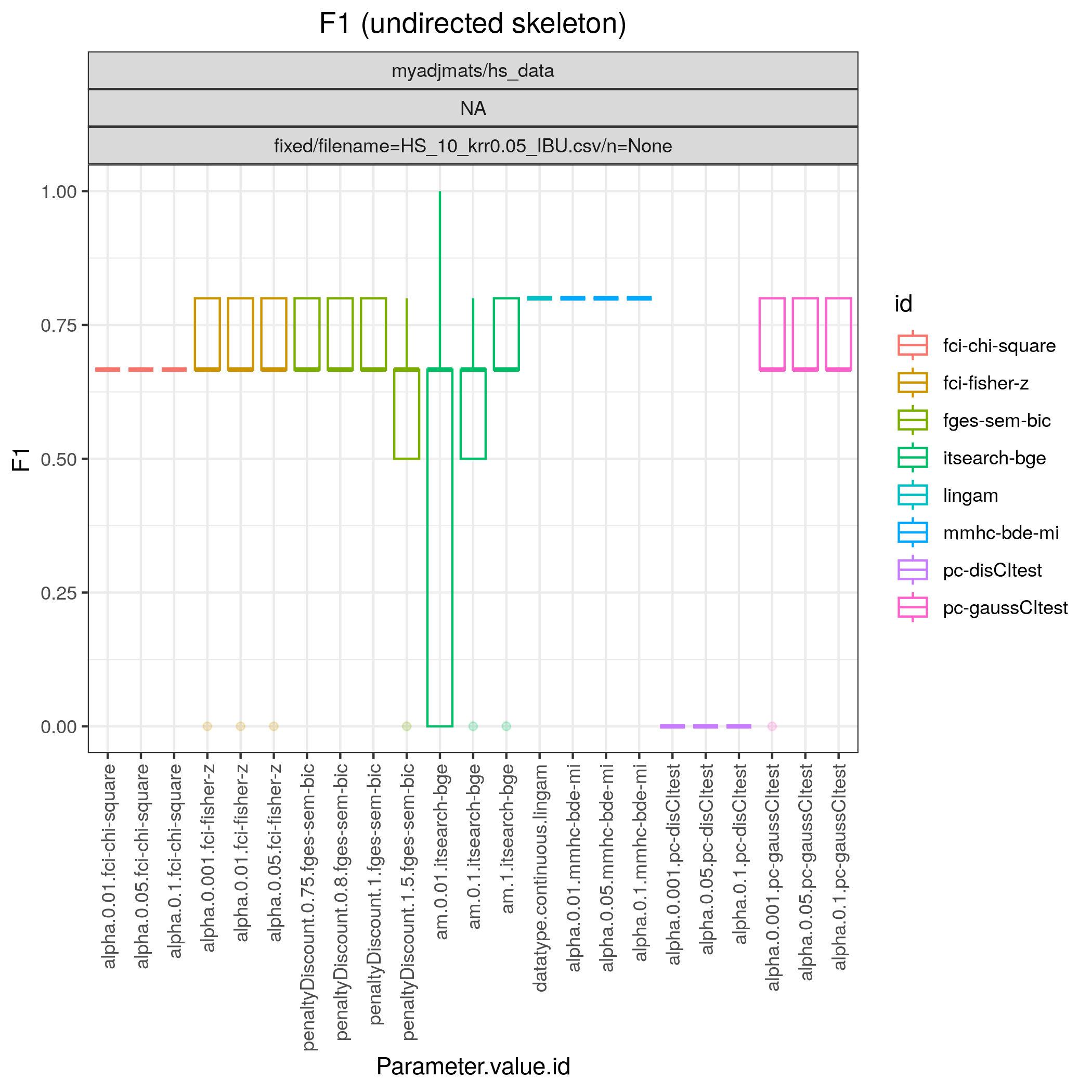}
	\caption{Human Stature data, $k$-RR C-wise IBU mechanism, max probability 0.05.}
\end{minipage}
\begin{minipage}{0.31\linewidth}
\centering
  \includegraphics[scale=0.34]{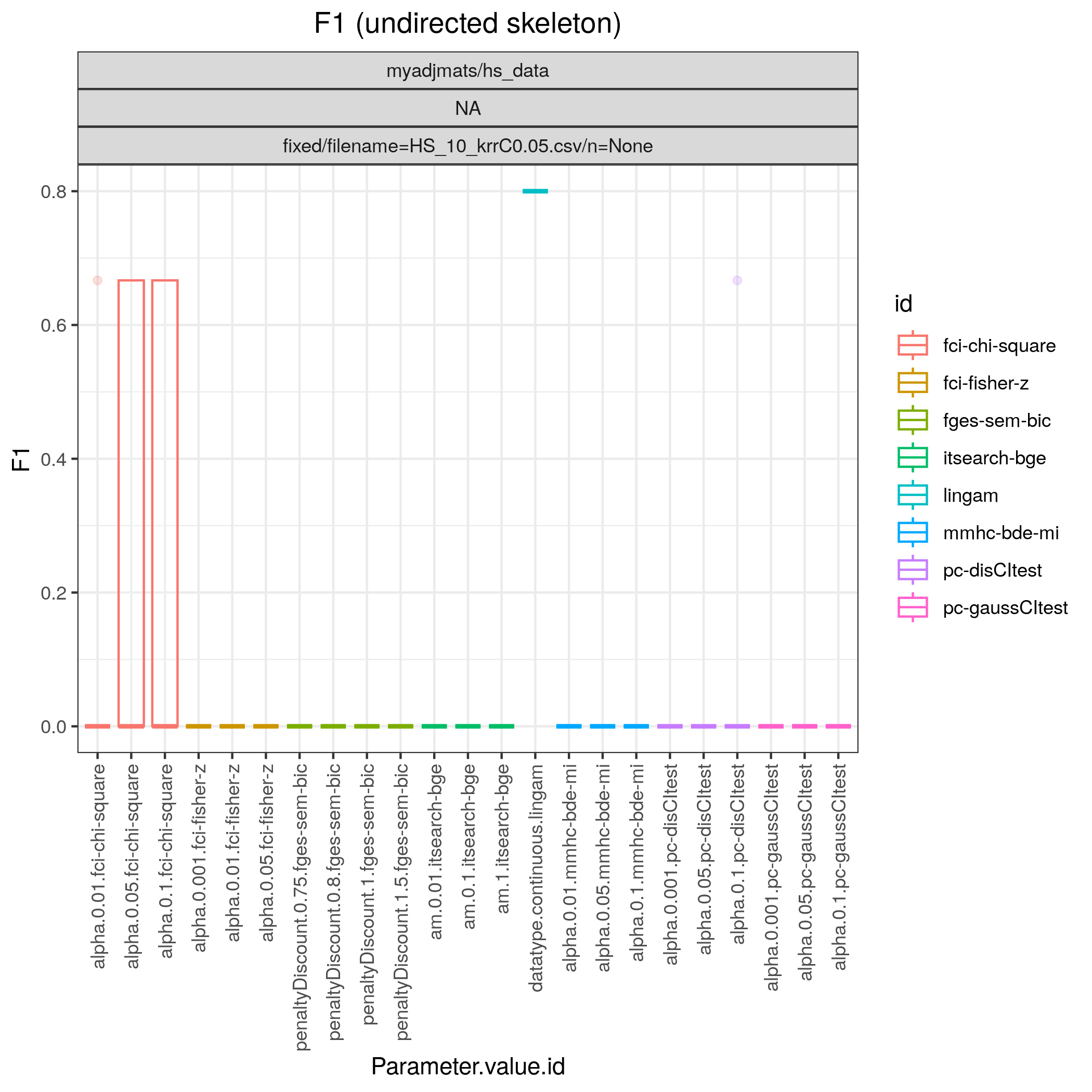}
	\caption{Human Stature data, $k$-RR Comb mechanism, max probability 0.05.}
\end{minipage}
\begin{minipage}{0.31\linewidth}
\centering
  \includegraphics[scale=0.34]{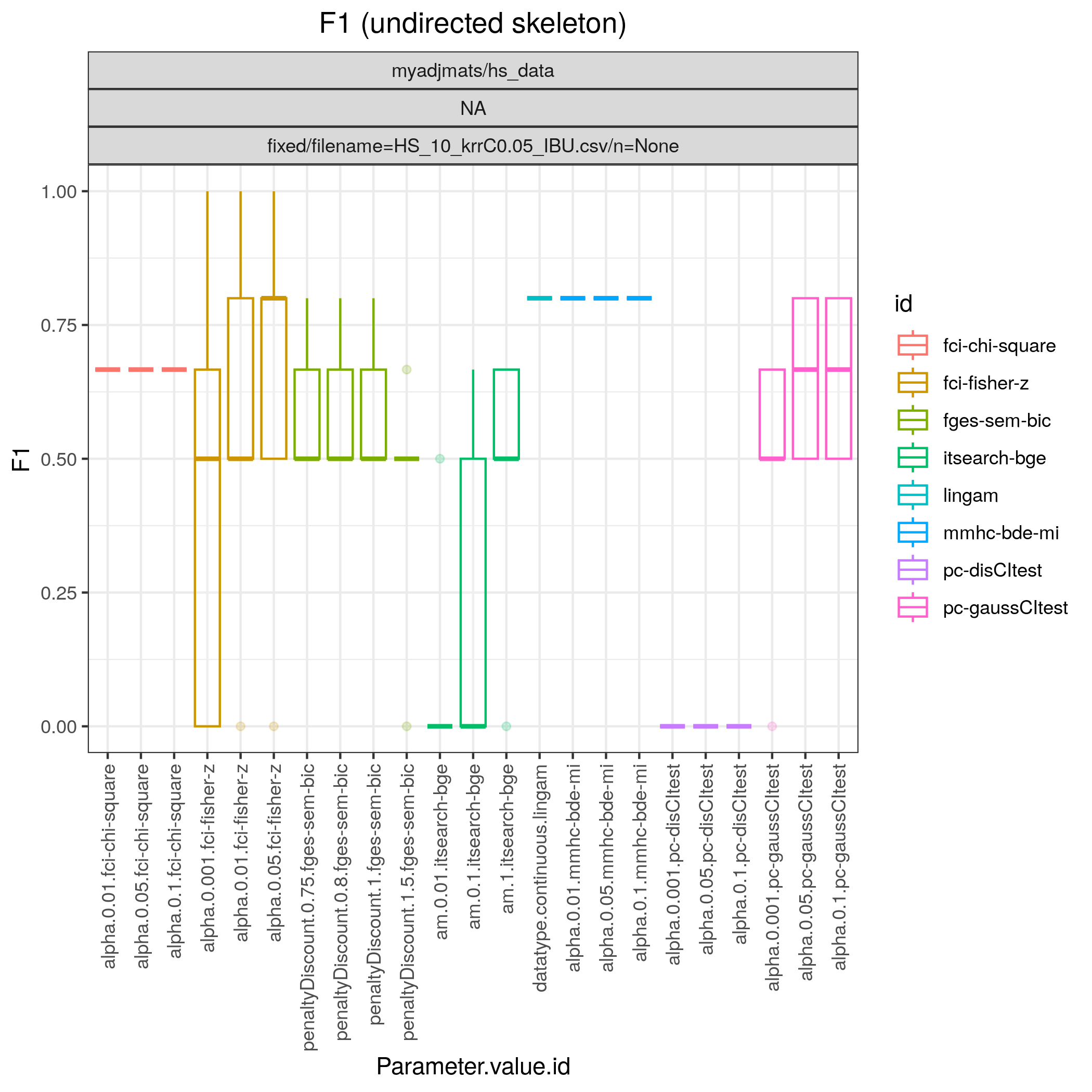}
	\caption{Human Stature data, $k$-RR Comb IBU mechanism, max probability 0.05.}
\end{minipage}
\end{figure}


\noindent
\begin{figure}[H]
\begin{minipage}{0.31\linewidth}
\centering
		\includegraphics[scale=0.34]{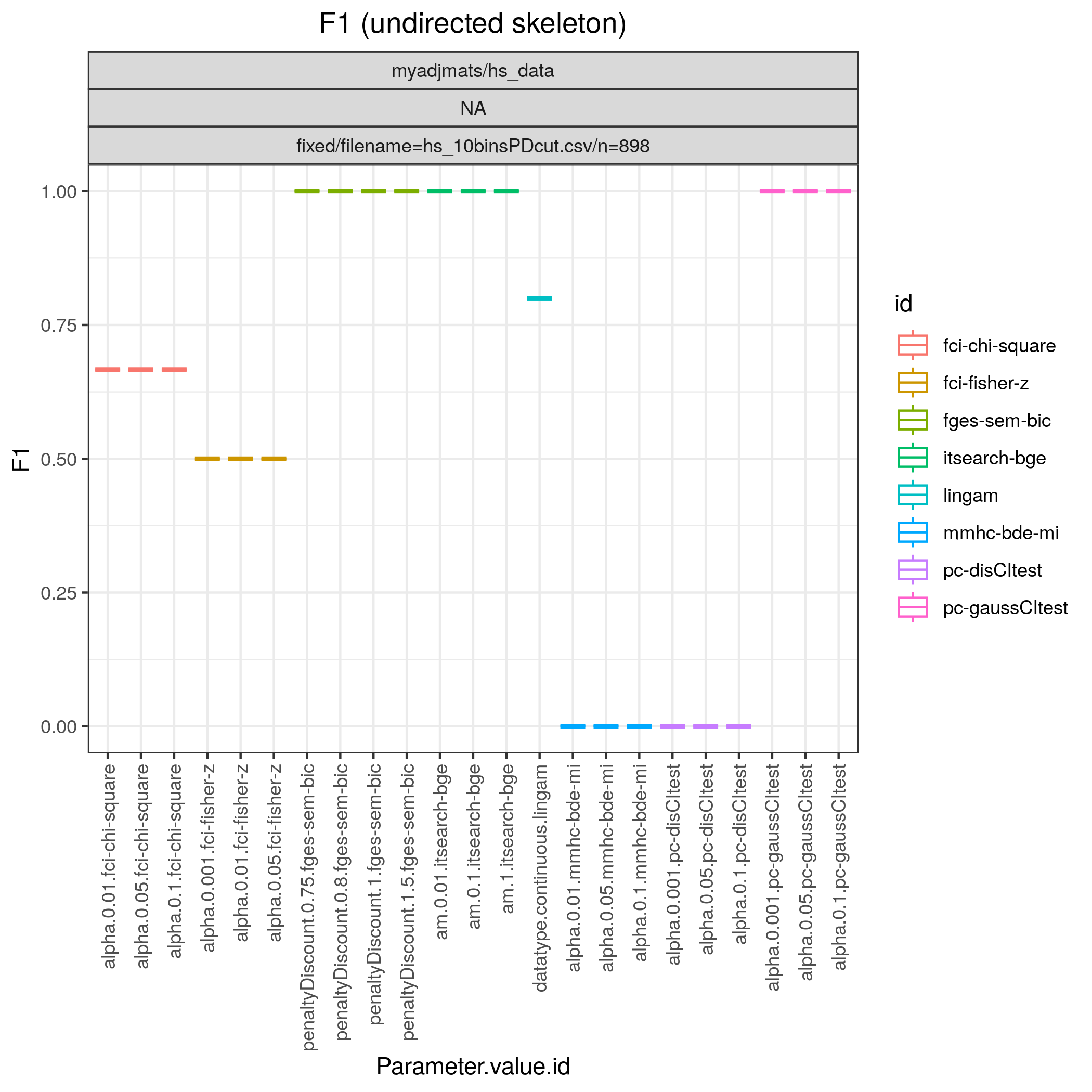}
	\caption{F1 Scores on the Human Stature data set. Discretized, no noise.}
\end{minipage}
\begin{minipage}{0.31\linewidth}
\centering
		\includegraphics[scale=0.34]{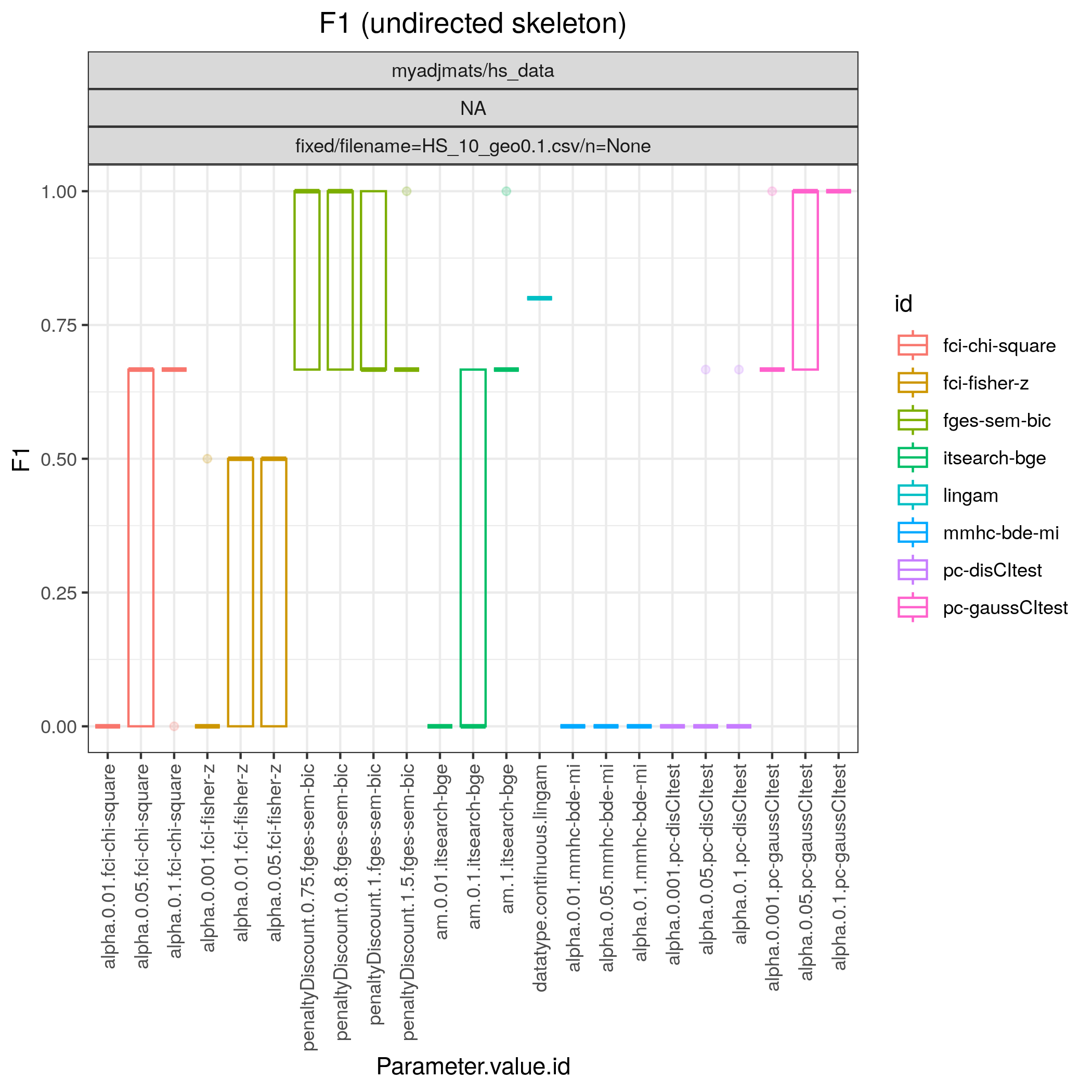}
	\caption{Human Stature data, Geo C-wise mechanism, max probability 0.1.}
\end{minipage}
\begin{minipage}{0.31\linewidth}
\centering
  \includegraphics[scale=0.34]{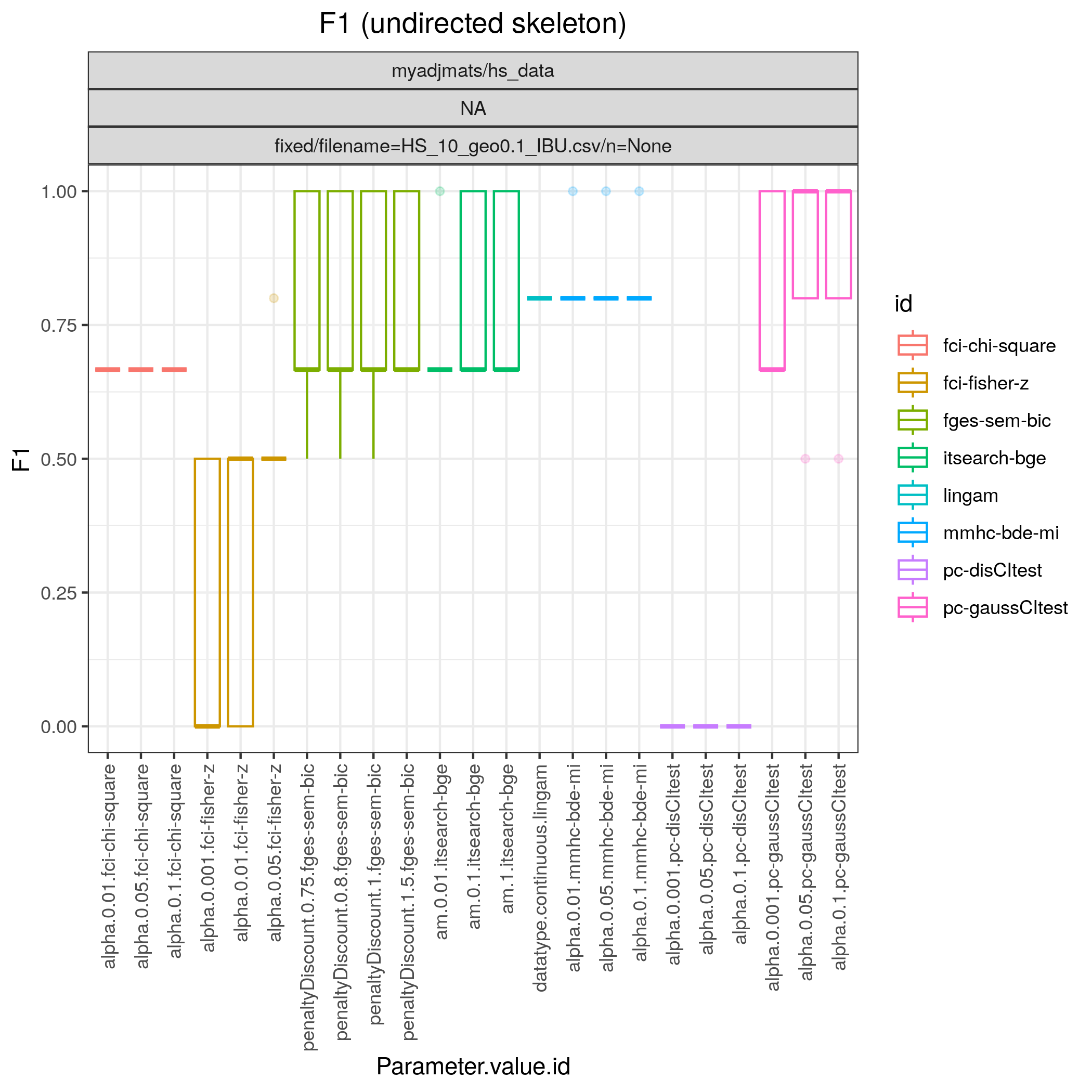}
	\caption{Human Stature data, Geo C-wise IBU mechanism, max probability 0.1.}
 \end{minipage}
\end{figure}

\begin{figure}[H]
    \centering
   \begin{minipage}{0.31\linewidth}
\centering
  \includegraphics[scale=0.34]{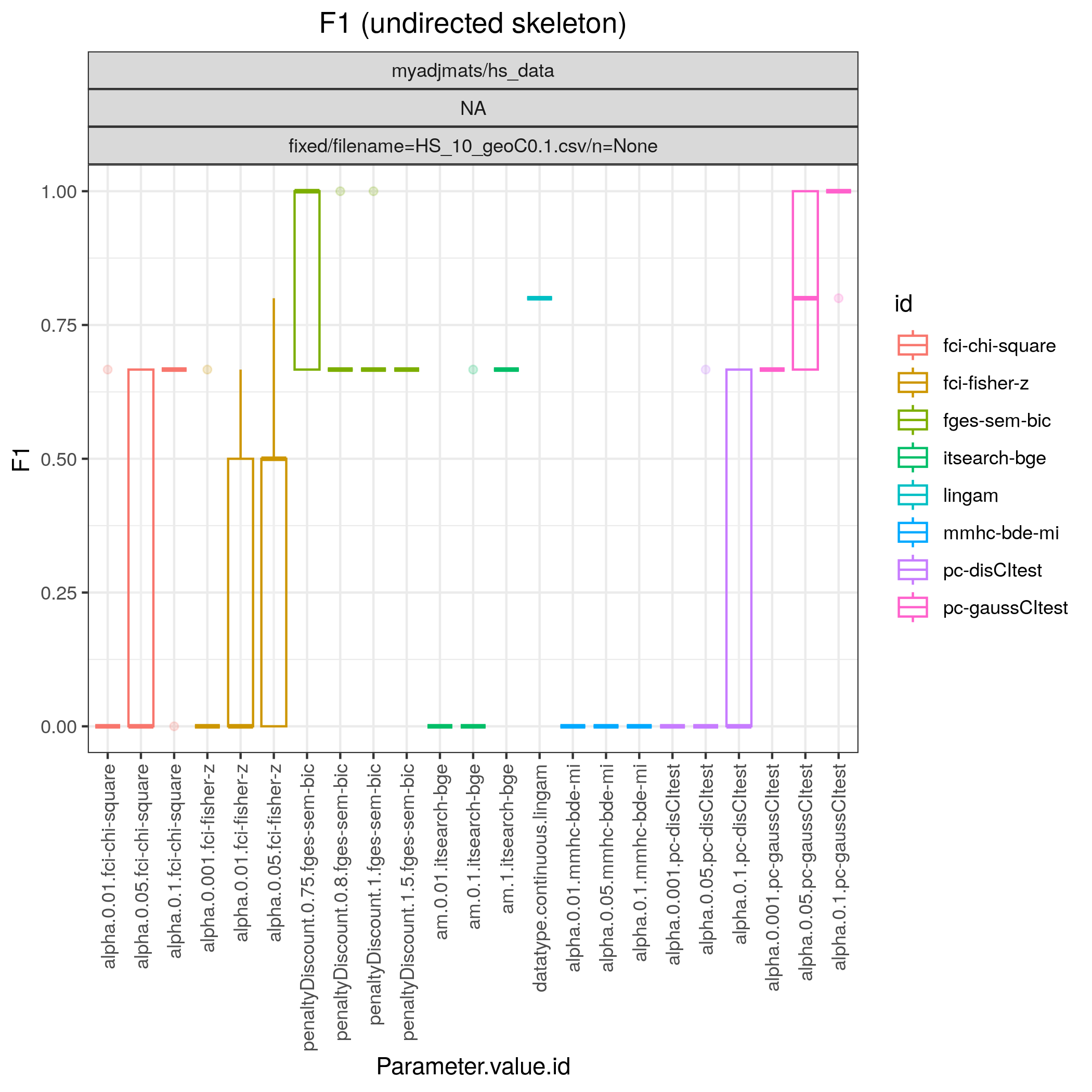}
	\caption{Human Stature data, Geo Comb mechanism, max probability 0.1.}
\end{minipage}
\begin{minipage}{0.31\linewidth}
\centering
  \includegraphics[scale=0.34]{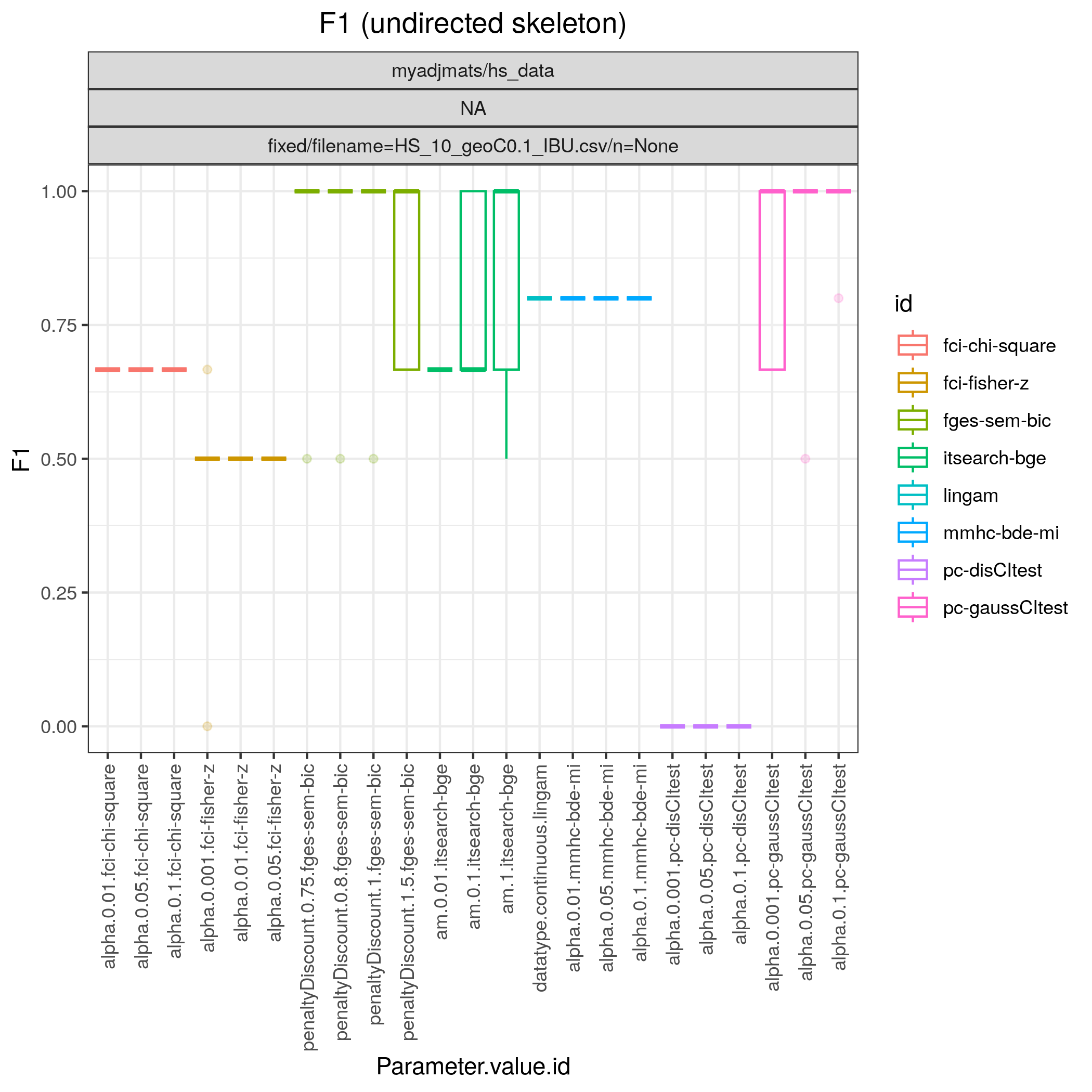}
	\caption{Human Stature data, Geo Comb IBU mechanism, max probability 0.1.}
\end{minipage}
\begin{minipage}{0.31\linewidth}
\centering
  \includegraphics[scale=0.34]{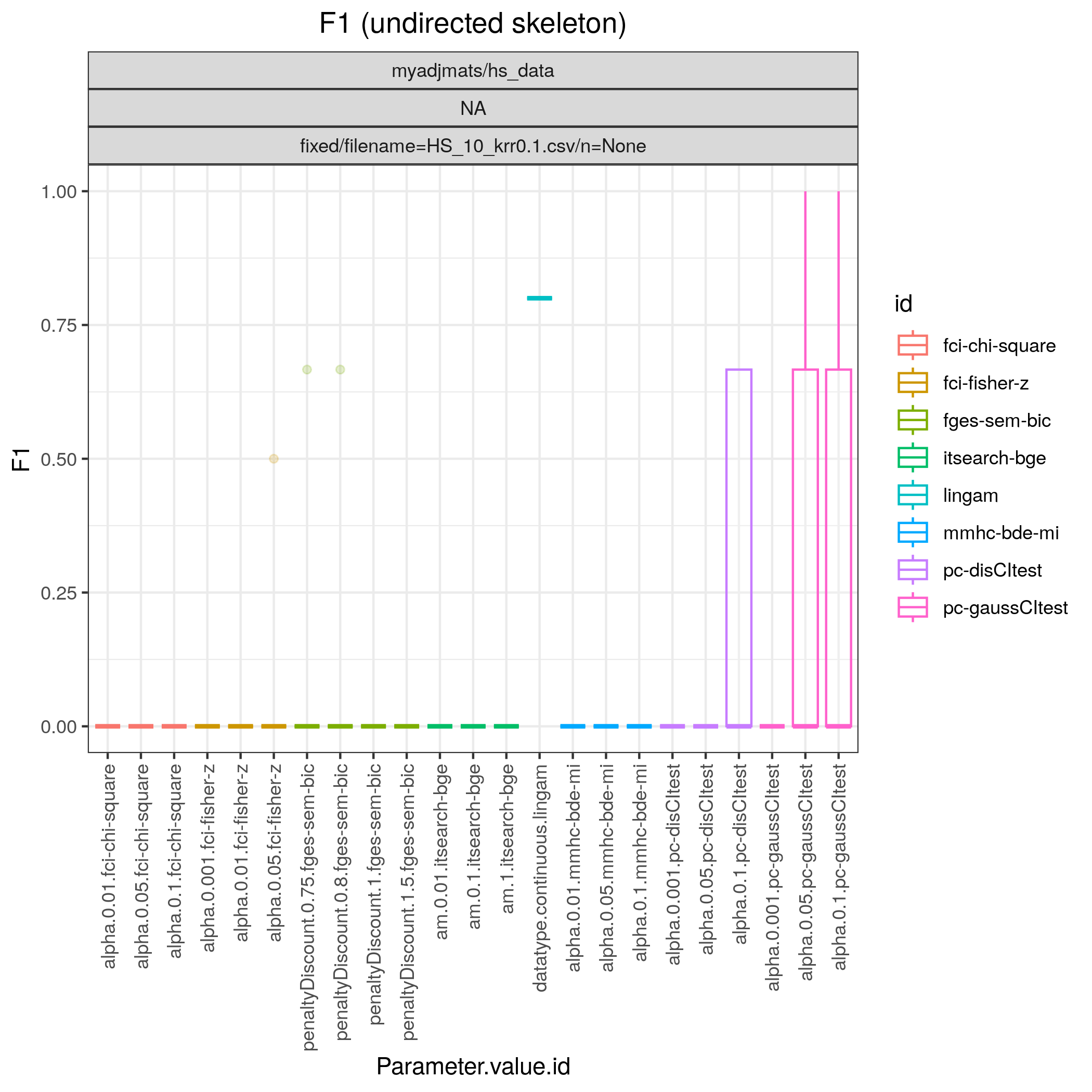}
	\caption{Human Stature data, $k$-RR C-wise mechanism, max probability 0.1.}
\end{minipage}
\end{figure}

\begin{figure}[H]
    \centering
   \begin{minipage}{0.31\linewidth}
\centering
  \includegraphics[scale=0.34]{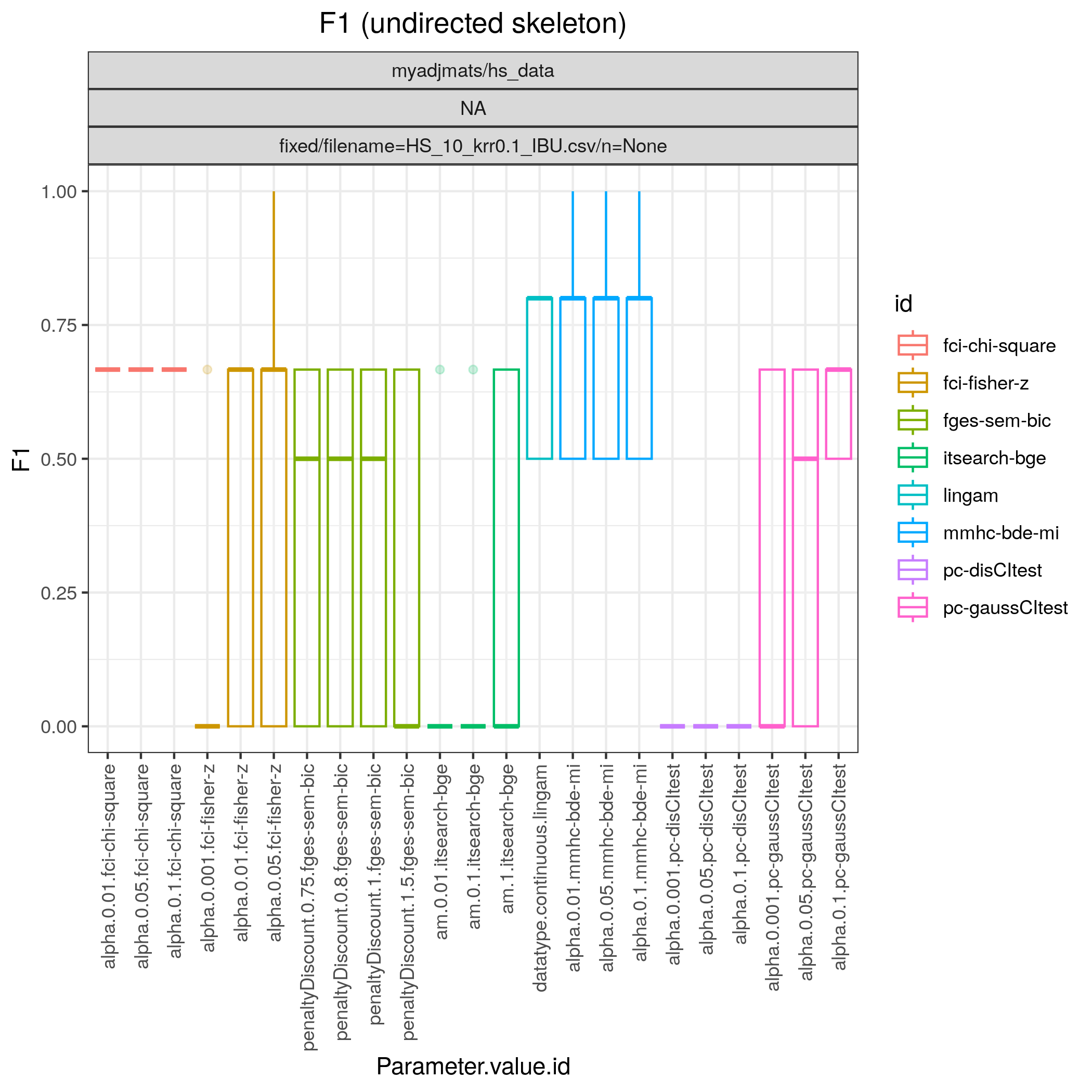}
	\caption{Human Stature data, $k$-RR C-wise IBU mechanism, max probability 0.1.}
\end{minipage}
\begin{minipage}{0.31\linewidth}
\centering
  \includegraphics[scale=0.34]{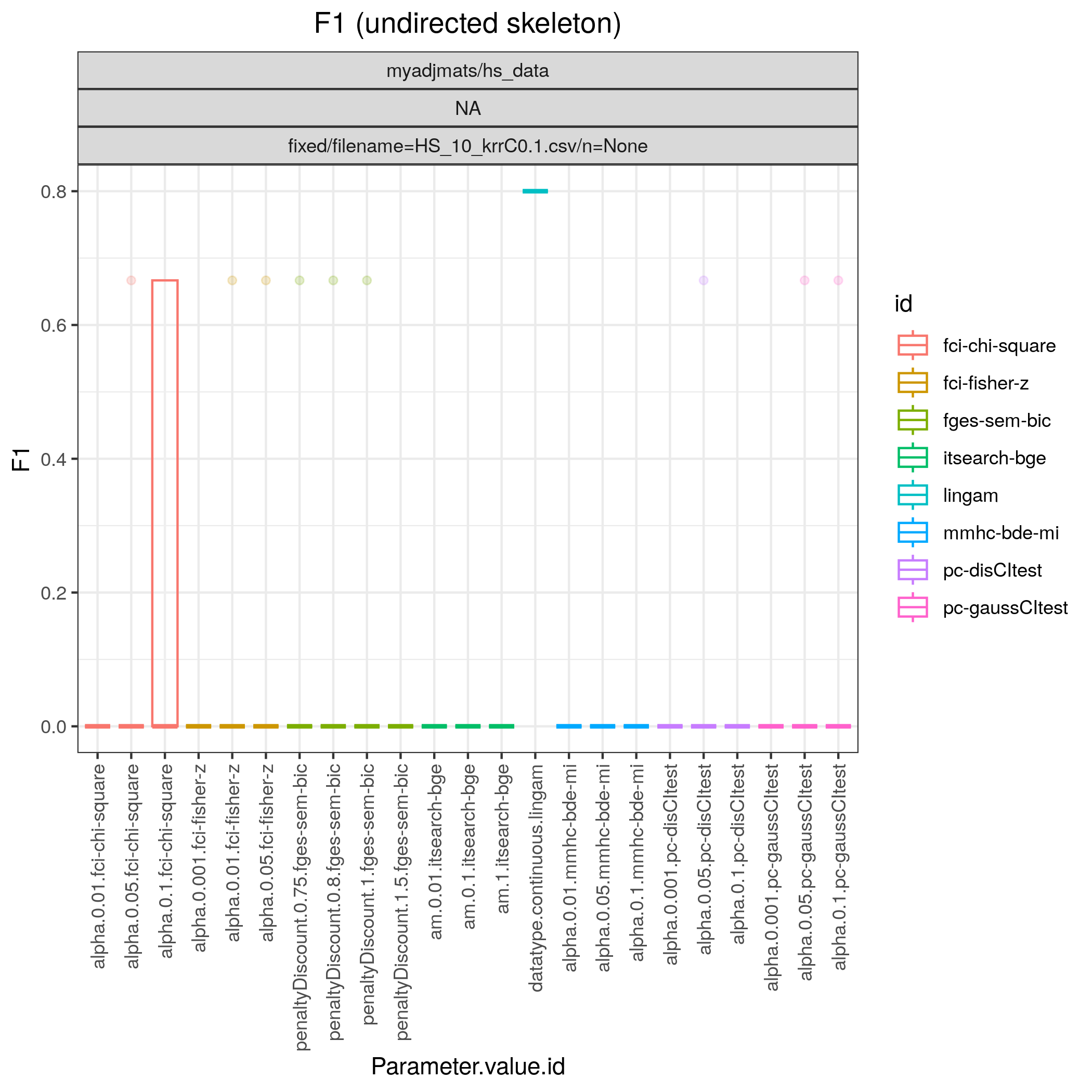}
	\caption{Human Stature data, $k$-RR Comb mechanism, max probability 0.1.}
\end{minipage}
\begin{minipage}{0.31\linewidth}
\centering
  \includegraphics[scale=0.34]{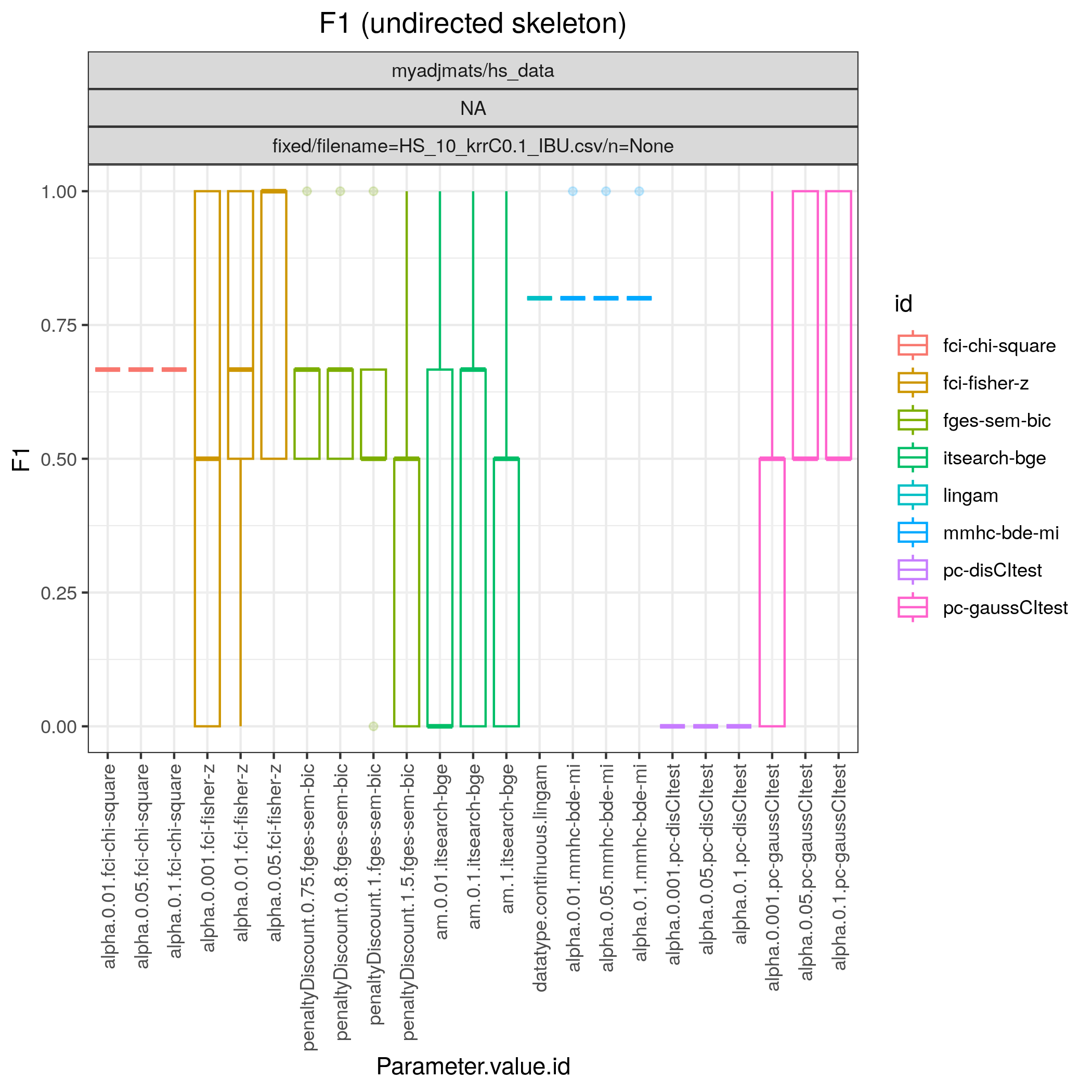}
	\caption{Human Stature data, $k$-RR Comb IBU mechanism, max probability 0.1.}
\end{minipage}
\end{figure}


\noindent
\begin{figure}[H]
\begin{minipage}{0.31\linewidth}
\centering
		\includegraphics[scale=0.34]{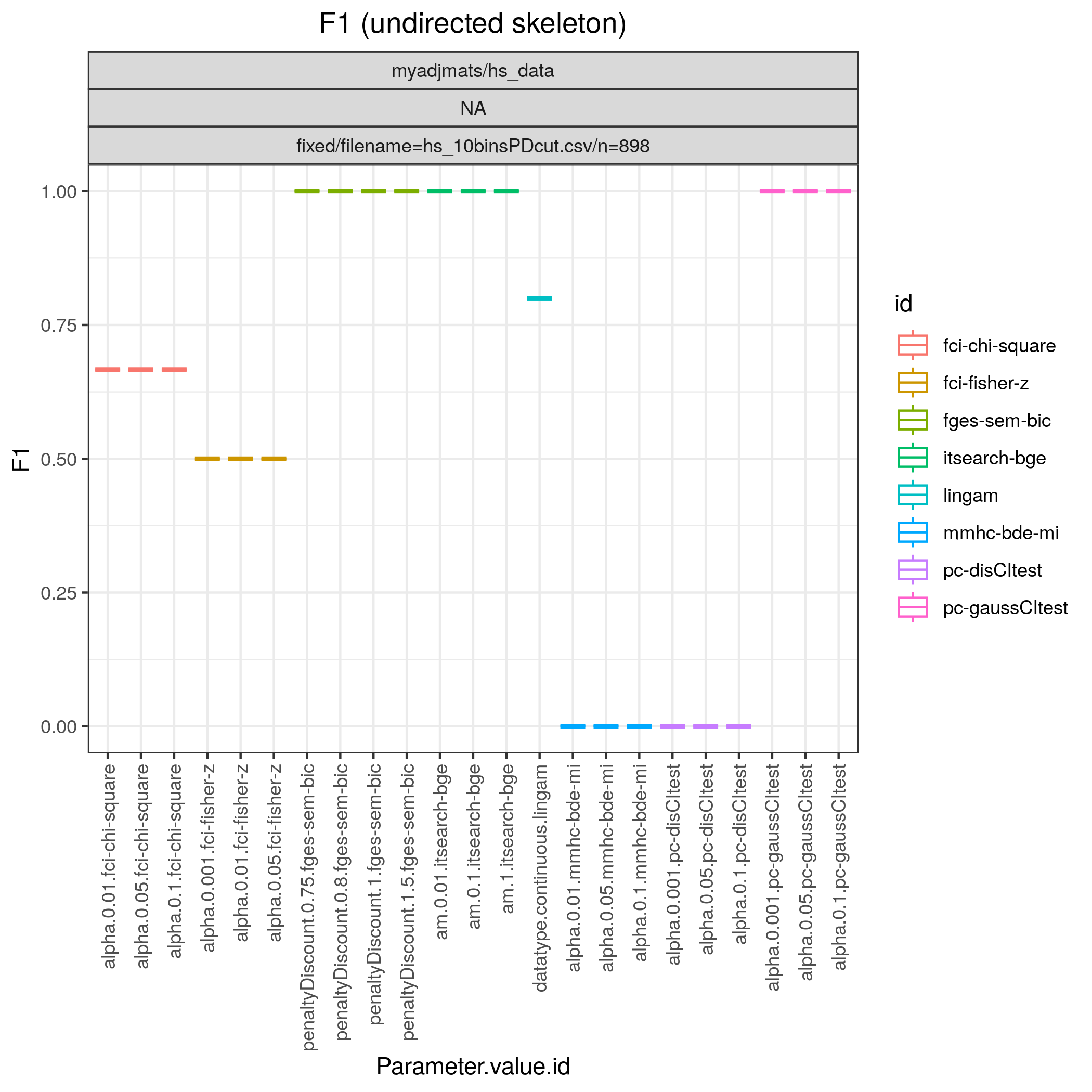}
	\caption{F1 Scores on the Human Stature data set. Discretized, no noise.}
\end{minipage}
\begin{minipage}{0.31\linewidth}
\centering
		\includegraphics[scale=0.34]{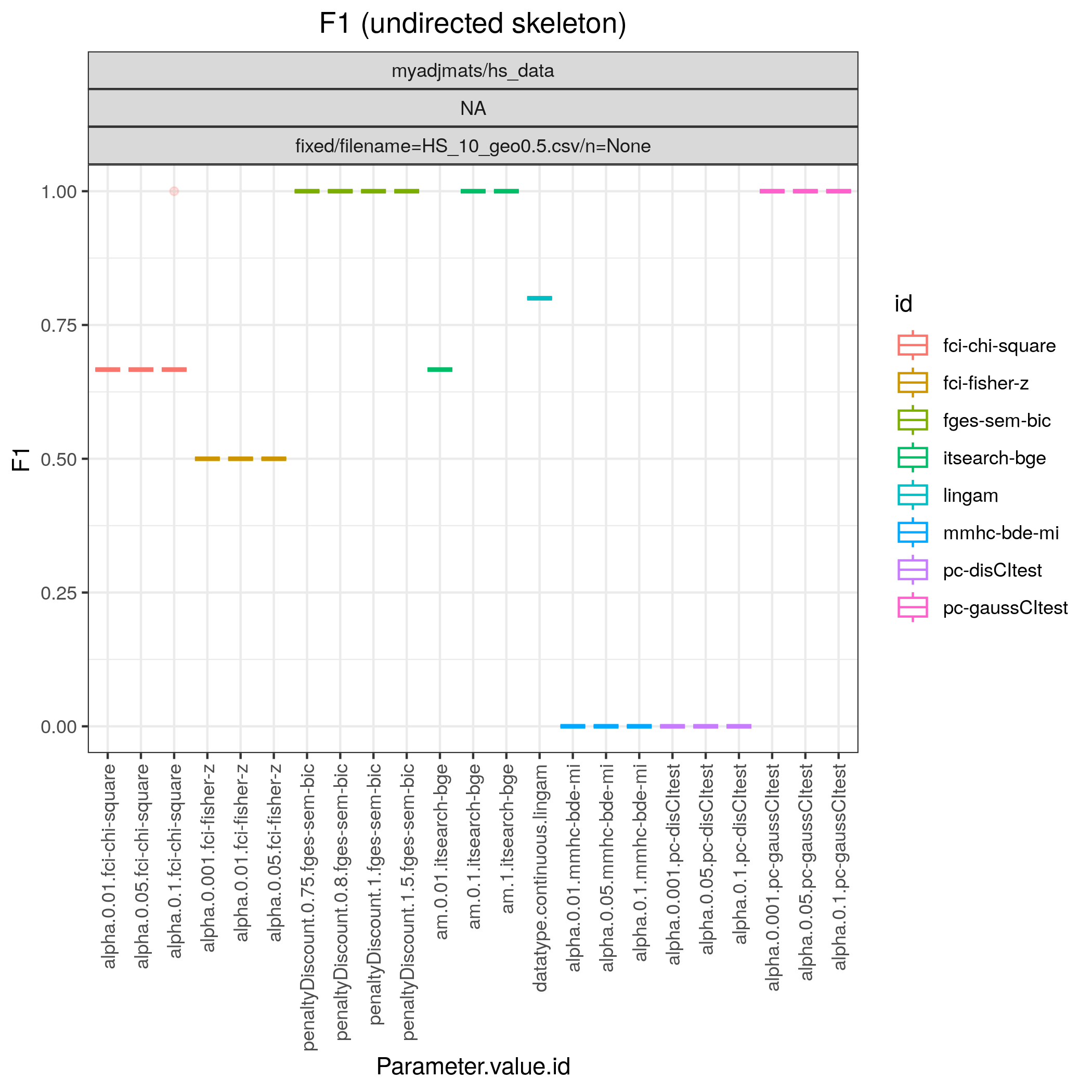}
	\caption{Human Stature data, Geo C-wise mechanism, max probability 0.5.}
\end{minipage}
\begin{minipage}{0.31\linewidth}
\centering
  \includegraphics[scale=0.34]{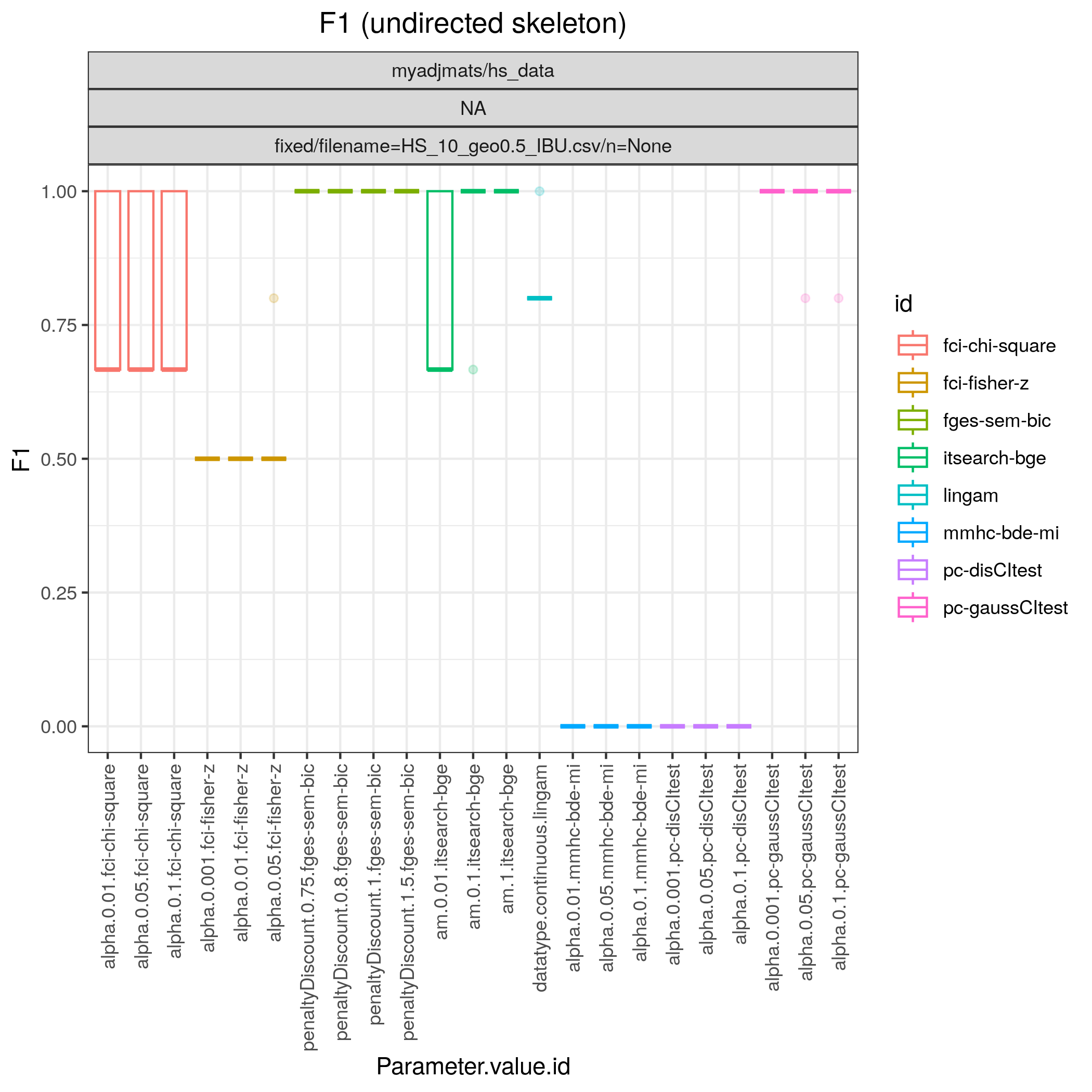}
	\caption{Human Stature data, Geo C-wise IBU mechanism, max probability 0.5.}
 \end{minipage}
\end{figure}

\begin{figure}[H]
    \centering
   \begin{minipage}{0.31\linewidth}
\centering
  \includegraphics[scale=0.34]{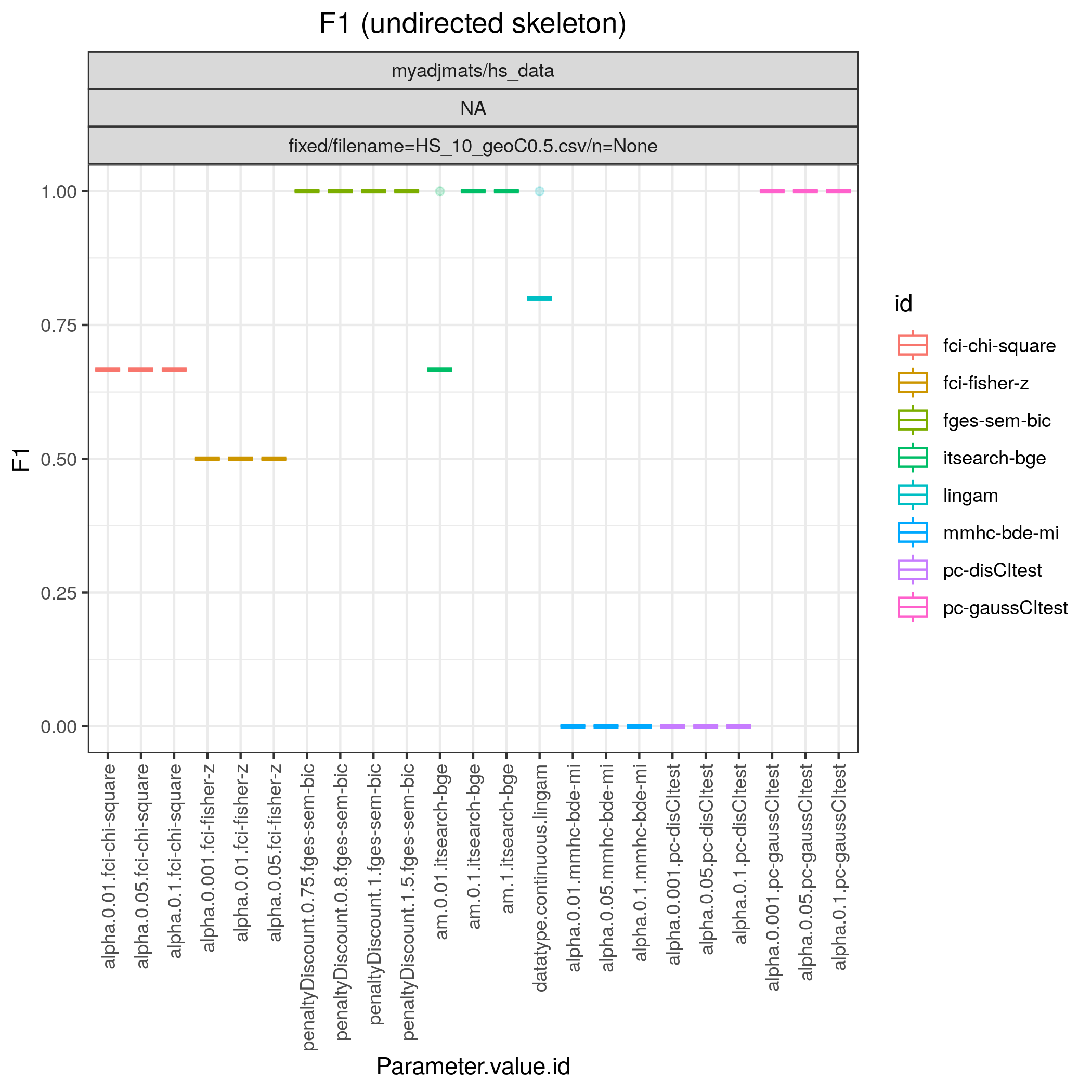}
	\caption{Human Stature data, Geo Comb mechanism, max probability 0.5.}
\end{minipage}
\begin{minipage}{0.31\linewidth}
\centering
  \includegraphics[scale=0.34]{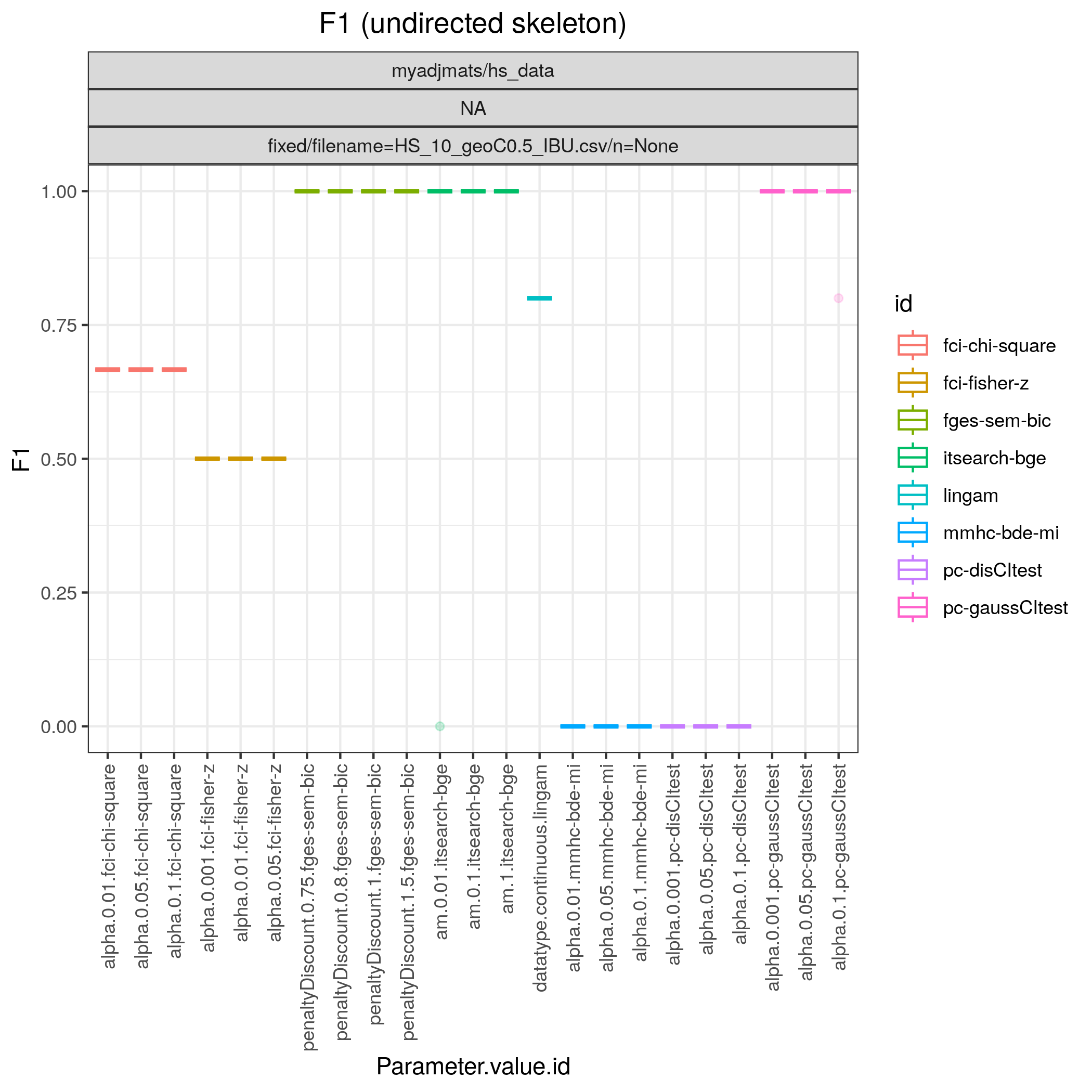}
	\caption{Human Stature data, Geo Comb IBU mechanism, max probability 0.5.}
\end{minipage}
\begin{minipage}{0.31\linewidth}
\centering
  \includegraphics[scale=0.34]{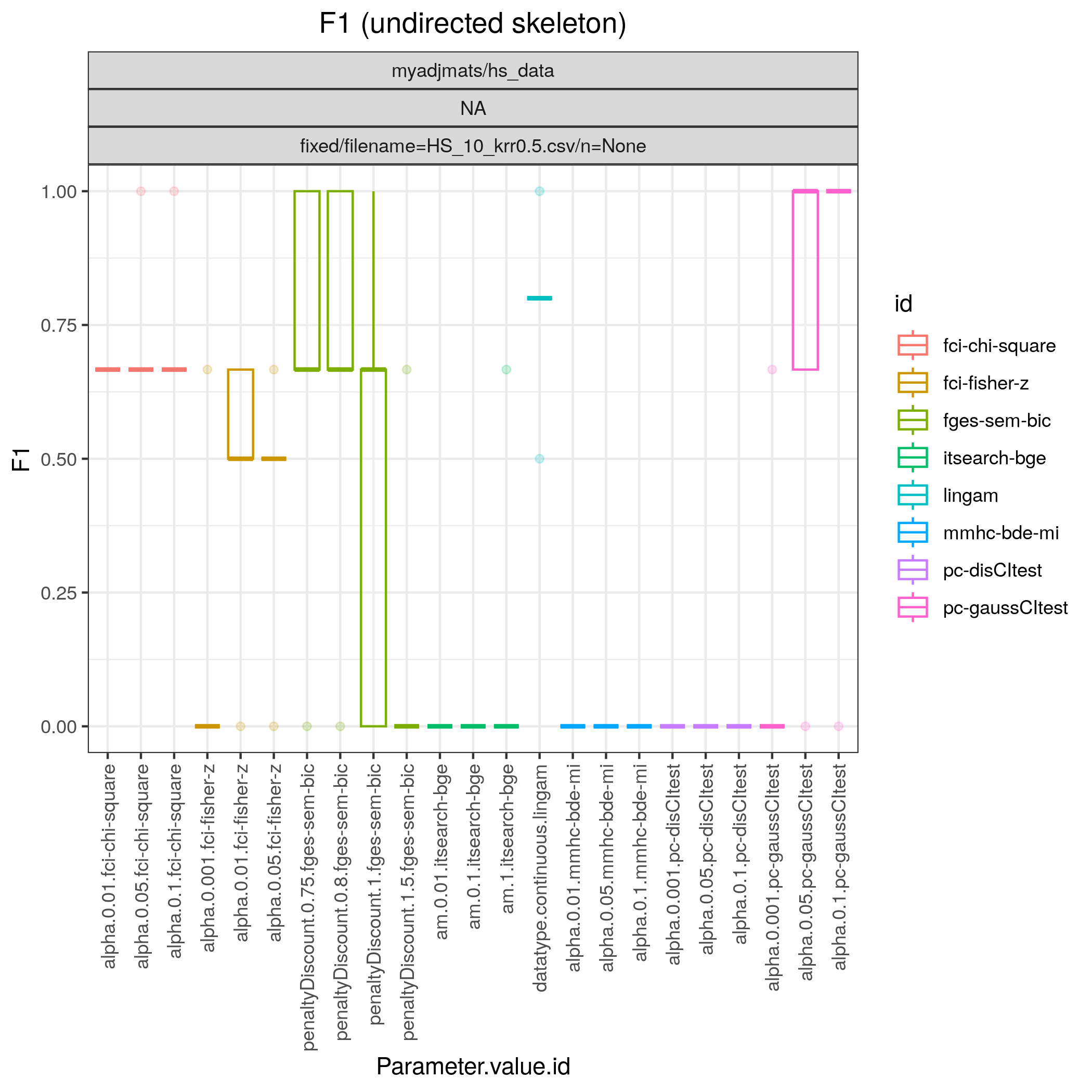}
	\caption{Human Stature data, $k$-RR C-wise mechanism, max probability 0.5.}
\end{minipage}
\end{figure}

\begin{figure}[H]
    \centering
   \begin{minipage}{0.31\linewidth}
\centering
  \includegraphics[scale=0.34]{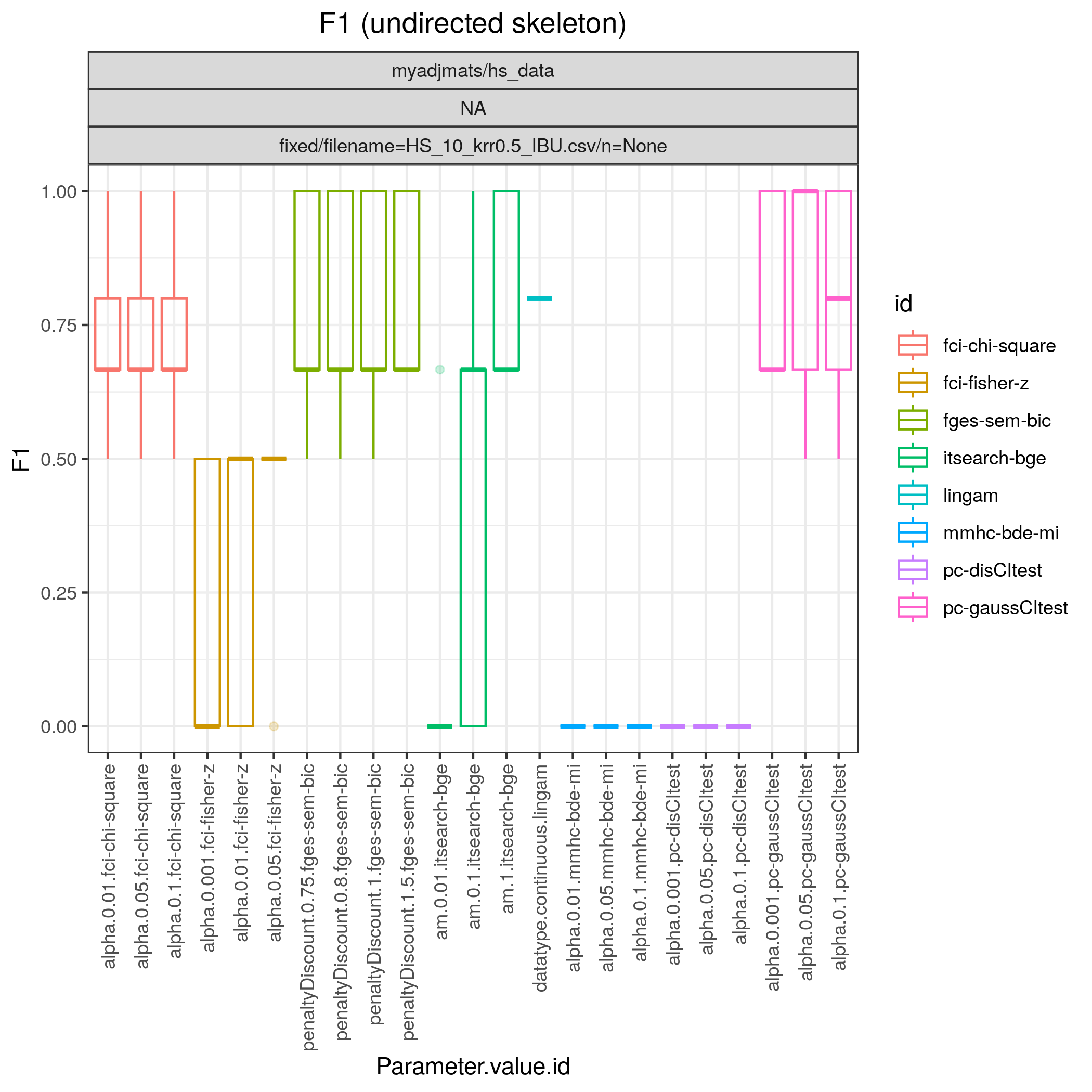}
	\caption{Human Stature data, $k$-RR C-wise IBU mechanism, max probability 0.5.}
\end{minipage}
\begin{minipage}{0.31\linewidth}
\centering
  \includegraphics[scale=0.34]{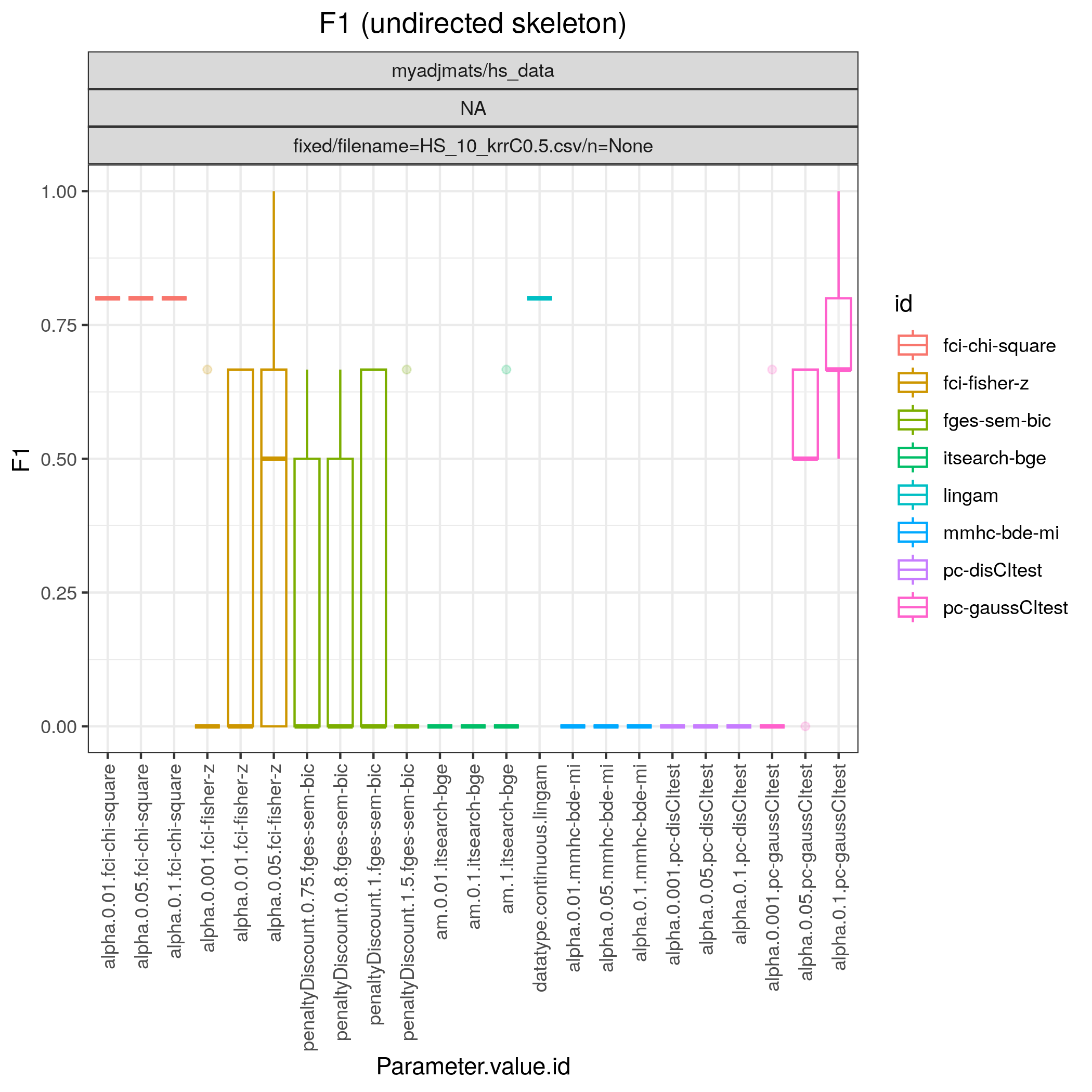}
	\caption{Human Stature data, $k$-RR Comb mechanism, max probability 0.5.}
\end{minipage}
\begin{minipage}{0.31\linewidth}
\centering
  \includegraphics[scale=0.34]{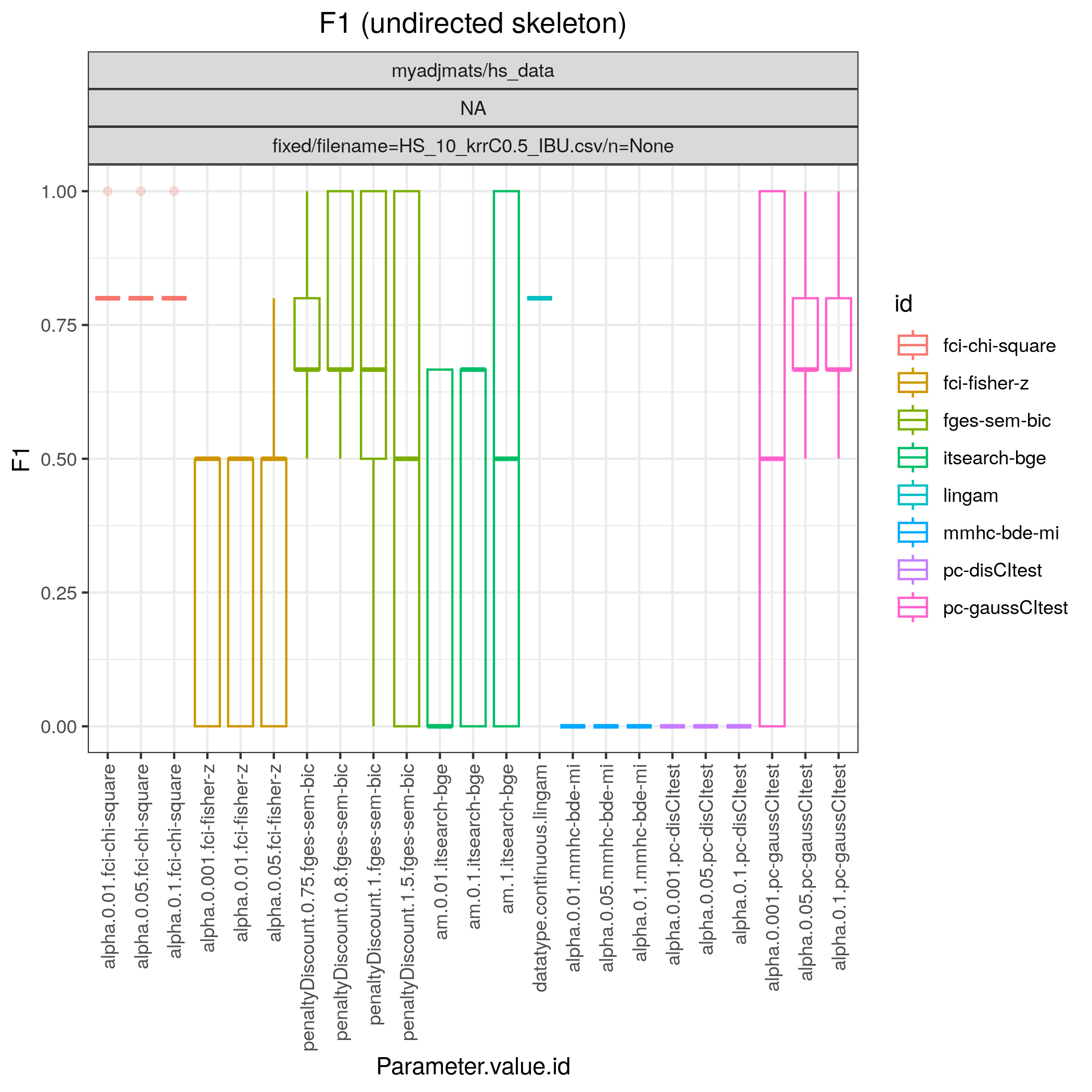}
	\caption{Human Stature data, $k$-RR Comb IBU mechanism, max probability 0.5.}
\end{minipage}
\end{figure}

 \subsection{SHD Score results Human Stature data set}

\noindent
\begin{figure}[H]
\begin{minipage}{0.31\linewidth}
\centering
		\includegraphics[scale=0.34]{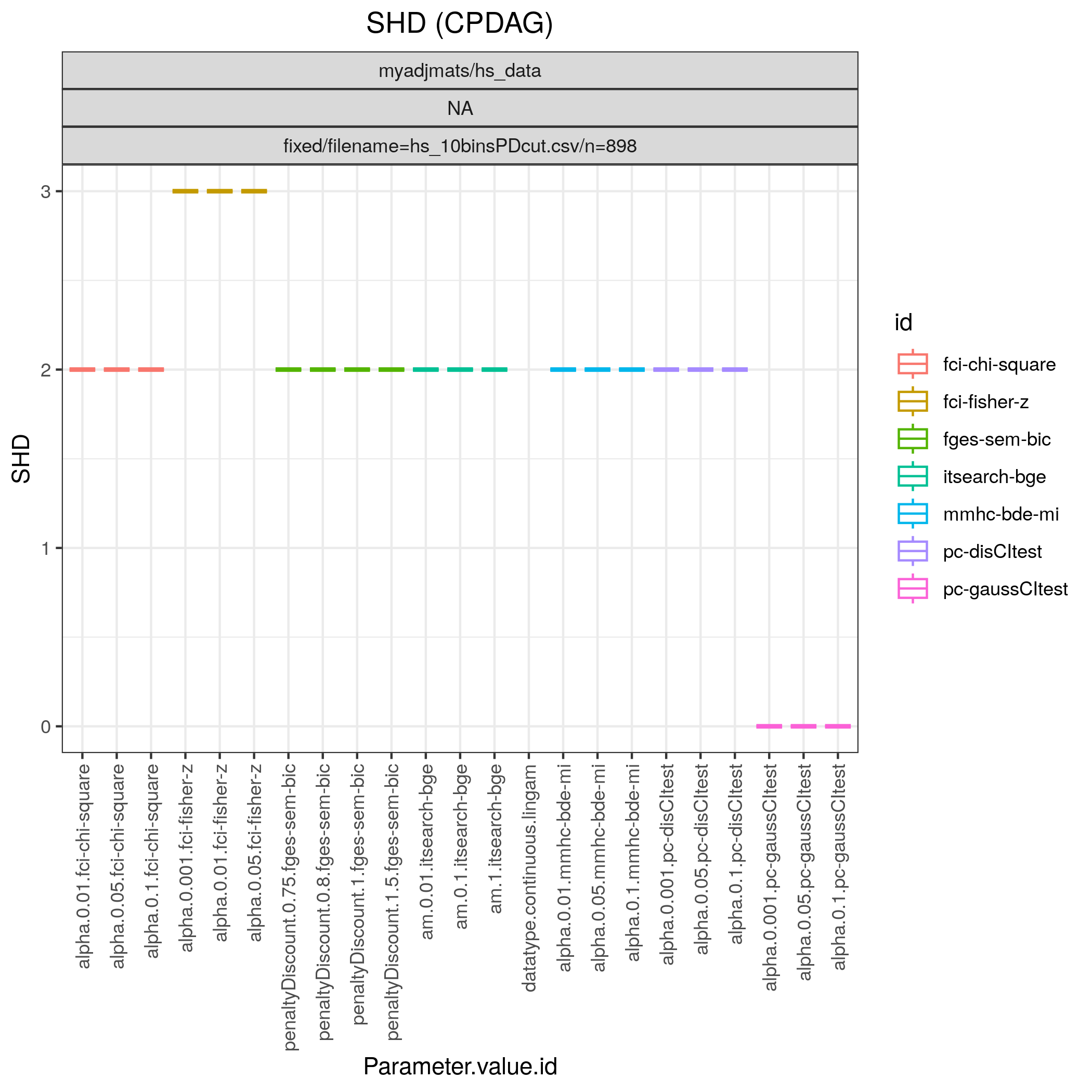}
	\caption{SHD Scores on the Human Stature data set. Discretized, no noise.}
\end{minipage}
\begin{minipage}{0.31\linewidth}
\centering
		\includegraphics[scale=0.34]{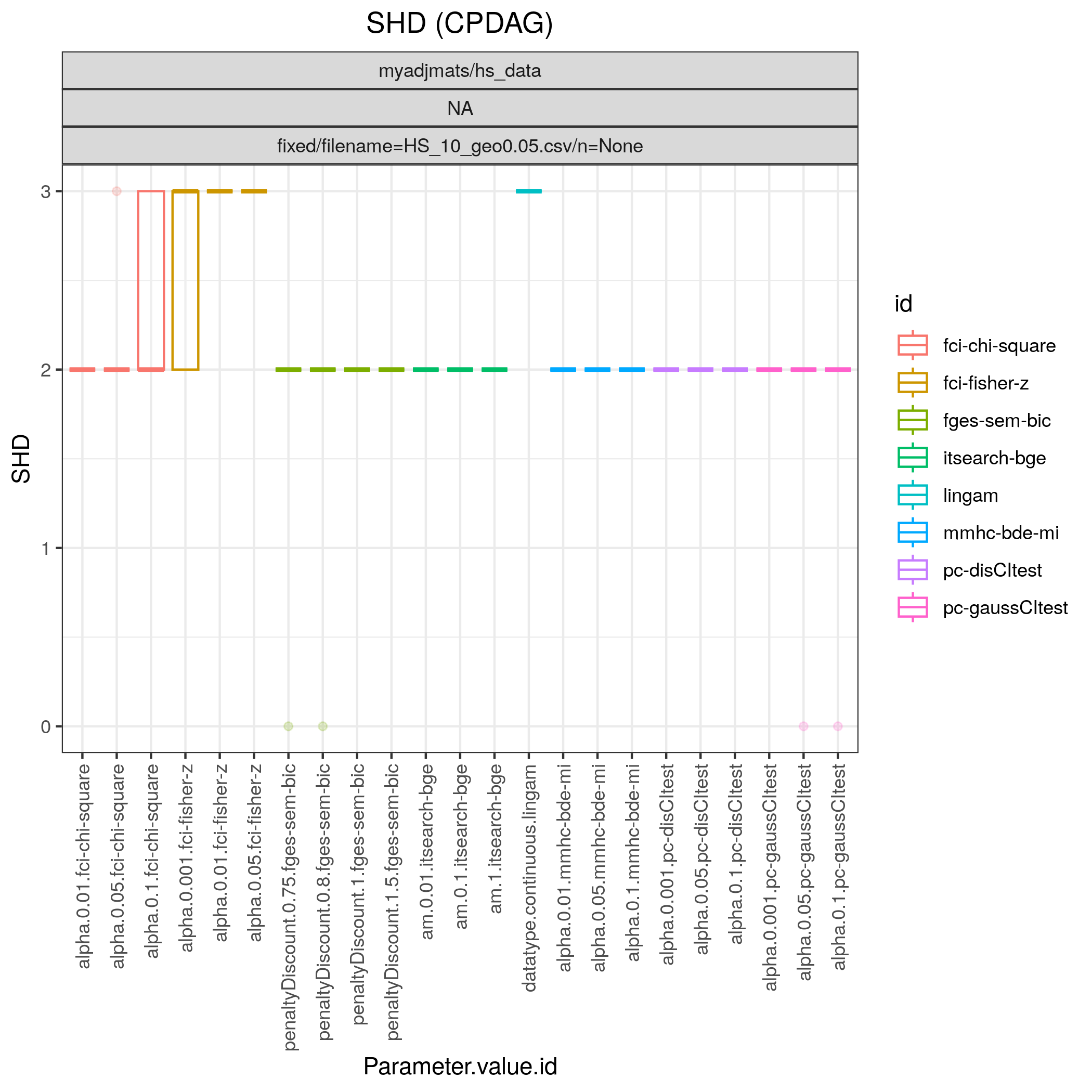}
	\caption{Human Stature data, Geo C-wise mechanism, max probability 0.05.}
\end{minipage}
\begin{minipage}{0.31\linewidth}
\centering
  \includegraphics[scale=0.34]{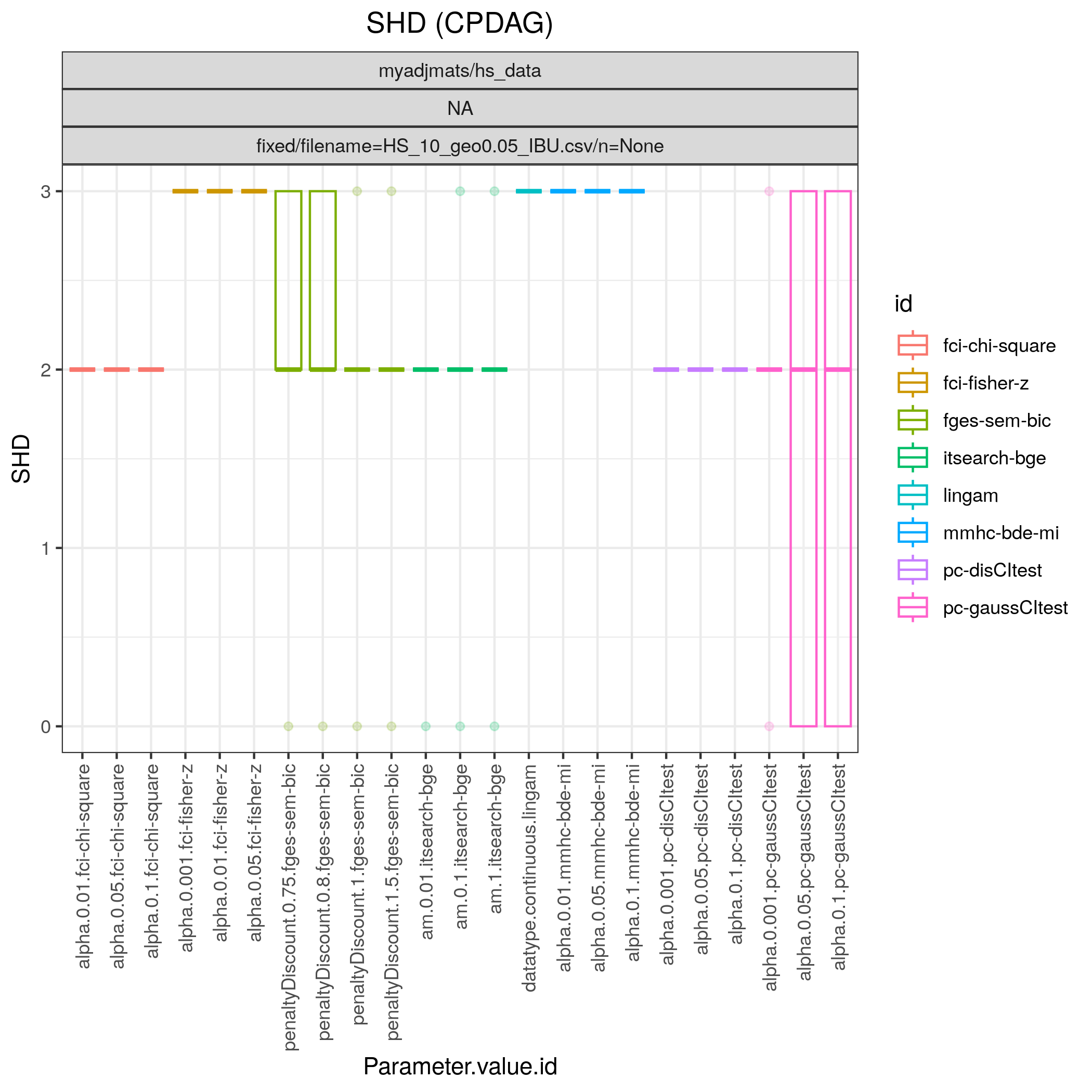}
	\caption{Human Stature data, Geo C-wise IBU mechanism, max probability 0.05.}
 \end{minipage}
\begin{minipage}{0.31\linewidth}
\centering
  \includegraphics[scale=0.34]{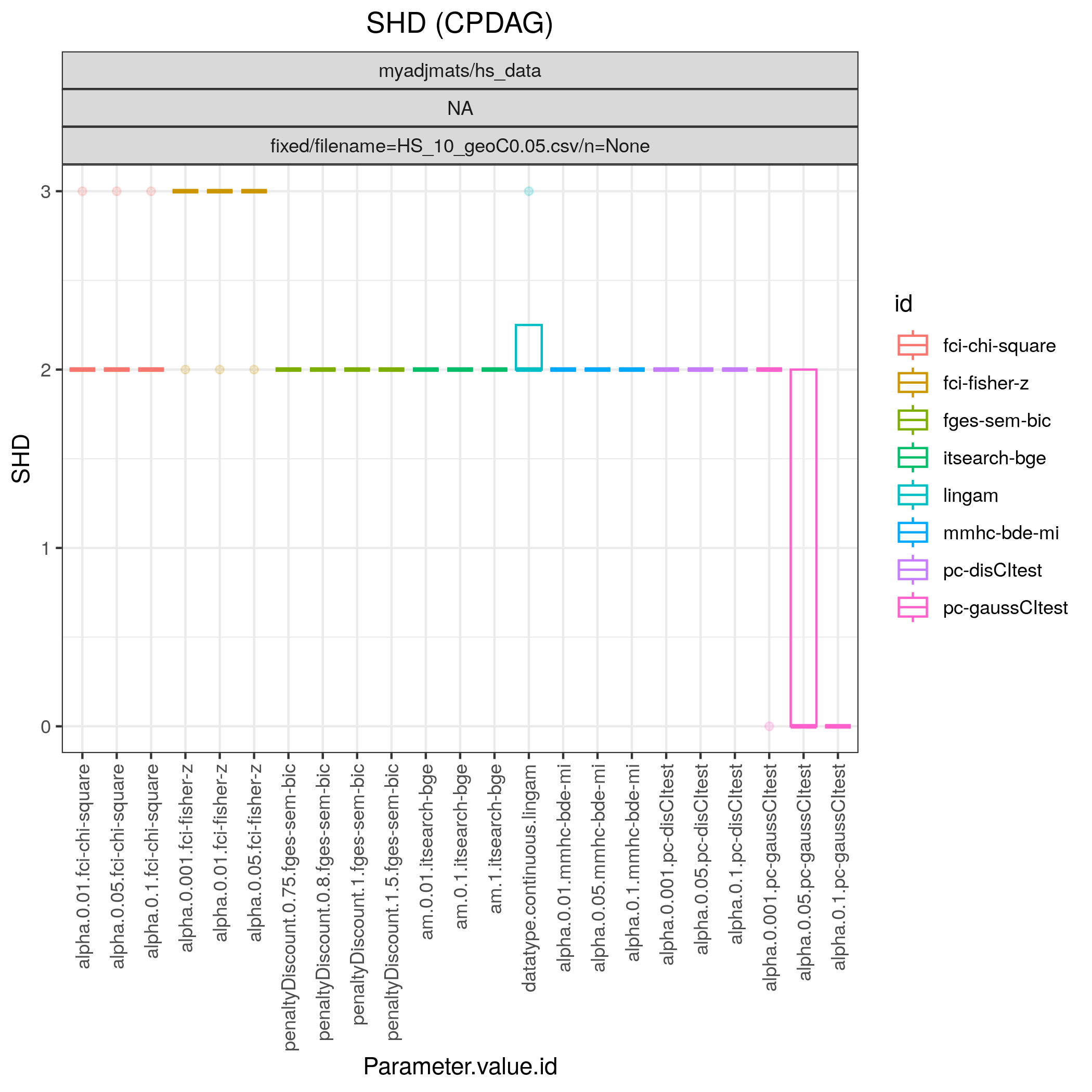}
	\caption{Human Stature data, Geo Comb mechanism, max probability 0.05.}
\end{minipage}
\begin{minipage}{0.31\linewidth}
\centering
  \includegraphics[scale=0.34]{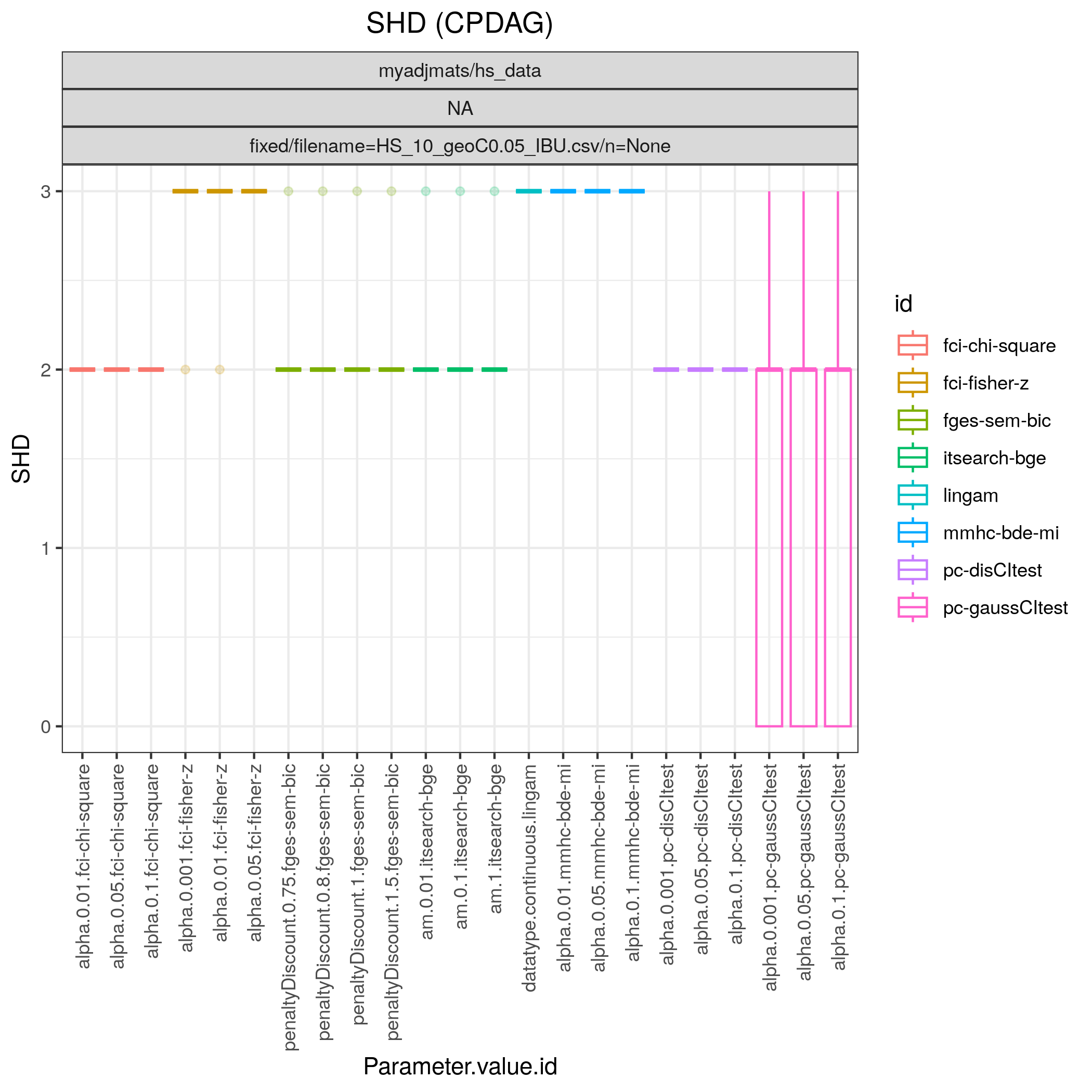}
	\caption{Human Stature data, Geo Comb IBU mechanism, max probability 0.05.}
\end{minipage}
\begin{minipage}{0.31\linewidth}
\centering
  \includegraphics[scale=0.34]{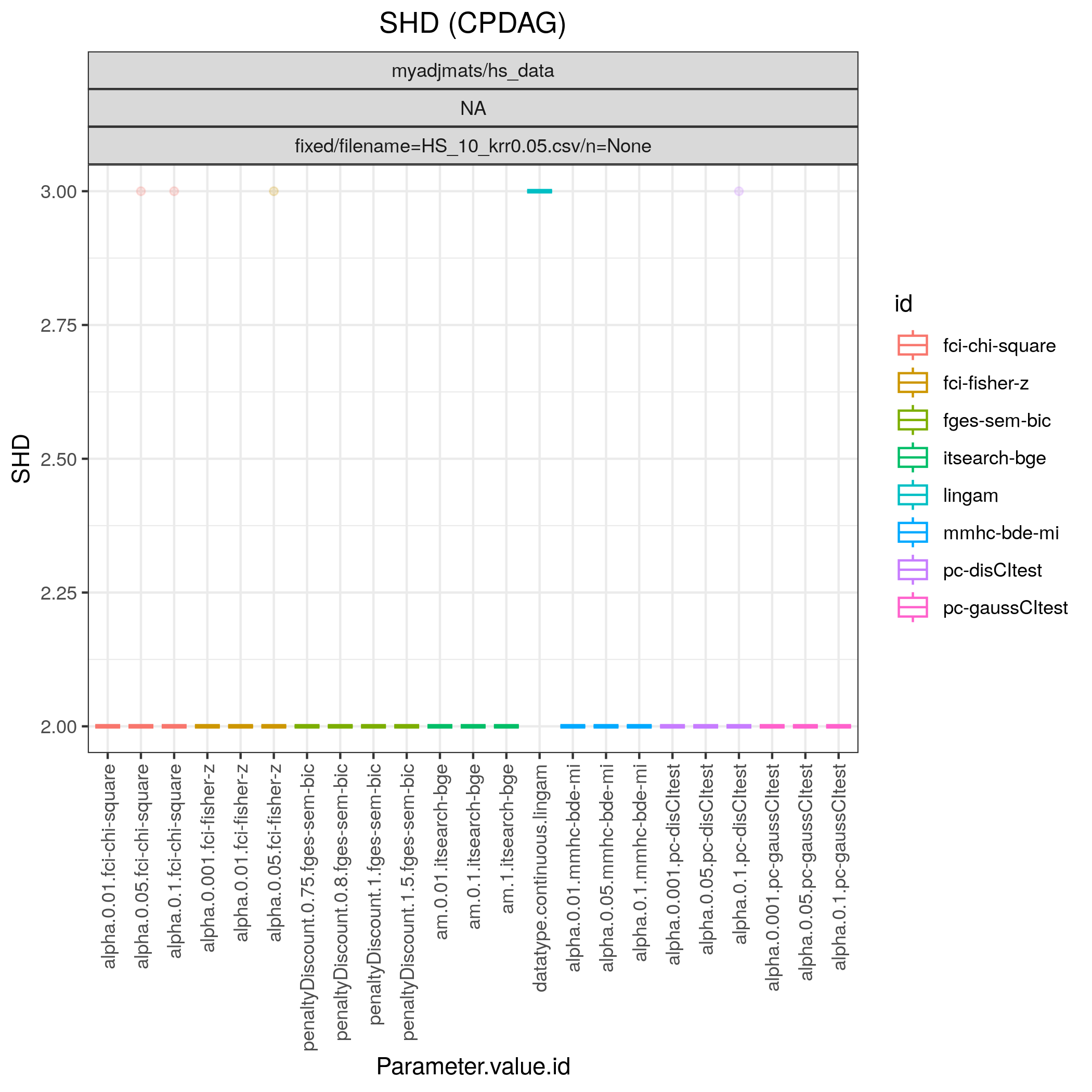}
	\caption{Human Stature data, $k$-RR C-wise mechanism, max probability 0.05.}
\end{minipage}

\end{figure}

\begin{figure}[H]
    \centering
   \begin{minipage}{0.31\linewidth}
\centering
  \includegraphics[scale=0.34]{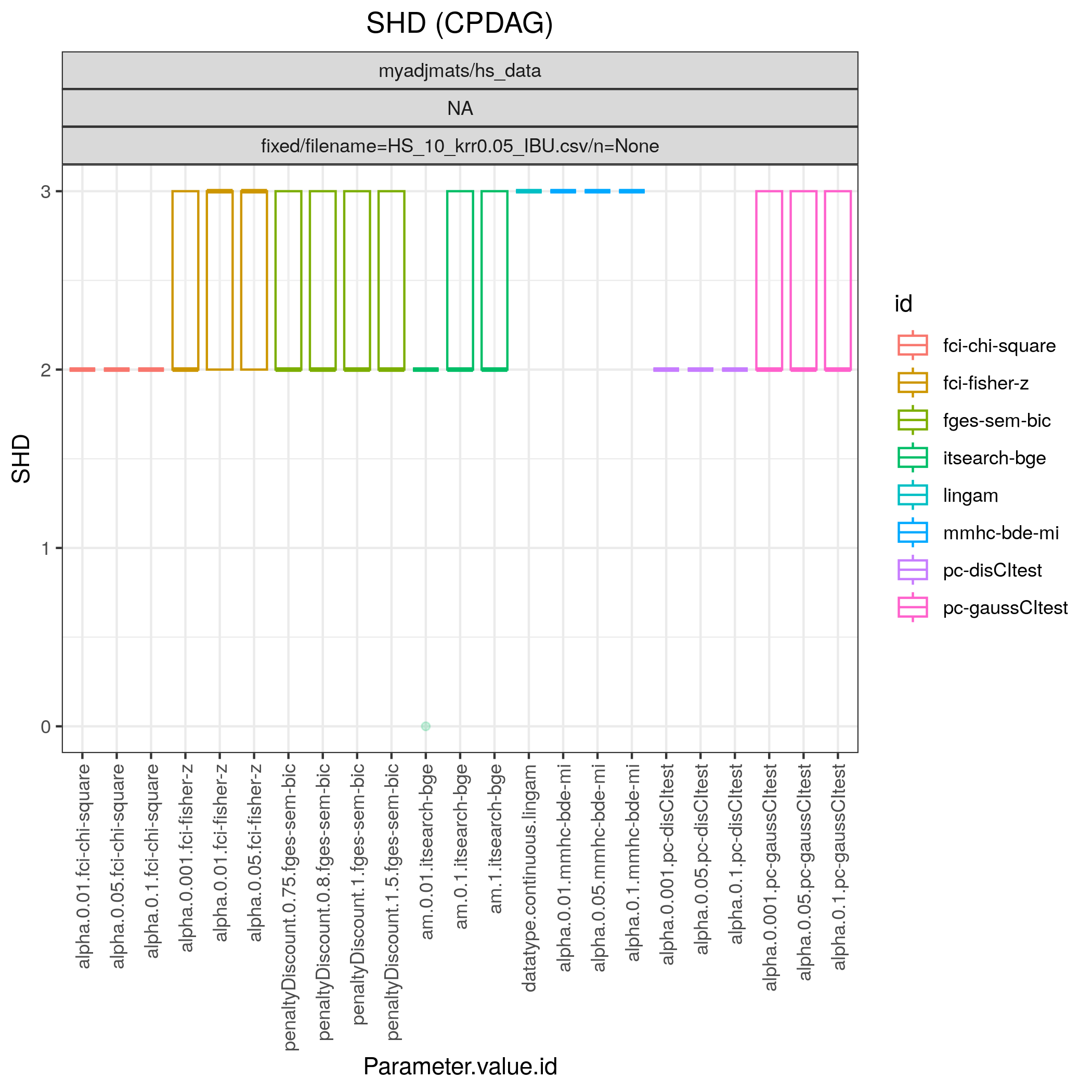}
	\caption{Human Stature data, $k$-RR C-wise IBU mechanism, max probability 0.05.}
\end{minipage}
\begin{minipage}{0.31\linewidth}
\centering
  \includegraphics[scale=0.34]{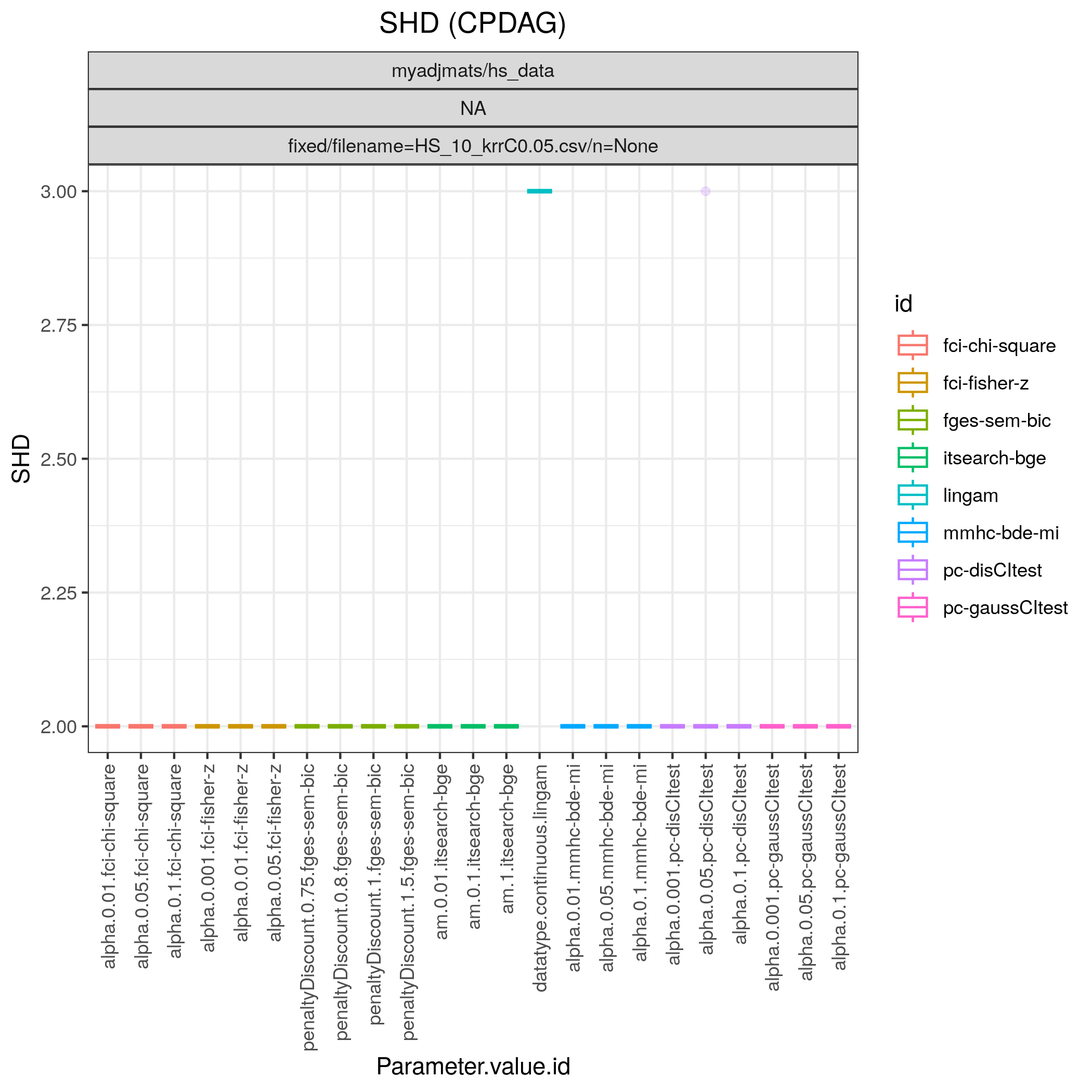}
	\caption{Human Stature data, $k$-RR Comb mechanism, max probability 0.05.}
\end{minipage}
\begin{minipage}{0.31\linewidth}
\centering
  \includegraphics[scale=0.34]{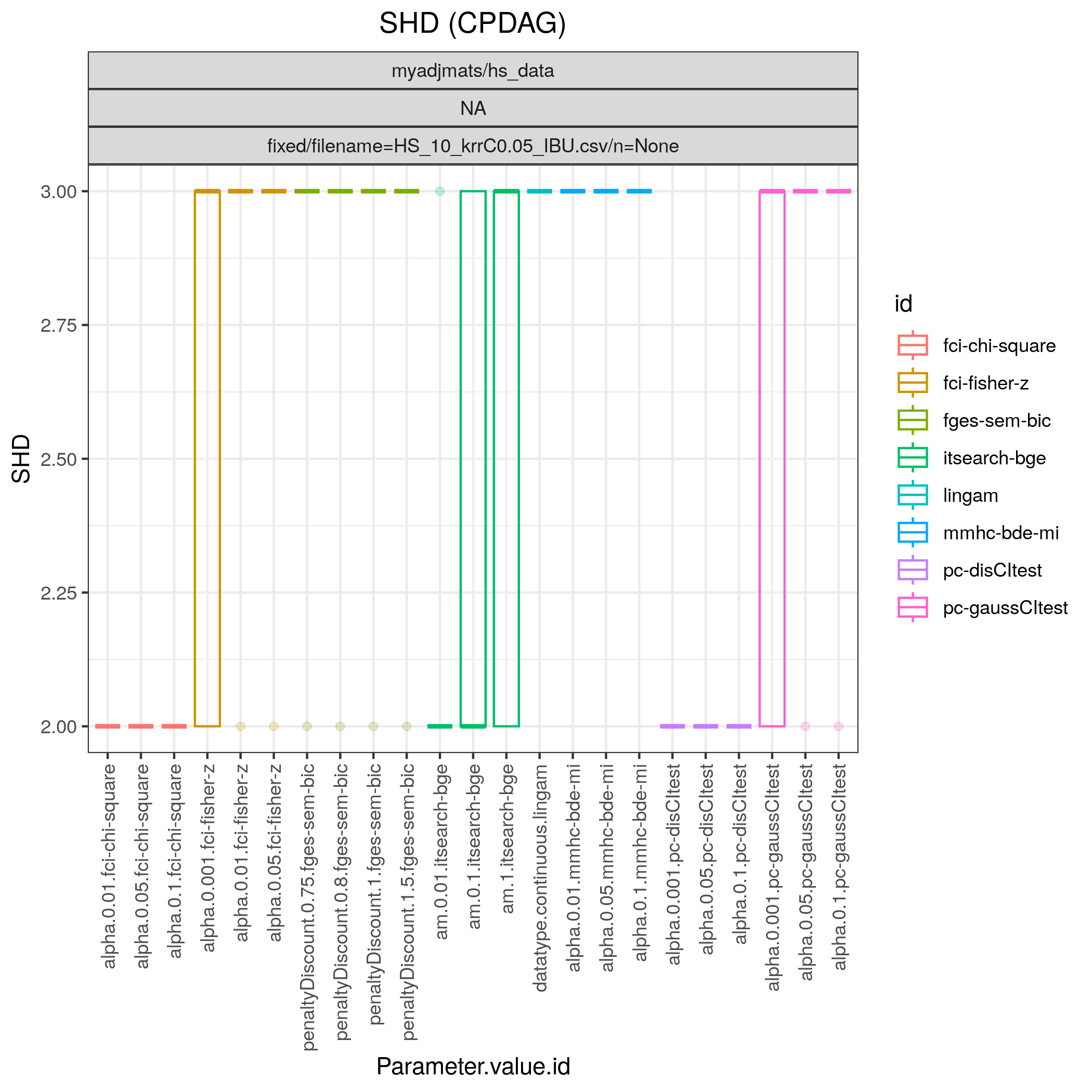}
	\caption{Human Stature data, $k$-RR Comb IBU mechanism, max probability 0.05.}
\end{minipage}
\end{figure}


\noindent
\begin{figure}[H]
\begin{minipage}{0.31\linewidth}
\centering
		\includegraphics[scale=0.34]{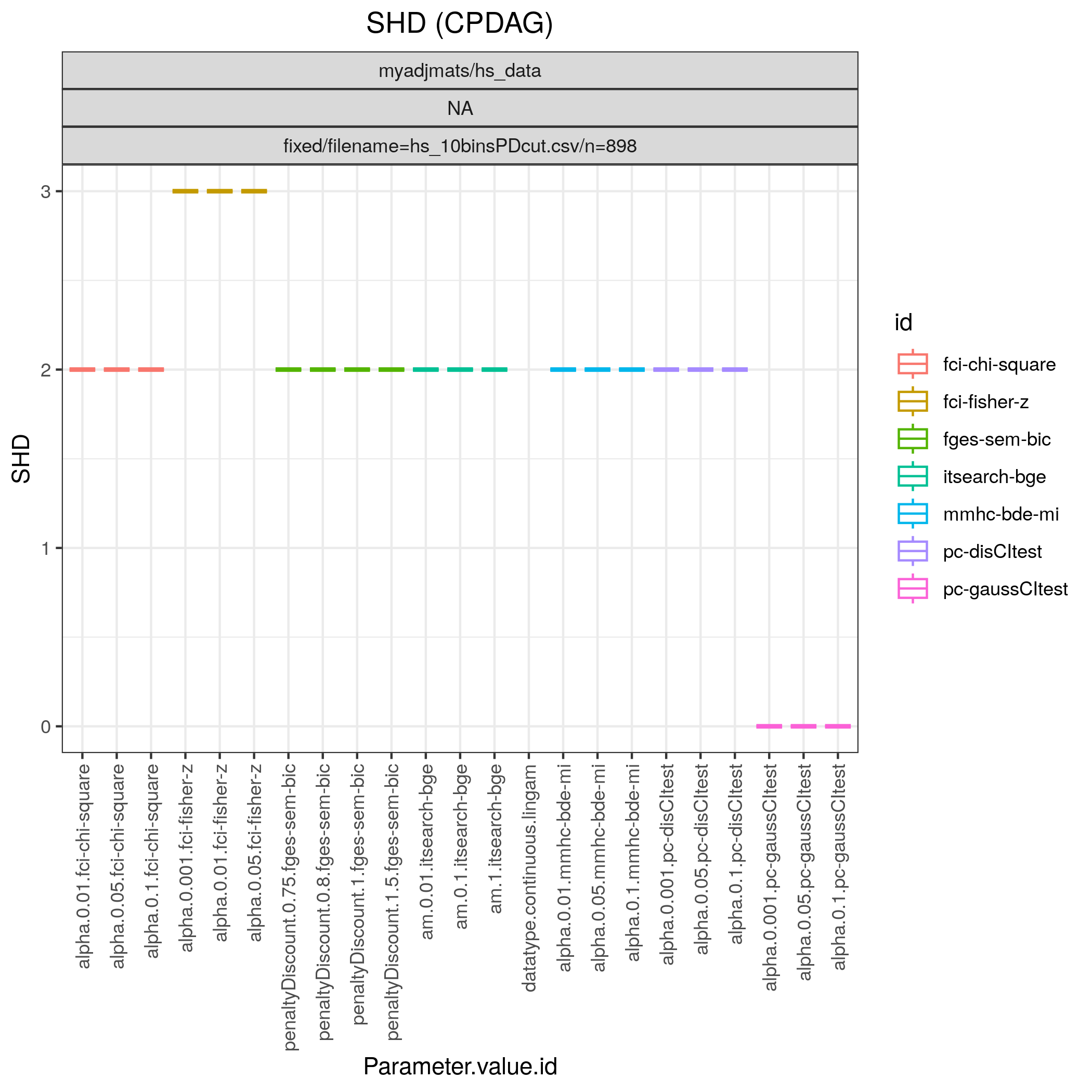}
	\caption{F1 Scores on the Human Stature data set. Discretized, no noise.}
\end{minipage}
\begin{minipage}{0.31\linewidth}
\centering
		\includegraphics[scale=0.34]{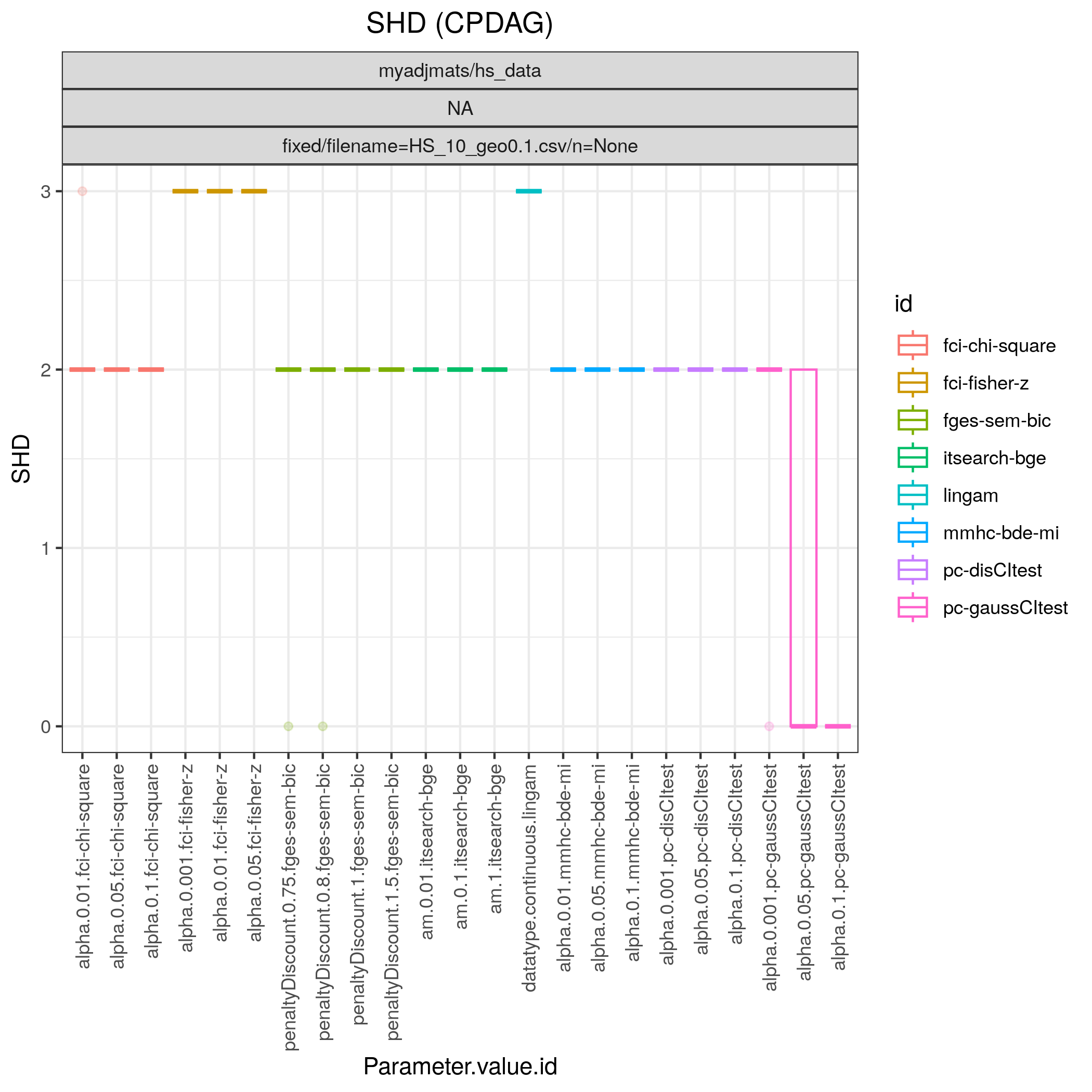}
	\caption{Human Stature data, Geo C-wise mechanism, max probability 0.1.}
\end{minipage}
\begin{minipage}{0.31\linewidth}
\centering
  \includegraphics[scale=0.34]{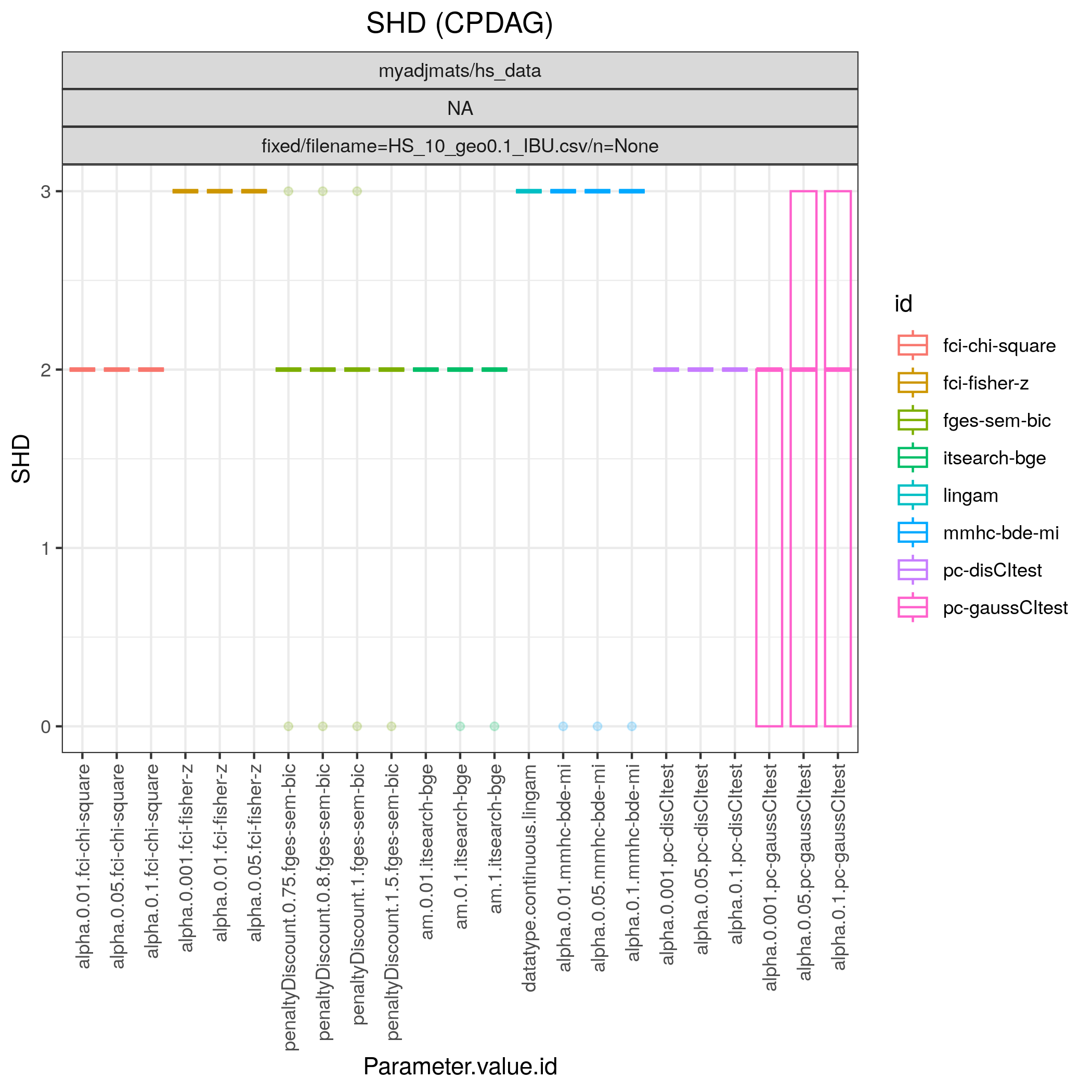}
	\caption{Human Stature data, Geo C-wise IBU mechanism, max probability 0.1.}
 \end{minipage}
\end{figure}

\begin{figure}[H]
    \centering
   \begin{minipage}{0.31\linewidth}
\centering
  \includegraphics[scale=0.34]{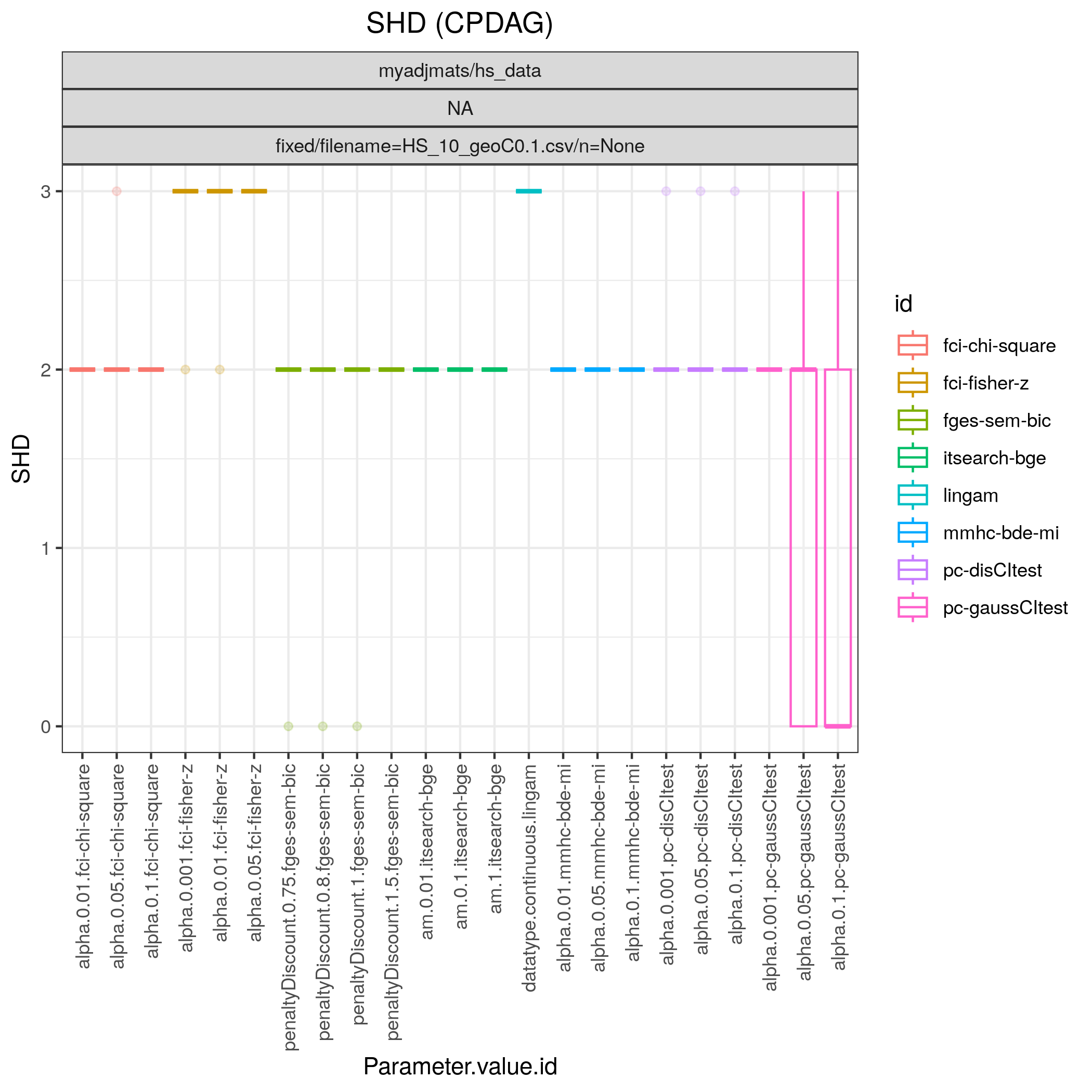}
	\caption{Human Stature data, Geo Comb mechanism, max probability 0.1.}
\end{minipage}
\begin{minipage}{0.31\linewidth}
\centering
  \includegraphics[scale=0.34]{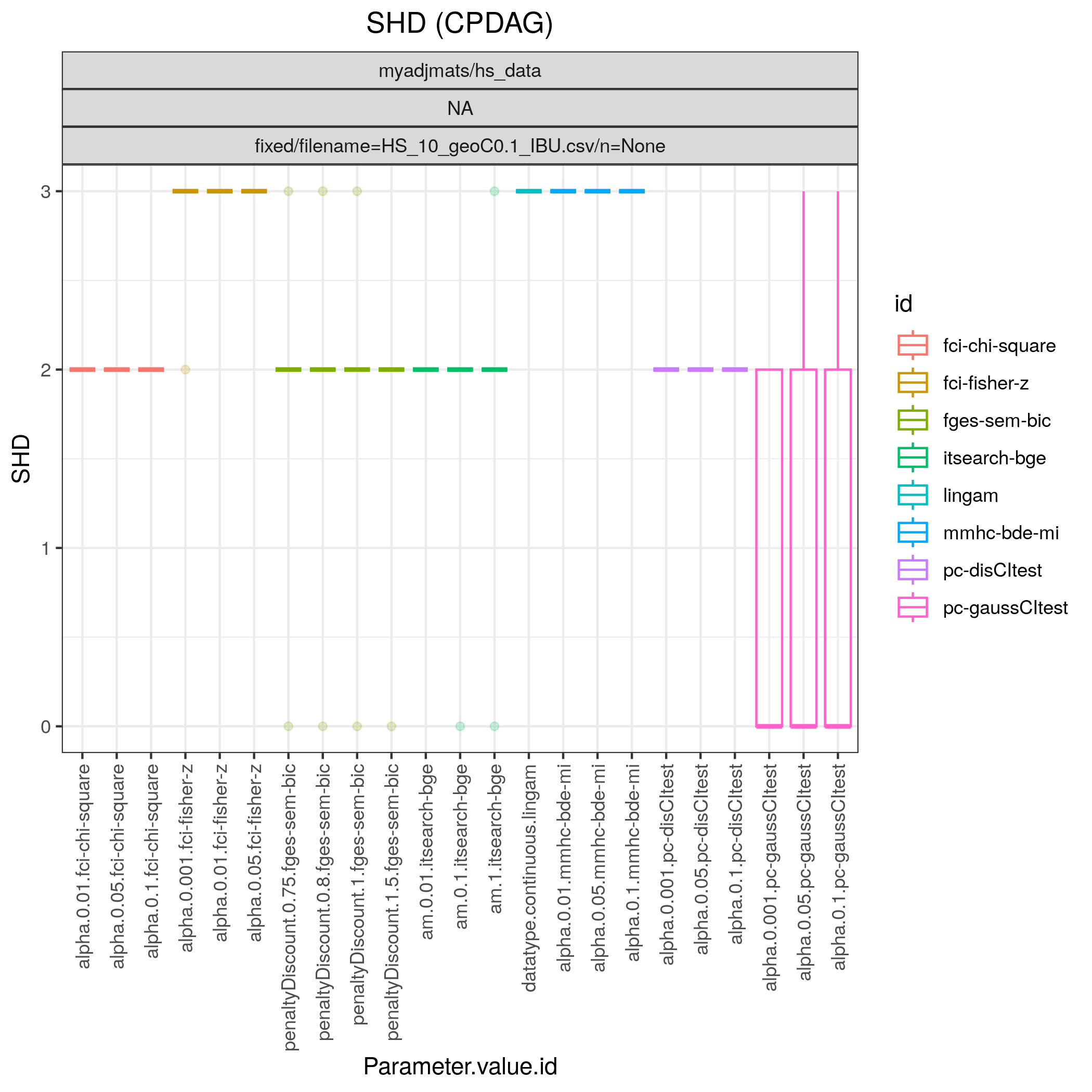}
	\caption{Human Stature data, Geo Comb IBU mechanism, max probability 0.1.}
\end{minipage}
\begin{minipage}{0.31\linewidth}
\centering
  \includegraphics[scale=0.34]{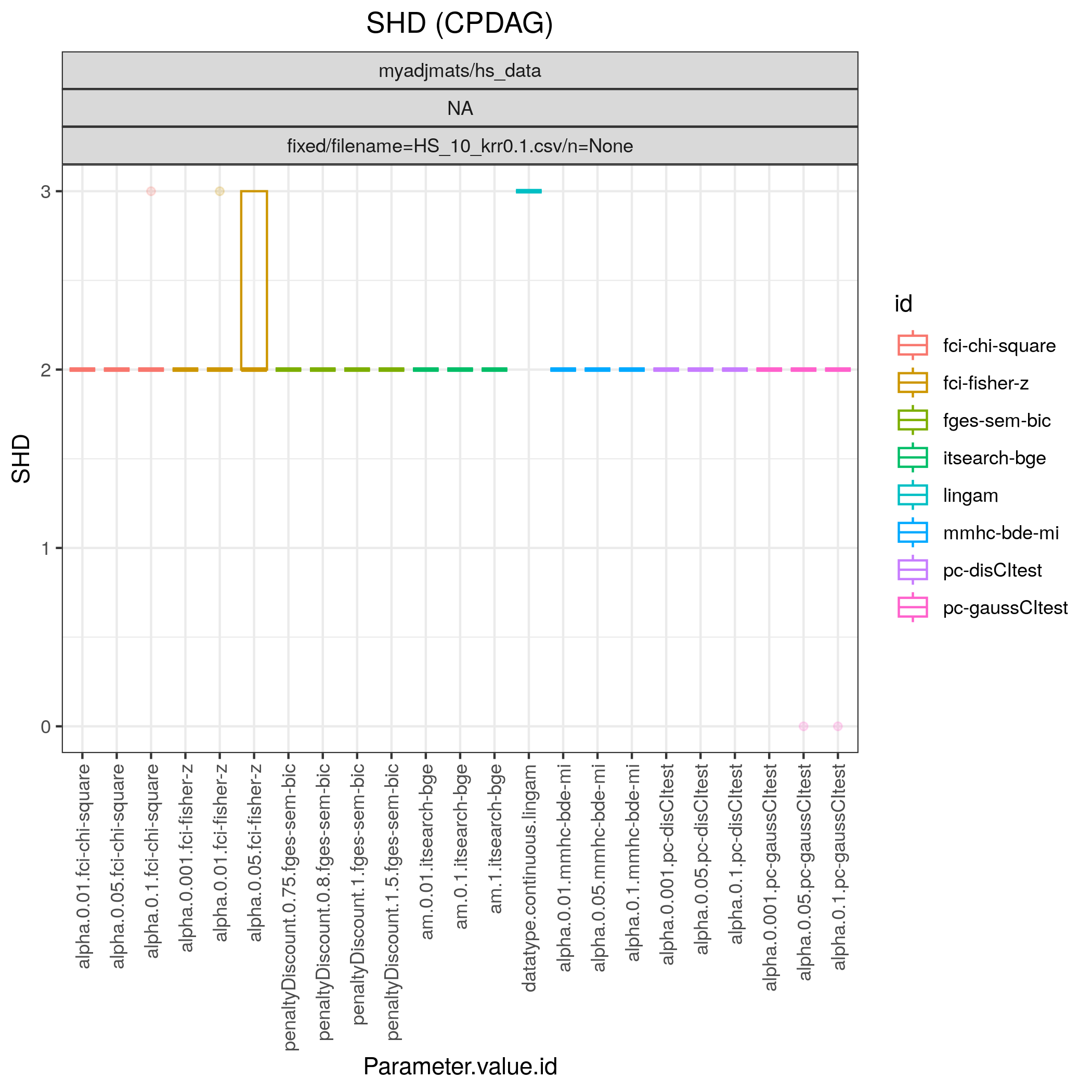}
	\caption{Human Stature data, $k$-RR C-wise mechanism, max probability 0.1.}
\end{minipage}
\end{figure}

\begin{figure}[H]
    \centering
   \begin{minipage}{0.31\linewidth}
\centering
  \includegraphics[scale=0.34]{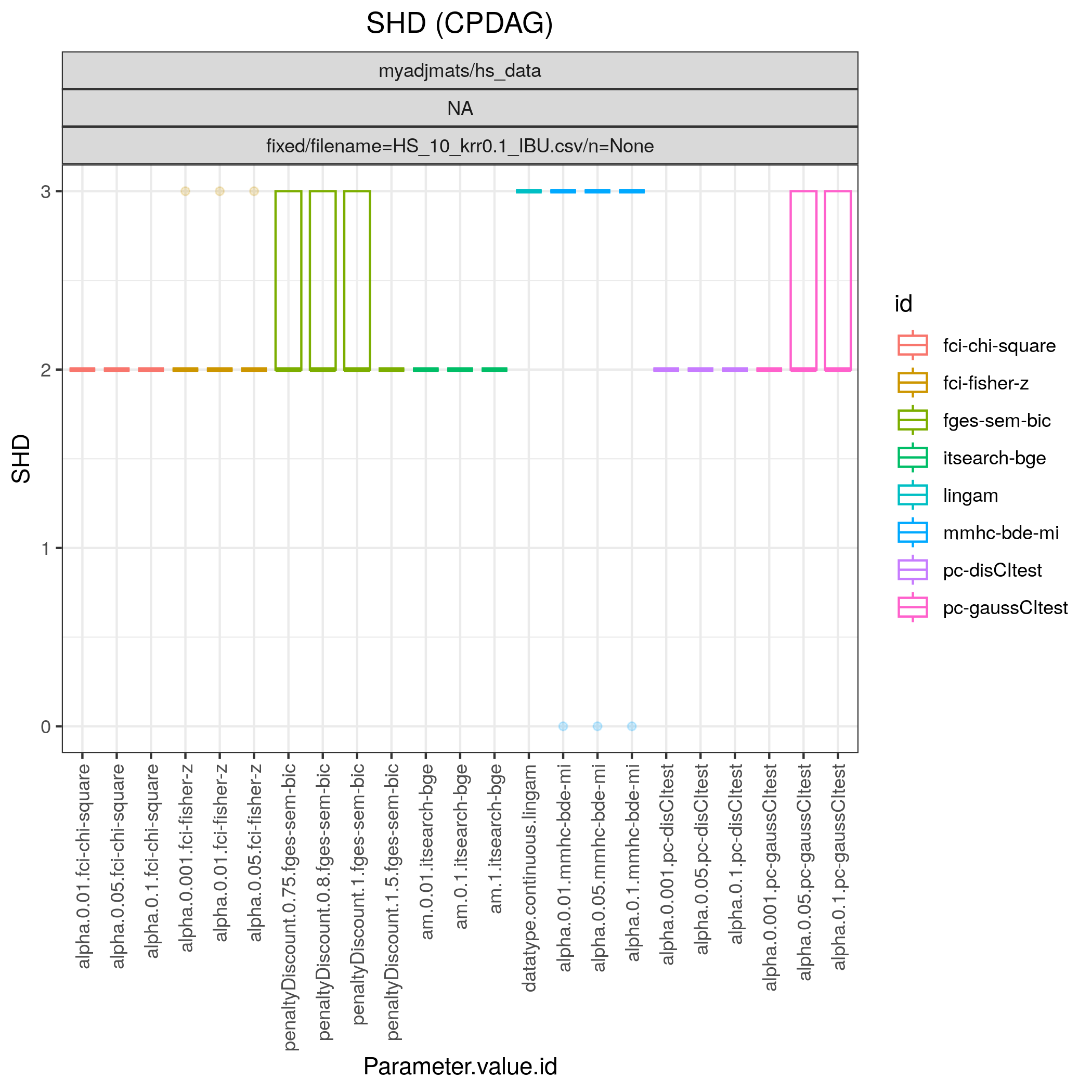}
	\caption{Human Stature data, $k$-RR C-wise IBU mechanism, max probability 0.1.}
\end{minipage}
\begin{minipage}{0.31\linewidth}
\centering
  \includegraphics[scale=0.34]{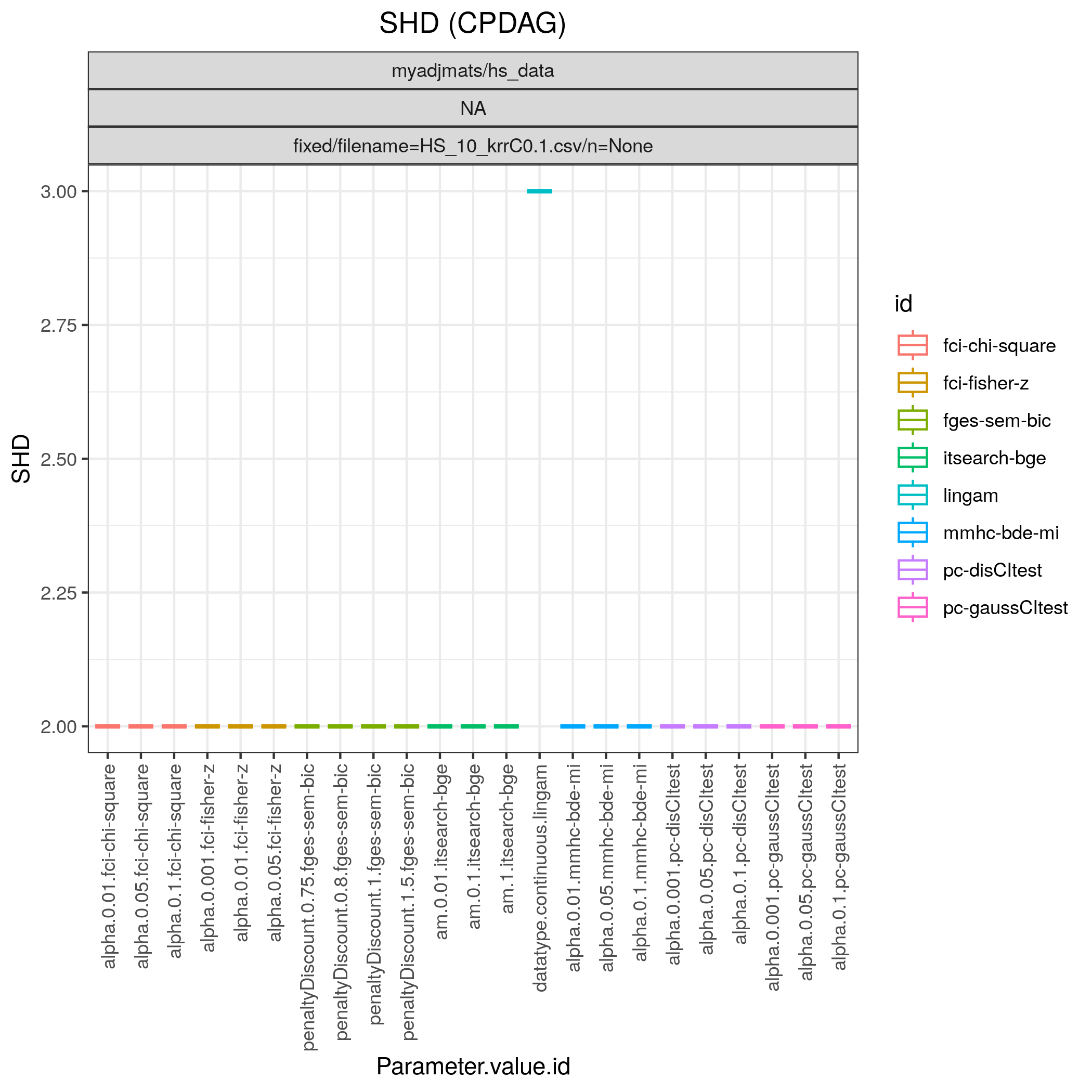}
	\caption{Human Stature data, $k$-RR Comb mechanism, max probability 0.1.}
\end{minipage}
\begin{minipage}{0.31\linewidth}
\centering
  \includegraphics[scale=0.34]{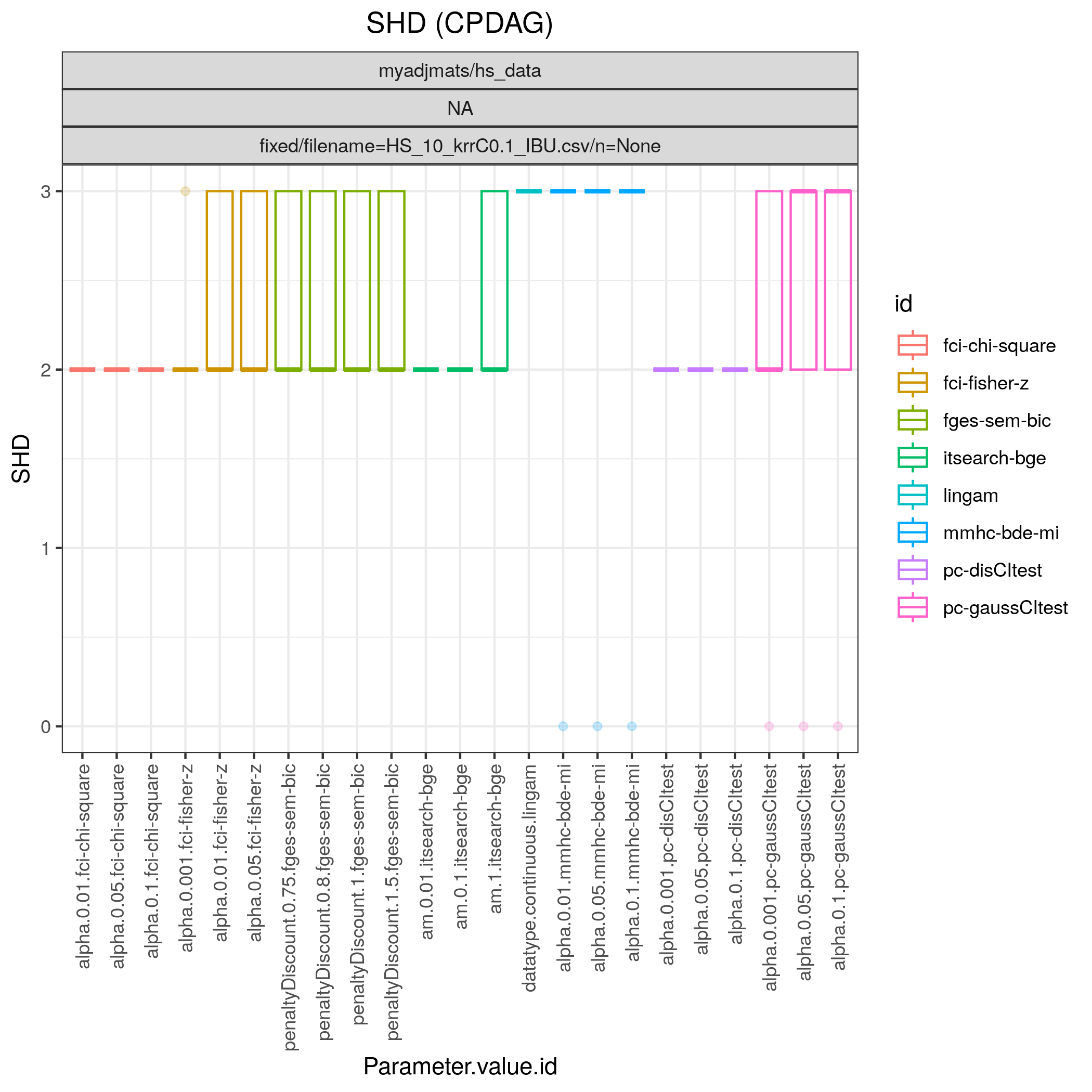}
	\caption{Human Stature data, $k$-RR Comb IBU mechanism, max probability 0.1.}
\end{minipage}
\end{figure}


\noindent
\begin{figure}[H]
\begin{minipage}{0.31\linewidth}
\centering
		\includegraphics[scale=0.34]{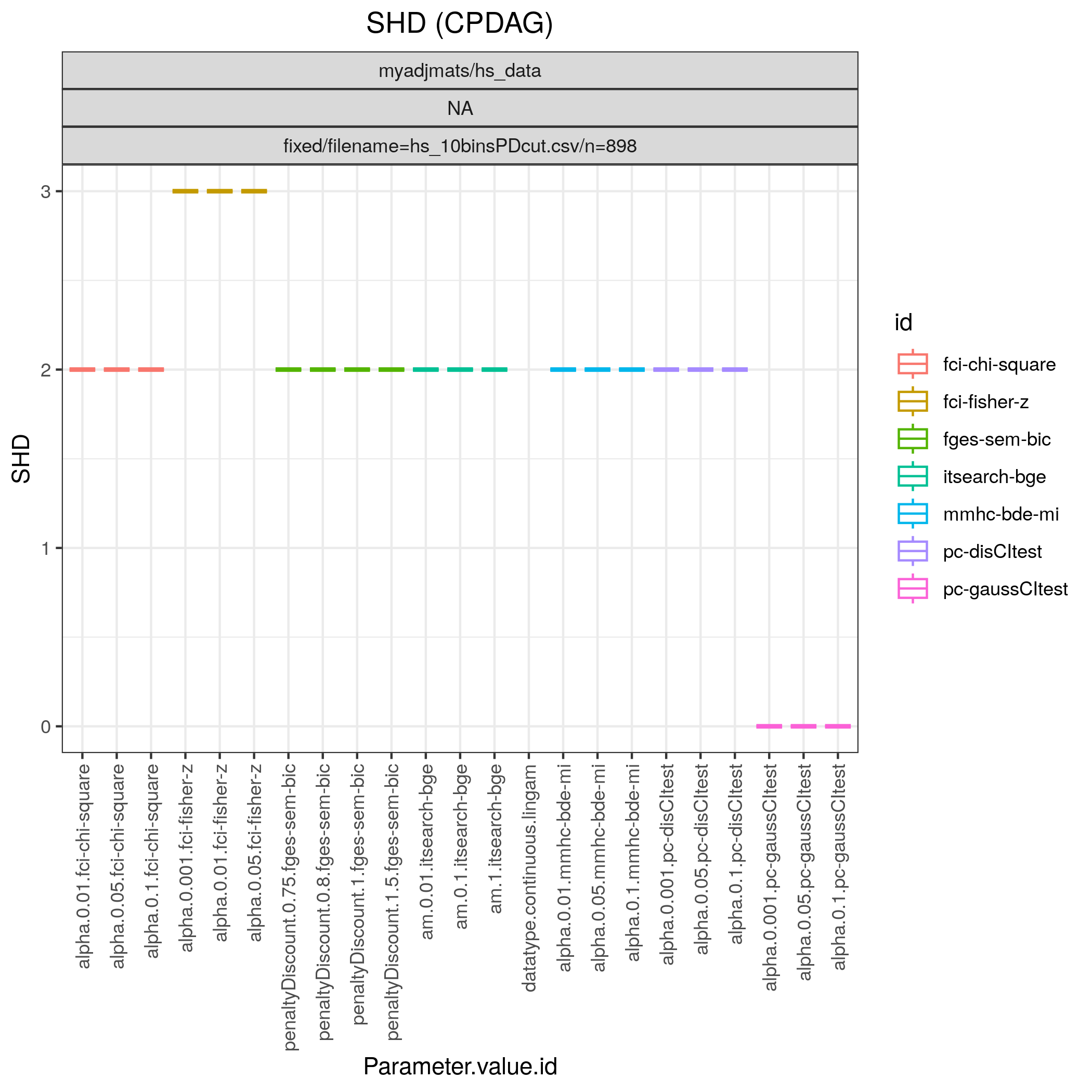}
	\caption{SHD Scores on the Human Stature data set. Discretized, no noise.}
\end{minipage}
\begin{minipage}{0.31\linewidth}
\centering
		\includegraphics[scale=0.34]{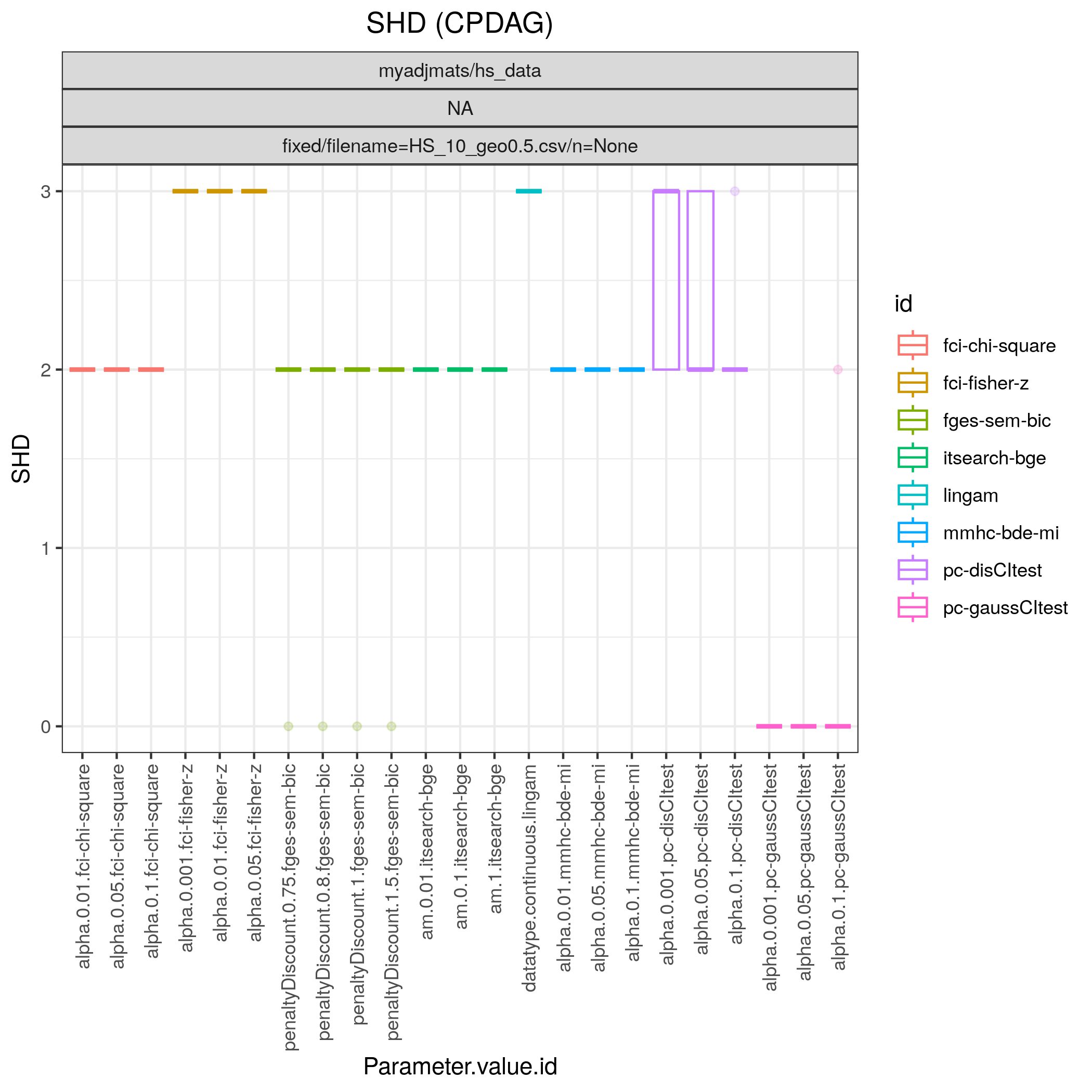}
	\caption{Human Stature data, Geo C-wise mechanism, max probability 0.5.}
\end{minipage}
\begin{minipage}{0.31\linewidth}
\centering
  \includegraphics[scale=0.34]{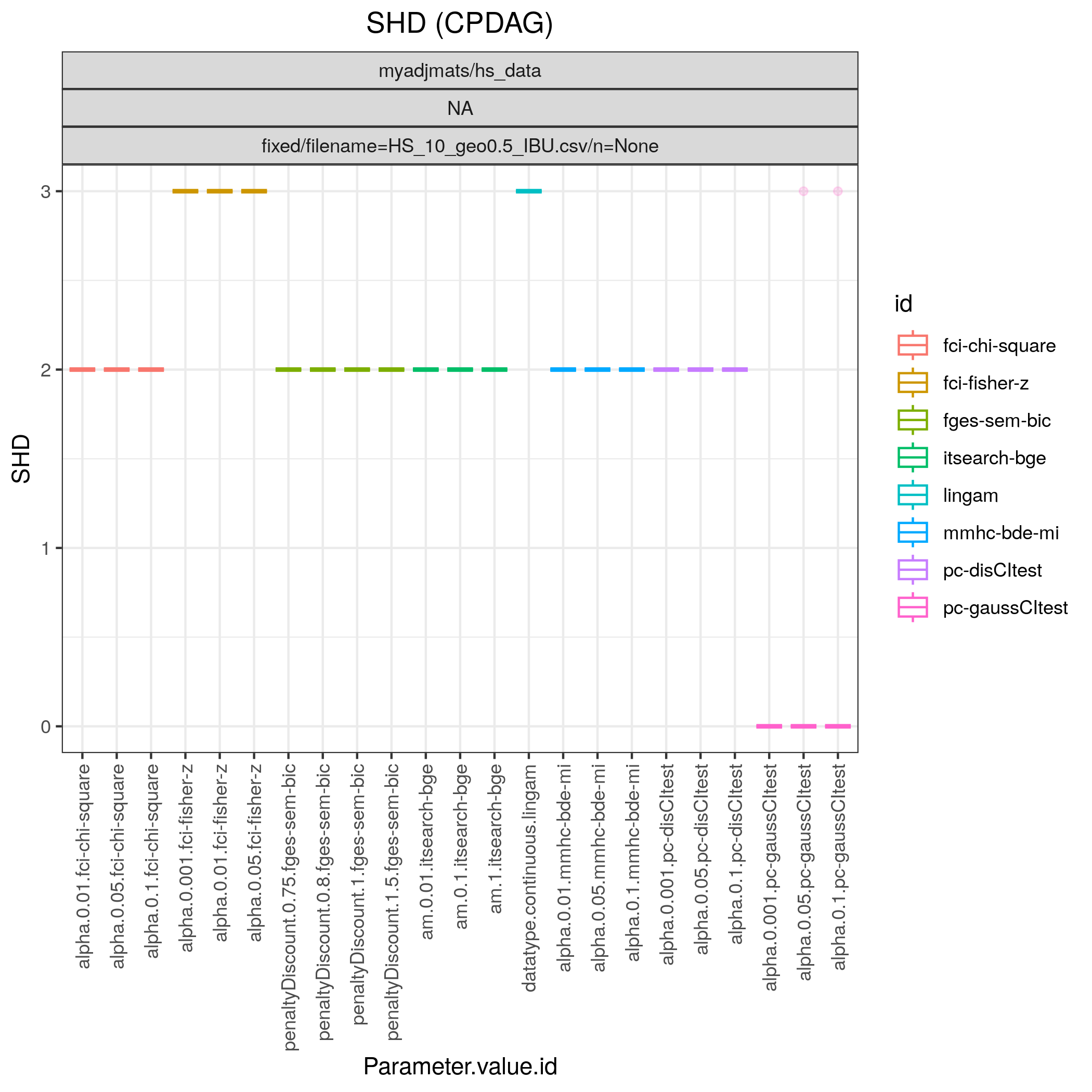}
	\caption{Human Stature data, Geo C-wise IBU mechanism, max probability 0.5.}
 \end{minipage}
\end{figure}

\begin{figure}[H]
    \centering
   \begin{minipage}{0.31\linewidth}
\centering
  \includegraphics[scale=0.34]{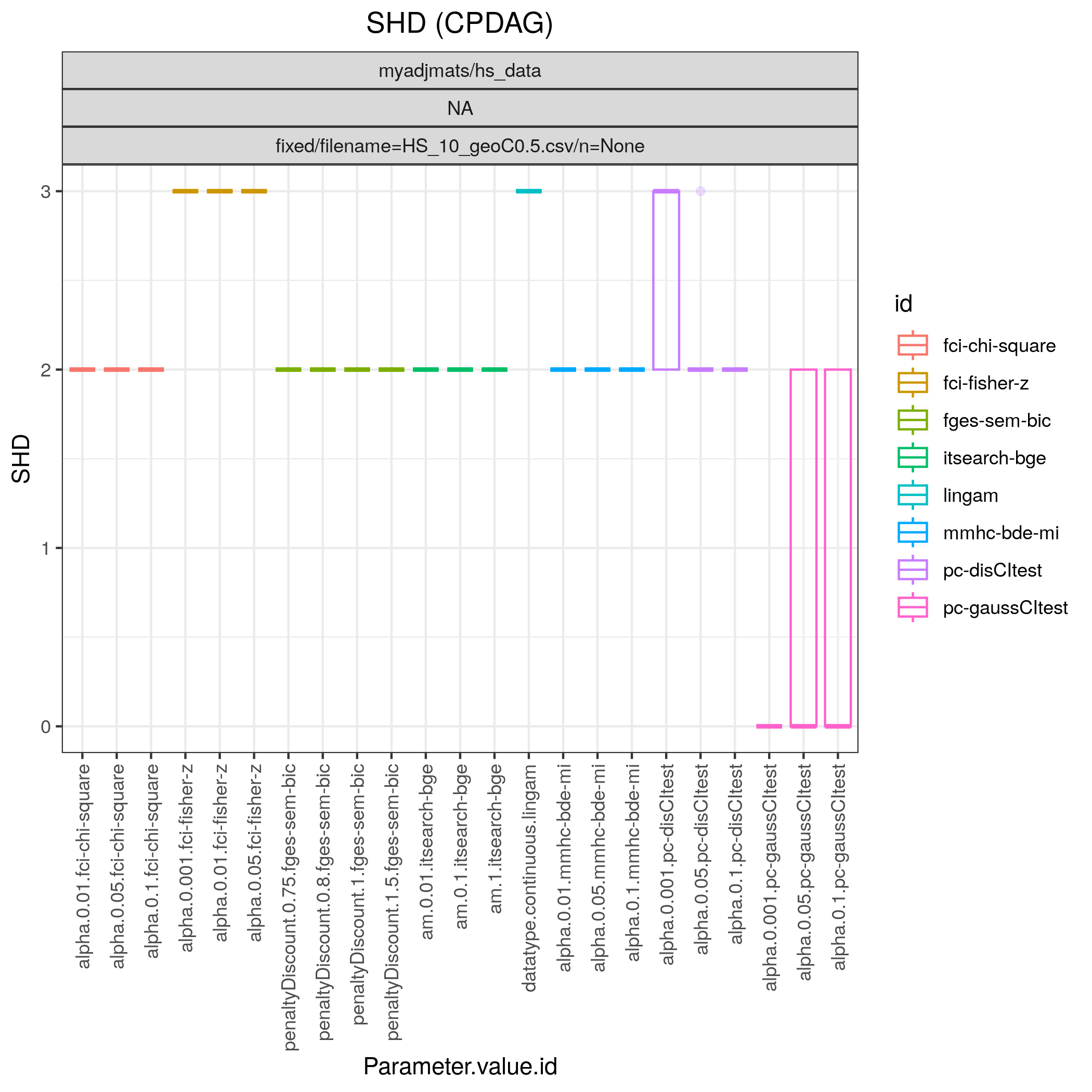}
	\caption{Human Stature data, Geo Comb mechanism, max probability 0.5.}
\end{minipage}
\begin{minipage}{0.31\linewidth}
\centering
  \includegraphics[scale=0.34]{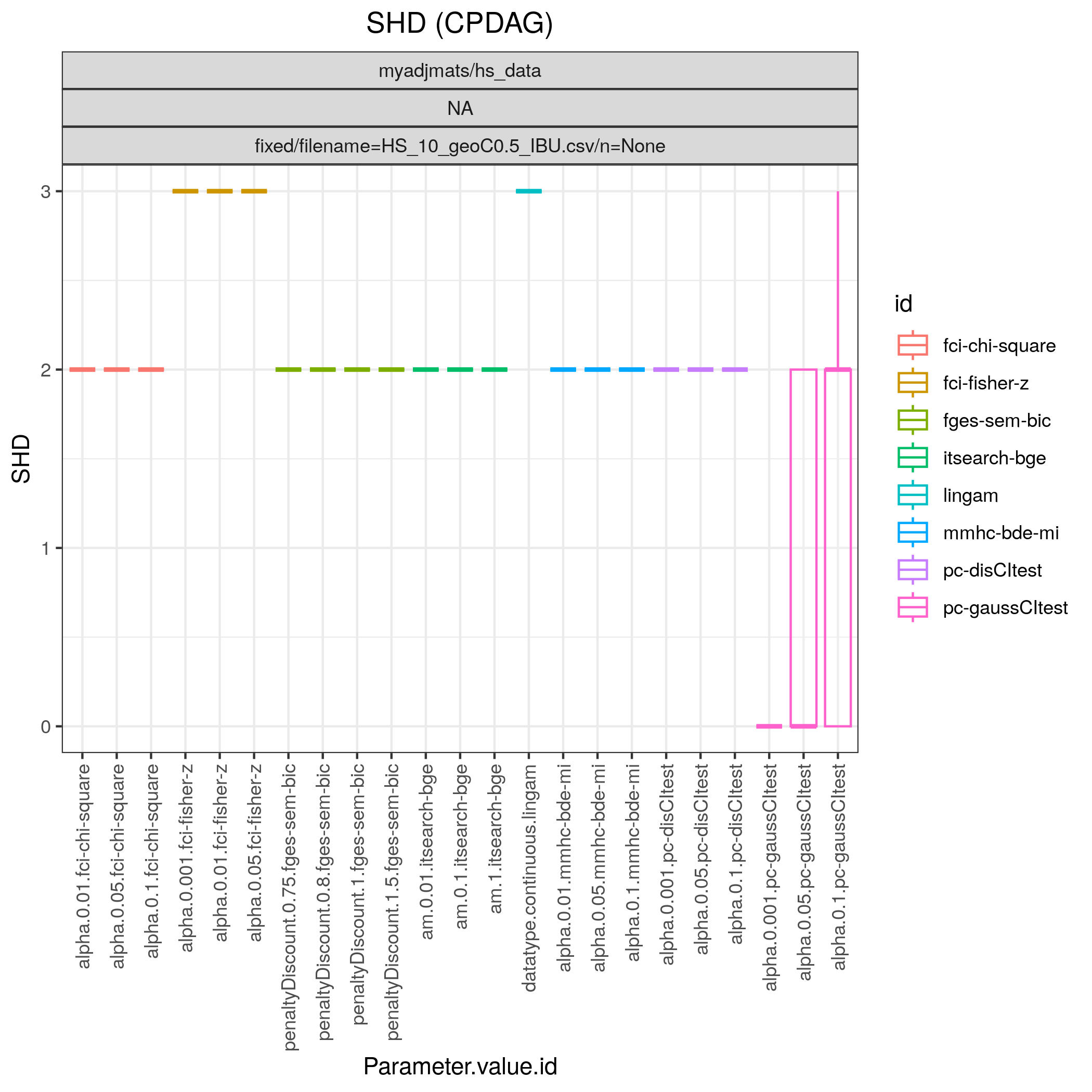}
	\caption{Human Stature data, Geo Comb IBU mechanism, max probability 0.5.}
\end{minipage}
\begin{minipage}{0.31\linewidth}
\centering
  \includegraphics[scale=0.34]{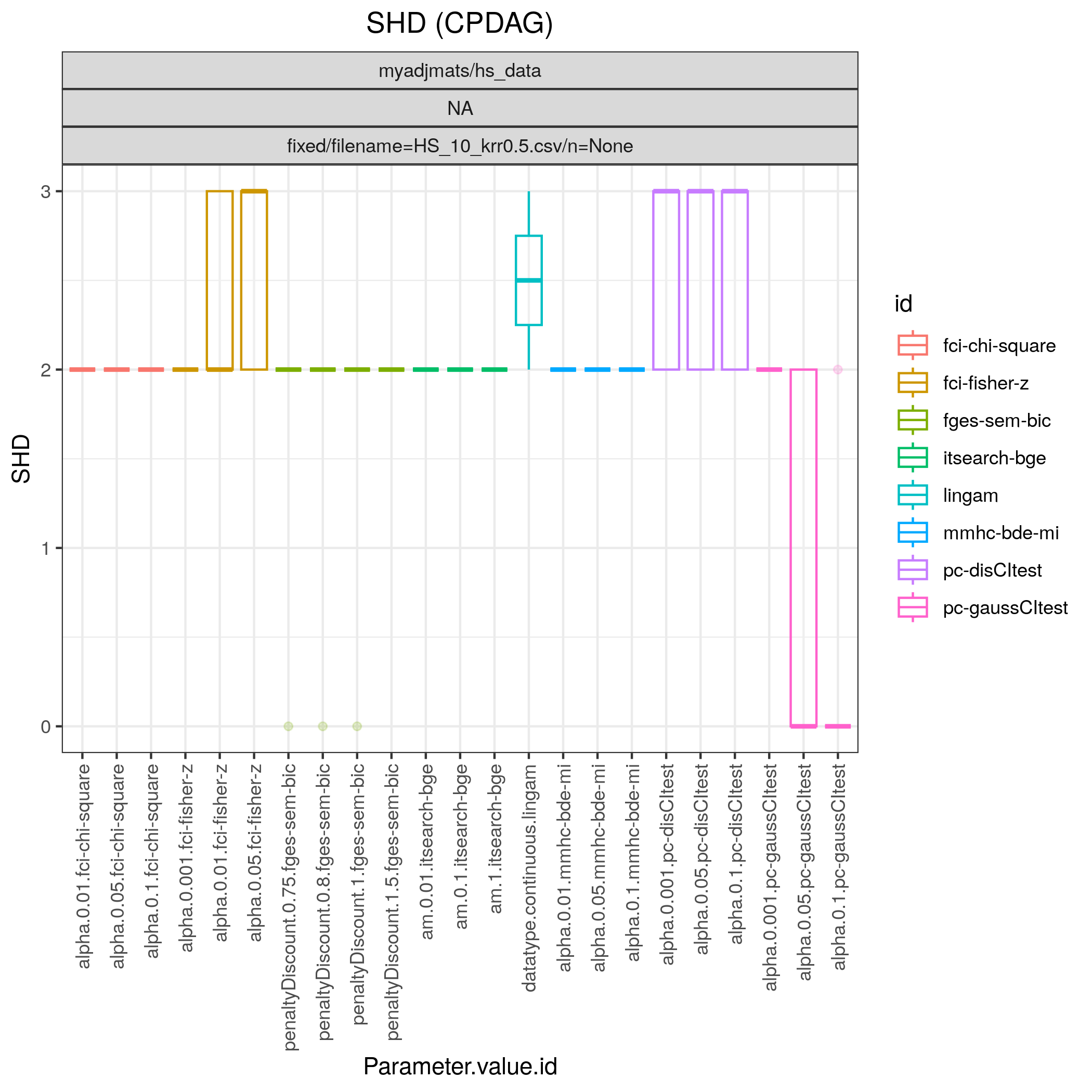}
	\caption{Human Stature data, $k$-RR C-wise mechanism, max probability 0.5.}
\end{minipage}
\end{figure}

\begin{figure}[H]
    \centering
   \begin{minipage}{0.31\linewidth}
\centering
  \includegraphics[scale=0.34]{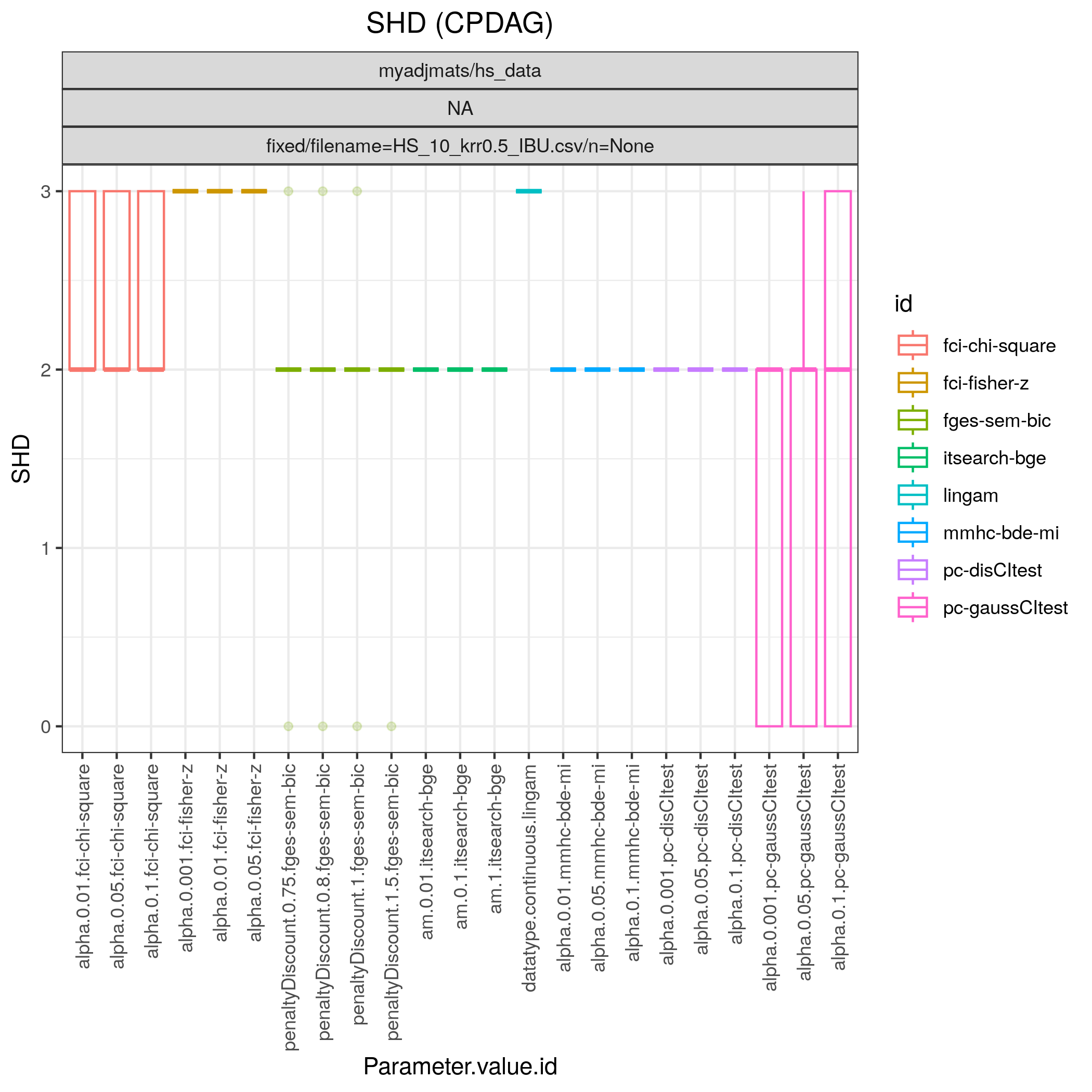}
	\caption{Human Stature data, $k$-RR C-wise IBU mechanism, max probability 0.5.}
\end{minipage}
\begin{minipage}{0.31\linewidth}
\centering
  \includegraphics[scale=0.34]{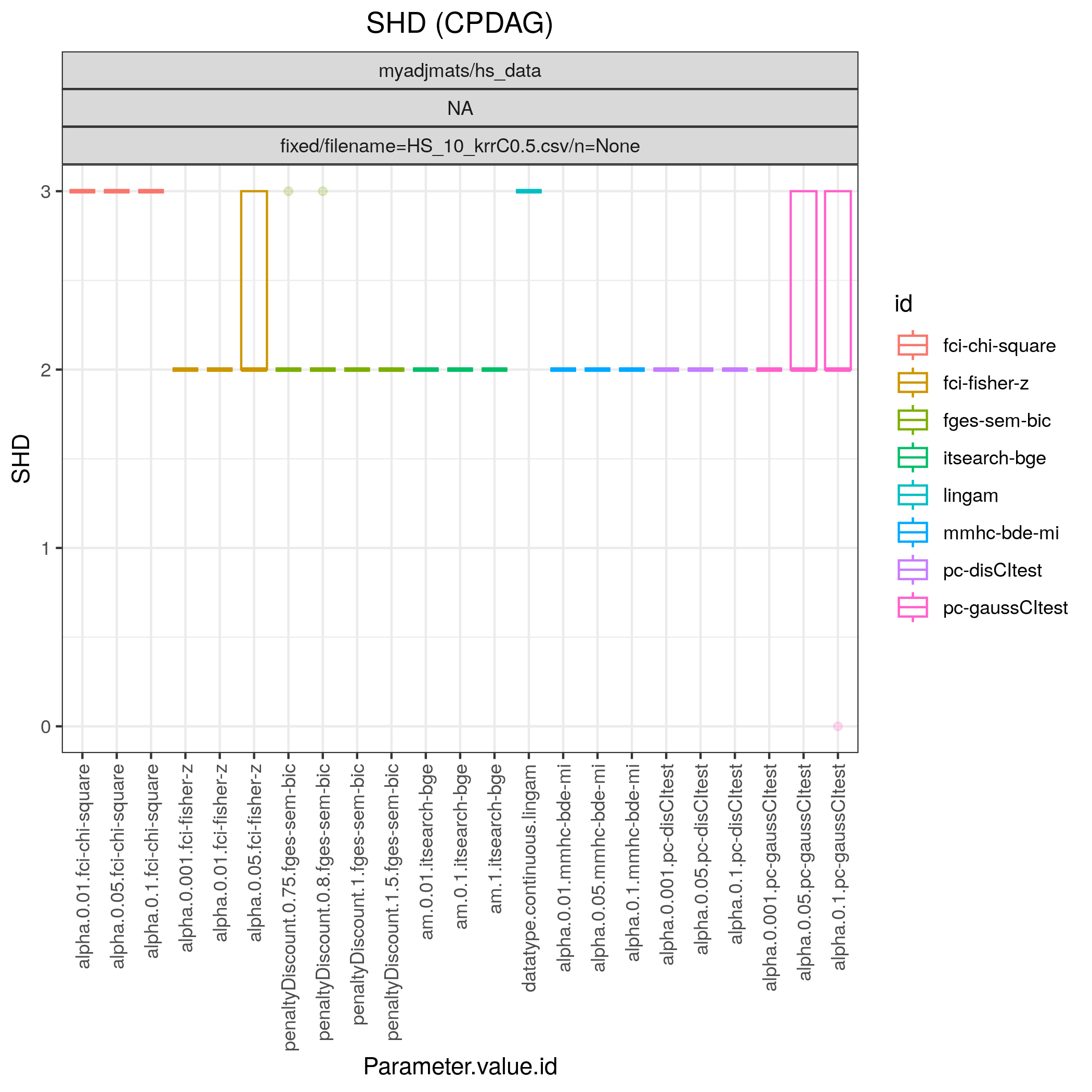}
	\caption{Human Stature data, $k$-RR Comb mechanism, max probability 0.5.}
\end{minipage}
\begin{minipage}{0.31\linewidth}
\centering
  \includegraphics[scale=0.34]{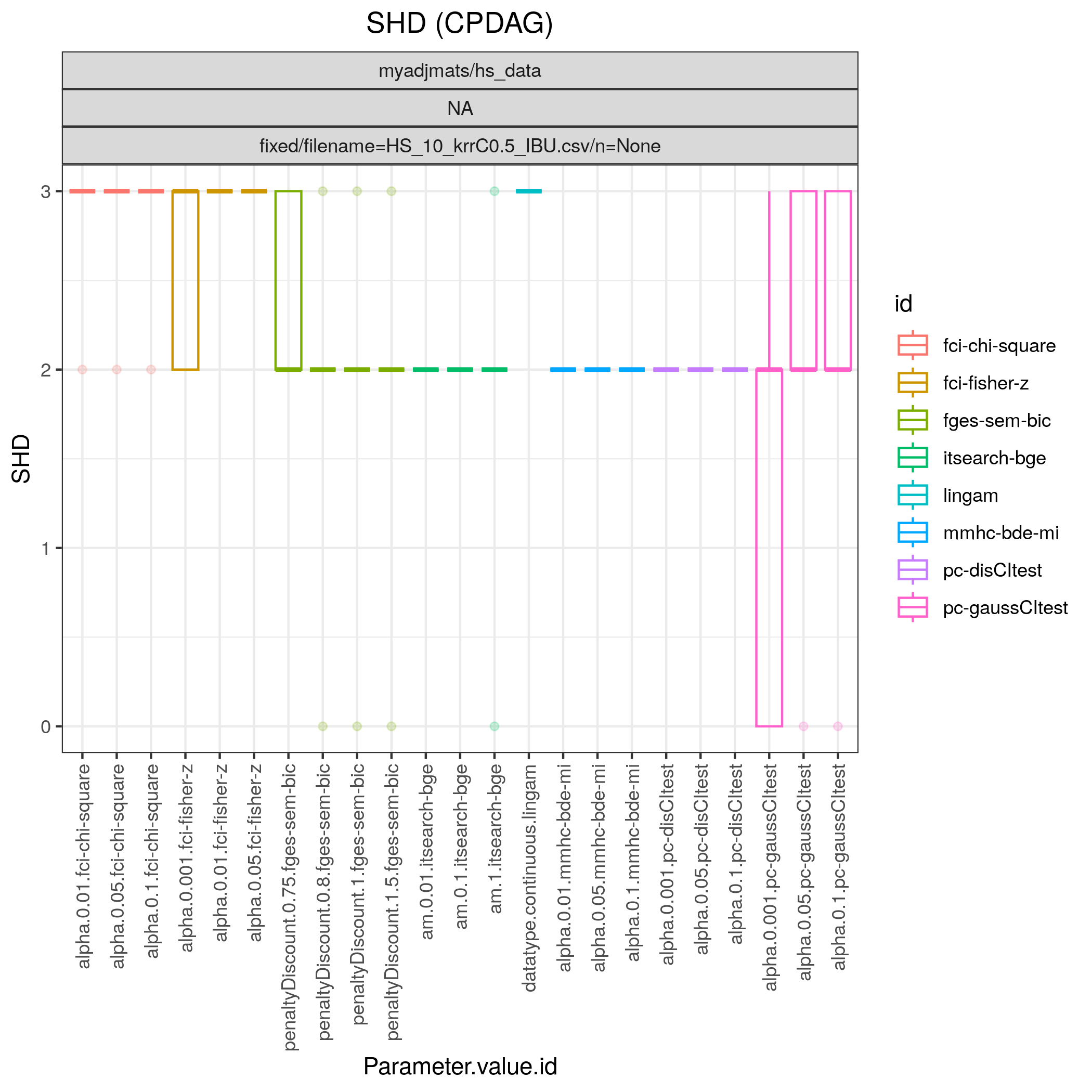}
	\caption{Human Stature data, $k$-RR Comb IBU mechanism, max probability 0.5.}
\end{minipage}
\end{figure}

 \subsection{F1 Score results Synthetic 5 nodes data set}

\begin{figure}[t]
    \centering
    \includegraphics[width=1\linewidth]{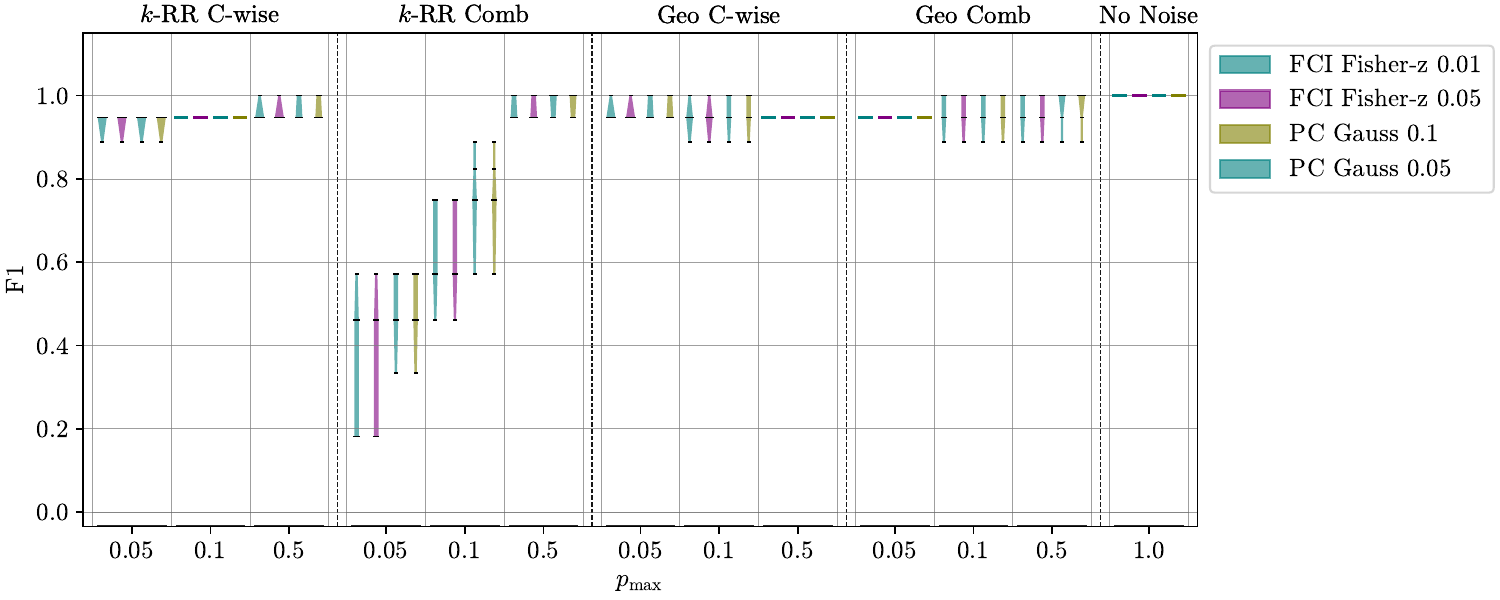}
    \vspace*{-10mm}
    \caption{Synthetic data, 5 nodes, F1.}
    \label{fig:SYNTH5_f1}
\end{figure}
\noindent
\begin{figure}[H]
\begin{minipage}{0.31\linewidth}
\centering
		\includegraphics[scale=0.34]{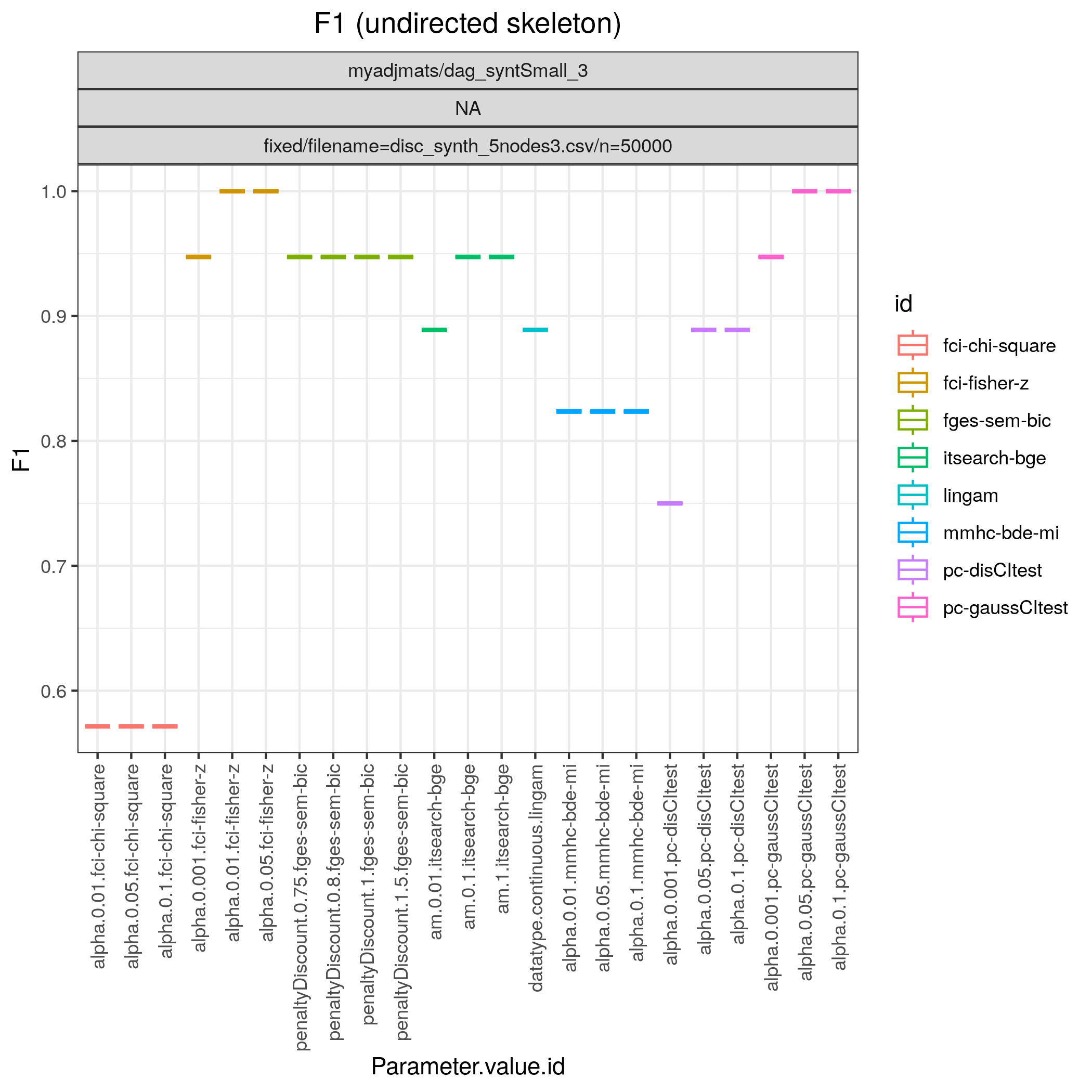}
	\caption{F1 Scores on the Synthetic 5 nodes data set. Discretized, no noise.}
\end{minipage}
\begin{minipage}{0.31\linewidth}
\centering
		\includegraphics[scale=0.34]{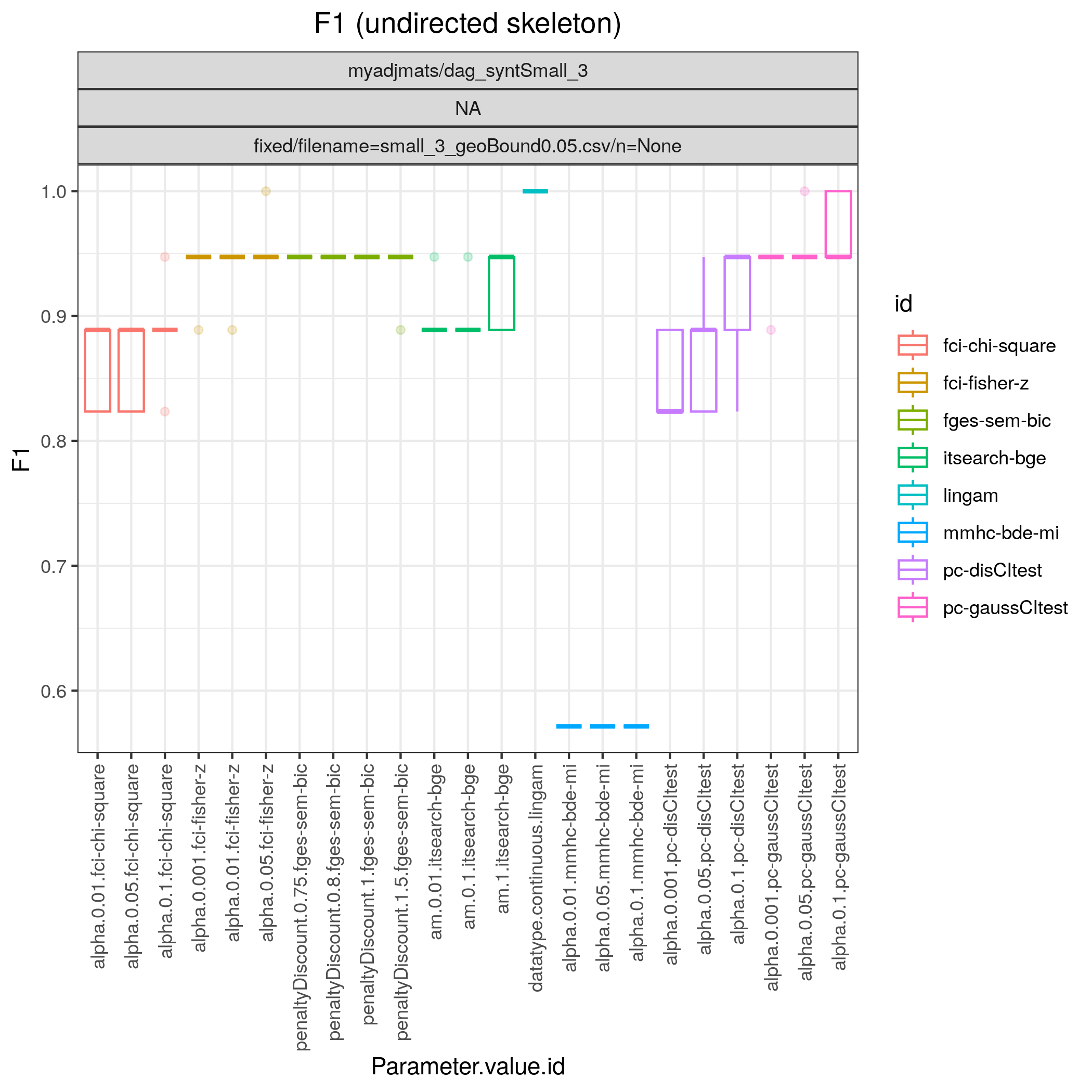}
	\caption{Synthetic 5 nodes data, Geo C-wise mechanism, max probability 0.05.}
\end{minipage}
\begin{minipage}{0.31\linewidth}
\centering
  \includegraphics[scale=0.34]{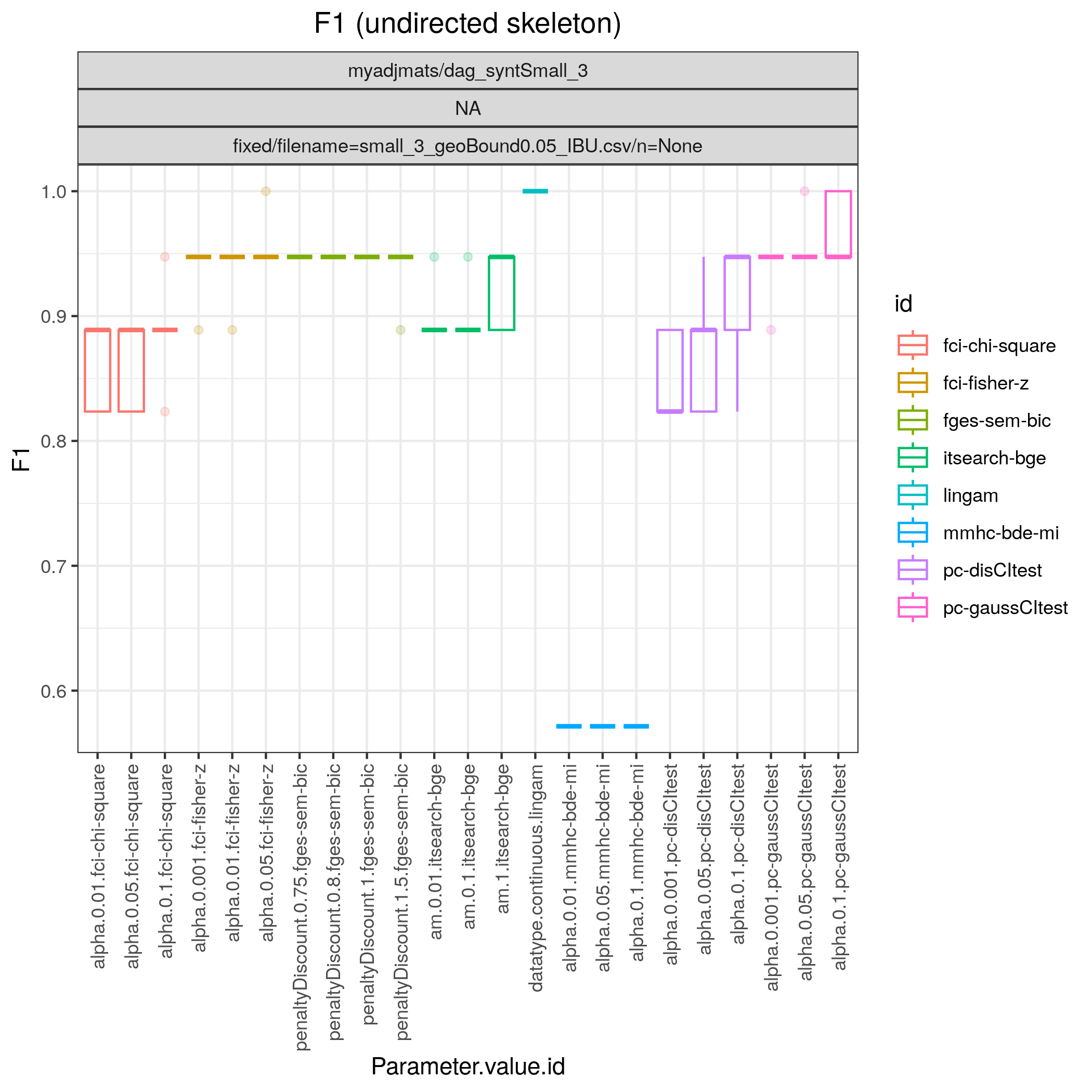}
	\caption{Synthetic 5 nodes data, Geo C-wise IBU mechanism, max probability 0.05.}
 \end{minipage}
\begin{minipage}{0.31\linewidth}
\centering
  \includegraphics[scale=0.34]{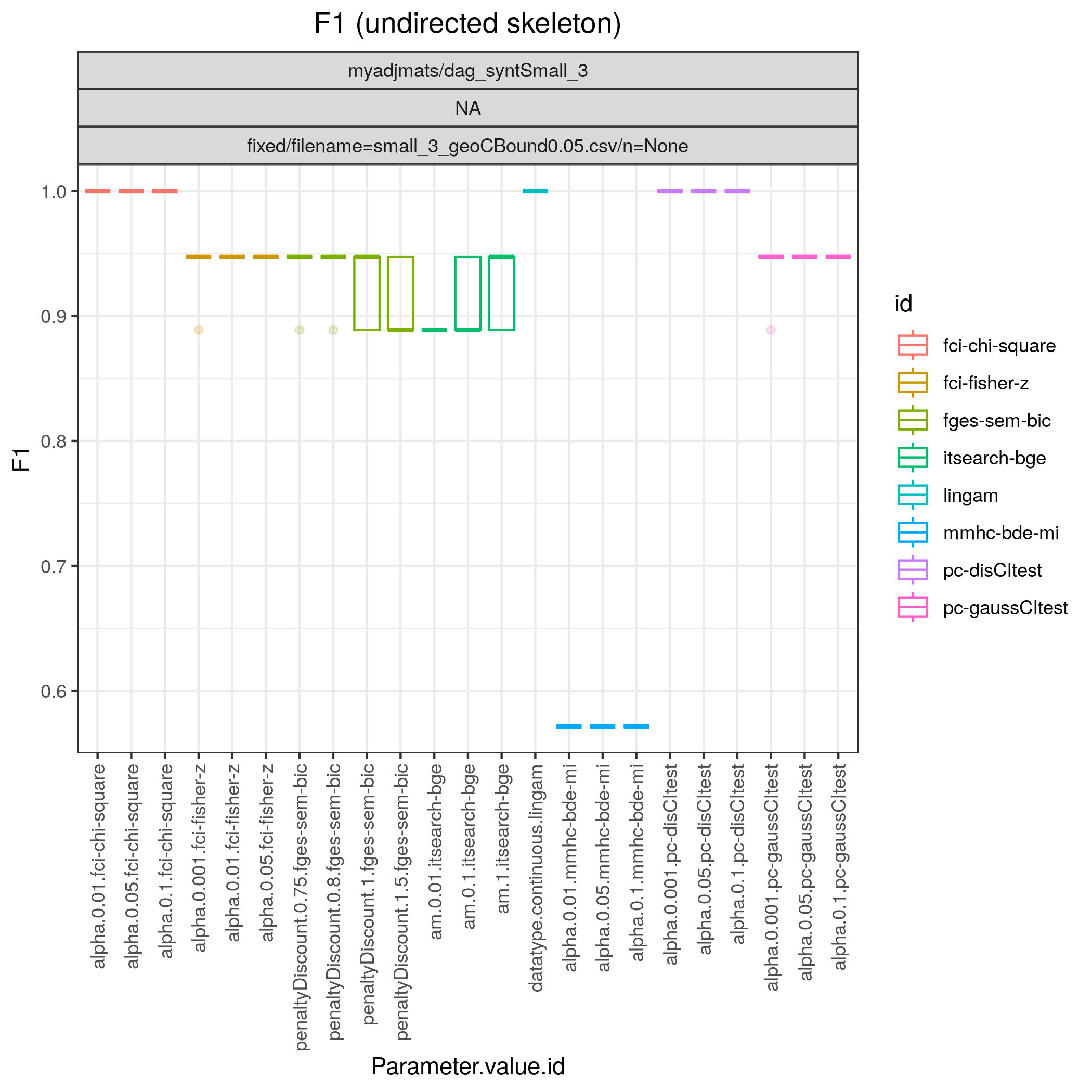}
	\caption{Synthetic 5 nodes data, Geo Comb mechanism, max probability 0.05.}
\end{minipage}
\begin{minipage}{0.31\linewidth}
\centering
  \includegraphics[scale=0.34]{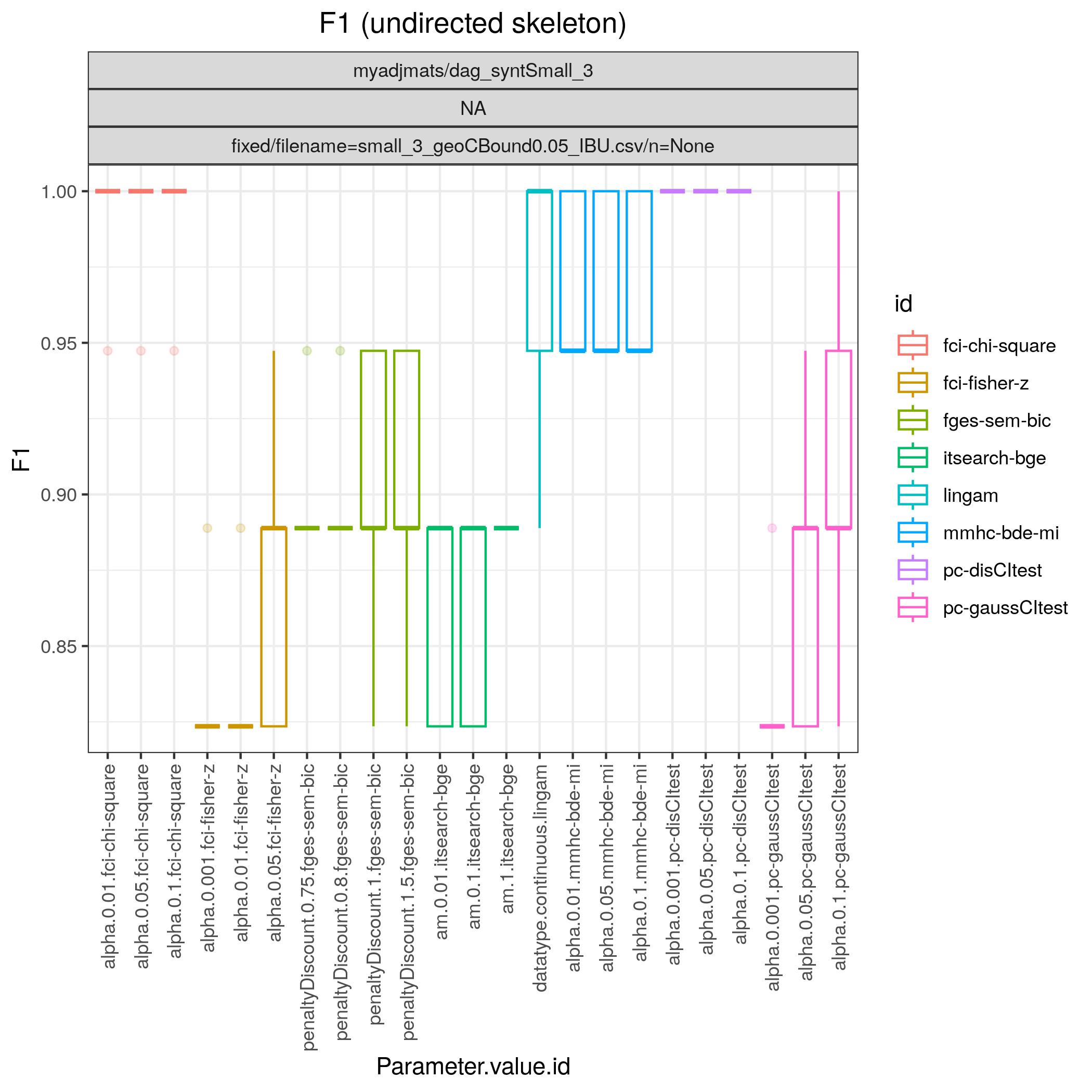}
	\caption{Synthetic 5 nodes data, Geo Comb IBU mechanism, max probability 0.05.}
\end{minipage}
\begin{minipage}{0.31\linewidth}
\centering
  \includegraphics[scale=0.34]{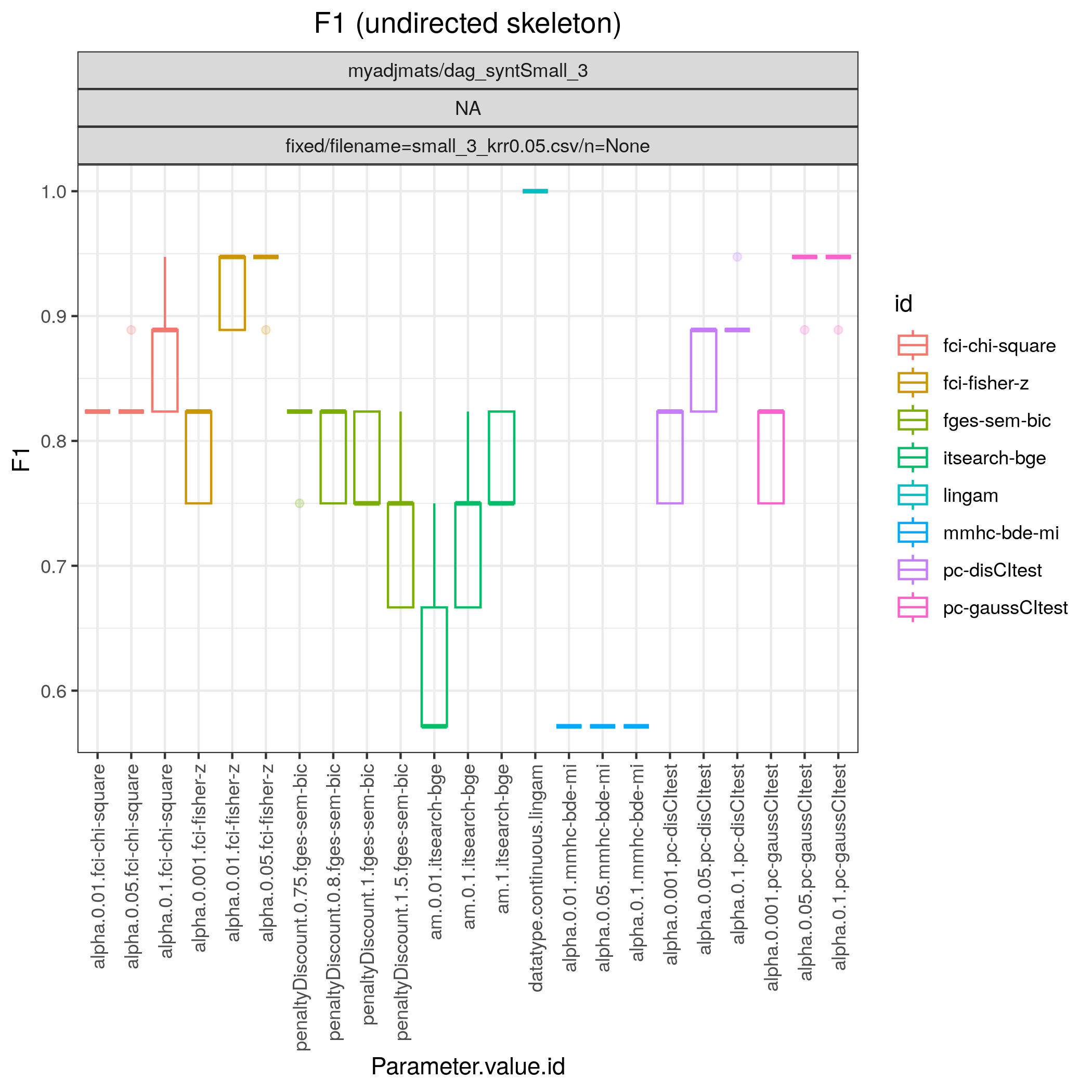}
	\caption{Synthetic 5 nodes data, $k$-RR C-wise mechanism, max probability 0.05.}
\end{minipage}

\end{figure}

\begin{figure}[H]
    \centering
   \begin{minipage}{0.31\linewidth}
\centering
  \includegraphics[scale=0.34]{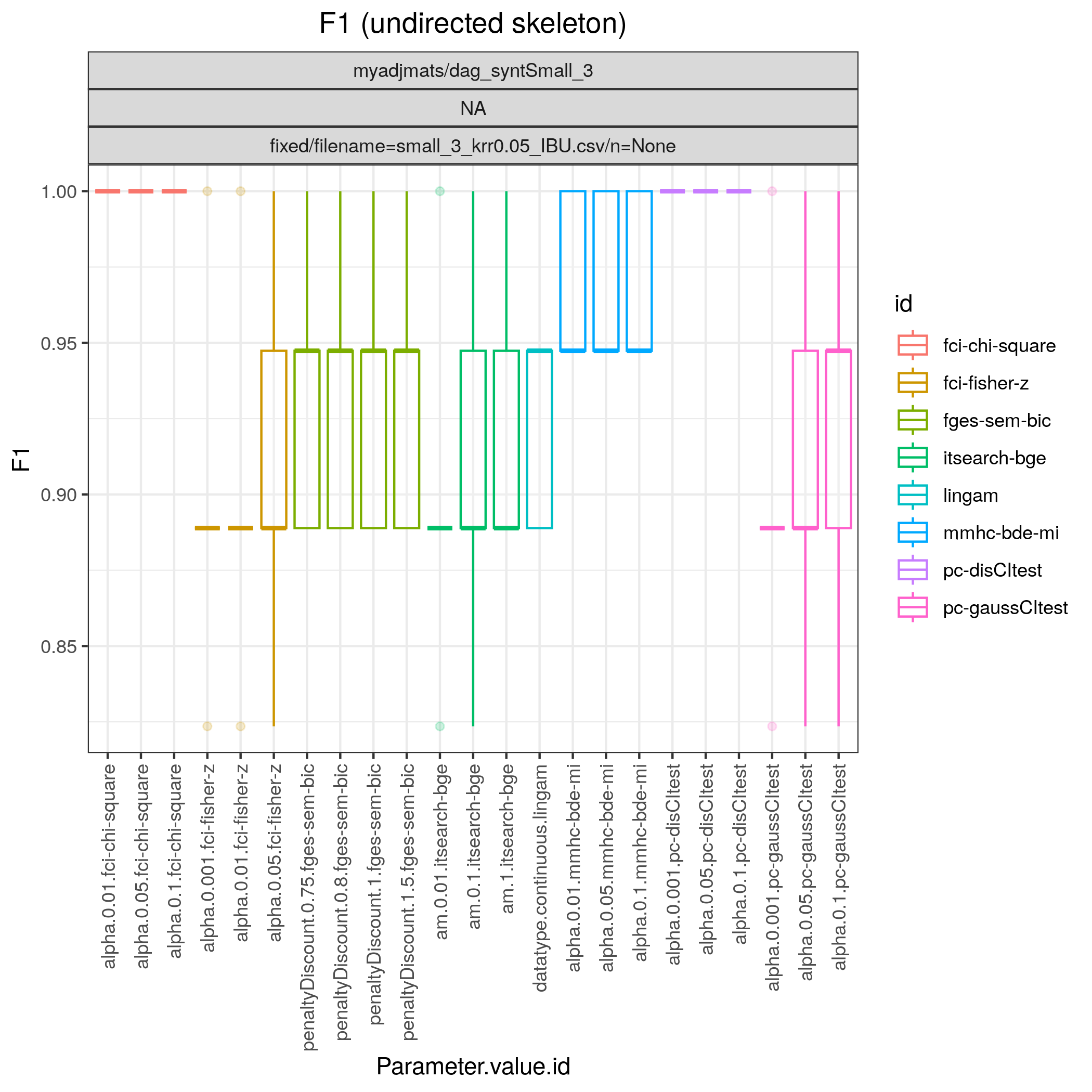}
	\caption{Synthetic 5 nodes data, $k$-RR C-wise IBU mechanism, max probability 0.05.}
\end{minipage}
\begin{minipage}{0.31\linewidth}
\centering
  \includegraphics[scale=0.34]{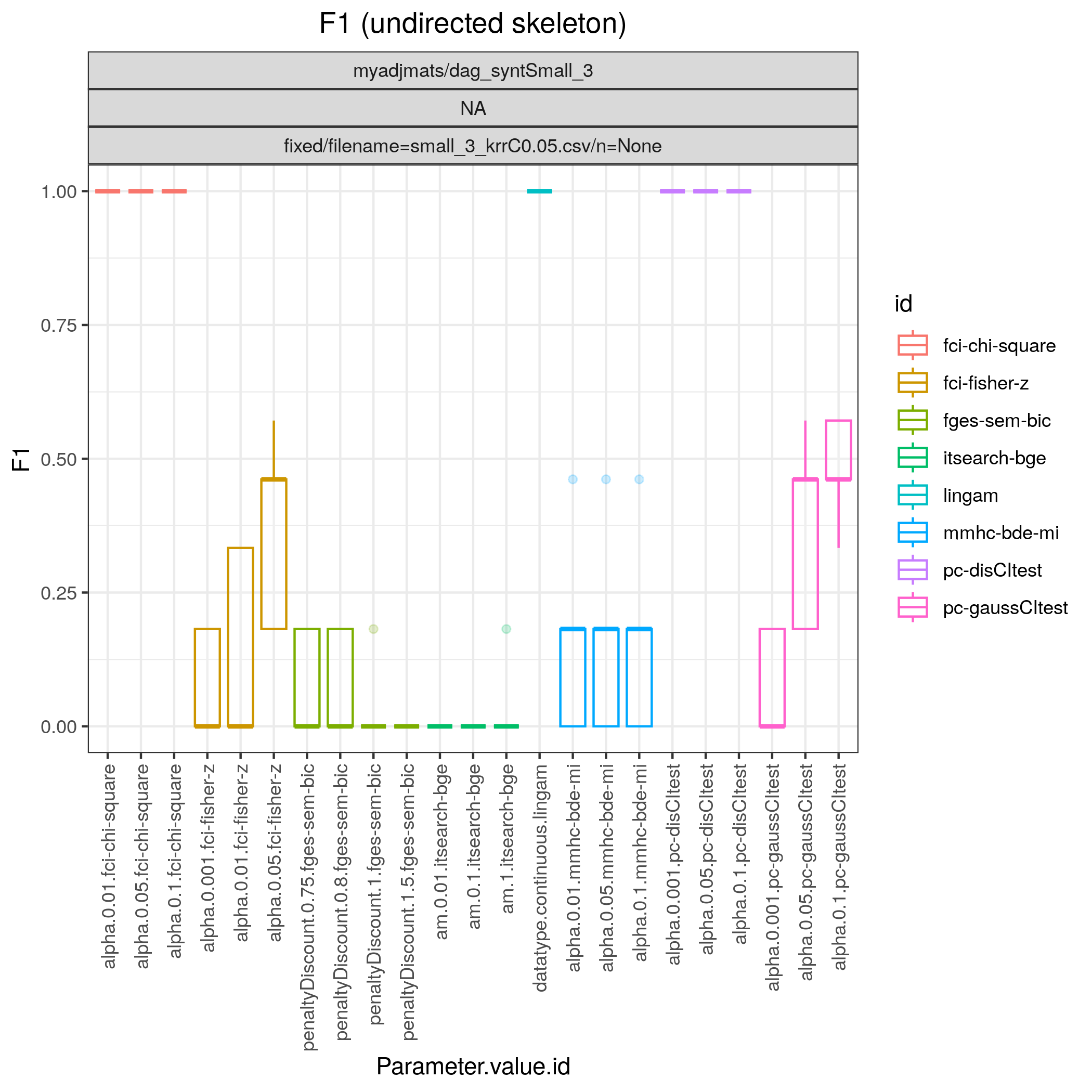}
	\caption{Synthetic 5 nodes data, $k$-RR Comb mechanism, max probability 0.05.}
\end{minipage}
\begin{minipage}{0.31\linewidth}
\centering
  \includegraphics[scale=0.34]{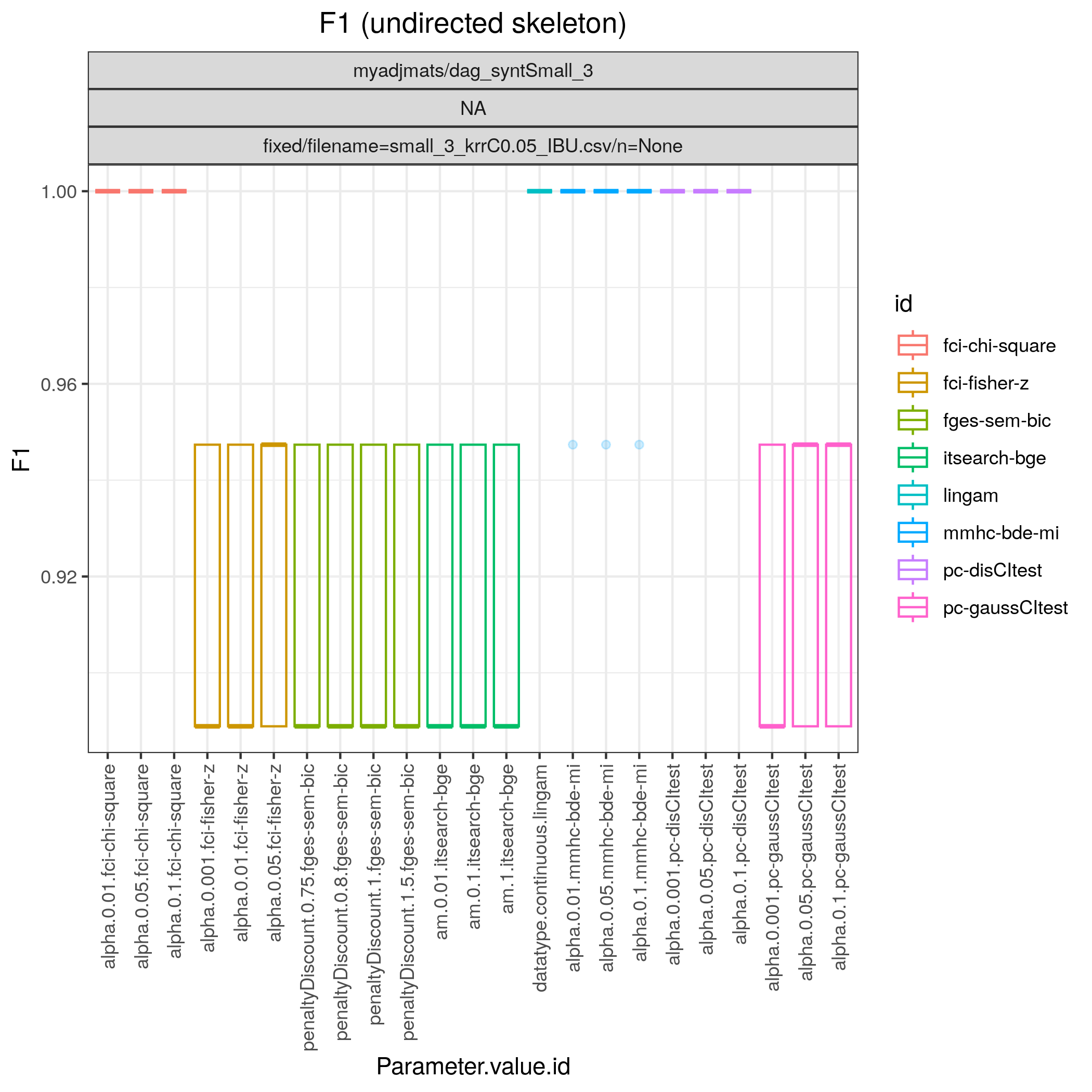}
	\caption{Synthetic 5 nodes data, $k$-RR Comb IBU mechanism, max probability 0.05.}
\end{minipage}
\end{figure}

\noindent
\begin{figure}[H]
\begin{minipage}{0.31\linewidth}
\centering
		\includegraphics[scale=0.34]{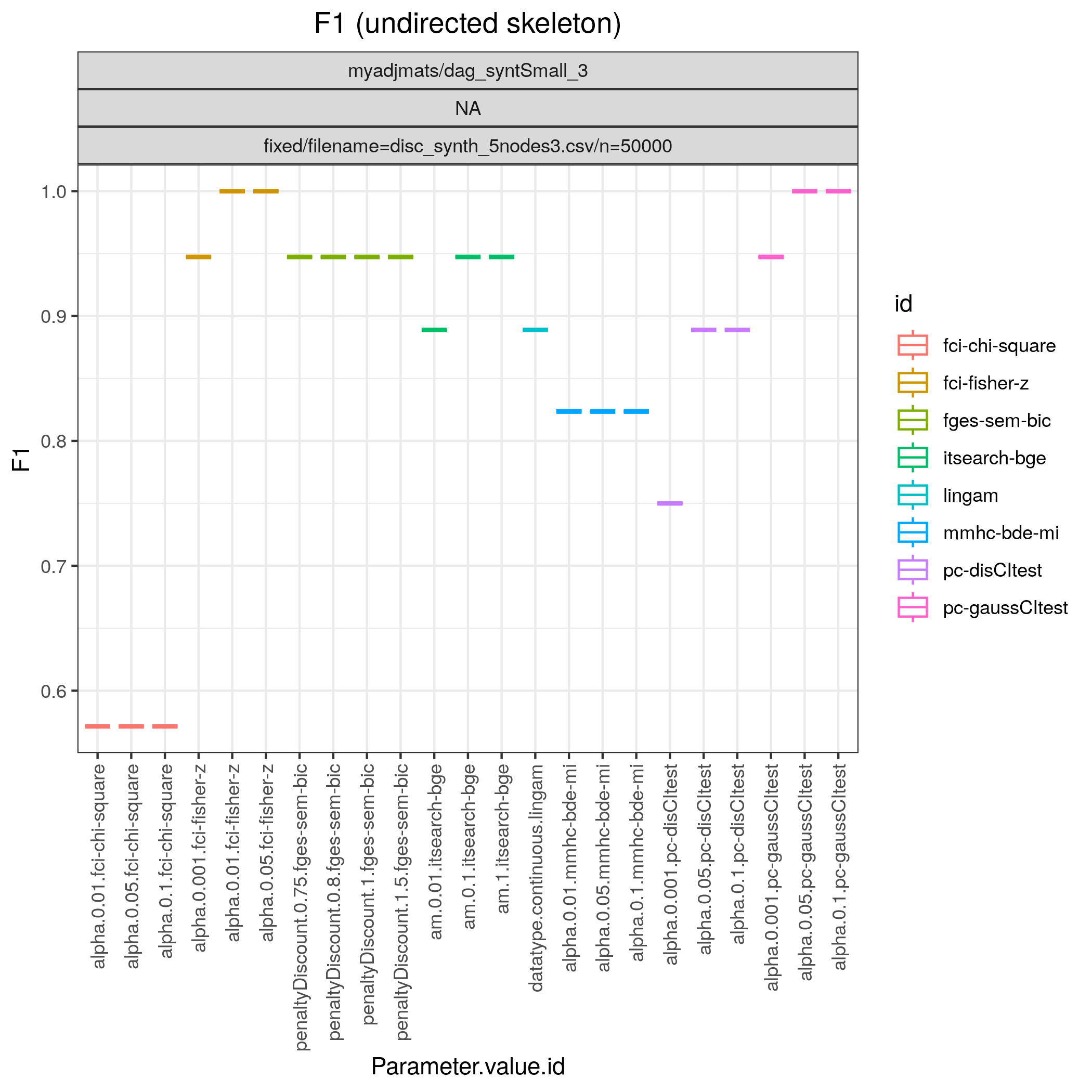}
	\caption{F1 Scores on the Synthetic 5 nodes data set. Discretized, no noise.}
\end{minipage}
\begin{minipage}{0.31\linewidth}
\centering
		\includegraphics[scale=0.34]{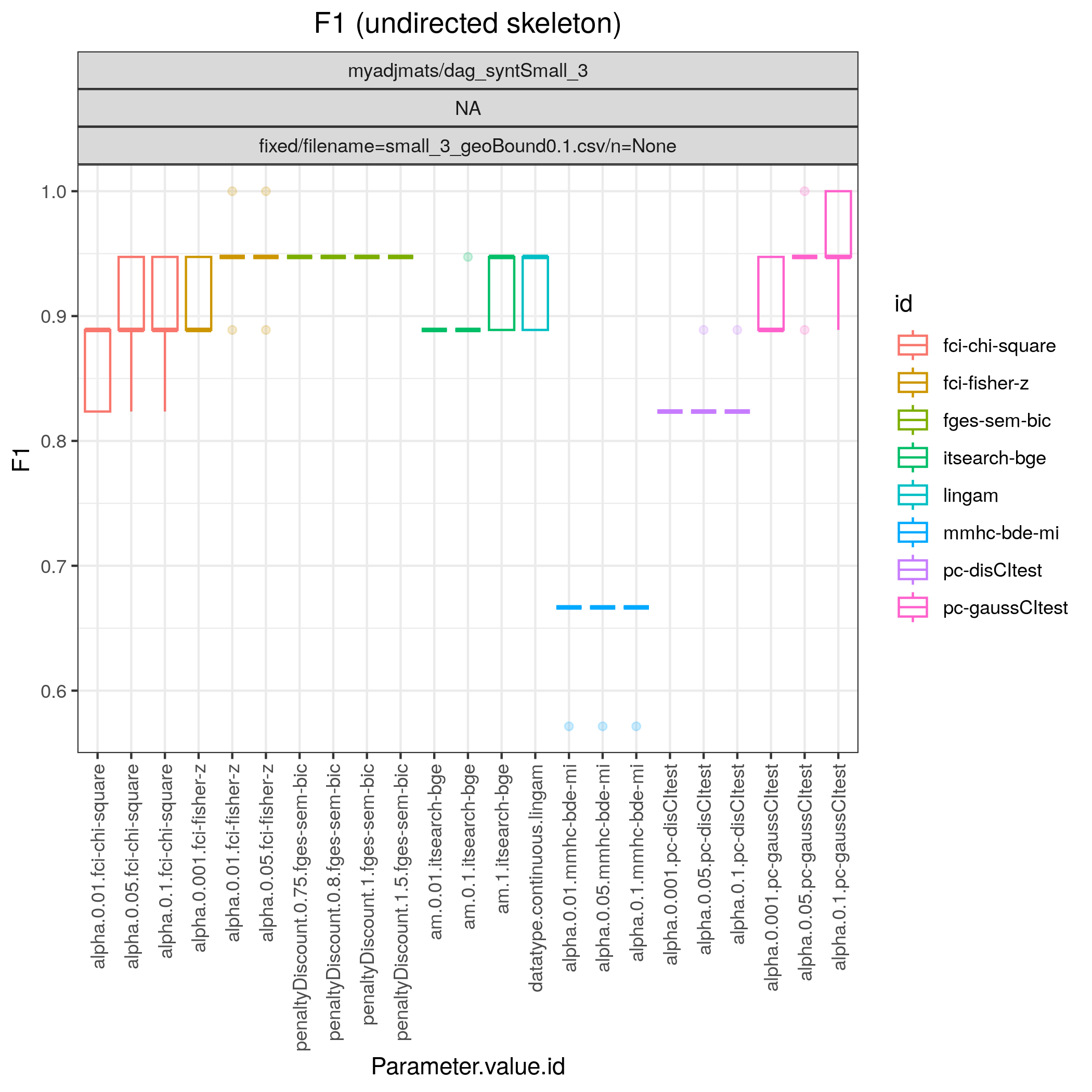}
	\caption{Synthetic 5 nodes data, Geo C-wise mechanism, max probability 0.1.}
\end{minipage}
\begin{minipage}{0.31\linewidth}
\centering
  \includegraphics[scale=0.34]{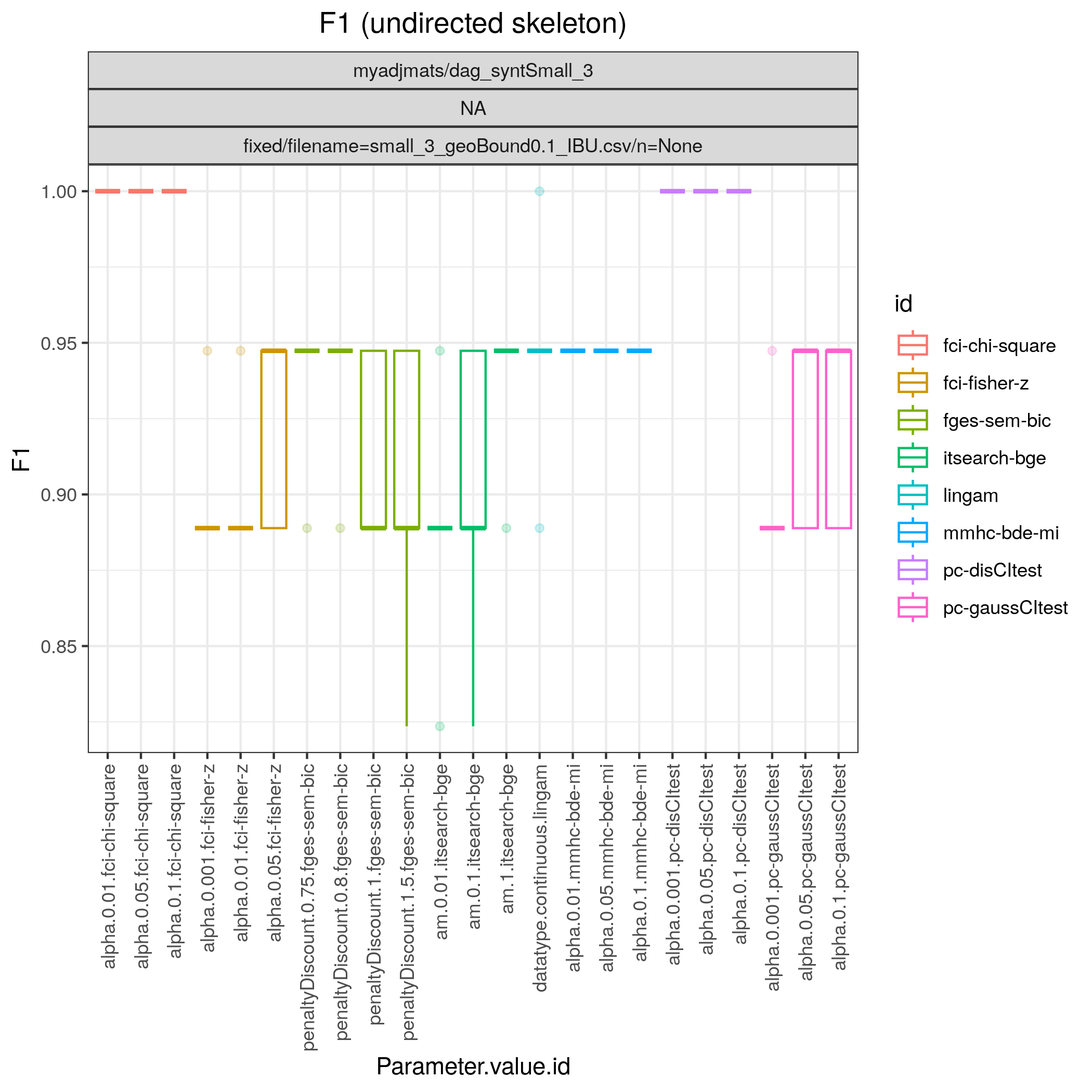}
	\caption{Synthetic 5 nodes data, Geo C-wise IBU mechanism, max probability 0.1.}
 \end{minipage}
\begin{minipage}{0.31\linewidth}
\centering
  \includegraphics[scale=0.34]{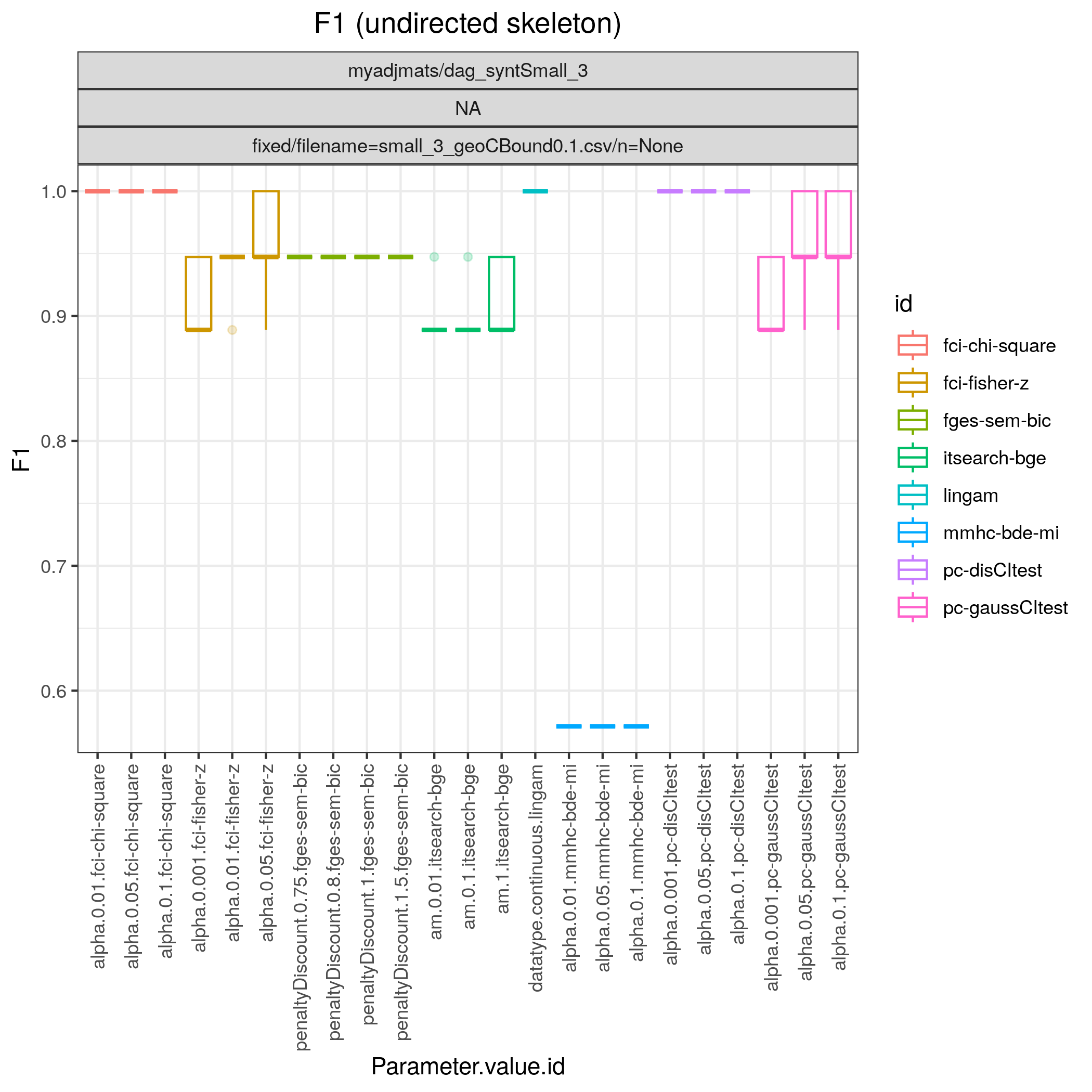}
	\caption{Synthetic 5 nodes data, Geo Comb mechanism, max probability 0.1.}
\end{minipage}
\begin{minipage}{0.31\linewidth}
\centering
  \includegraphics[scale=0.34]{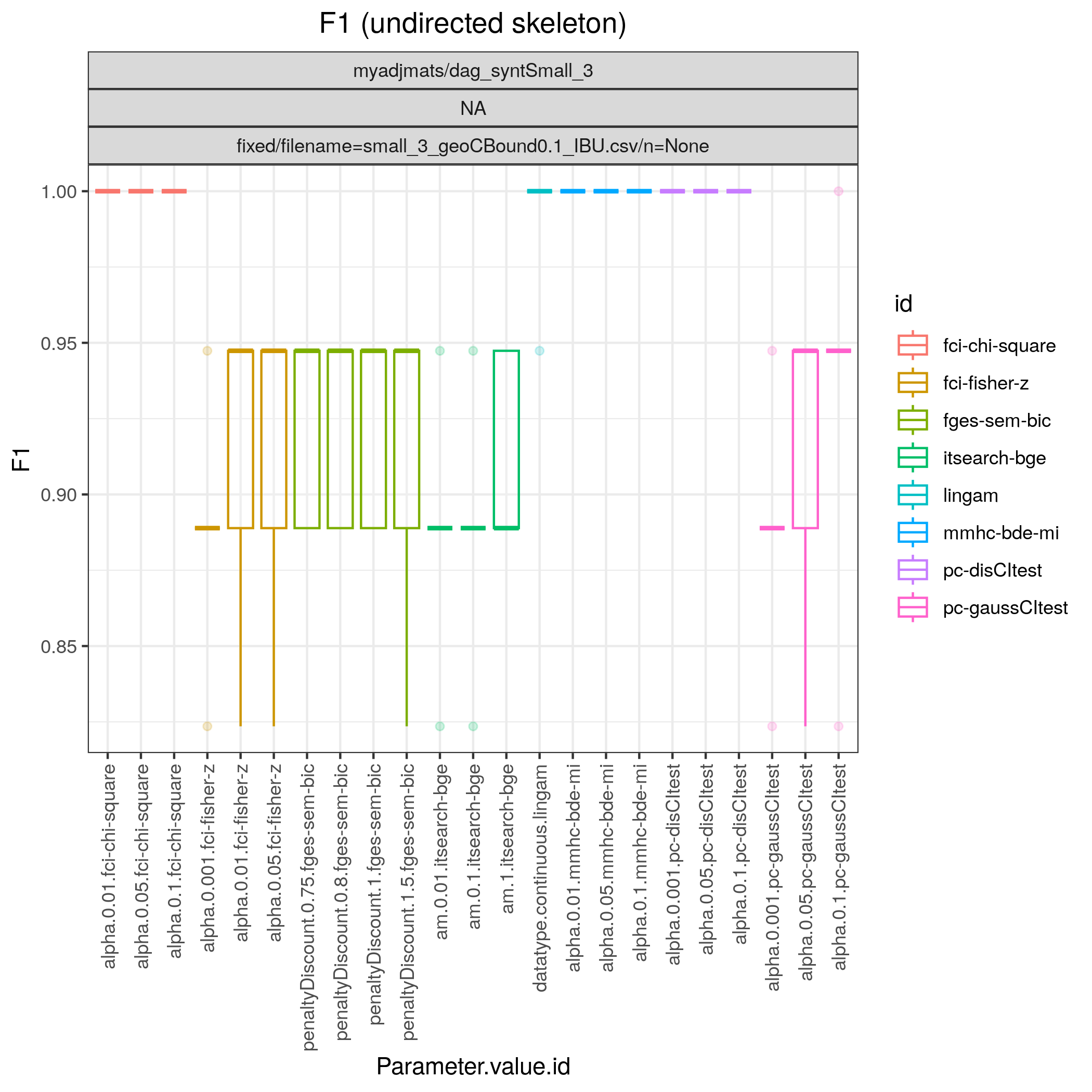}
	\caption{Synthetic 5 nodes data, Geo Comb IBU mechanism, max probability 0.1.}
\end{minipage}
\begin{minipage}{0.31\linewidth}
\centering
  \includegraphics[scale=0.34]{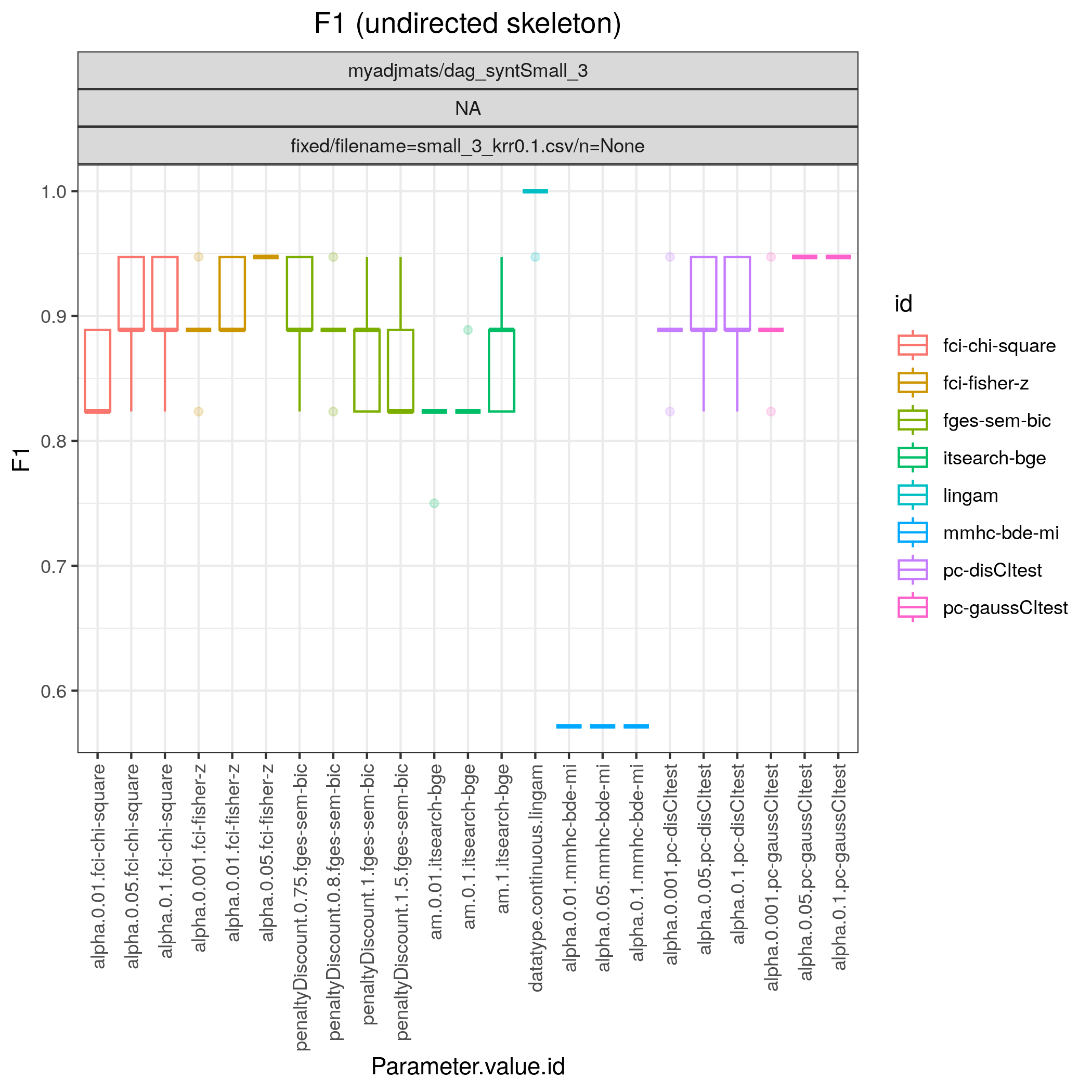}
	\caption{Synthetic 5 nodes data, $k$-RR C-wise mechanism, max probability 0.1.}
\end{minipage}

\end{figure}

\begin{figure}[H]
    \centering
   \begin{minipage}{0.31\linewidth}
\centering
  \includegraphics[scale=0.34]{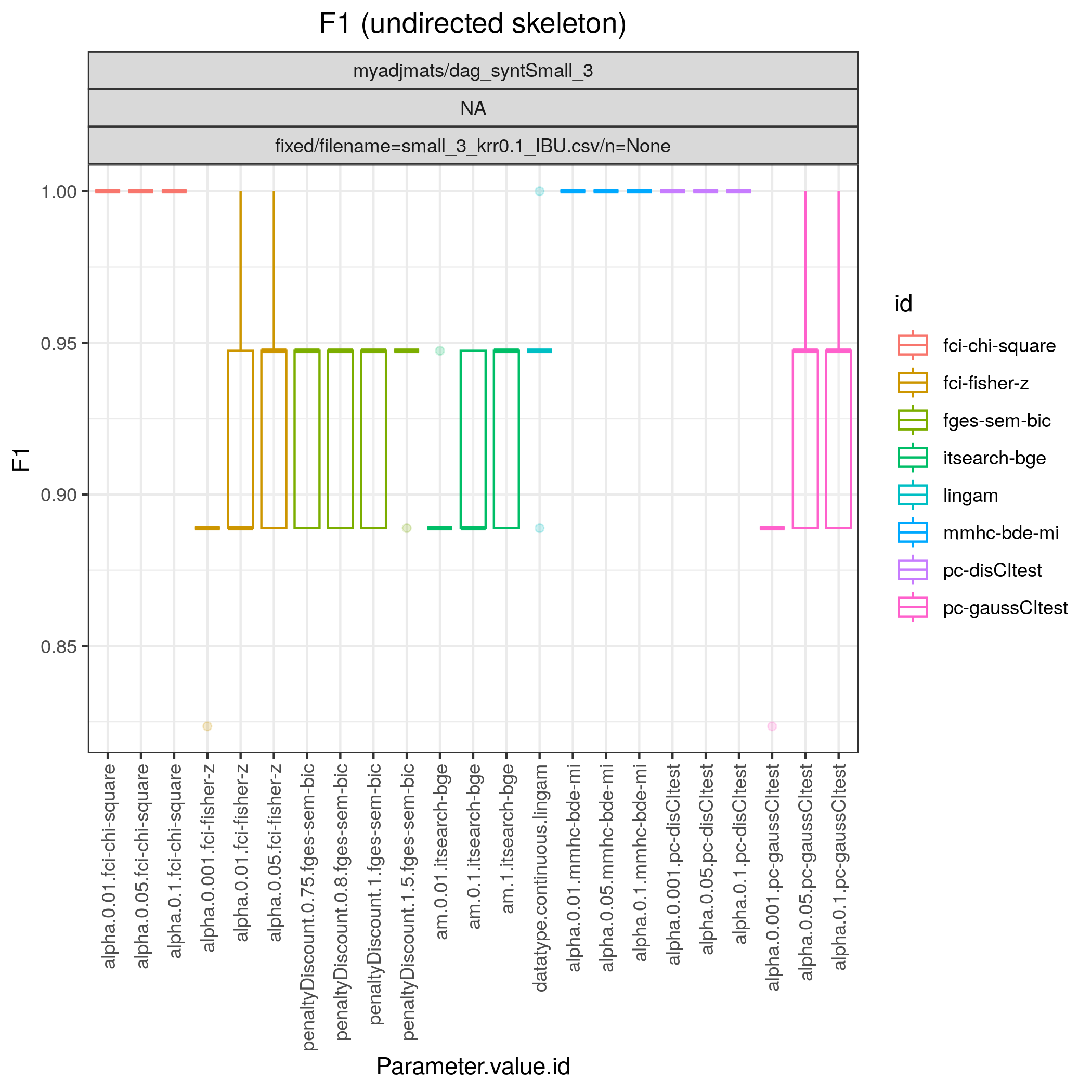}
	\caption{Synthetic 5 nodes data, $k$-RR C-wise IBU mechanism, max probability 0.1.}
\end{minipage}
\begin{minipage}{0.31\linewidth}
\centering
  \includegraphics[scale=0.34]{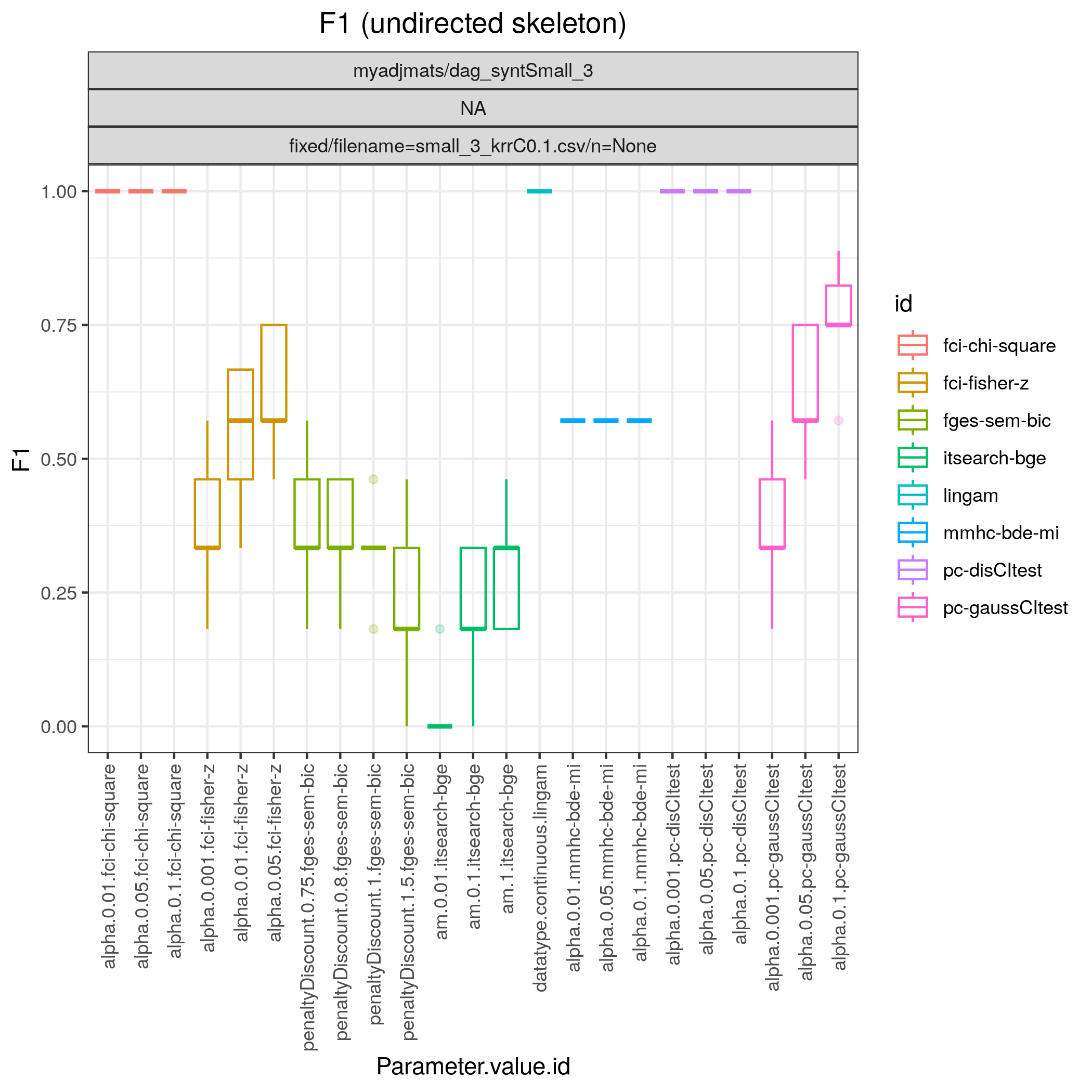}
	\caption{Synthetic 5 nodes data, $k$-RR Comb mechanism, max probability 0.1.}
\end{minipage}
\begin{minipage}{0.31\linewidth}
\centering
  \includegraphics[scale=0.34]{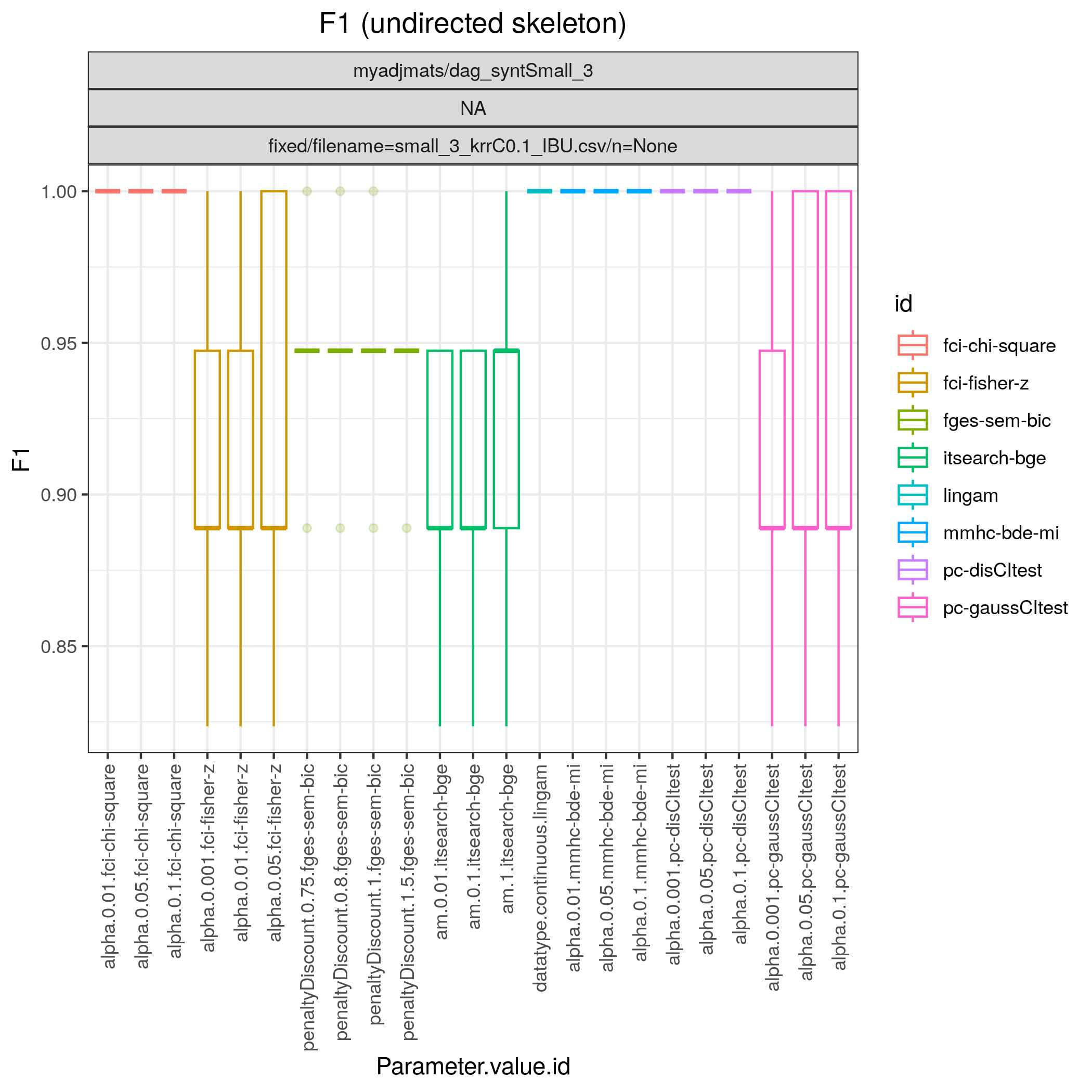}
	\caption{Synthetic 5 nodes data, $k$-RR Comb IBU mechanism, max probability 0.1.}
\end{minipage}
\end{figure}

\noindent
\begin{figure}[H]
\begin{minipage}{0.31\linewidth}
\centering
		\includegraphics[scale=0.34]{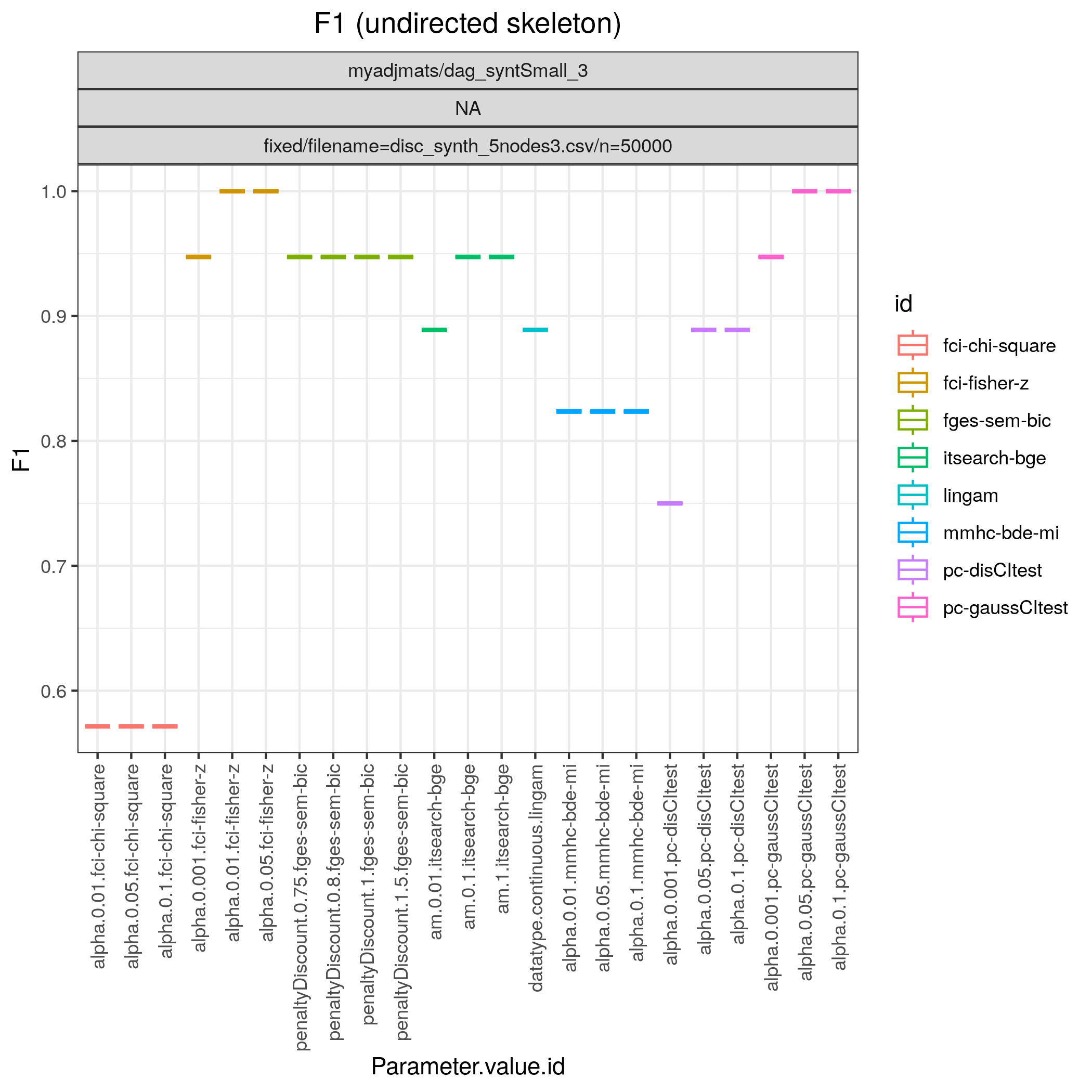}
	\caption{F1 Scores on the Synthetic 5 nodes data set. Discretized, no noise.}
\end{minipage}
\begin{minipage}{0.31\linewidth}
\centering
		\includegraphics[scale=0.34]{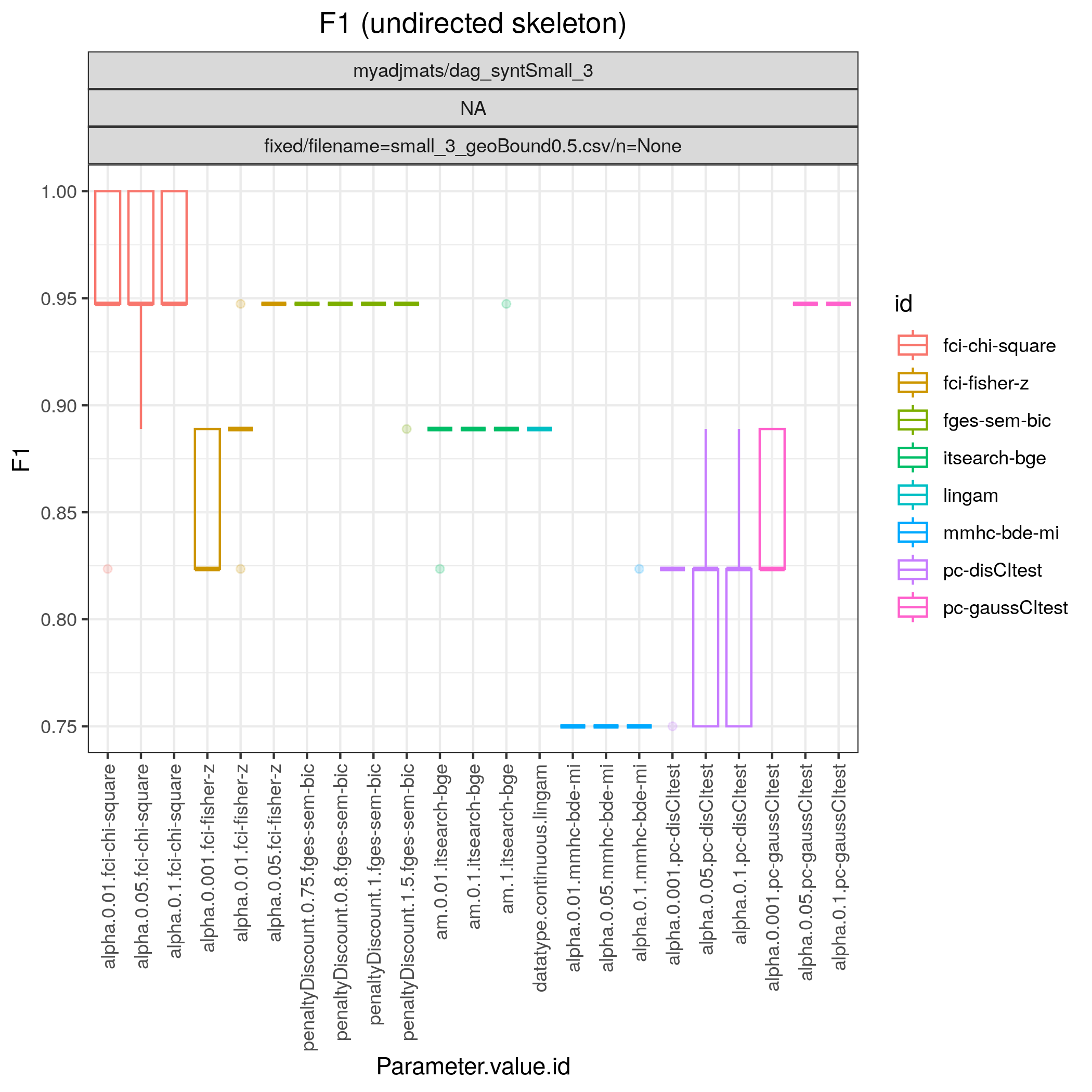}
	\caption{Synthetic 5 nodes data, Geo C-wise mechanism, max probability 0.5.}
\end{minipage}
\begin{minipage}{0.31\linewidth}
\centering
  \includegraphics[scale=0.34]{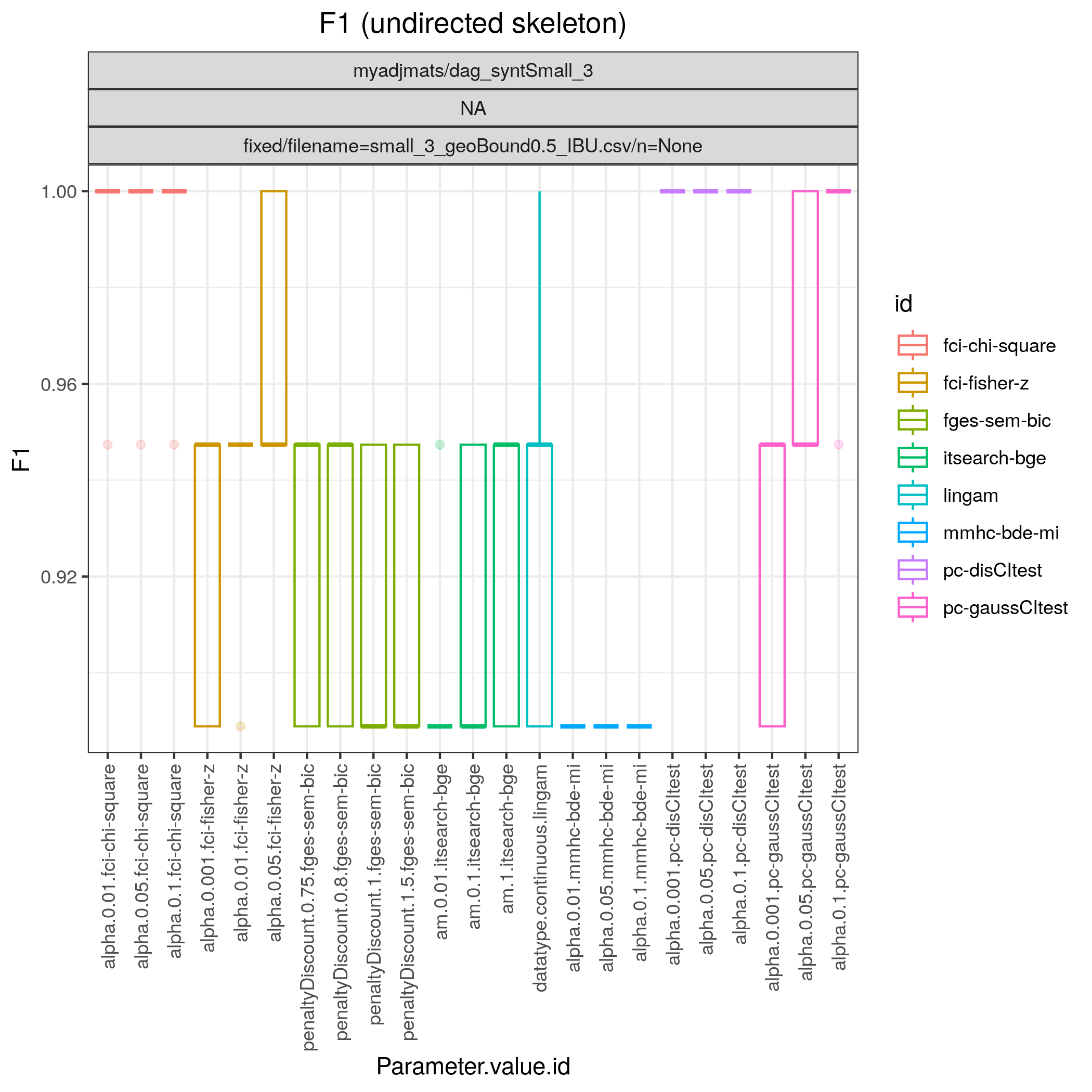}
	\caption{Synthetic 5 nodes data, Geo C-wise IBU mechanism, max probability 0.5.}
 \end{minipage}
\begin{minipage}{0.31\linewidth}
\centering
  \includegraphics[scale=0.34]{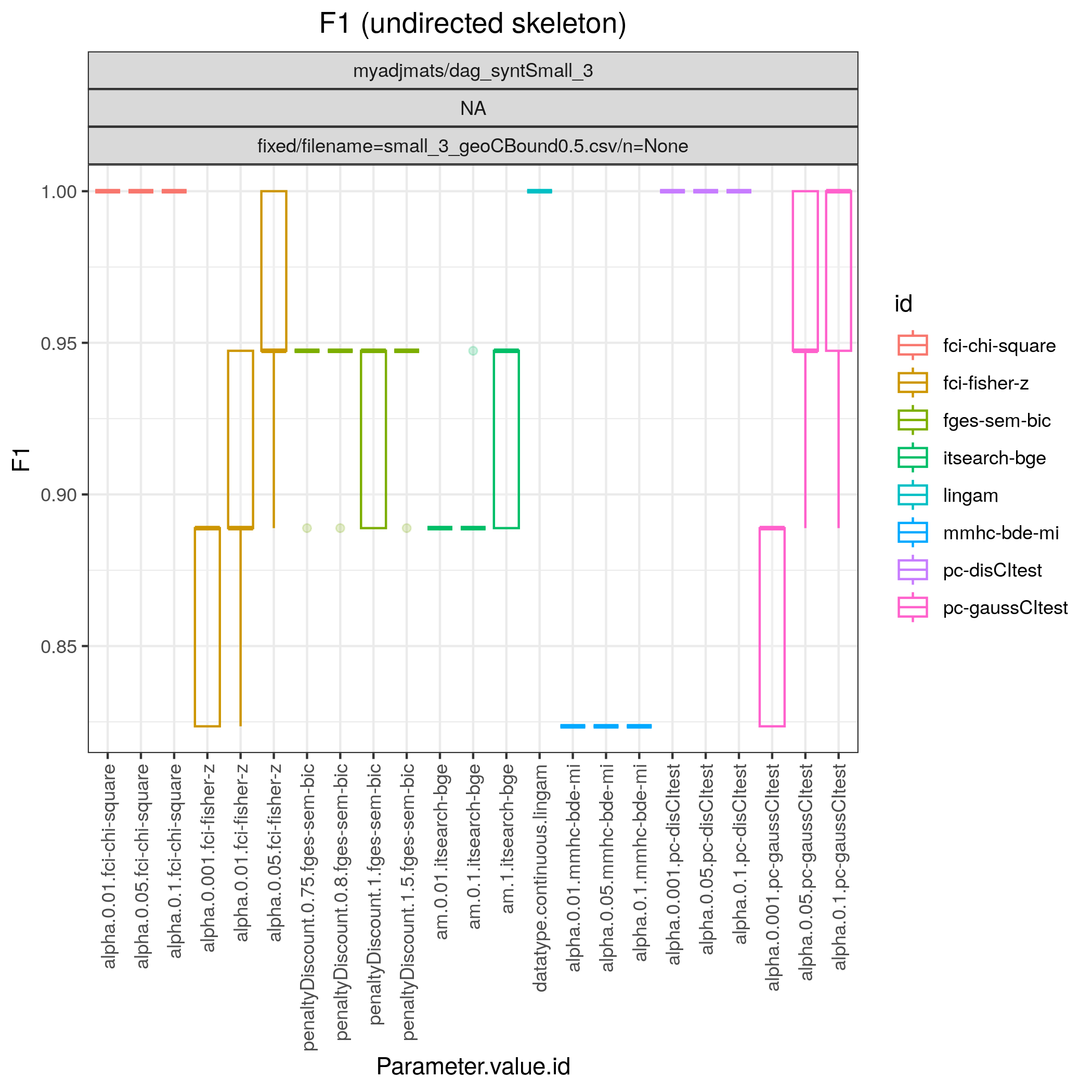}
	\caption{Synthetic 5 nodes data, Geo Comb mechanism, max probability 0.5.}
\end{minipage}
\begin{minipage}{0.31\linewidth}
\centering
  \includegraphics[scale=0.34]{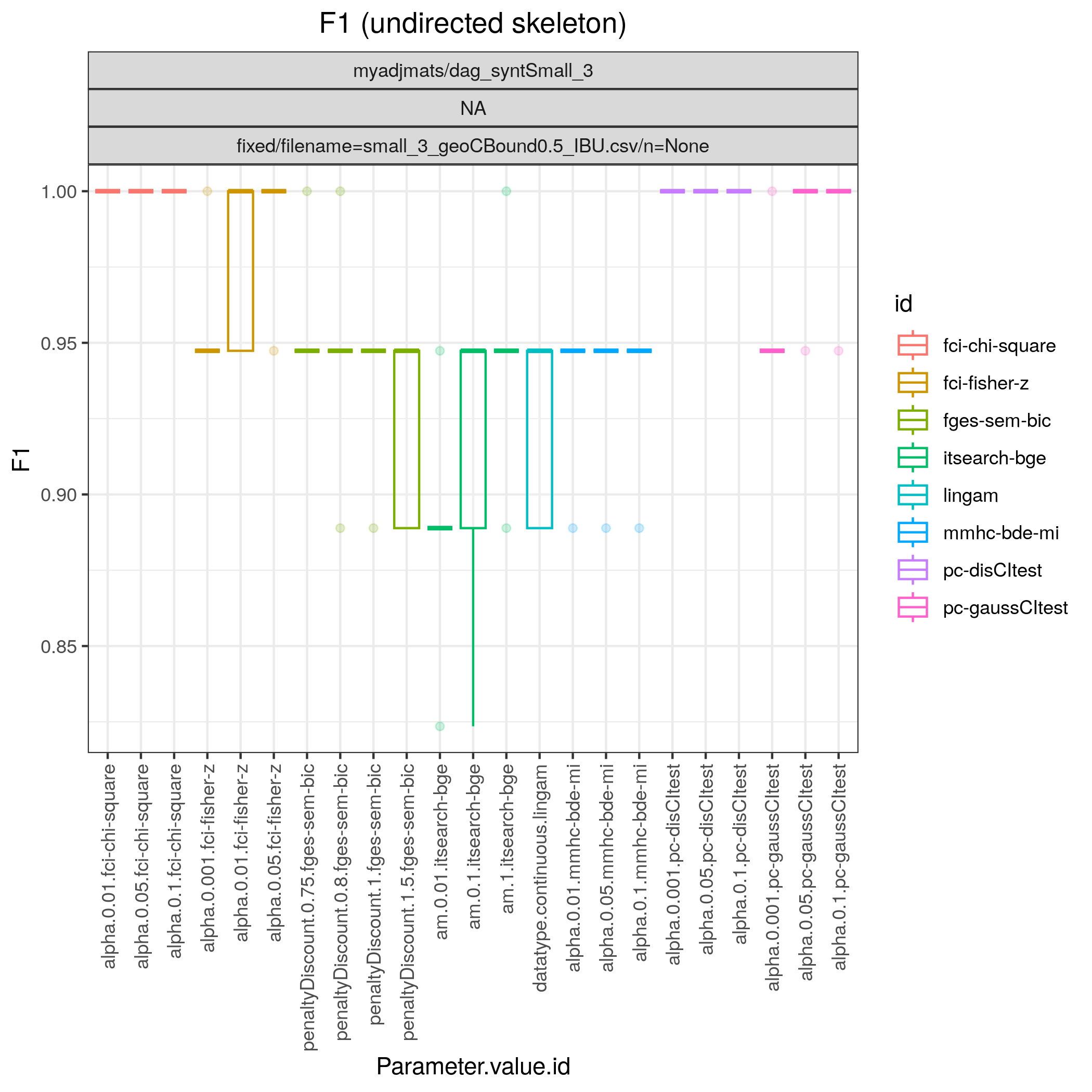}
	\caption{Synthetic 5 nodes data, Geo Comb IBU mechanism, max probability 0.5.}
\end{minipage}
\begin{minipage}{0.31\linewidth}
\centering
  \includegraphics[scale=0.34]{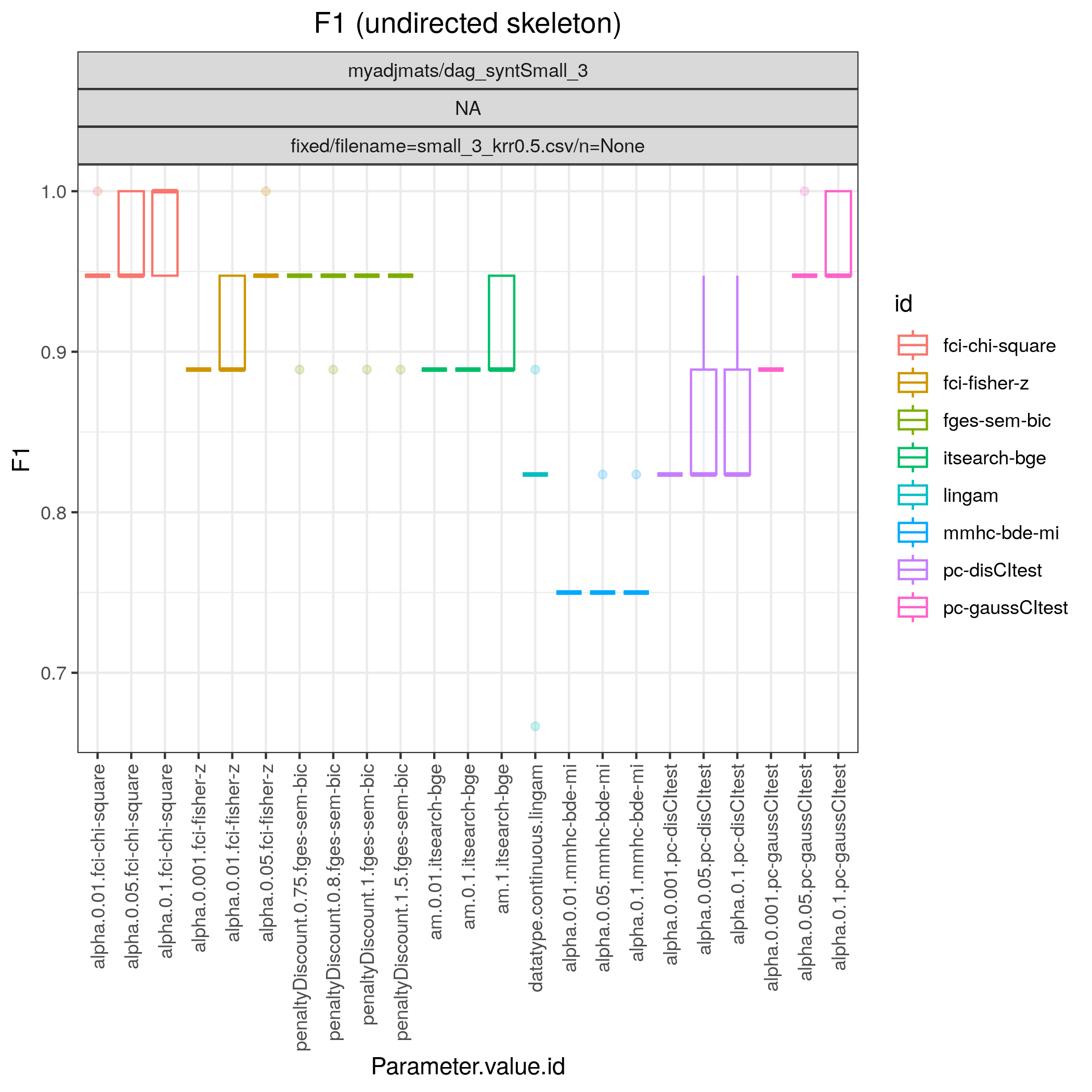}
	\caption{Synthetic 5 nodes data, $k$-RR C-wise mechanism, max probability 0.5.}
\end{minipage}

\end{figure}

\begin{figure}[H]
    \centering
   \begin{minipage}{0.31\linewidth}
\centering
  \includegraphics[scale=0.34]{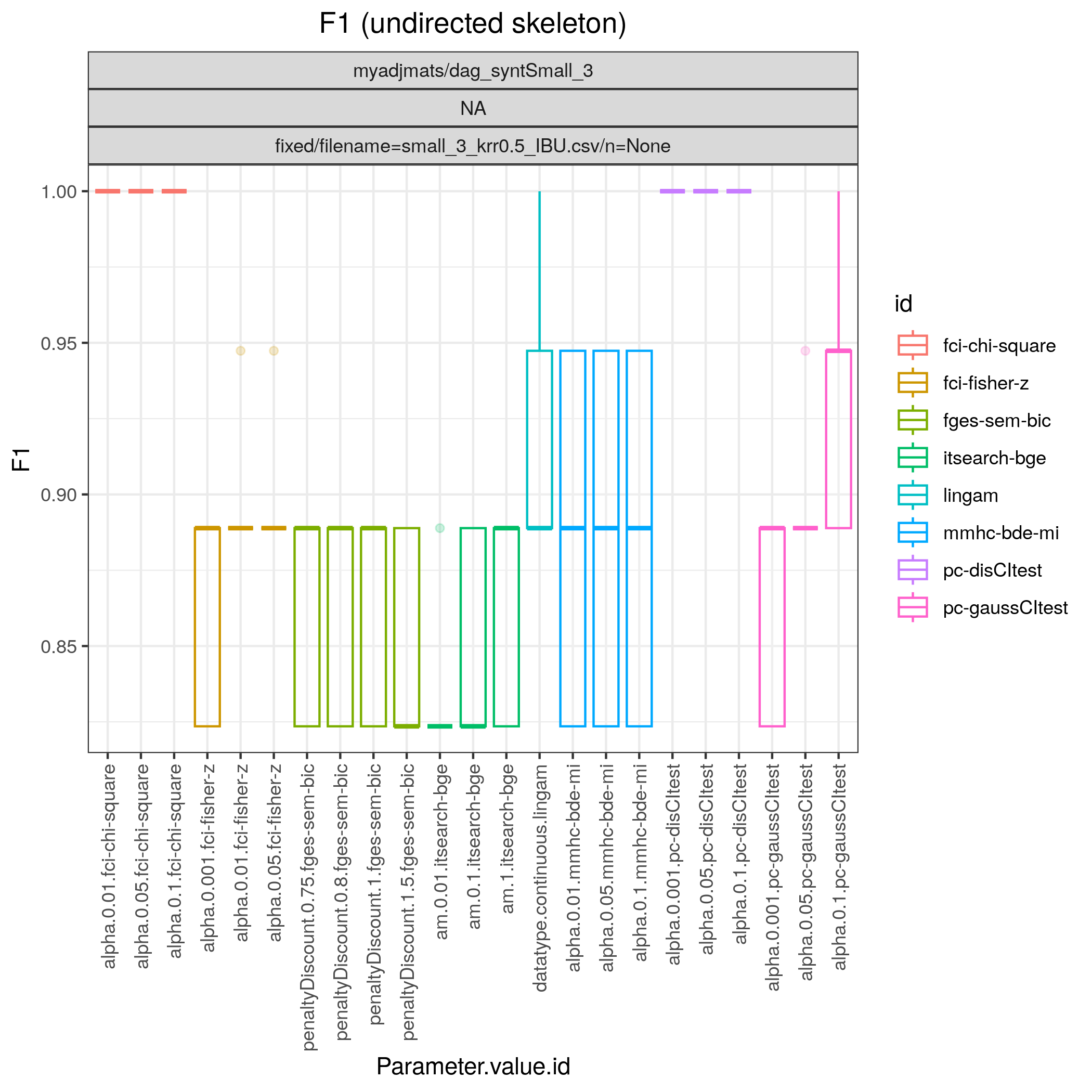}
	\caption{Synthetic 5 nodes data, $k$-RR C-wise IBU mechanism, max probability 0.5.}
\end{minipage}
\begin{minipage}{0.31\linewidth}
\centering
  \includegraphics[scale=0.34]{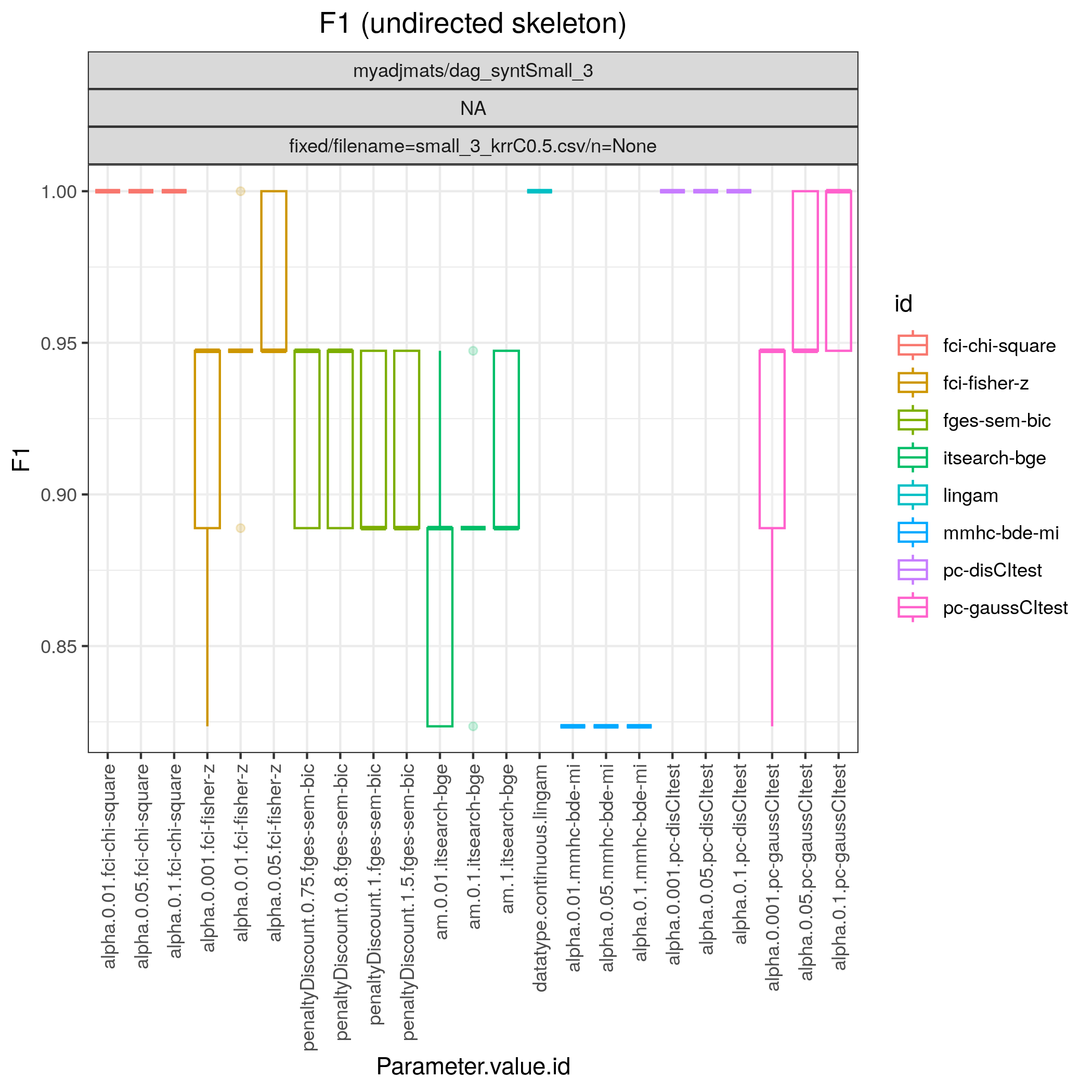}
	\caption{Synthetic 5 nodes data, $k$-RR Comb mechanism, max probability 0.5.}
\end{minipage}
\begin{minipage}{0.31\linewidth}
\centering
  \includegraphics[scale=0.34]{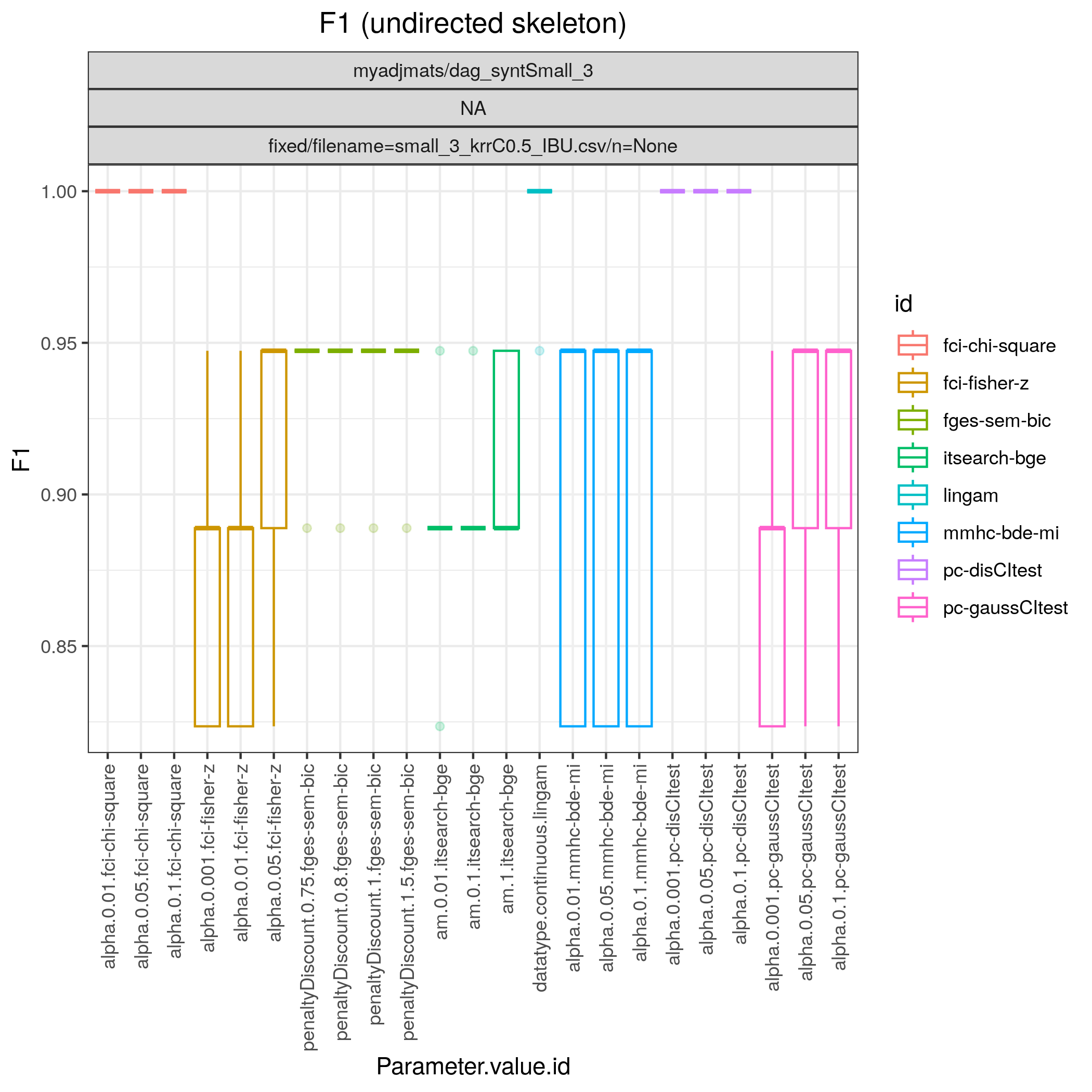}
	\caption{Synthetic 5 nodes data, $k$-RR Comb IBU mechanism, max probability 0.5.}
\end{minipage}
\end{figure}

\subsection{SHD Score results Synthetic 5 nodes data set}
\noindent
\begin{figure}[H]
\begin{minipage}{0.31\linewidth}
\centering
		\includegraphics[scale=0.34]{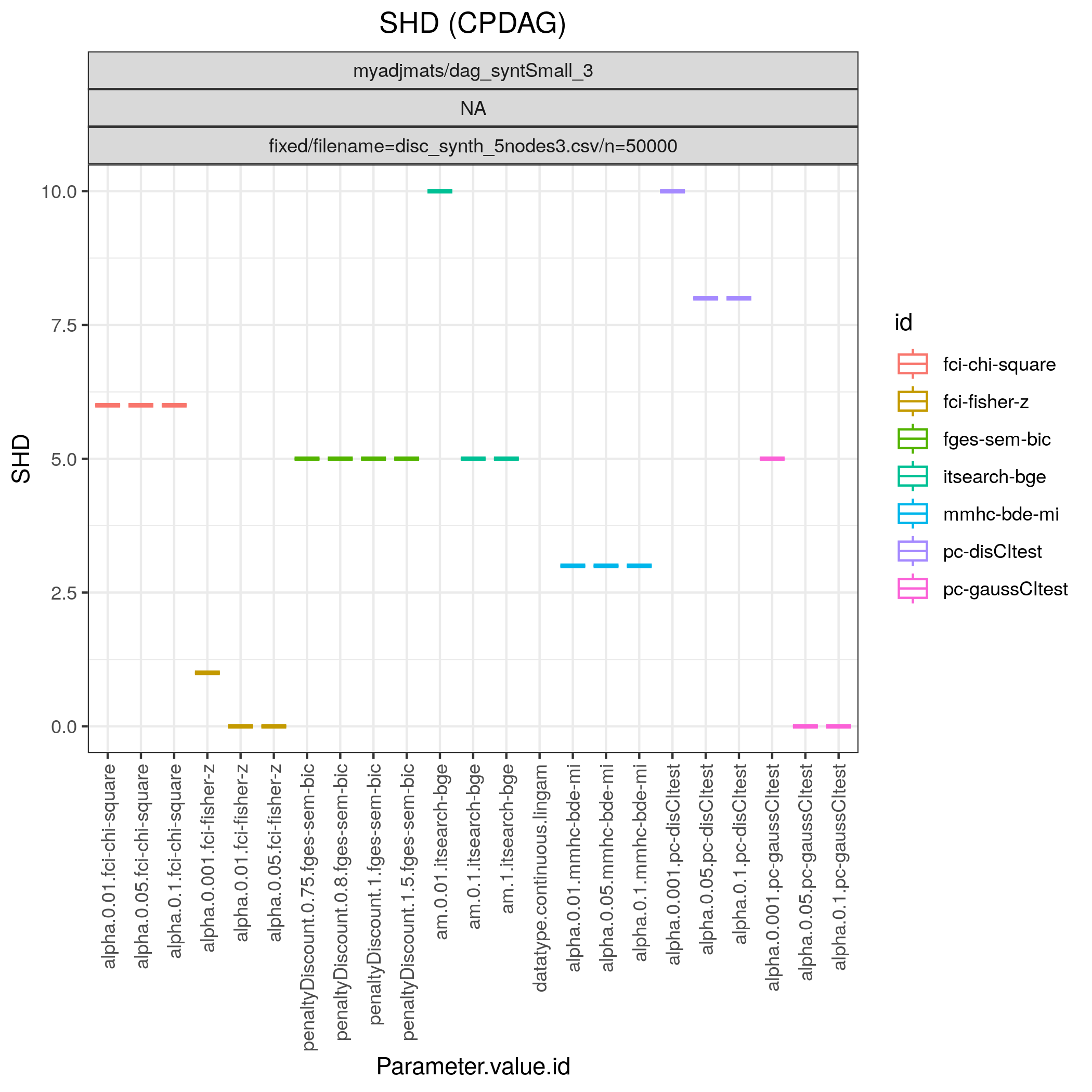}
	\caption{SHD Scores on the Synthetic 5 nodes data set. Discretized, no noise.}
\end{minipage}
\begin{minipage}{0.31\linewidth}
\centering
		\includegraphics[scale=0.34]{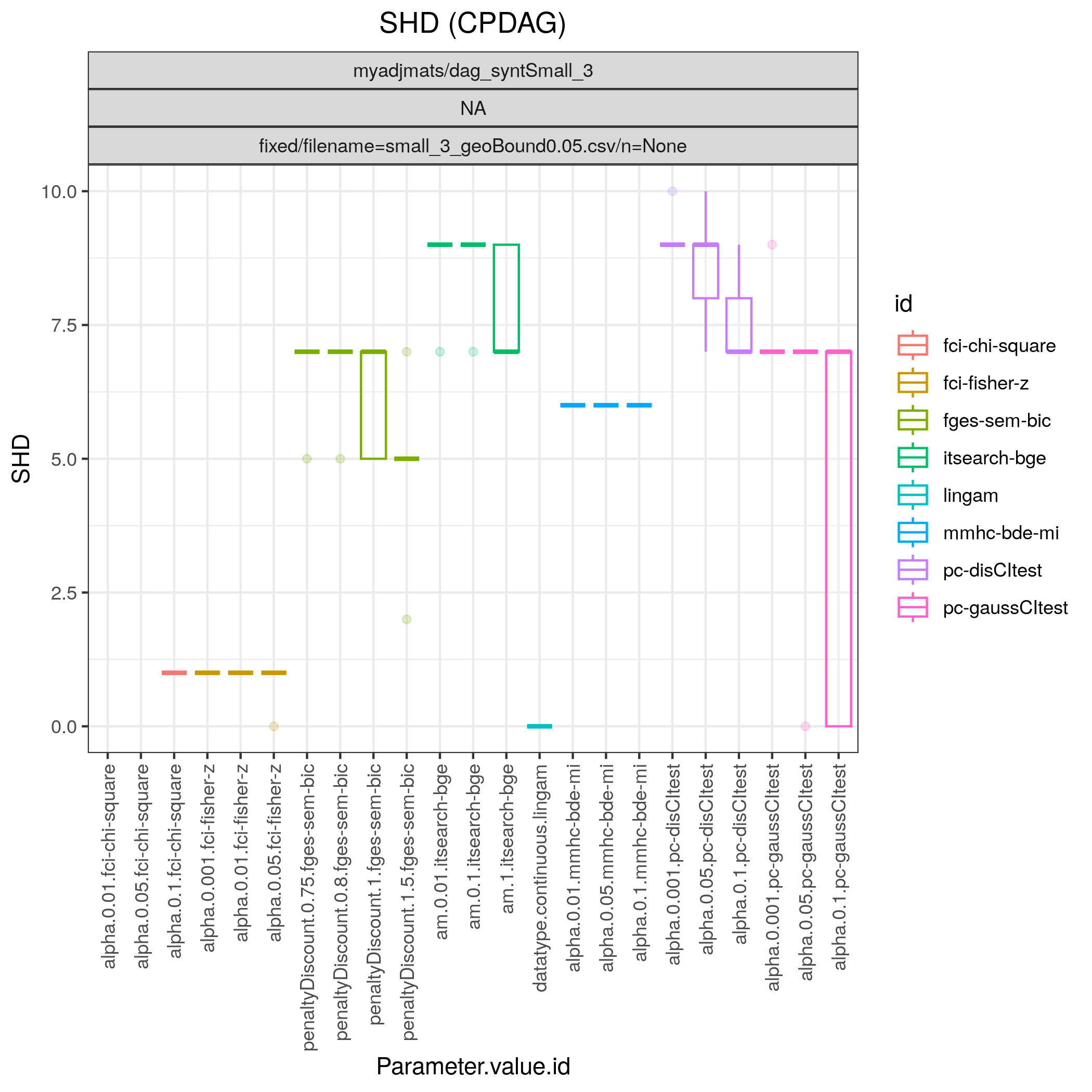}
	\caption{Synthetic 5 nodes data, Geo C-wise mechanism, max probability 0.05.}
\end{minipage}
\begin{minipage}{0.31\linewidth}
\centering
  \includegraphics[scale=0.34]{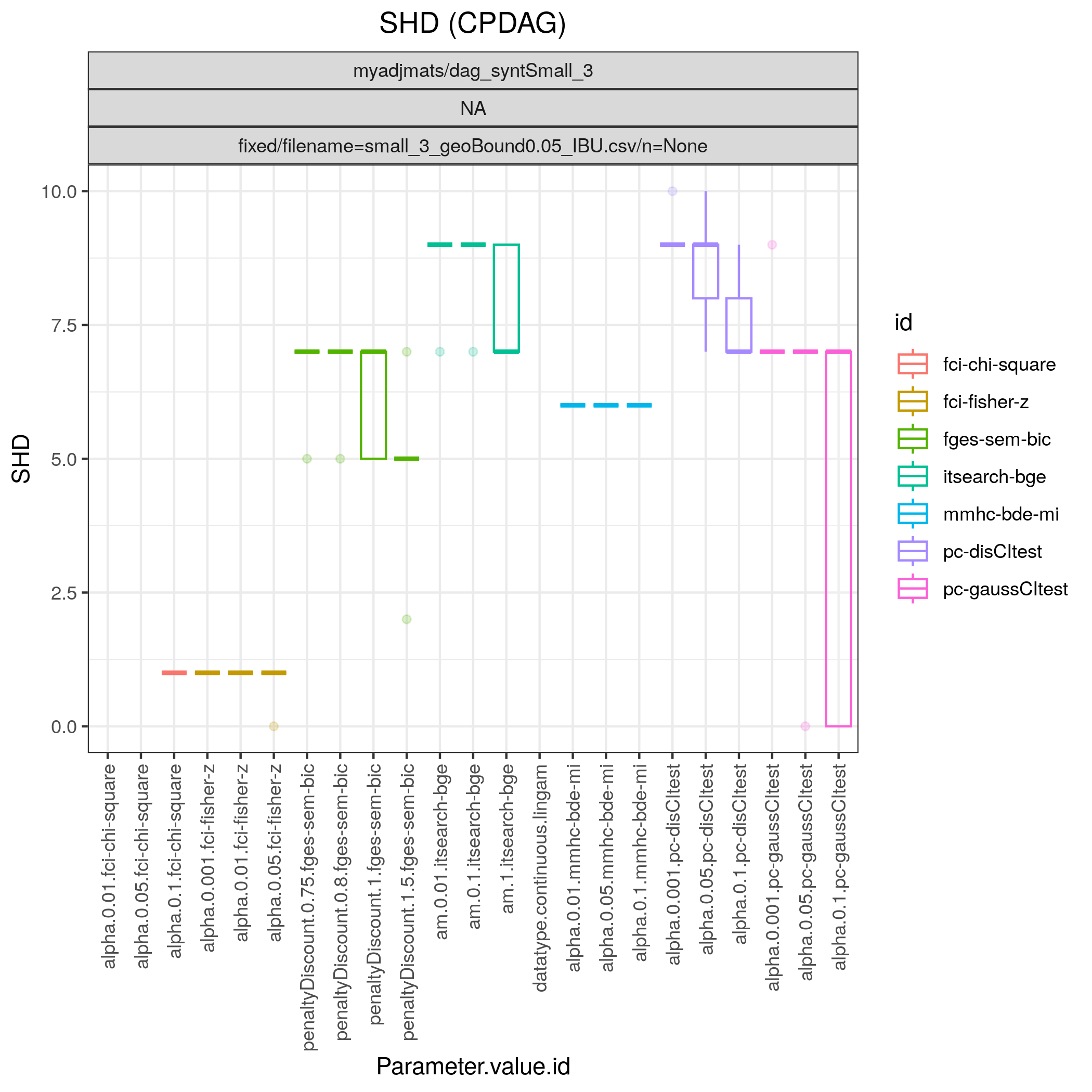}
	\caption{Synthetic 5 nodes data, Geo C-wise IBU mechanism, max probability 0.05.}
 \end{minipage}
\begin{minipage}{0.31\linewidth}
\centering
  \includegraphics[scale=0.34]{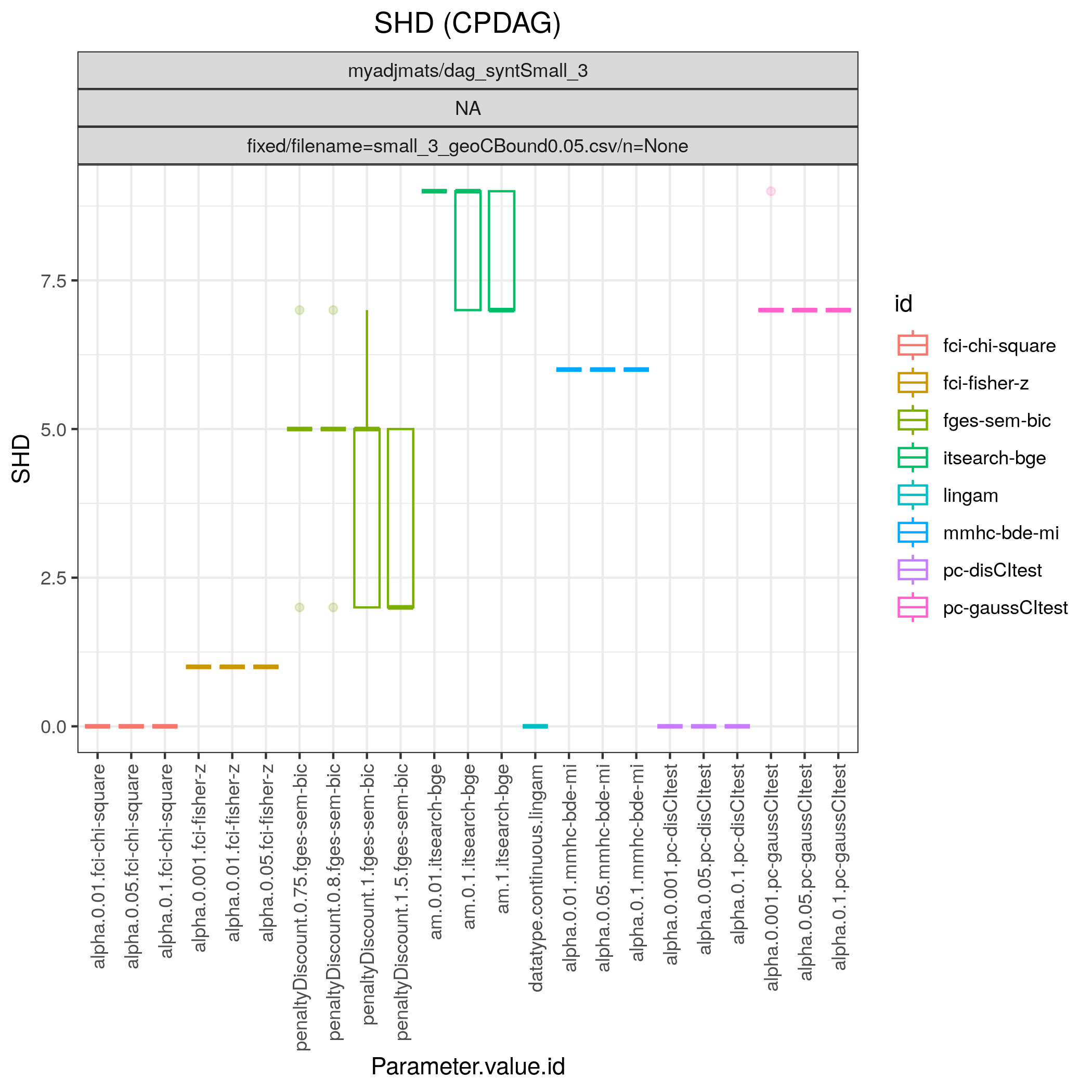}
	\caption{Synthetic 5 nodes data, Geo Comb mechanism, max probability 0.05.}
\end{minipage}
\begin{minipage}{0.31\linewidth}
\centering
  \includegraphics[scale=0.34]{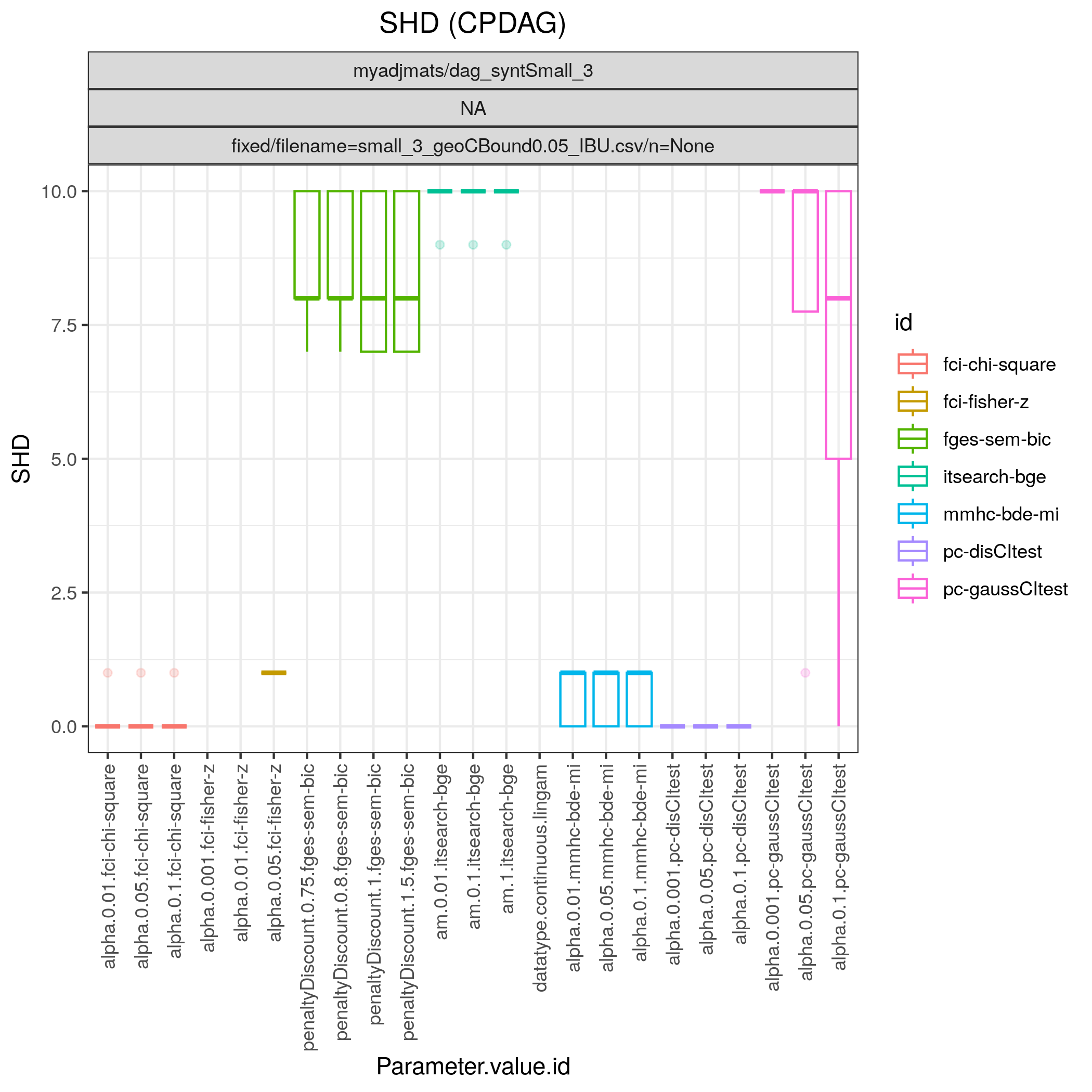}
	\caption{Synthetic 5 nodes data, Geo Comb IBU mechanism, max probability 0.05.}
\end{minipage}
\begin{minipage}{0.31\linewidth}
\centering
  \includegraphics[scale=0.34]{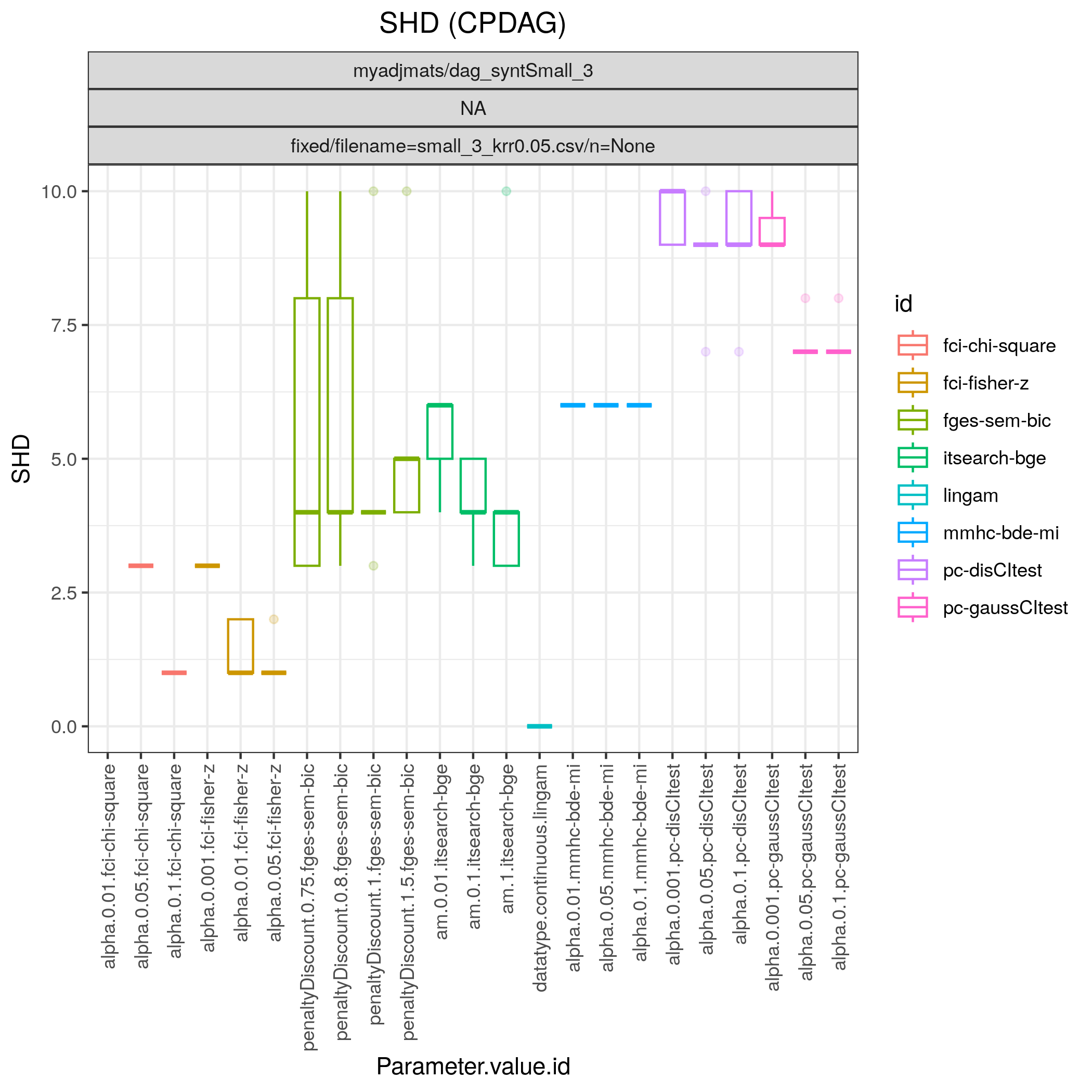}
	\caption{Synthetic 5 nodes data, $k$-RR C-wise mechanism, max probability 0.05.}
\end{minipage}

\end{figure}

\begin{figure}[H]
    \centering
   \begin{minipage}{0.31\linewidth}
\centering
  \includegraphics[scale=0.34]{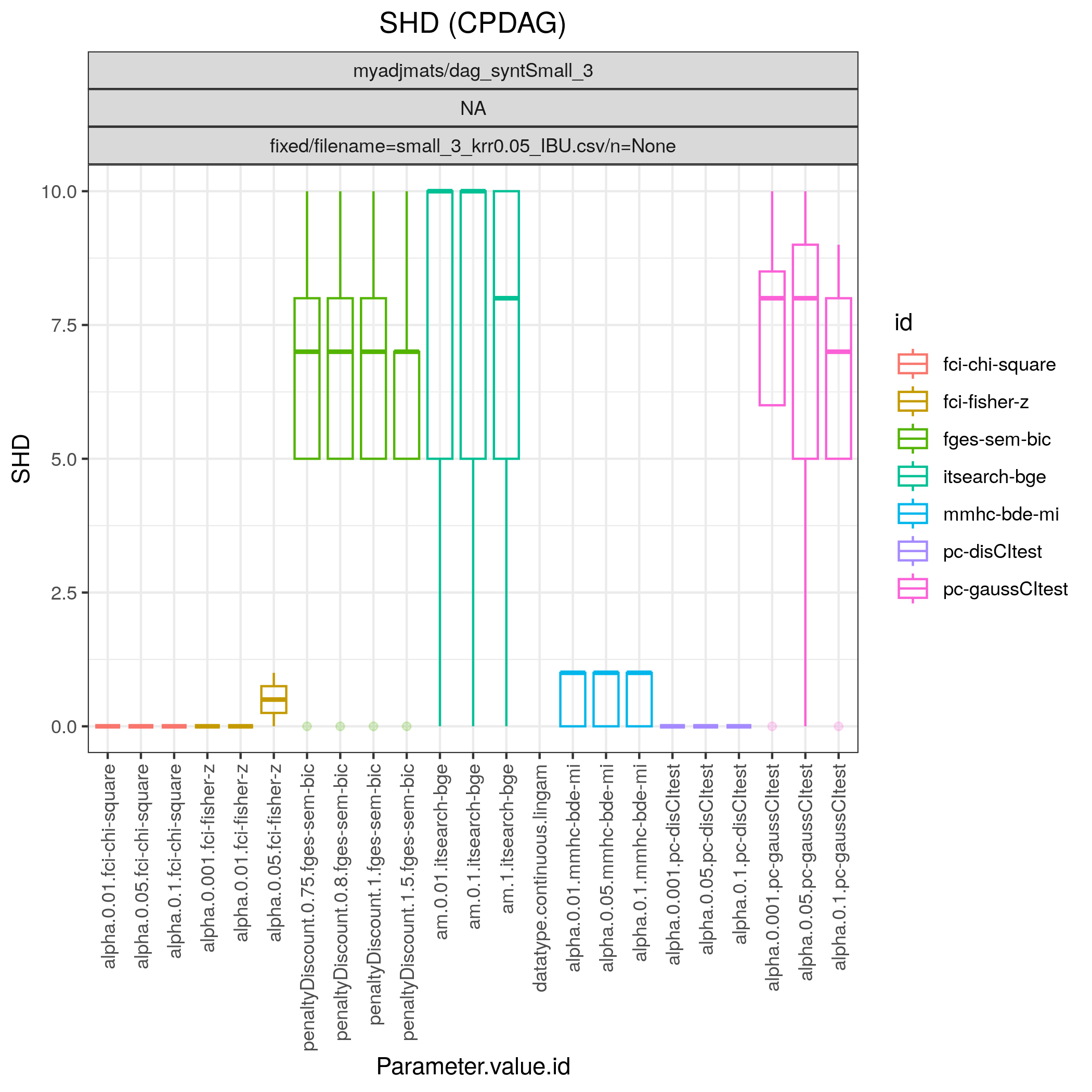}
	\caption{Synthetic 5 nodes data, $k$-RR C-wise IBU mechanism, max probability 0.05.}
\end{minipage}
\begin{minipage}{0.31\linewidth}
\centering
  \includegraphics[scale=0.34]{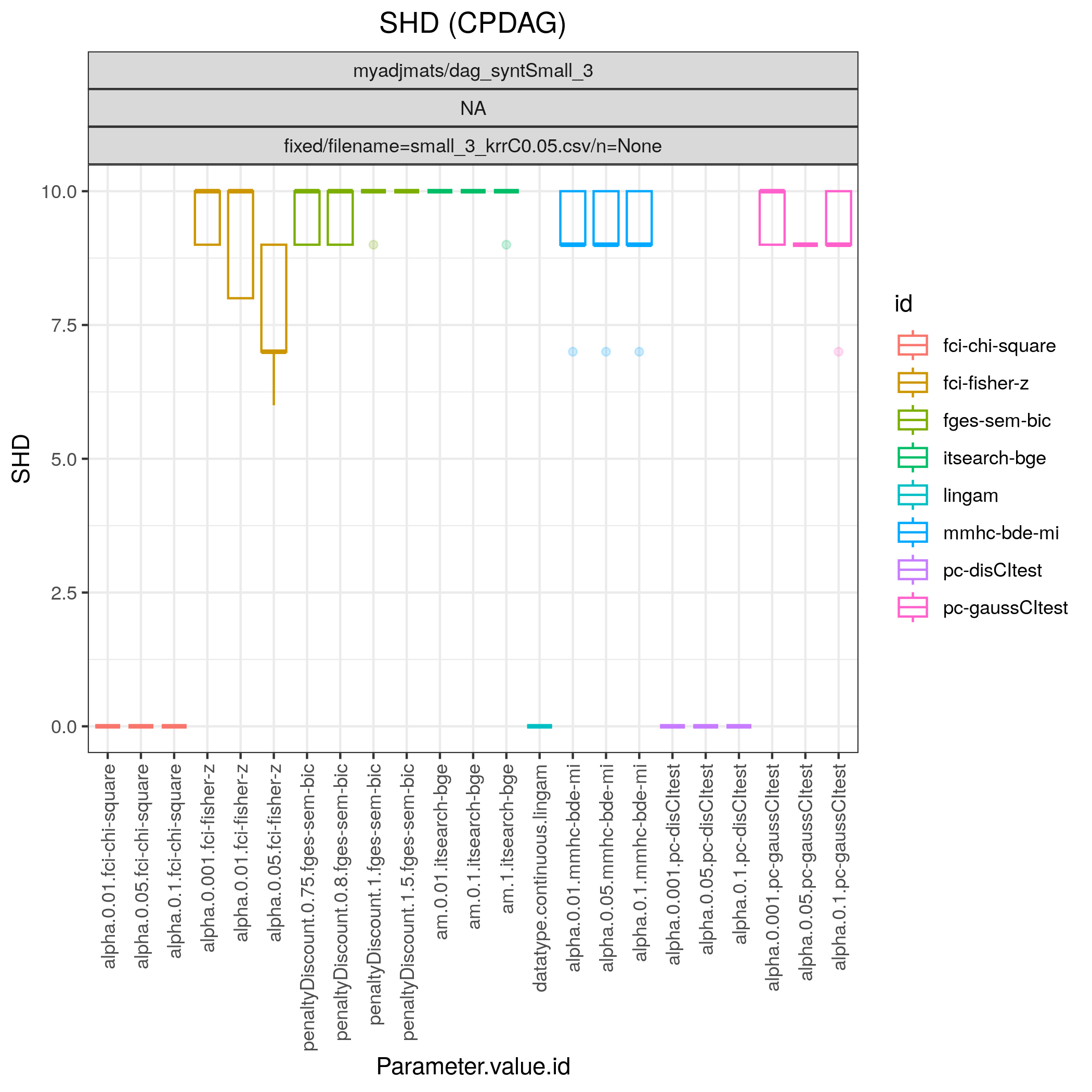}
	\caption{Synthetic 5 nodes data, $k$-RR Comb mechanism, max probability 0.05.}
\end{minipage}
\begin{minipage}{0.31\linewidth}
\centering
  \includegraphics[scale=0.34]{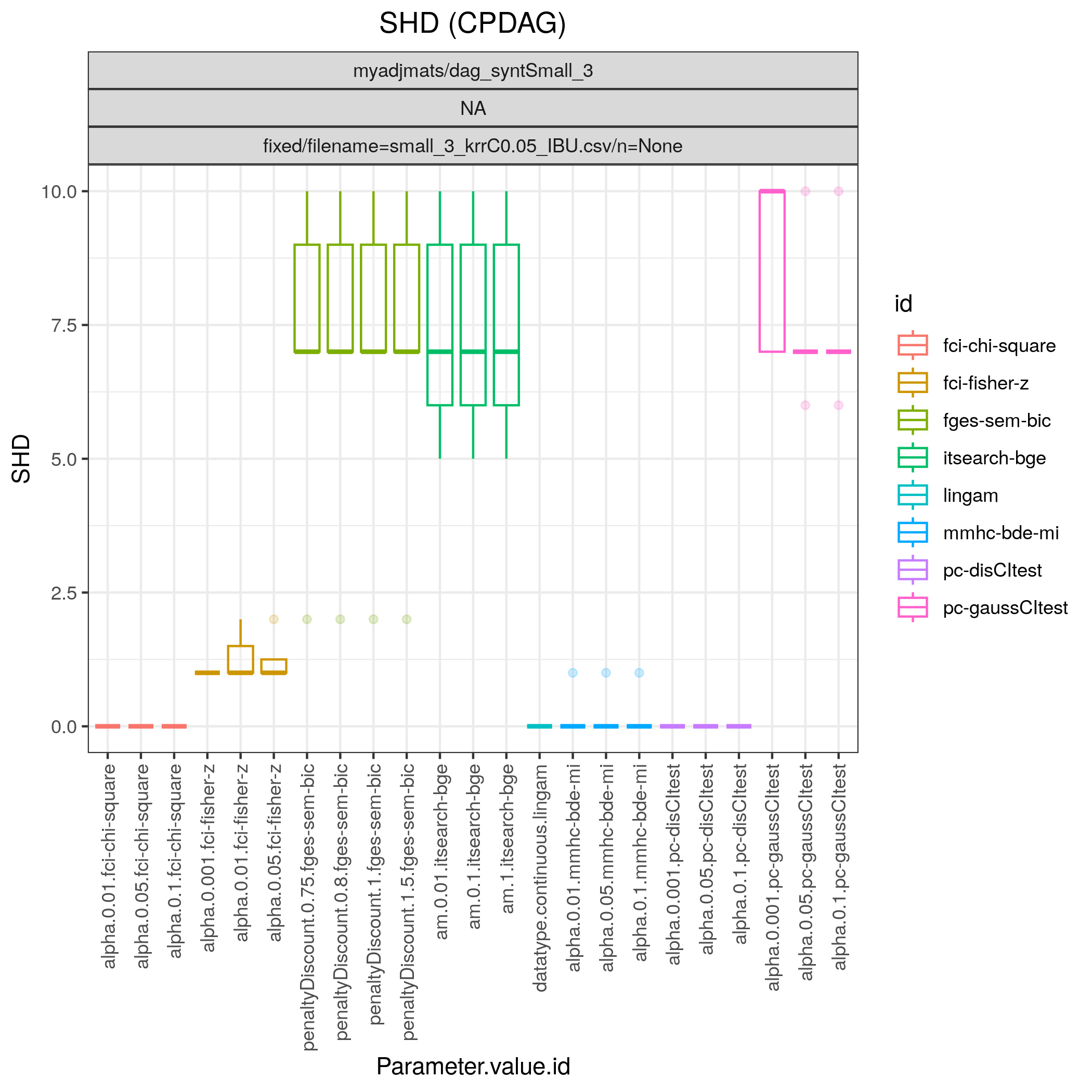}
	\caption{Synthetic 5 nodes data, $k$-RR Comb IBU mechanism, max probability 0.05.}
\end{minipage}
\end{figure}
\noindent
\begin{figure}[H]
\begin{minipage}{0.31\linewidth}
\centering
		\includegraphics[scale=0.34]{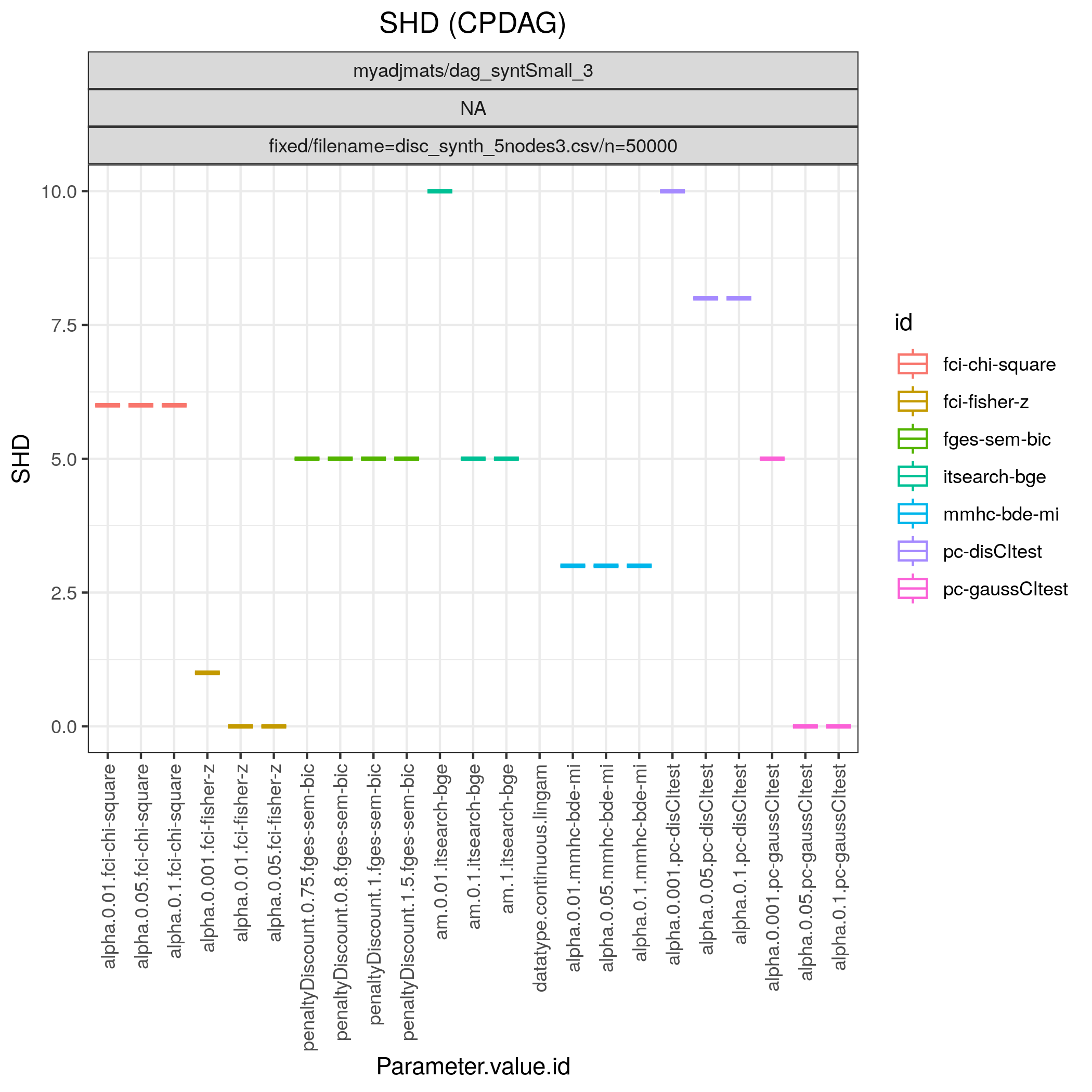}
	\caption{SHD Scores on the Synthetic 5 nodes data set. Discretized, no noise.}
\end{minipage}
\begin{minipage}{0.31\linewidth}
\centering
		\includegraphics[scale=0.34]{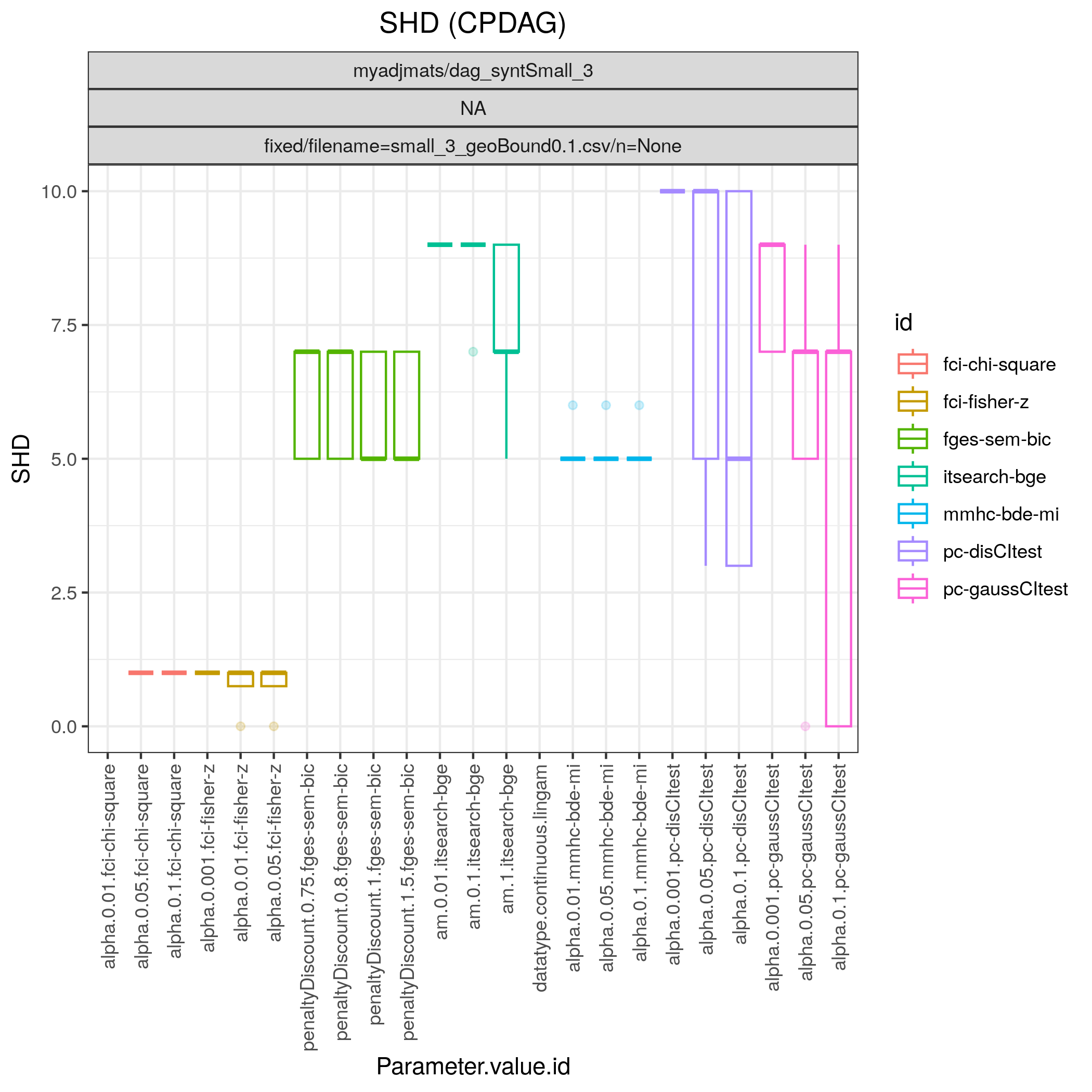}
	\caption{Synthetic 5 nodes data, Geo C-wise mechanism, max probability 0.1.}
\end{minipage}
\begin{minipage}{0.31\linewidth}
\centering
  \includegraphics[scale=0.34]{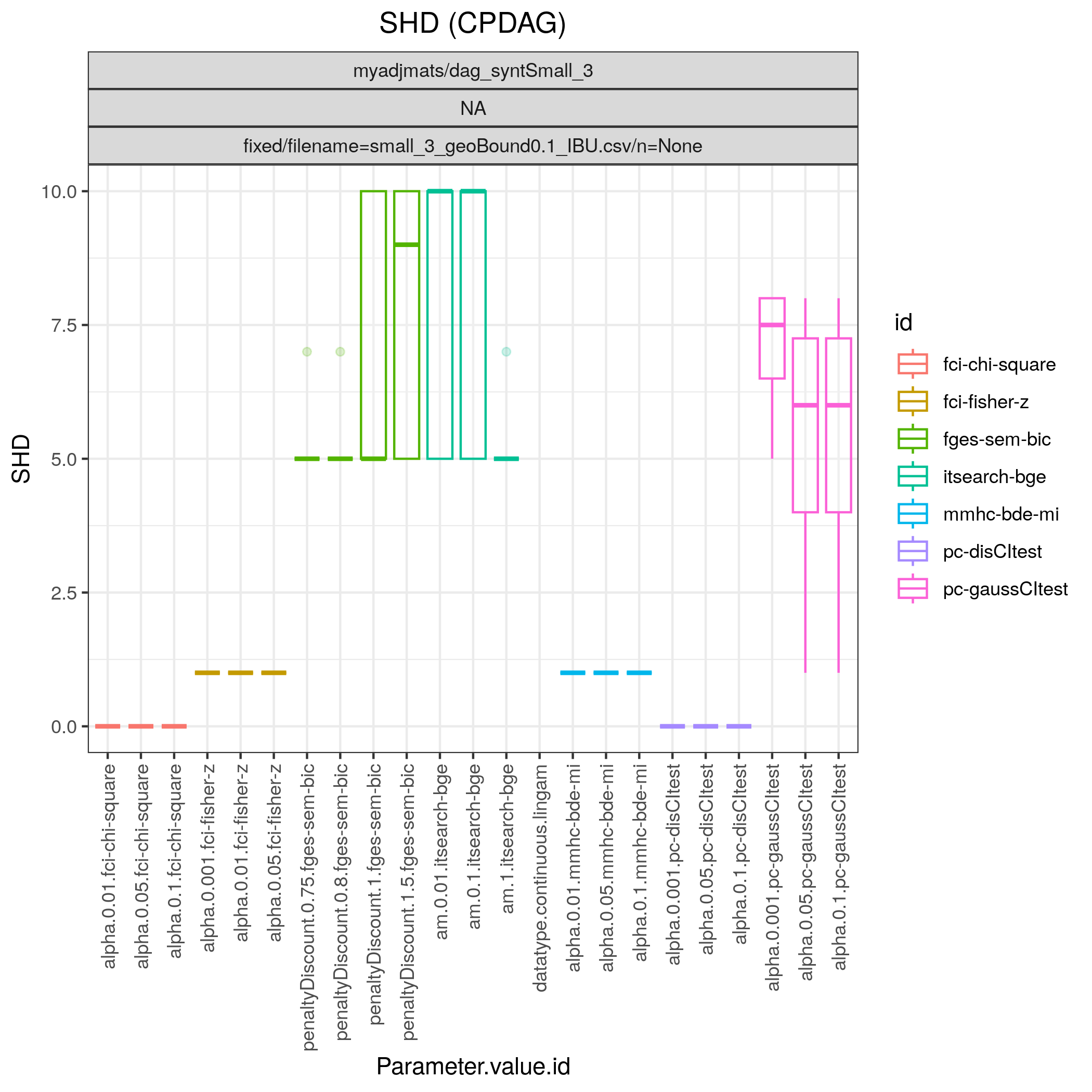}
	\caption{Synthetic 5 nodes data, Geo C-wise IBU mechanism, max probability 0.1.}
 \end{minipage}
\begin{minipage}{0.31\linewidth}
\centering
  \includegraphics[scale=0.34]{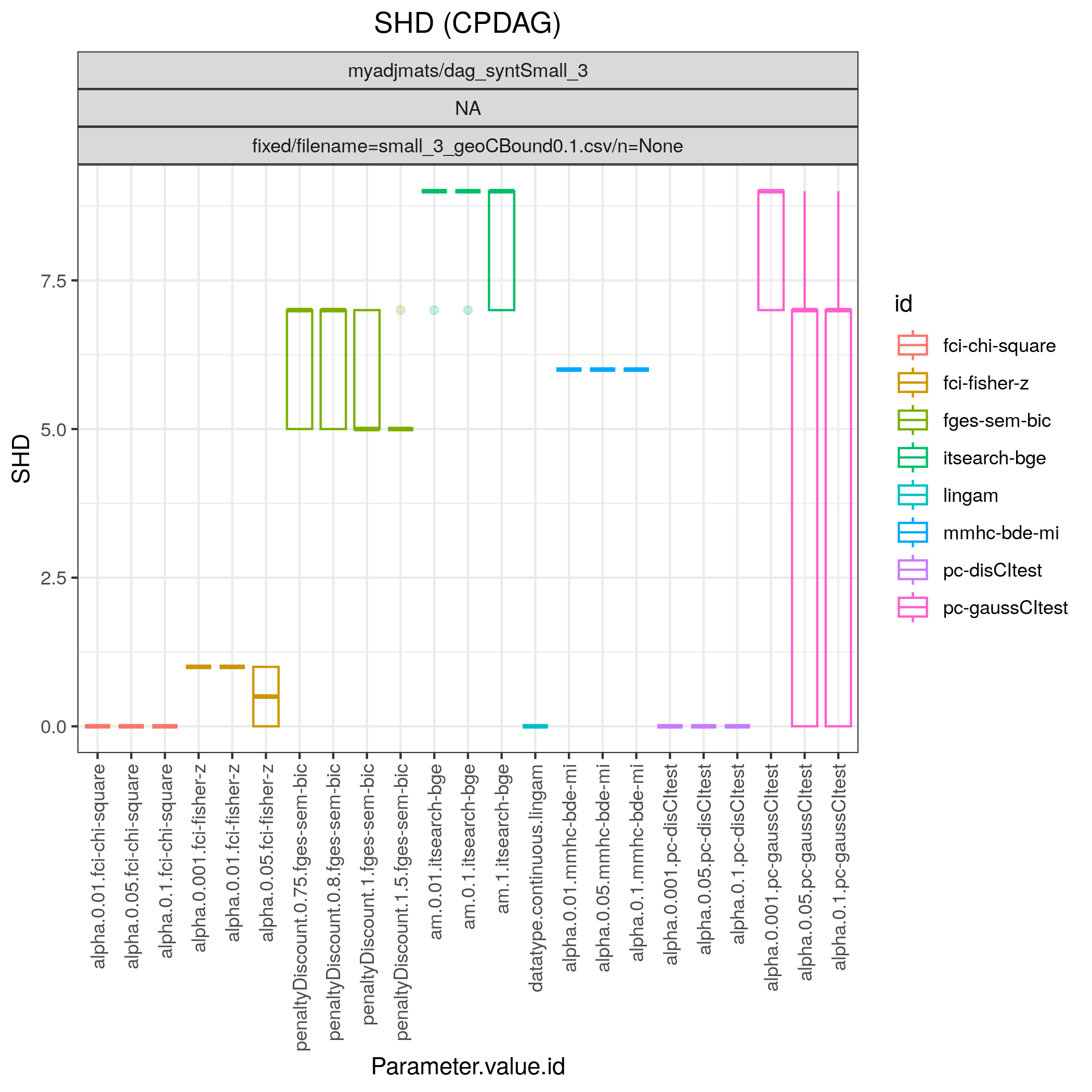}
	\caption{Synthetic 5 nodes data, Geo Comb mechanism, max probability 0.1.}
\end{minipage}
\begin{minipage}{0.31\linewidth}
\centering
  \includegraphics[scale=0.34]{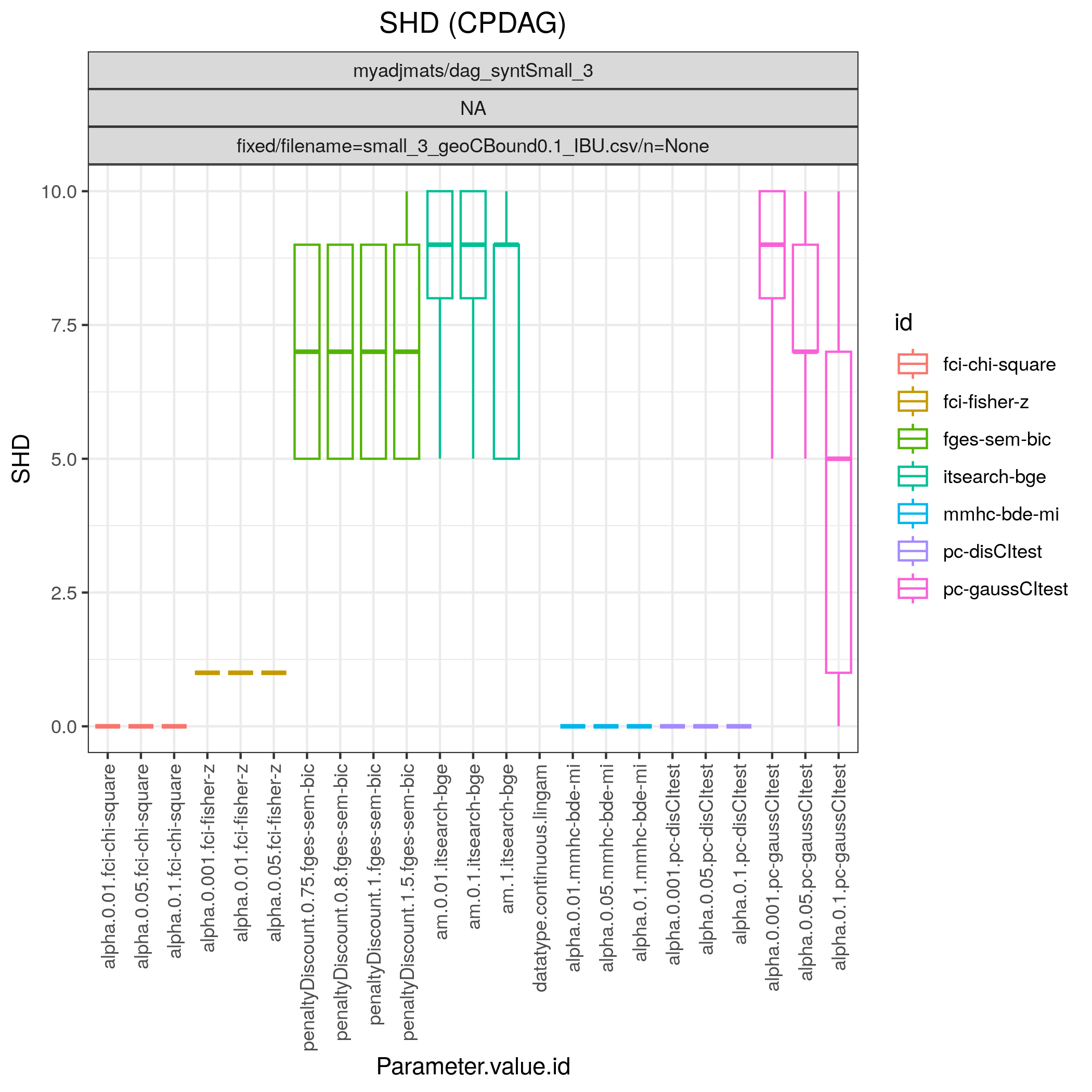}
	\caption{Synthetic 5 nodes data, Geo Comb IBU mechanism, max probability 0.1.}
\end{minipage}
\begin{minipage}{0.31\linewidth}
\centering
  \includegraphics[scale=0.34]{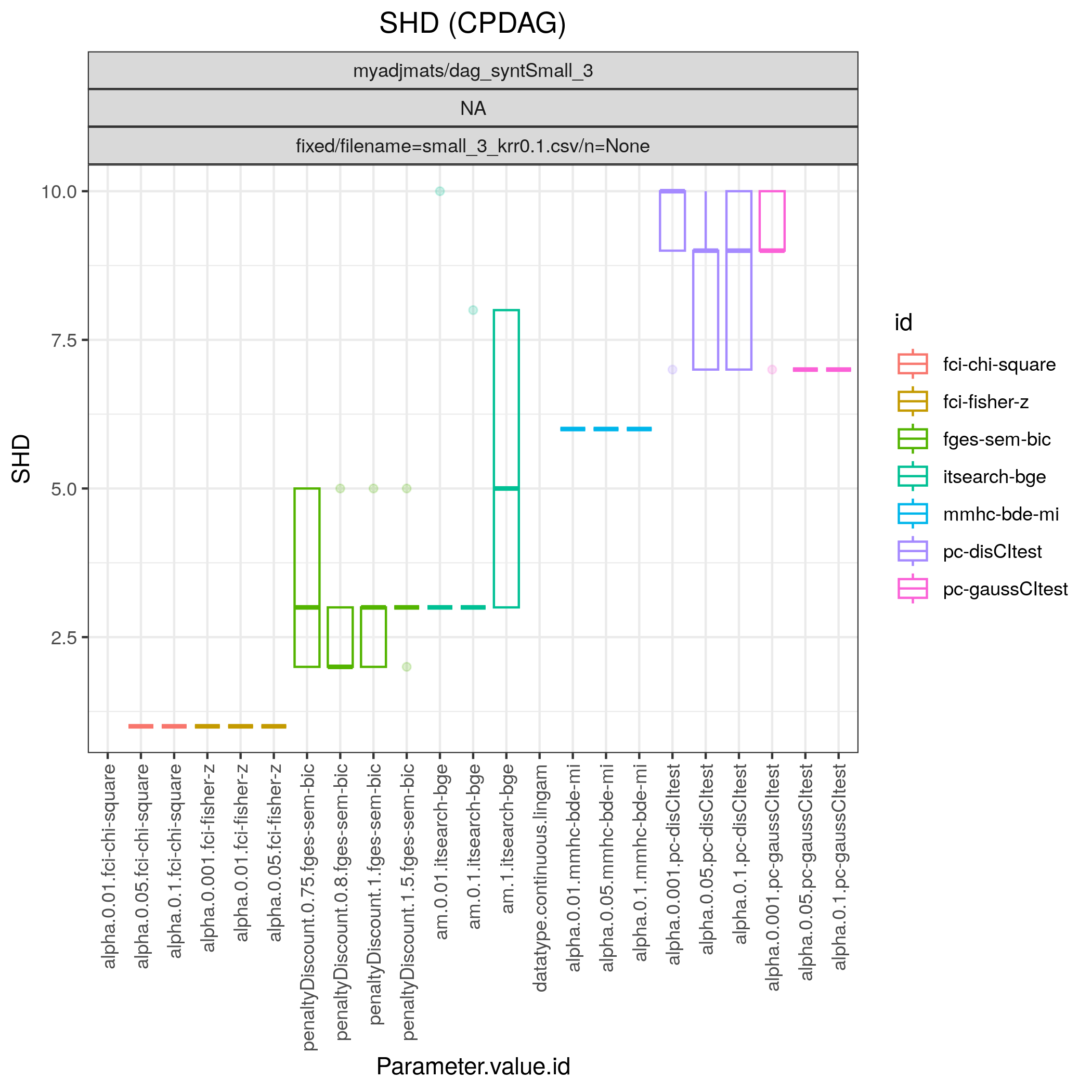}
	\caption{Synthetic 5 nodes data, $k$-RR C-wise mechanism, max probability 0.1.}
\end{minipage}

\end{figure}

\begin{figure}[H]
    \centering
   \begin{minipage}{0.31\linewidth}
\centering
  \includegraphics[scale=0.34]{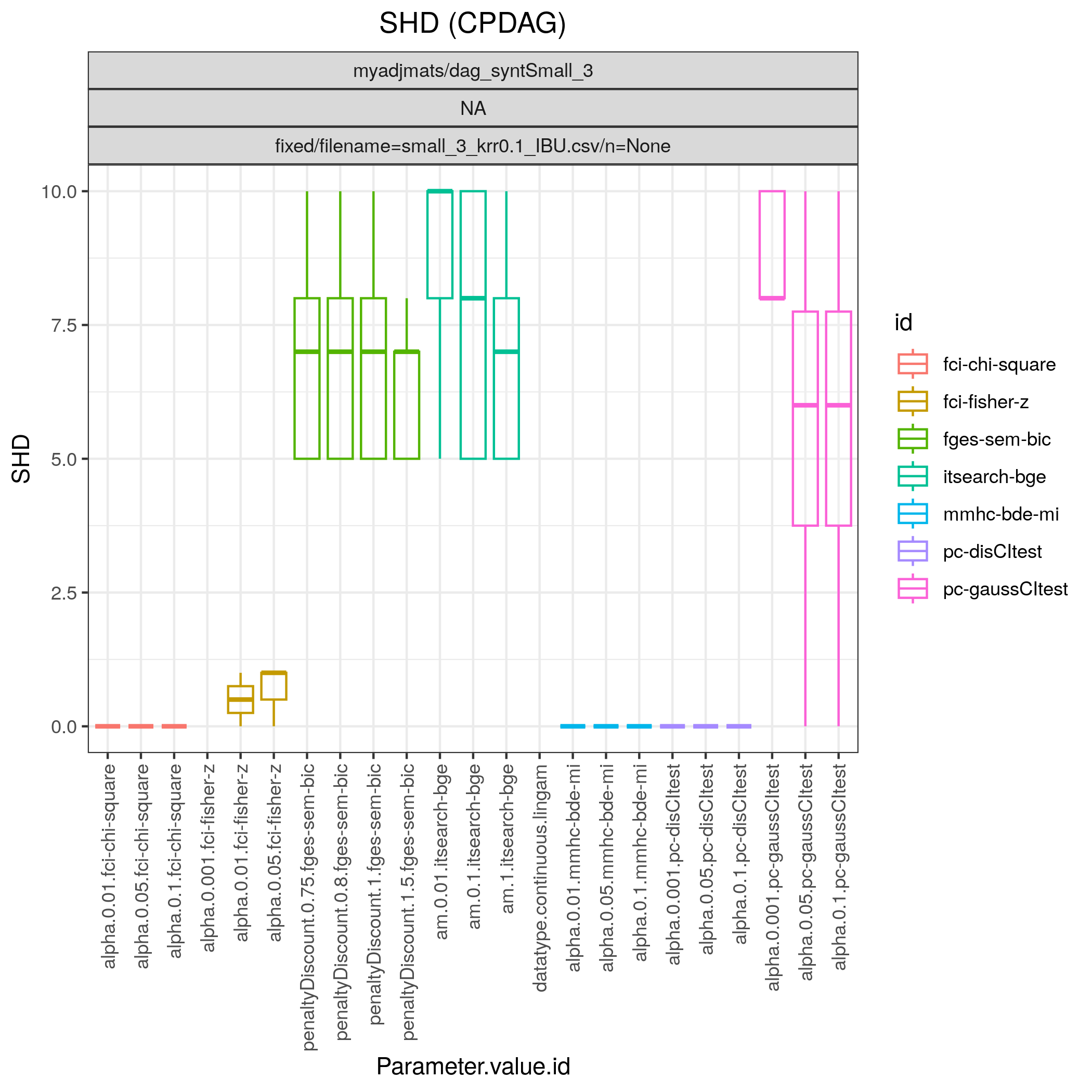}
	\caption{Synthetic 5 nodes data, $k$-RR C-wise IBU mechanism, max probability 0.1.}
\end{minipage}
\begin{minipage}{0.31\linewidth}
\centering
  \includegraphics[scale=0.34]{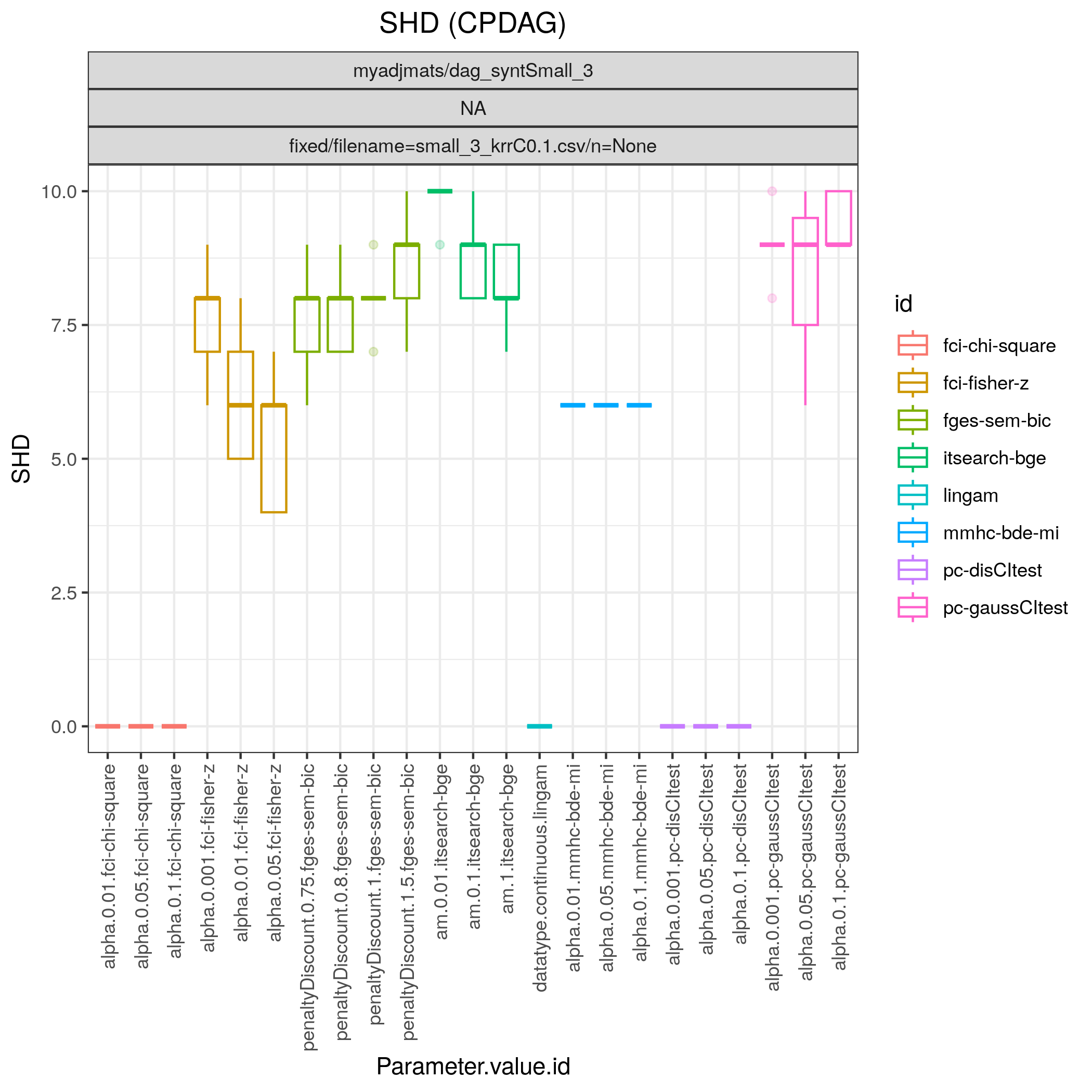}
	\caption{Synthetic 5 nodes data, $k$-RR Comb mechanism, max probability 0.1.}
\end{minipage}
\begin{minipage}{0.31\linewidth}
\centering
  \includegraphics[scale=0.34]{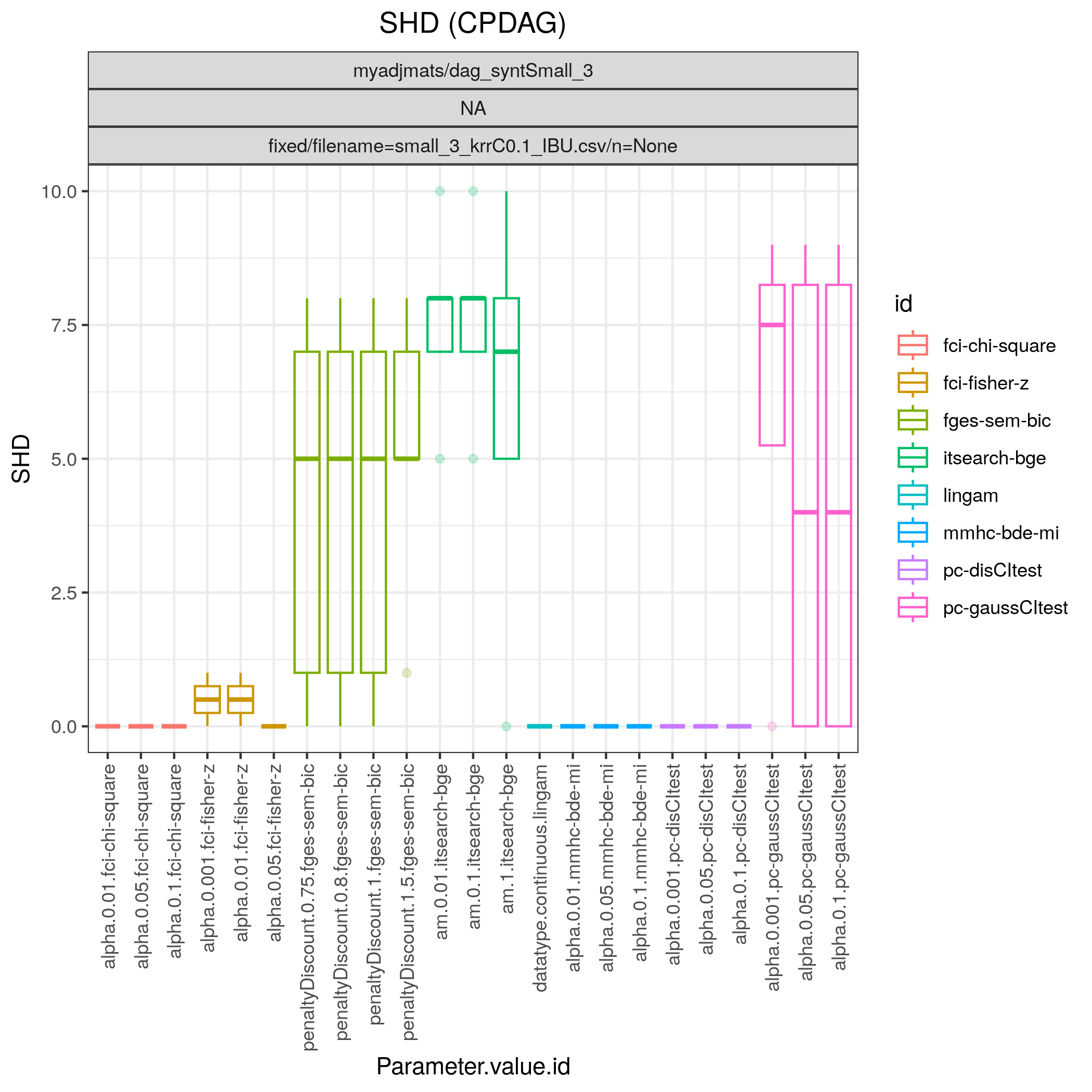}
	\caption{Synthetic 5 nodes data, $k$-RR Comb IBU mechanism, max probability 0.1.}
\end{minipage}
\end{figure}

\noindent
\begin{figure}[H]
\begin{minipage}{0.31\linewidth}
\centering
		\includegraphics[scale=0.34]{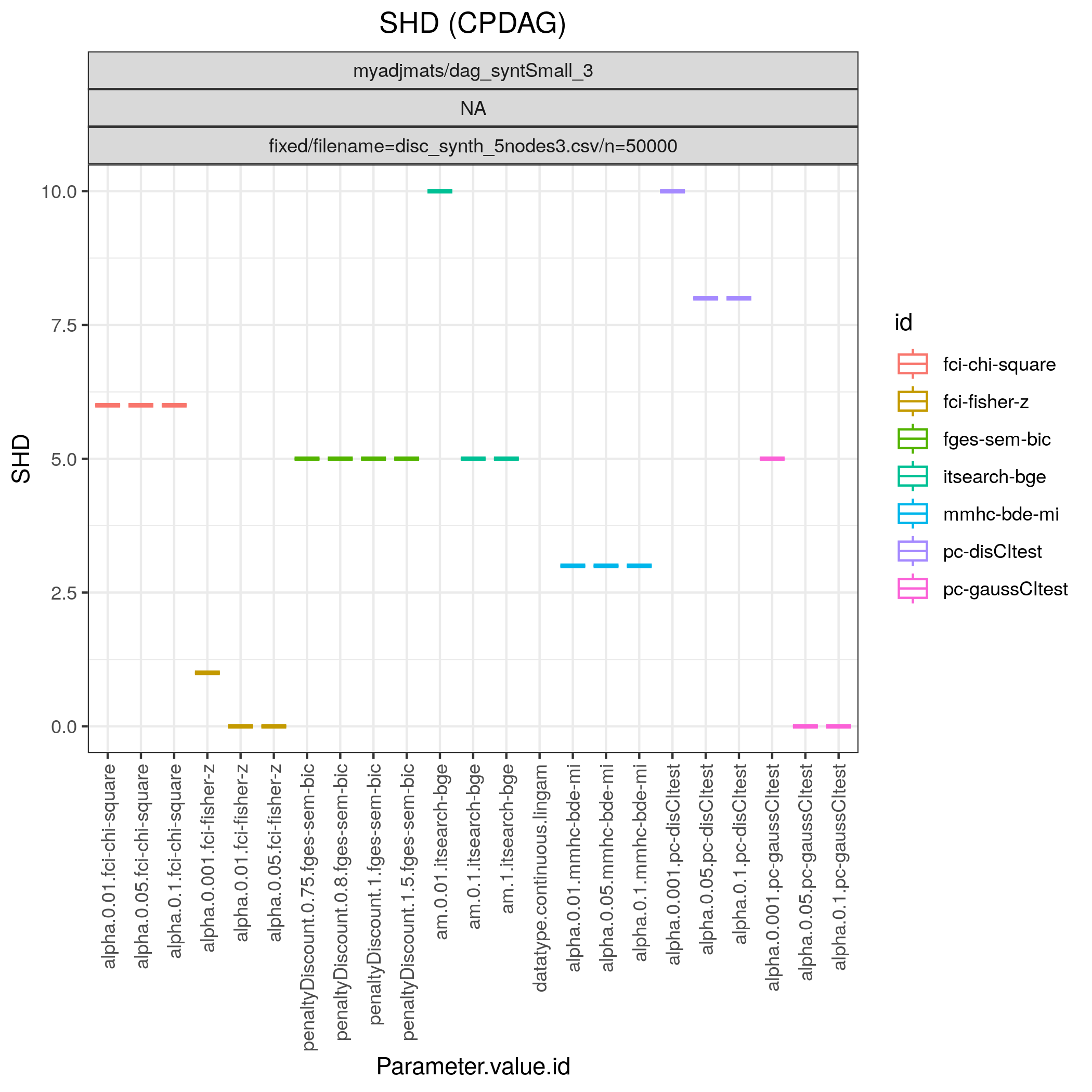}
	\caption{SHD Scores on the Synthetic 5 nodes data set. Discretized, no noise.}
\end{minipage}
\begin{minipage}{0.31\linewidth}
\centering
		\includegraphics[scale=0.34]{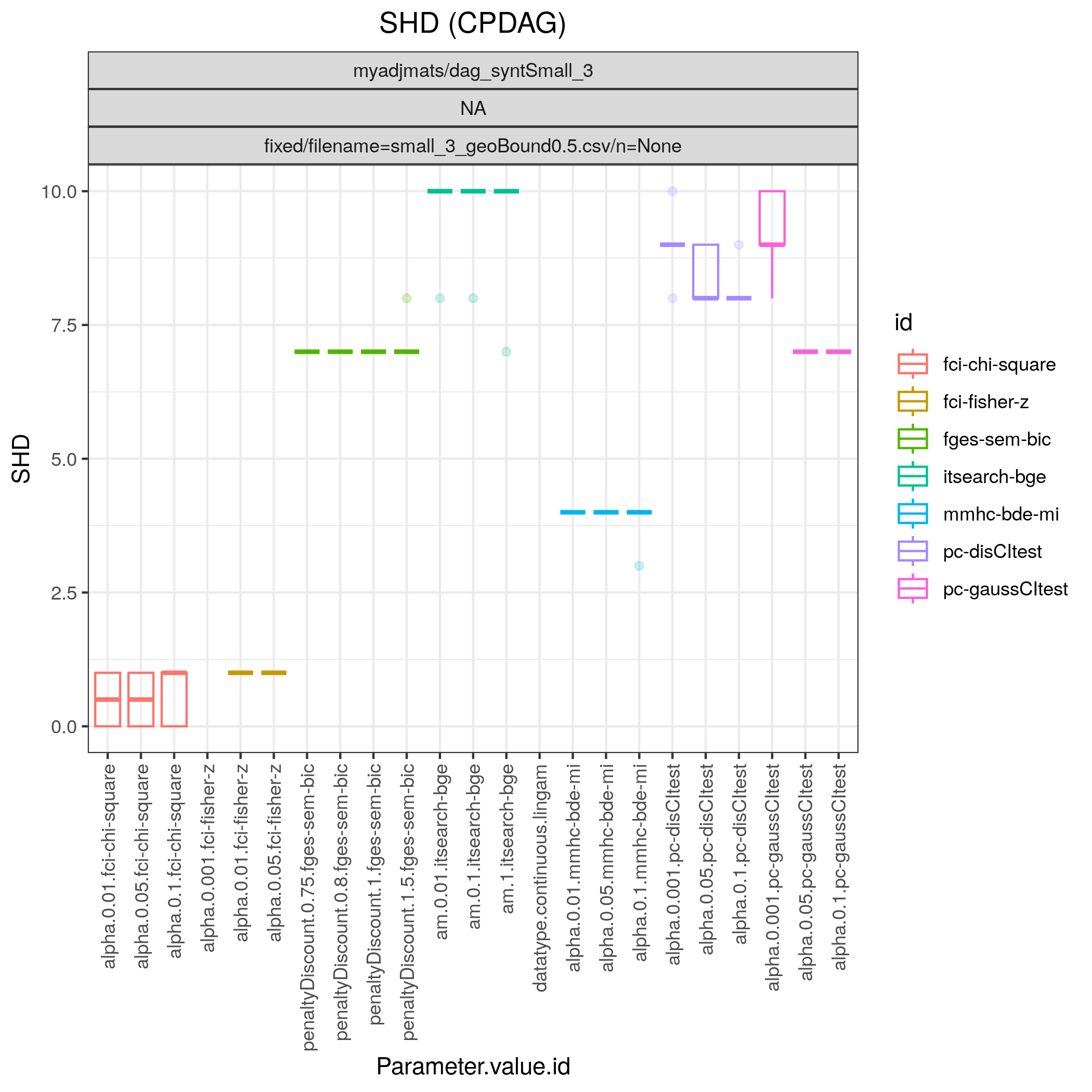}
	\caption{Synthetic 5 nodes data, Geo C-wise mechanism, max probability 0.5.}
\end{minipage}
\begin{minipage}{0.31\linewidth}
\centering
  \includegraphics[scale=0.34]{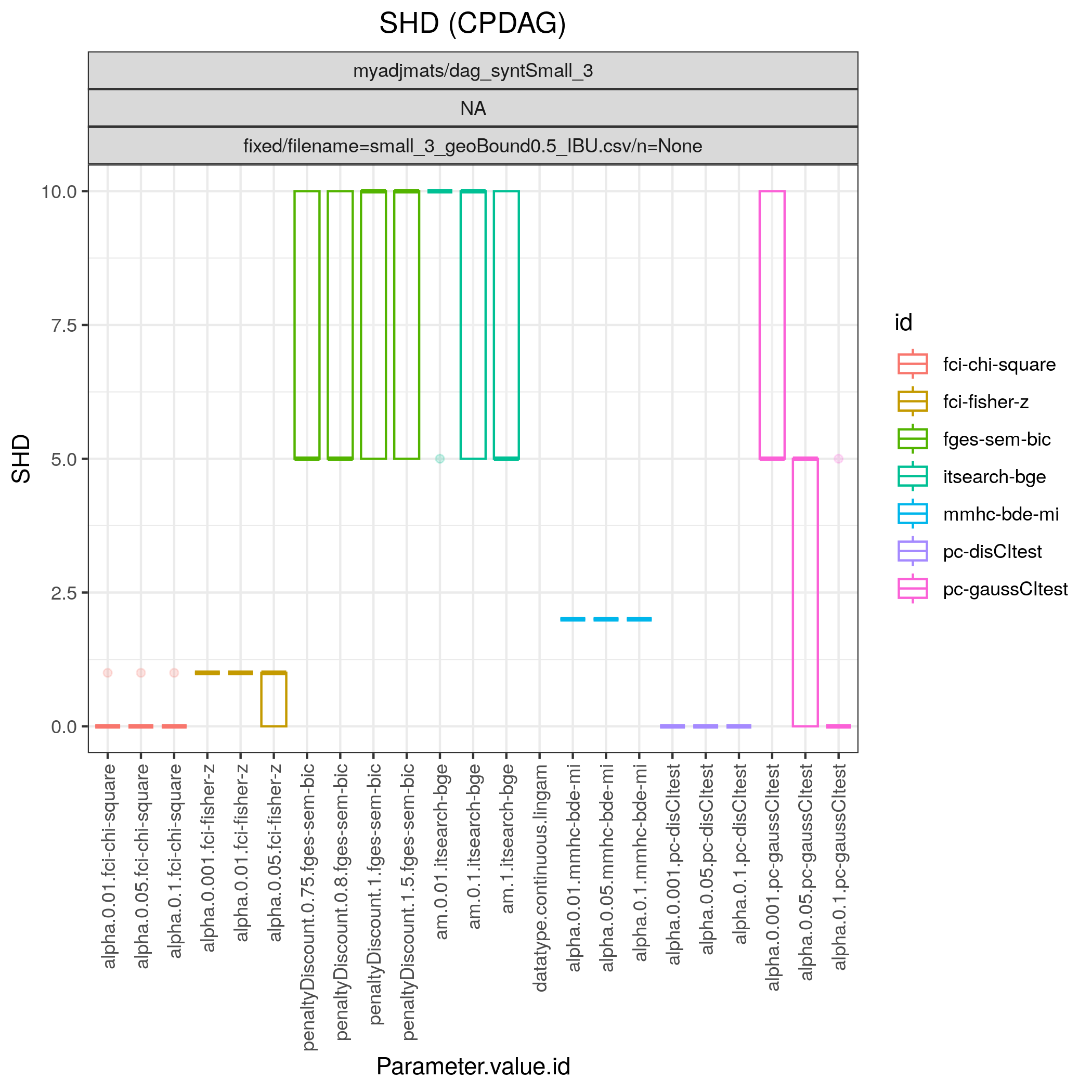}
	\caption{Synthetic 5 nodes data, Geo C-wise IBU mechanism, max probability 0.5.}
 \end{minipage}
\begin{minipage}{0.31\linewidth}
\centering
  \includegraphics[scale=0.34]{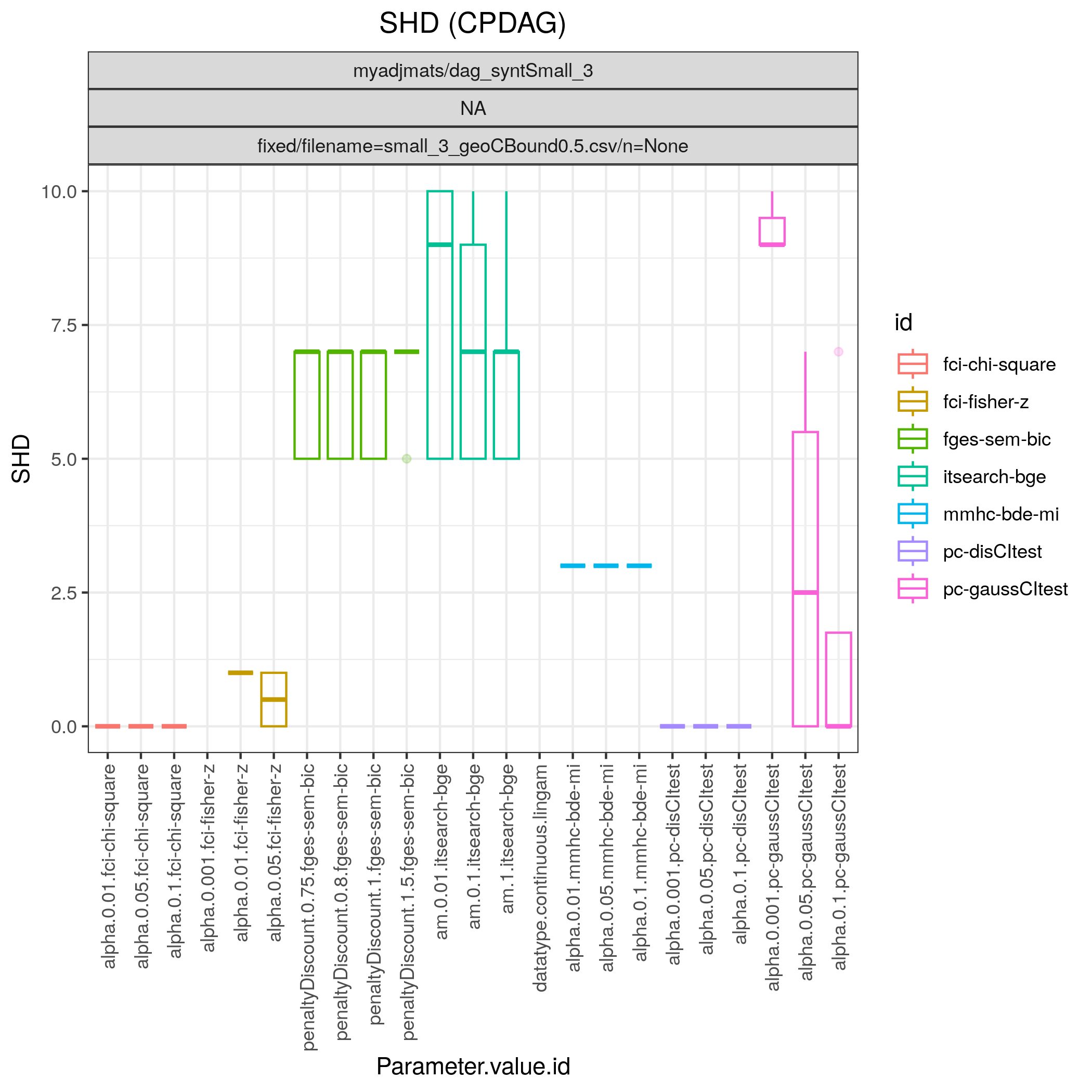}
	\caption{Synthetic 5 nodes data, Geo Comb mechanism, max probability 0.5.}
\end{minipage}
\begin{minipage}{0.31\linewidth}
\centering
  \includegraphics[scale=0.34]{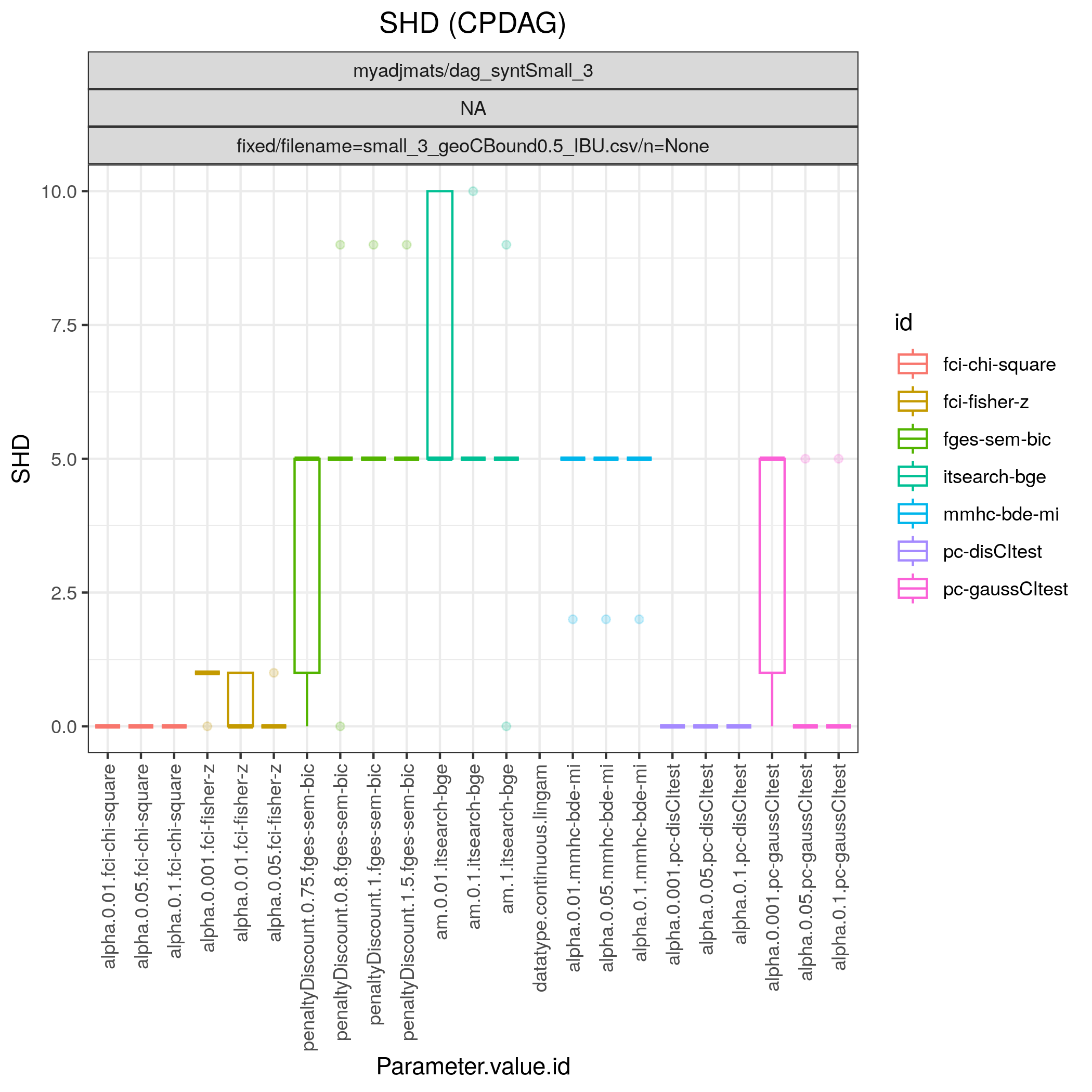}
	\caption{Synthetic 5 nodes data, Geo Comb IBU mechanism, max probability 0.5.}
\end{minipage}
\begin{minipage}{0.31\linewidth}
\centering
  \includegraphics[scale=0.34]{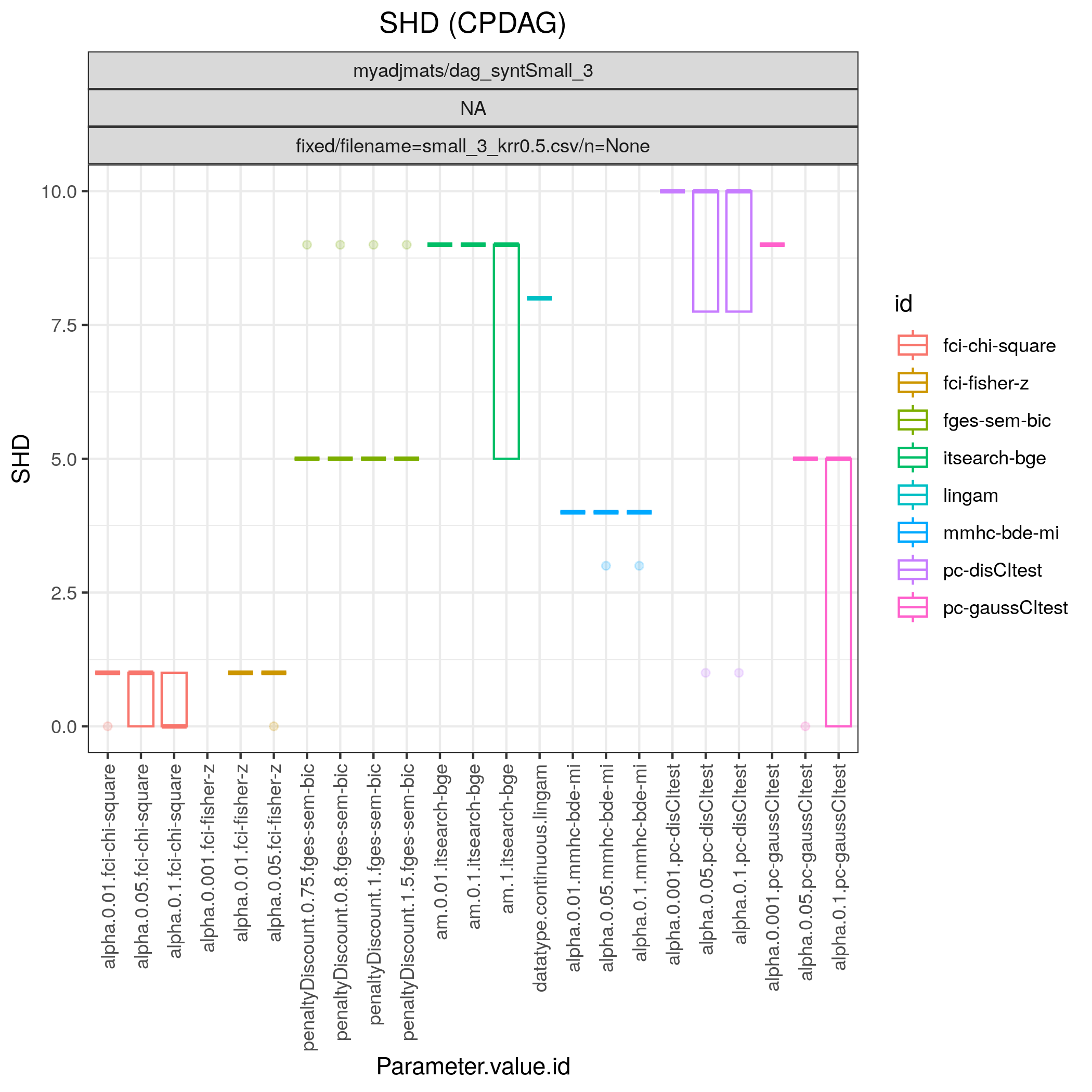}
	\caption{Synthetic 5 nodes data, $k$-RR C-wise mechanism, max probability 0.5.}
\end{minipage}

\end{figure}

\begin{figure}[H]
    \centering
   \begin{minipage}{0.31\linewidth}
\centering
  \includegraphics[scale=0.34]{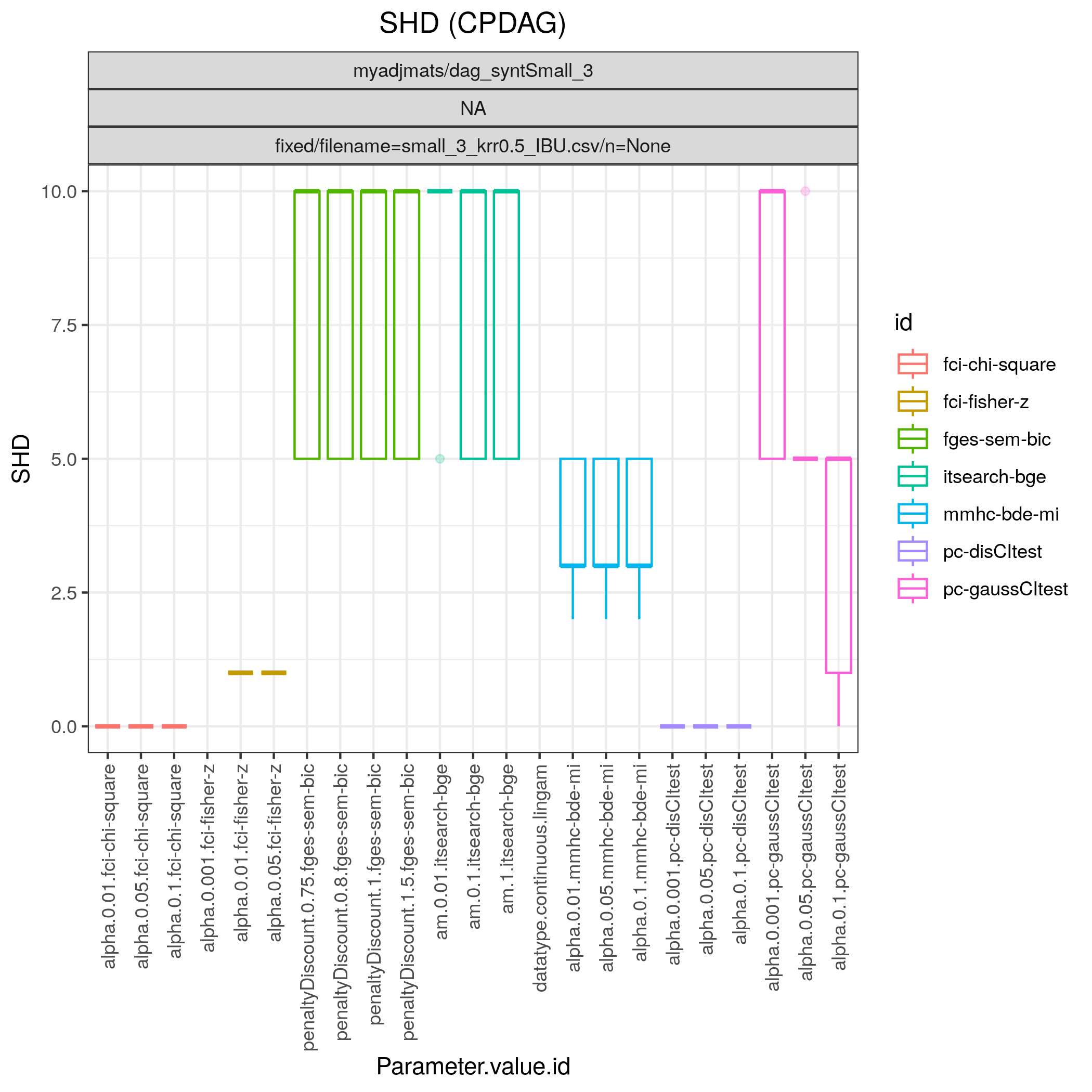}
	\caption{Synthetic 5 nodes data, $k$-RR C-wise IBU mechanism, max probability 0.5.}
\end{minipage}
\begin{minipage}{0.31\linewidth}
\centering
  \includegraphics[scale=0.34]{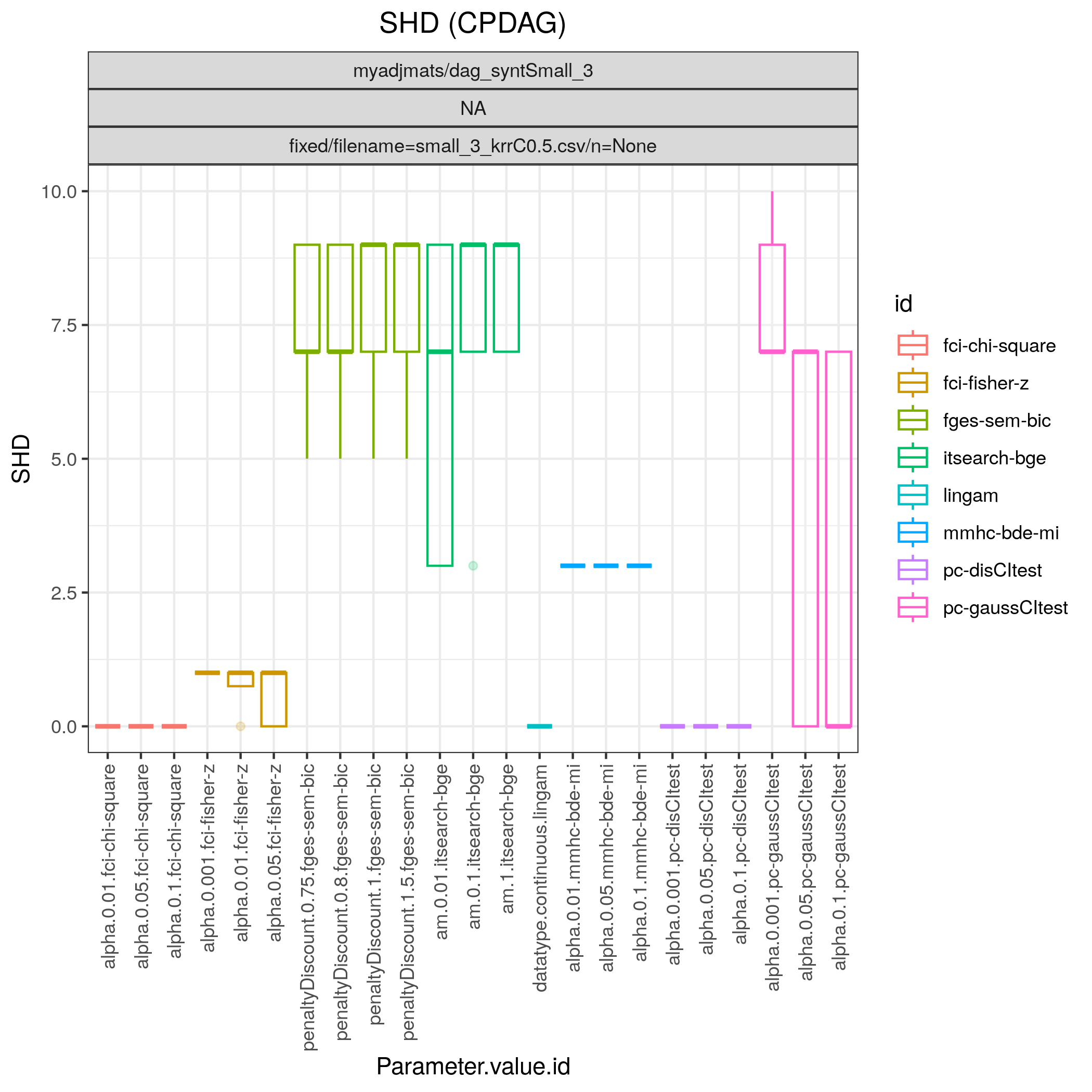}
	\caption{Synthetic 5 nodes data, $k$-RR Comb mechanism, max probability 0.5.}
\end{minipage}
\begin{minipage}{0.31\linewidth}
\centering
  \includegraphics[scale=0.34]{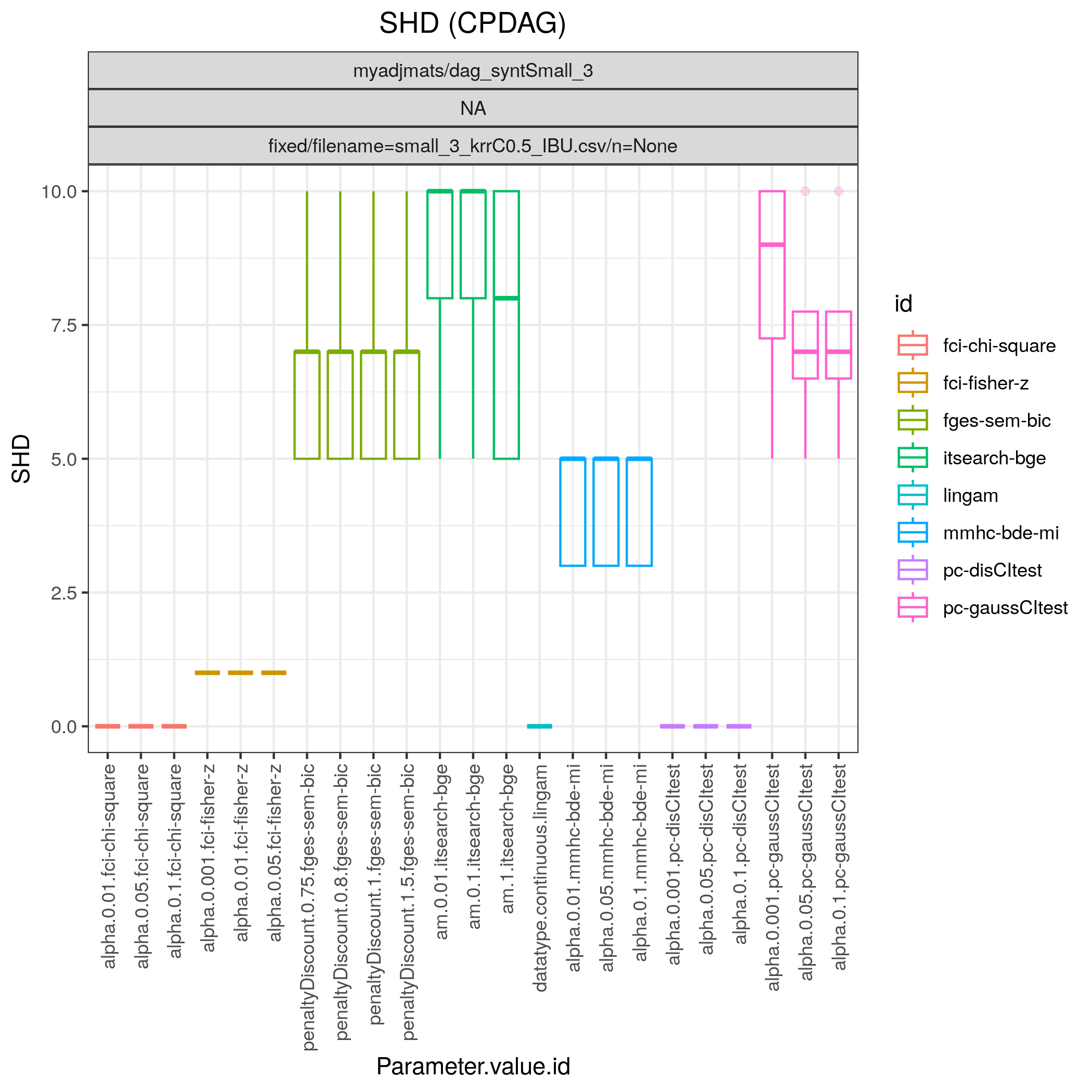}
	\caption{Synthetic 5 nodes data, $k$-RR Comb IBU mechanism, max probability 0.5.}
\end{minipage}
\end{figure}


 \subsection{F1 Score results Synthetic 10 nodes data set}
 \begin{figure}[t]
    \centering
    \includegraphics[width=1\linewidth]{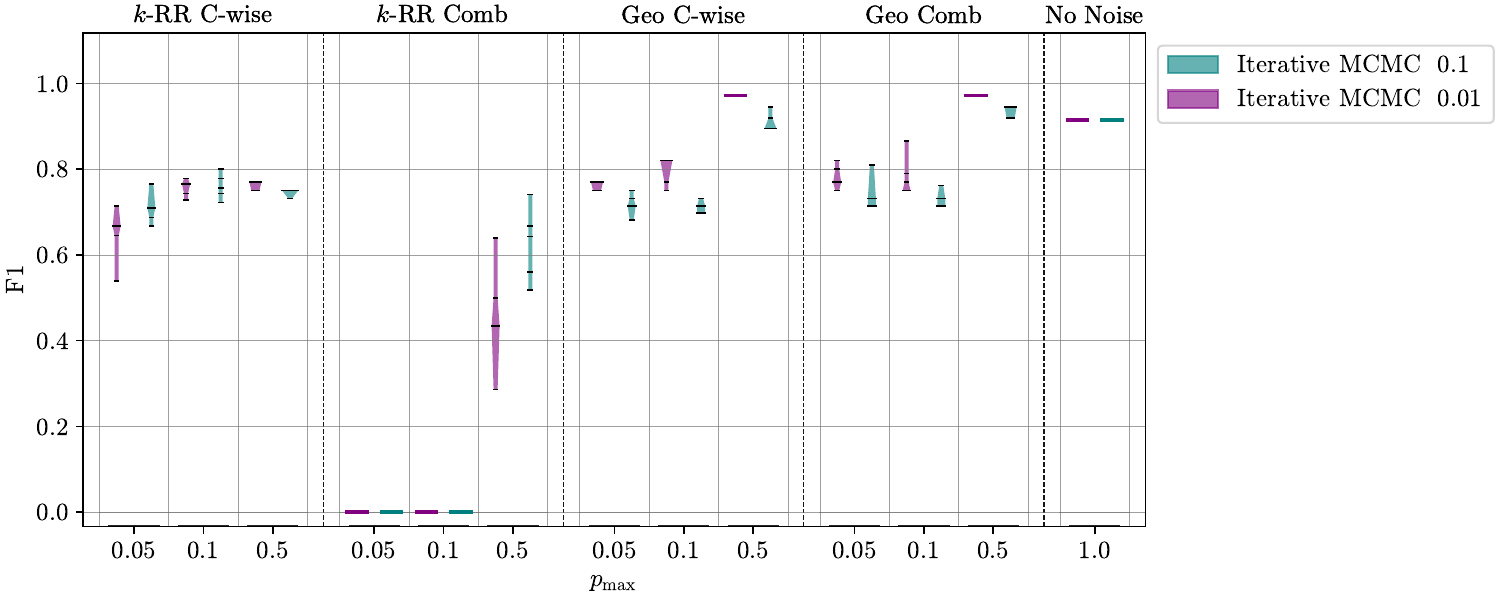}
    \vspace*{-10mm}
    \caption{Synthetic data, 10 nodes, F1.}
    \label{fig:SYNTH10_f1}
\end{figure}
\noindent
\begin{figure}[H]
\begin{minipage}{0.31\linewidth}
\centering
		\includegraphics[scale=0.34]{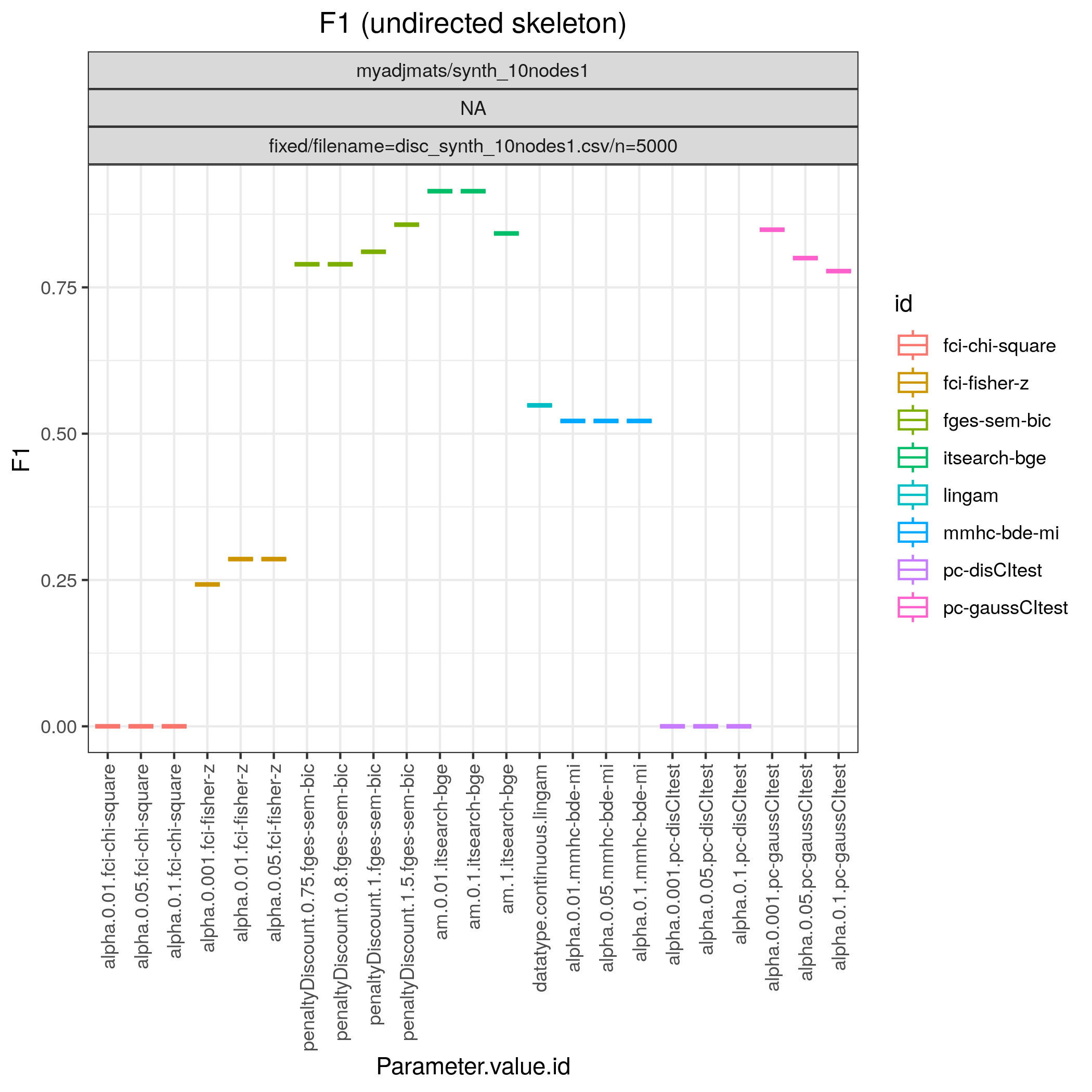}
	\caption{F1 Scores on the Synthetic 10 nodes data set. Discretized, no noise.}
\end{minipage}
\begin{minipage}{0.31\linewidth}
\centering
		\includegraphics[scale=0.34]{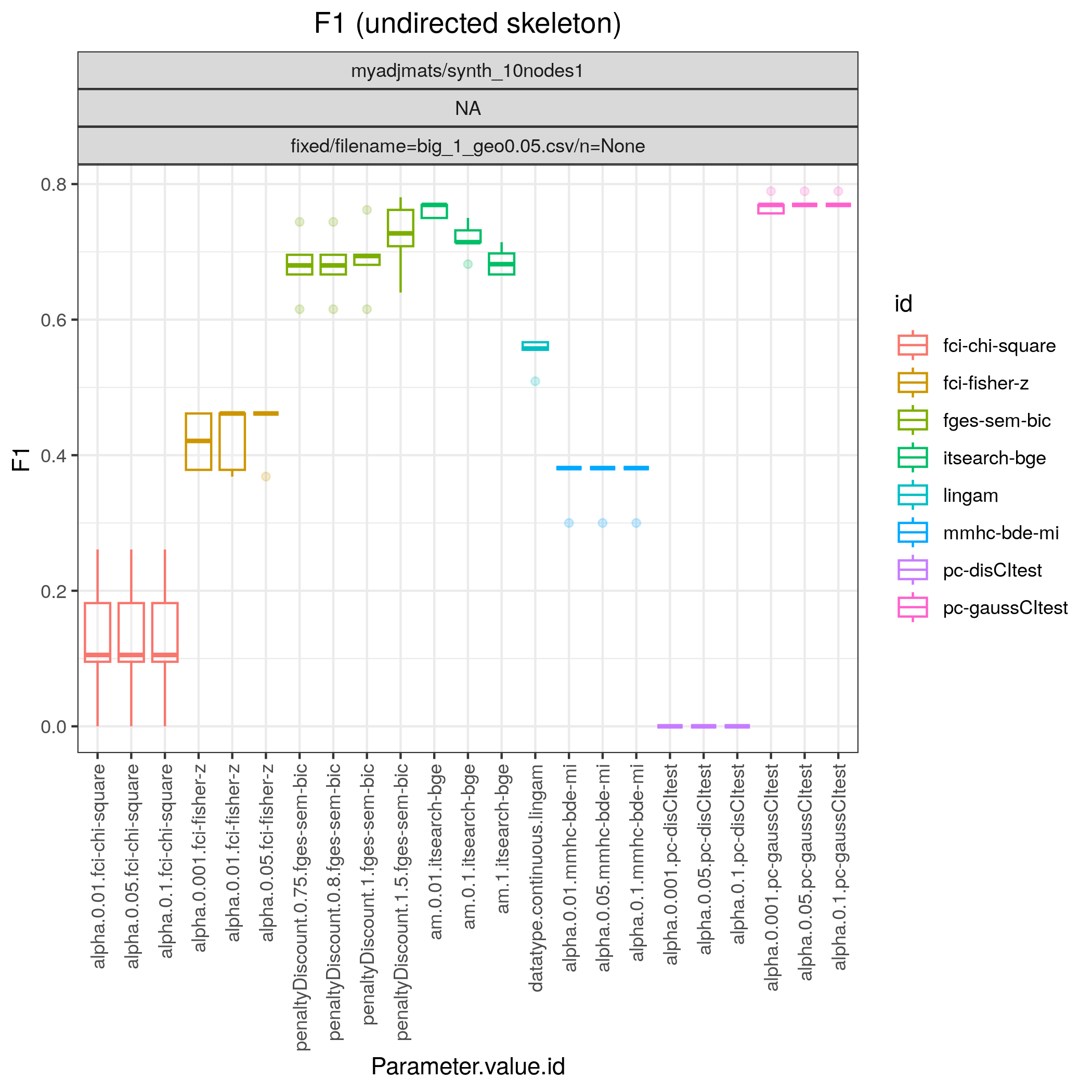}
	\caption{Synthetic 10 nodes data, Geo C-wise mechanism, max probability 0.05.}
\end{minipage}
\begin{minipage}{0.31\linewidth}
\centering
  \includegraphics[scale=0.34]{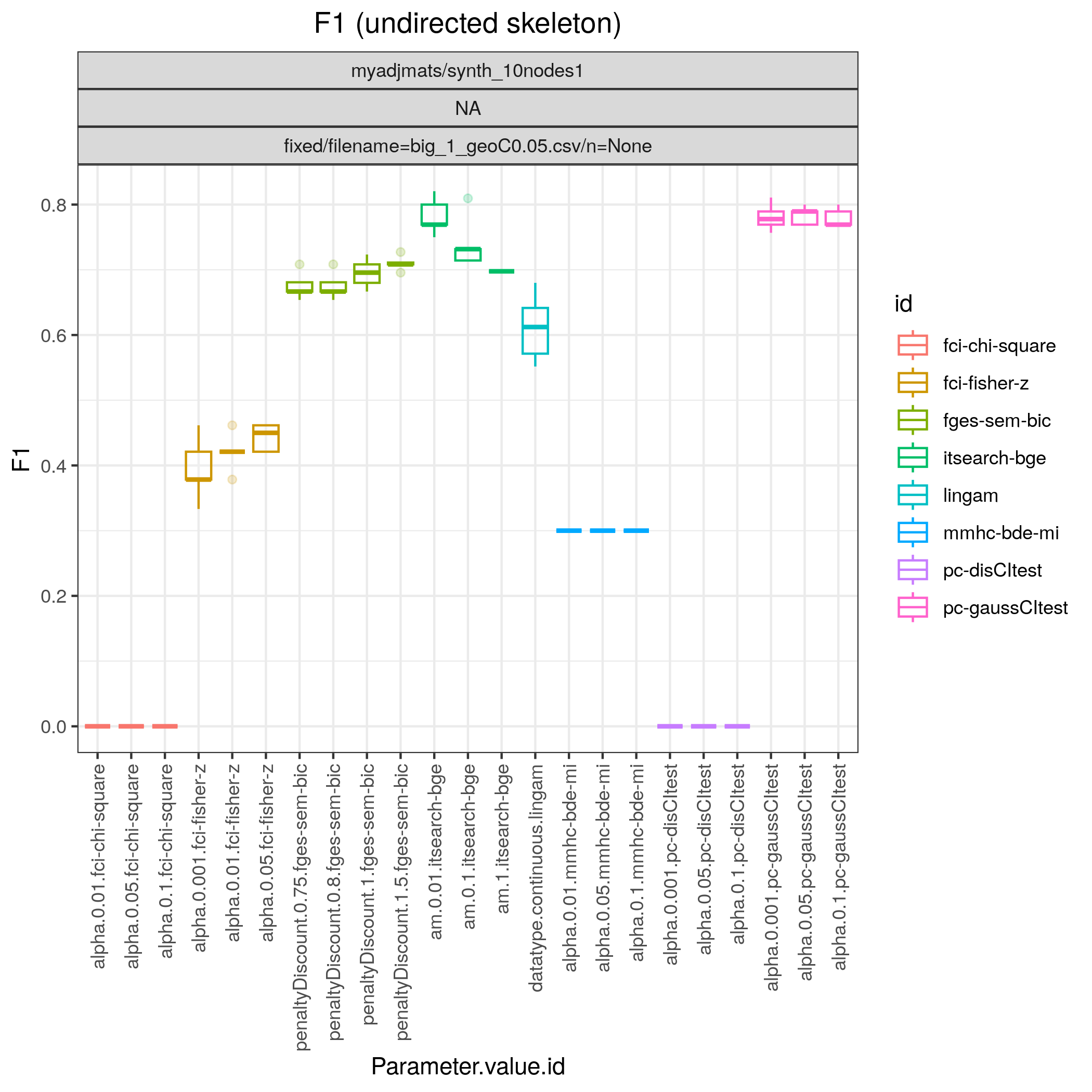}
	\caption{Synthetic 10 nodes data, Geo Comb mechanism, max probability 0.05.}
 \end{minipage}
    \begin{minipage}{0.31\linewidth}
\centering
  \includegraphics[scale=0.34]{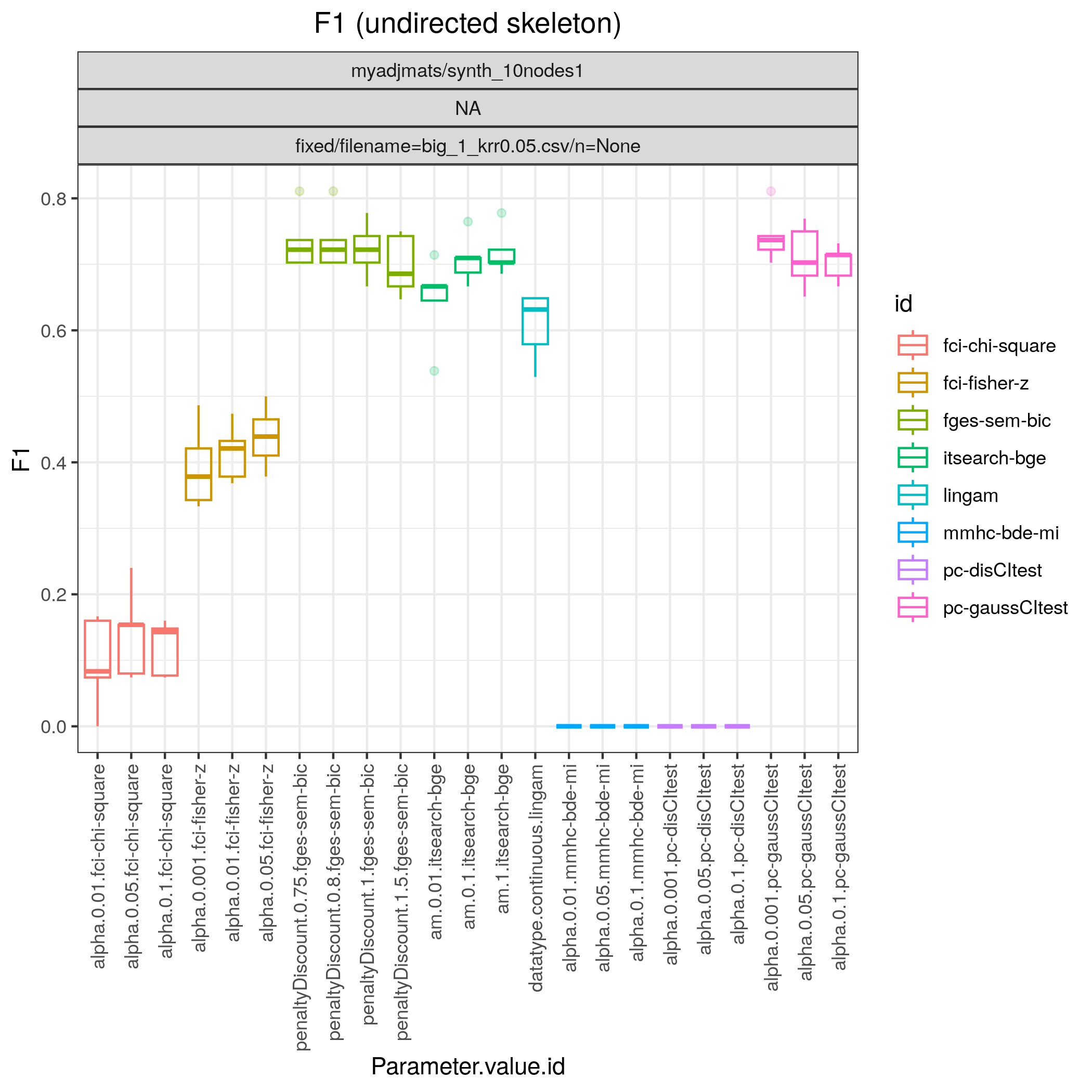}
	\caption{Synthetic 10 nodes data, $k$-RR C-wise mechanism, max probability 0.05.}
\end{minipage}
\begin{minipage}{0.31\linewidth}
\centering
  \includegraphics[scale=0.34]{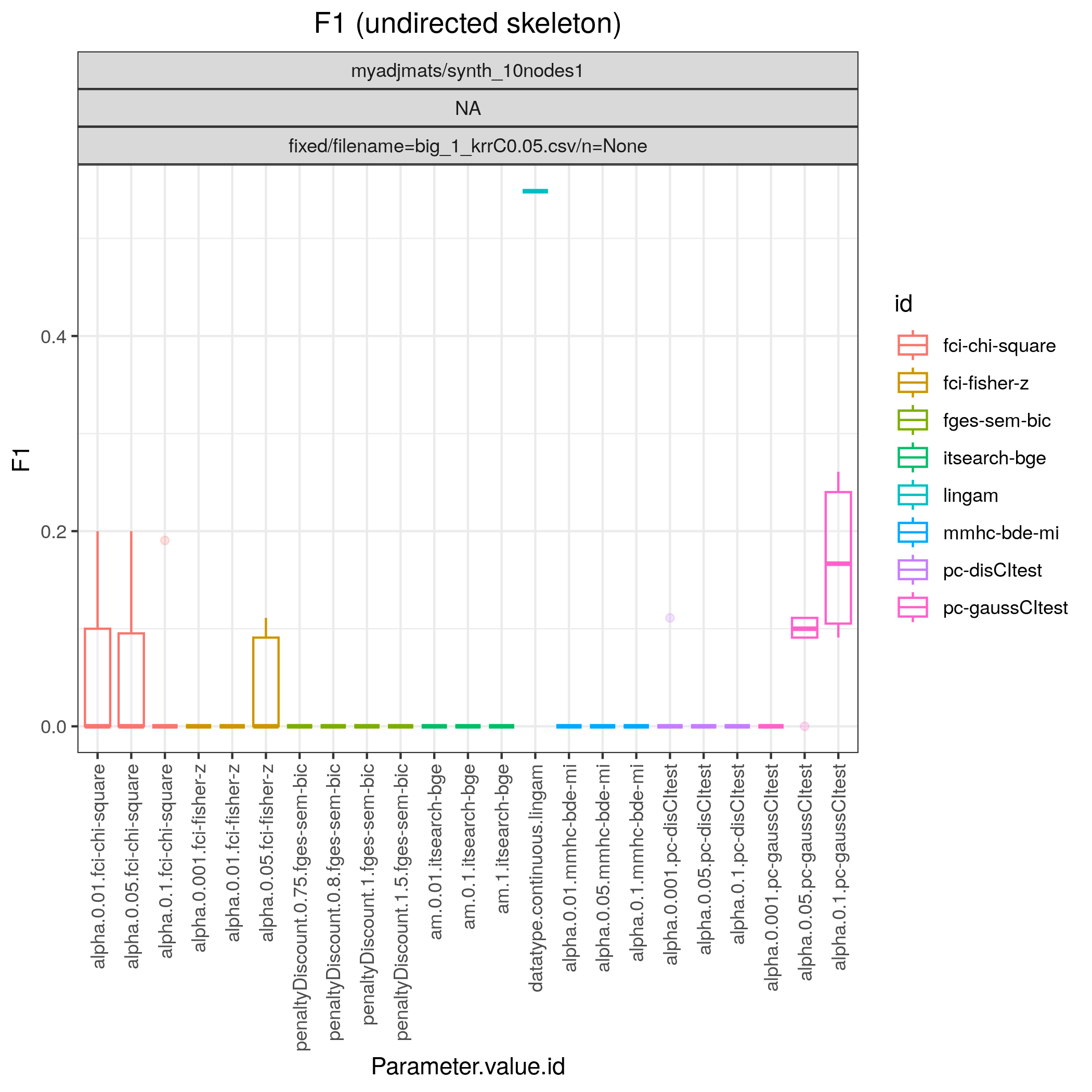}
	\caption{Synthetic 10 nodes data, $k$-RR Comb mechanism, max probability 0.05.}
\end{minipage}
\end{figure}


\noindent
\begin{figure}[H]
\begin{minipage}{0.31\linewidth}
\centering
		\includegraphics[scale=0.34]{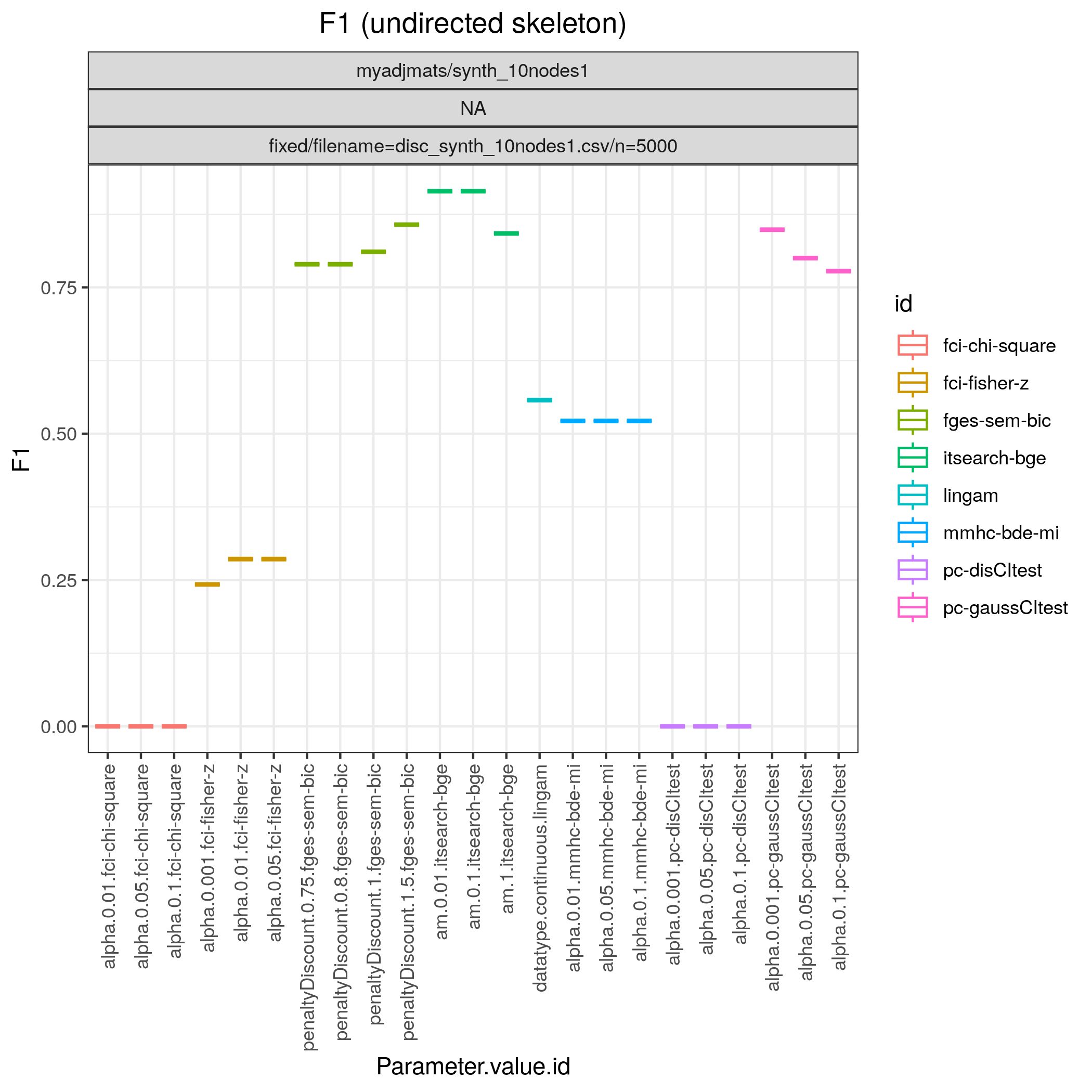}
	\caption{F1 Scores on the Synthetic 10 nodes data set. Discretized, no noise.}
\end{minipage}
\begin{minipage}{0.31\linewidth}
\centering
		\includegraphics[scale=0.34]{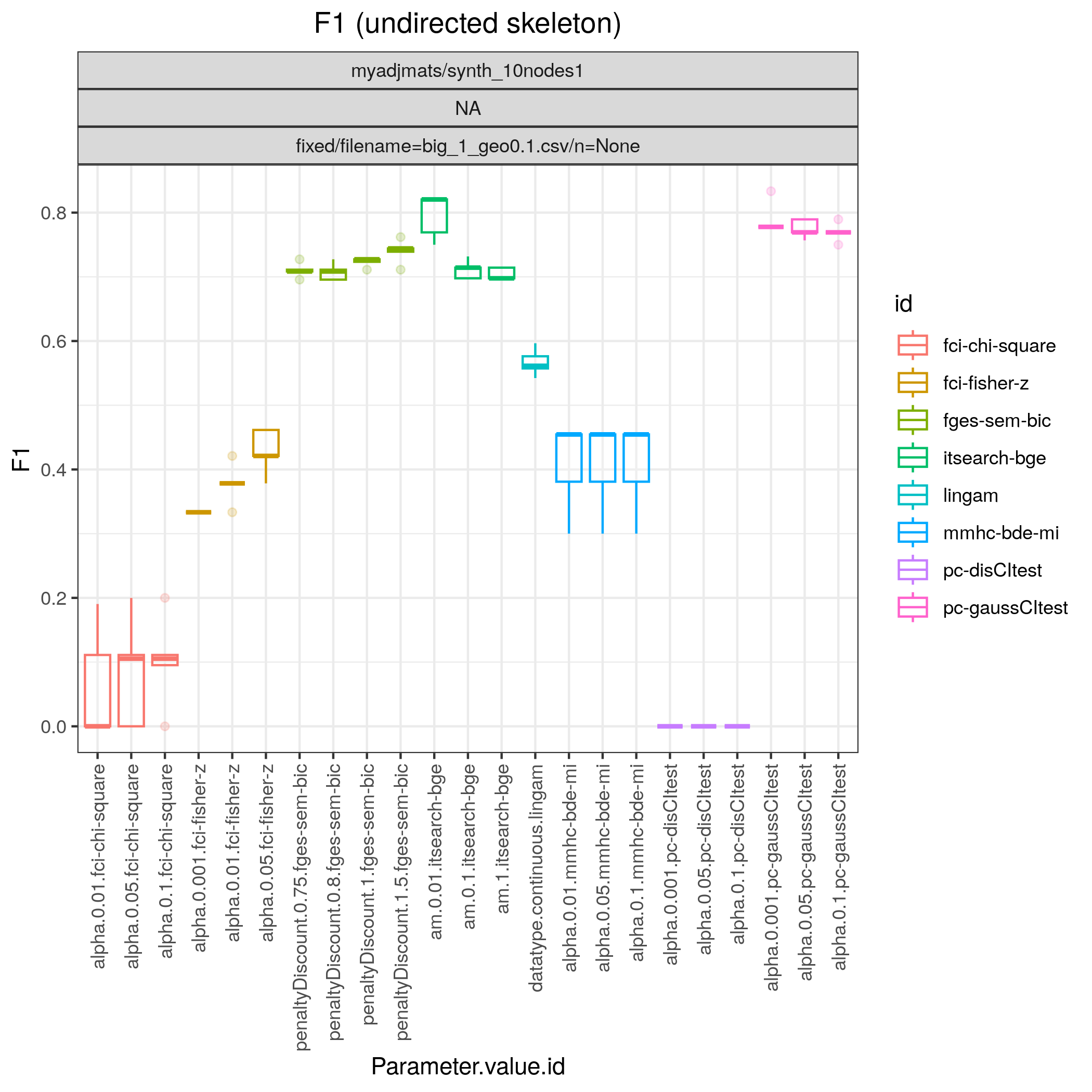}
	\caption{Synthetic 10 nodes data, Geo C-wise mechanism, max probability 0.1.}
\end{minipage}
\begin{minipage}{0.31\linewidth}
\centering
  \includegraphics[scale=0.34]{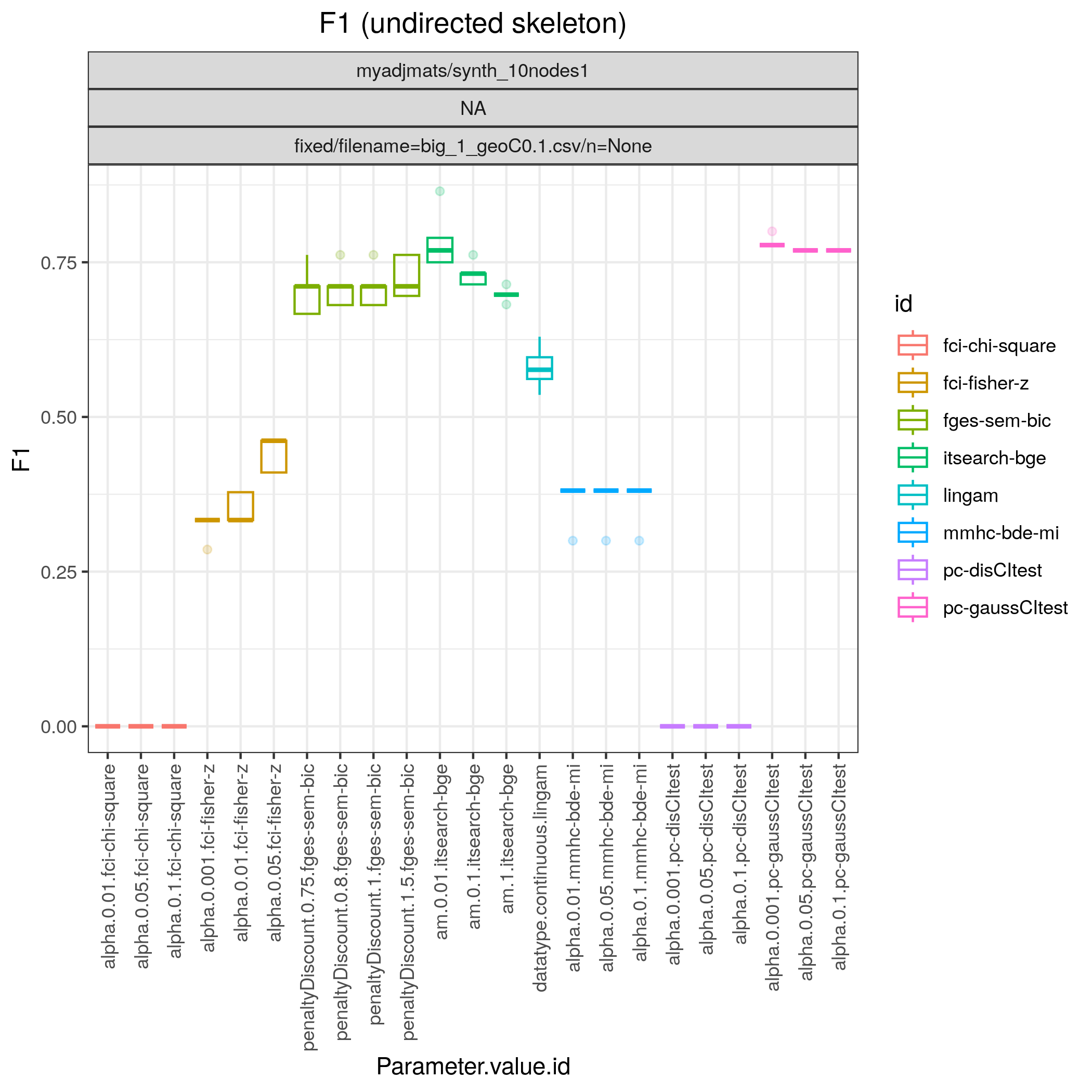}
	\caption{Synthetic 10 nodes data, Geo Comb mechanism, max probability 0.1.}
 \end{minipage}
    \begin{minipage}{0.31\linewidth}
\centering
  \includegraphics[scale=0.34]{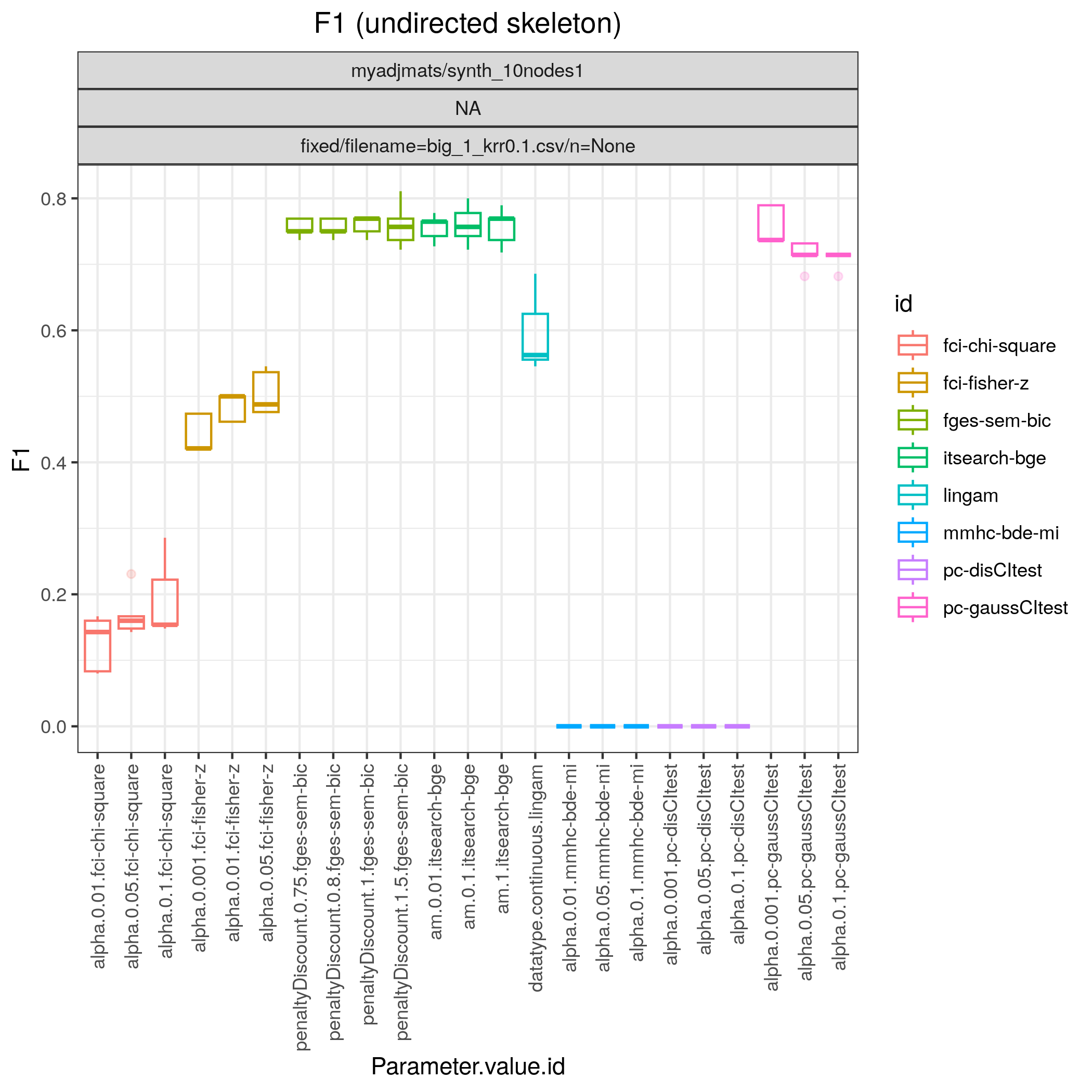}
	\caption{Synthetic 10 nodes data, $k$-RR C-wise mechanism, max probability 0.1.}
\end{minipage}
\begin{minipage}{0.31\linewidth}
\centering
  \includegraphics[scale=0.34]{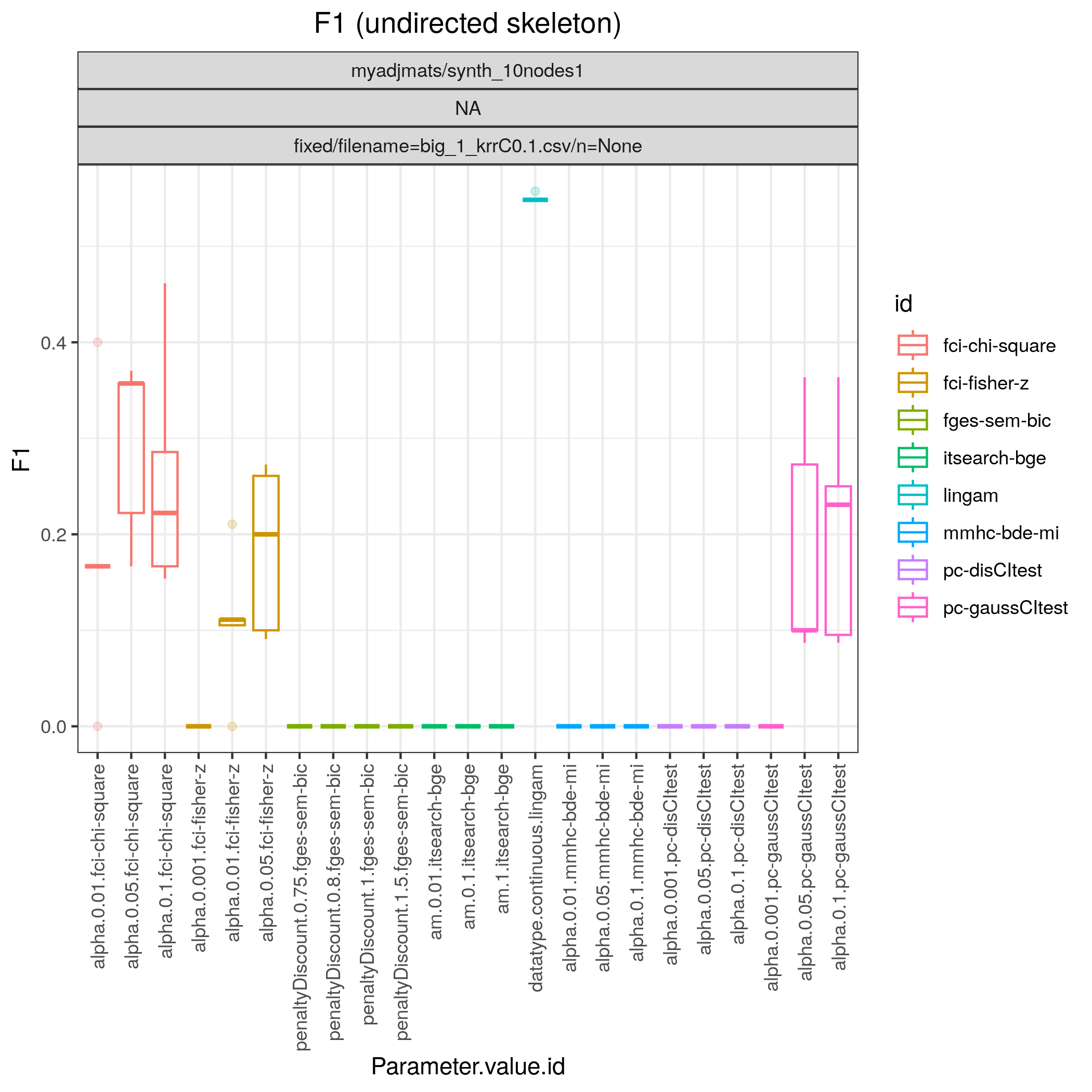}
	\caption{Synthetic 10 nodes data, $k$-RR Comb mechanism, max probability 0.1.}
\end{minipage}
\end{figure}


\noindent
\begin{figure}[H]
\begin{minipage}{0.31\linewidth}
\centering
		\includegraphics[scale=0.34]{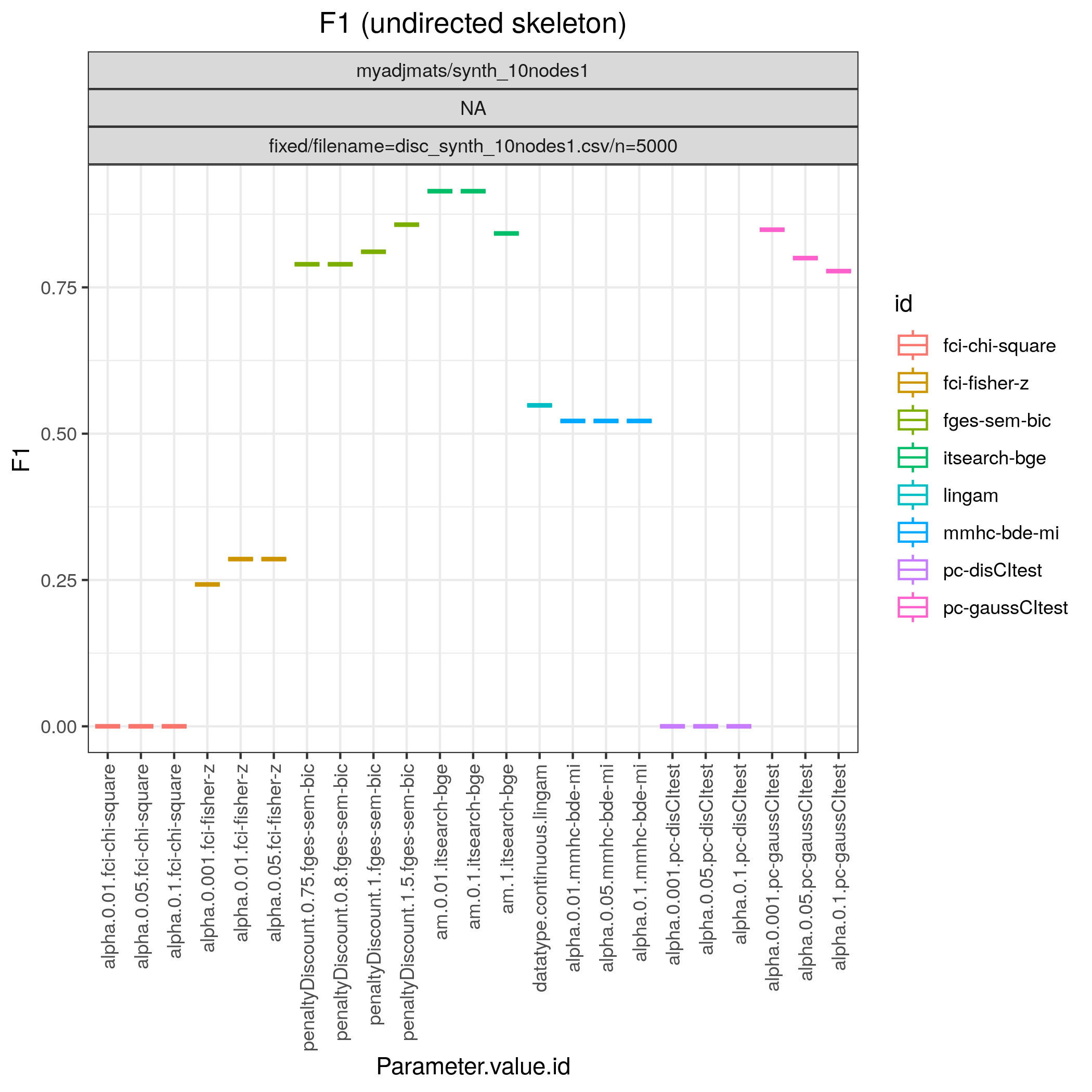}
	\caption{F1 Scores on the Synthetic 10 nodes data set. Discretized, no noise.}
\end{minipage}
\begin{minipage}{0.31\linewidth}
\centering
		\includegraphics[scale=0.34]{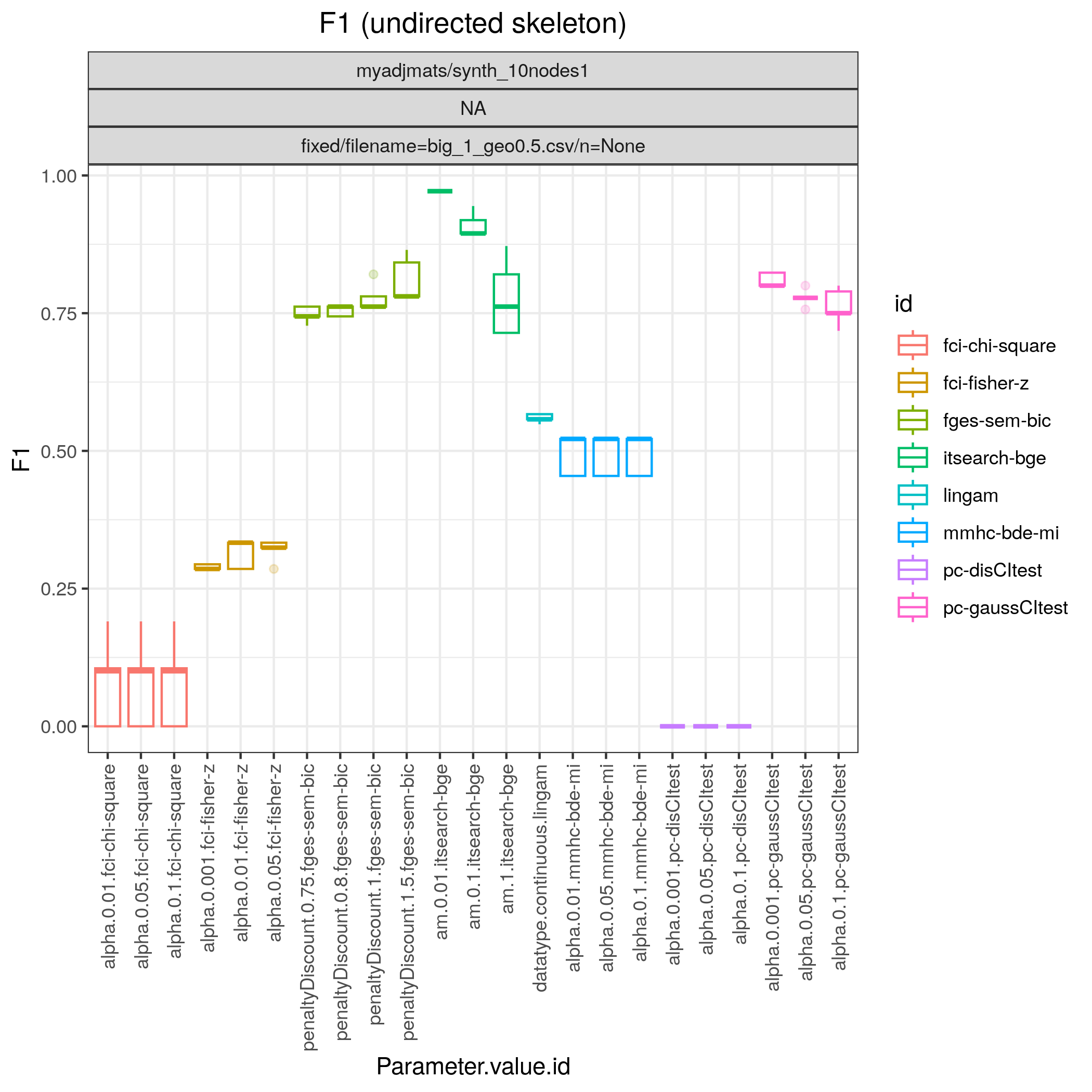}
	\caption{Synthetic 10 nodes data, Geo C-wise mechanism, max probability 0.5.}
\end{minipage}
\begin{minipage}{0.31\linewidth}
\centering
  \includegraphics[scale=0.34]{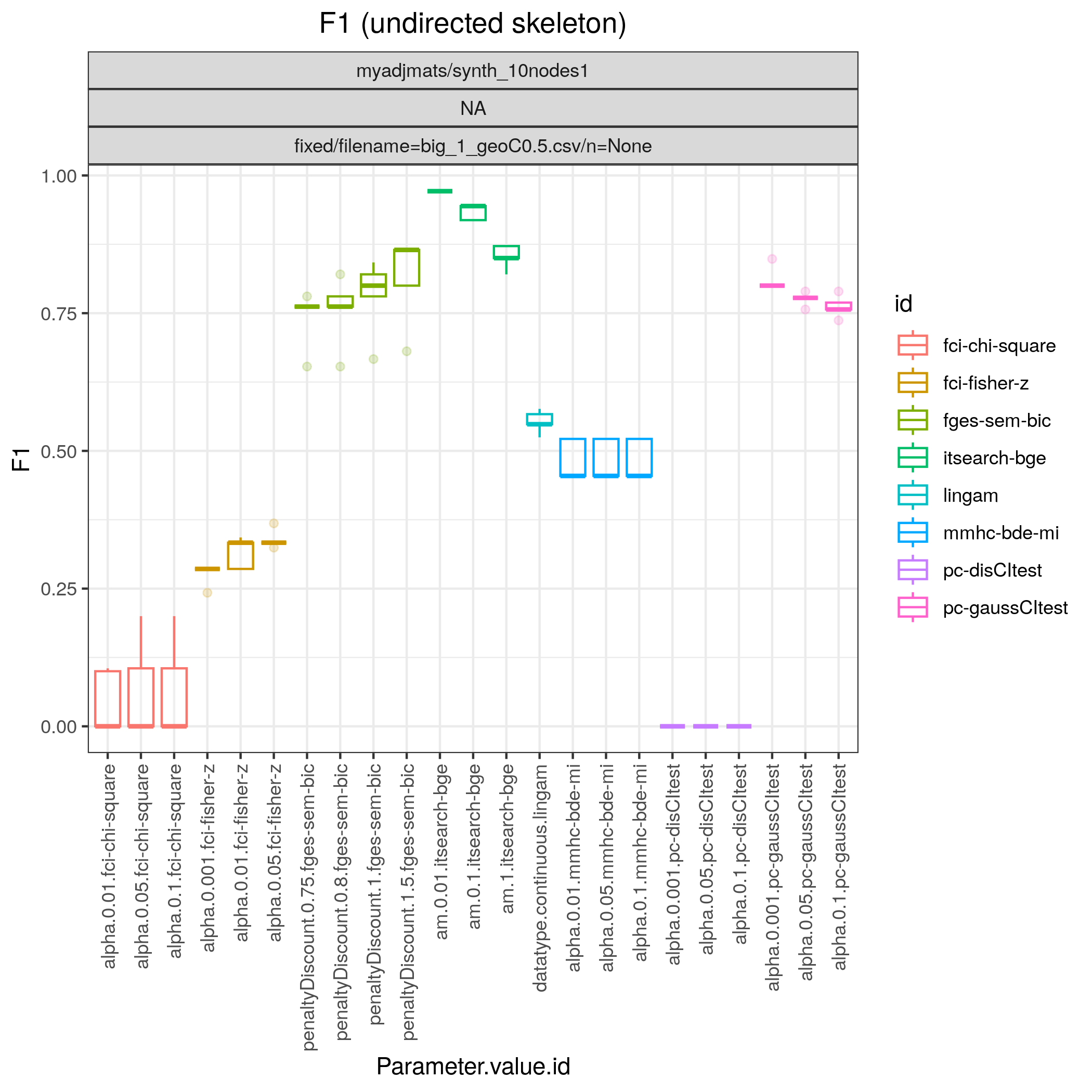}
	\caption{Synthetic 10 nodes data, Geo Comb mechanism, max probability 0.5.}
 \end{minipage}
    \begin{minipage}{0.31\linewidth}
\centering
  \includegraphics[scale=0.34]{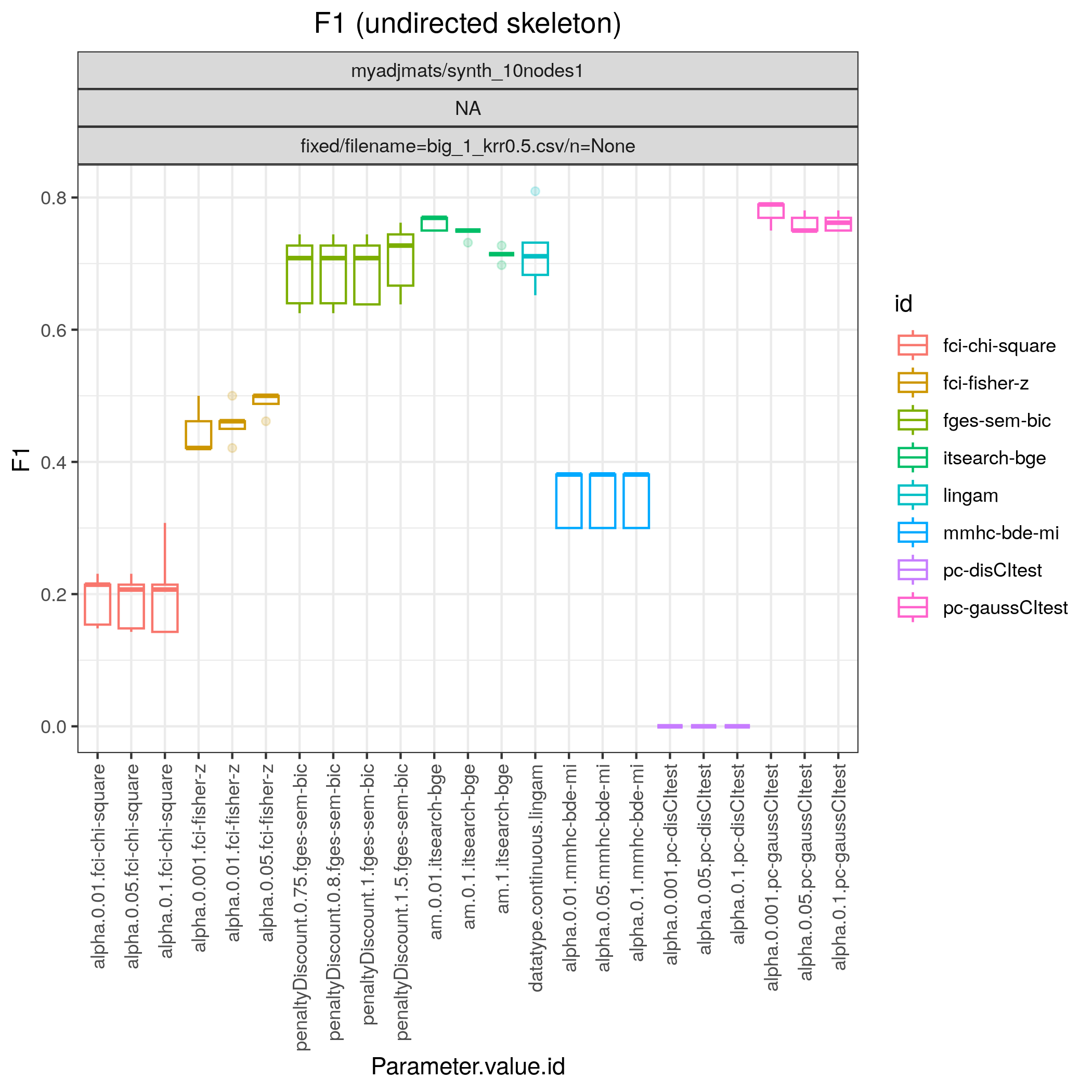}
	\caption{Synthetic 10 nodes data, $k$-RR C-wise mechanism, max probability 0.5.}
\end{minipage}
\begin{minipage}{0.31\linewidth}
\centering
  \includegraphics[scale=0.34]{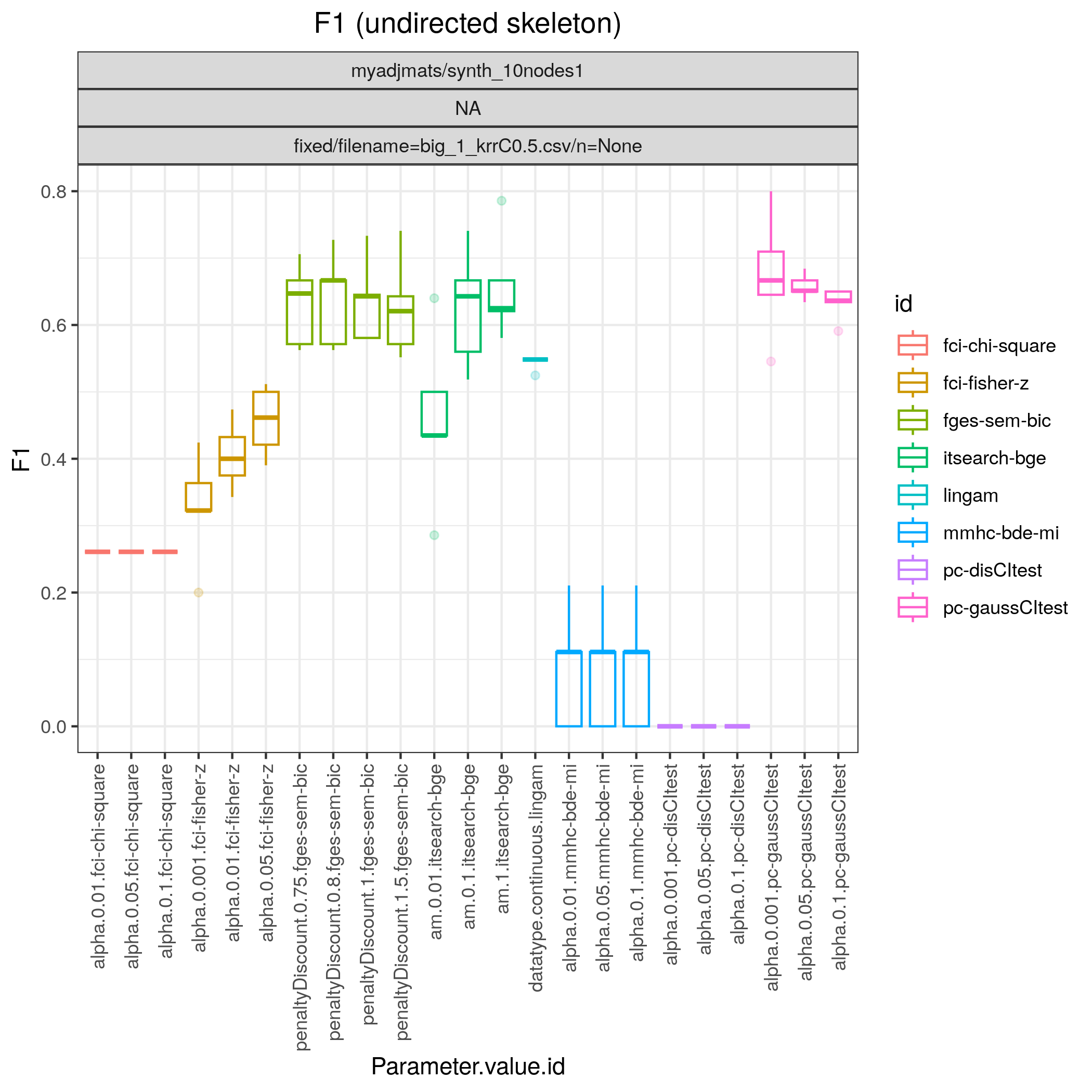}
	\caption{Synthetic 10 nodes data, $k$-RR Comb mechanism, max probability 0.5.}
\end{minipage}
\end{figure}

 \subsection{SHD Score results Synthetic 10 nodes data set}


\noindent
\begin{figure}[H]
\begin{minipage}{0.31\linewidth}
\centering
		\includegraphics[scale=0.34]{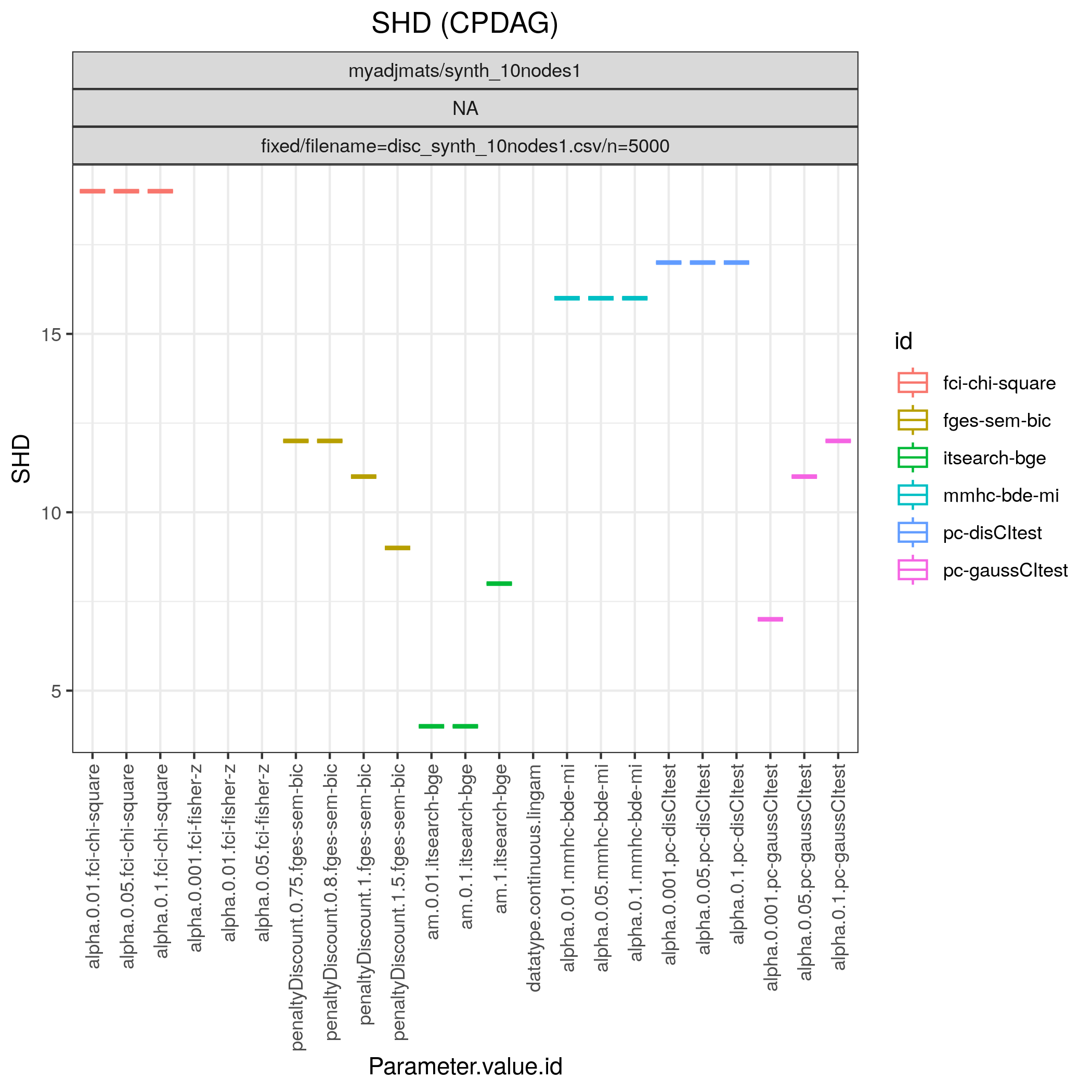}
	\caption{SHD Scores on the Synthetic 10 nodes data set. Discretized, no noise.}
\end{minipage}
\begin{minipage}{0.31\linewidth}
\centering
		\includegraphics[scale=0.34]{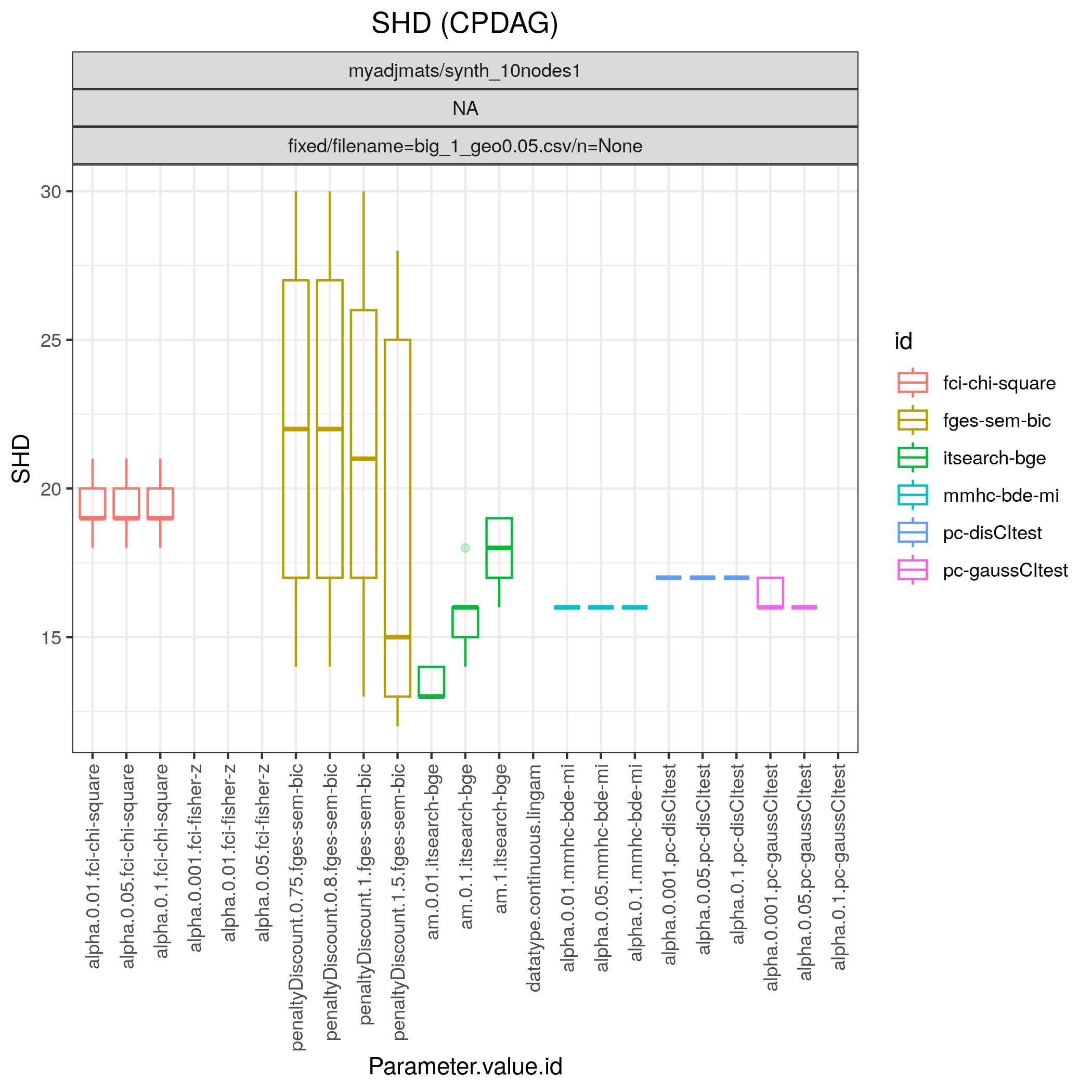}
	\caption{Synthetic 10 nodes data, Geo C-wise mechanism, max probability 0.05.}
\end{minipage}
\begin{minipage}{0.31\linewidth}
\centering
  \includegraphics[scale=0.34]{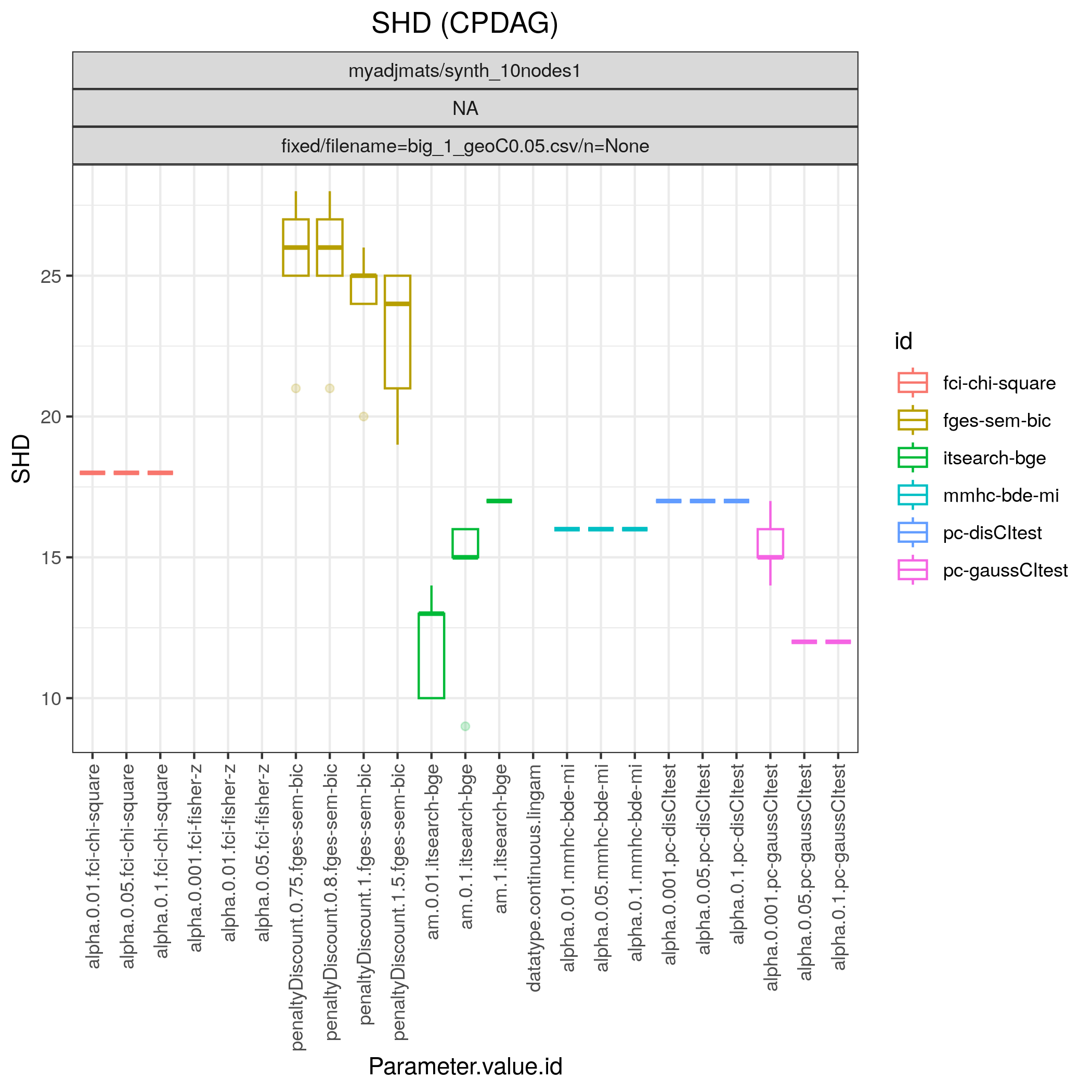}
	\caption{Synthetic 10 nodes data, Geo Comb mechanism, max probability 0.05.}
 \end{minipage}
    \begin{minipage}{0.31\linewidth}
\centering
  \includegraphics[scale=0.34]{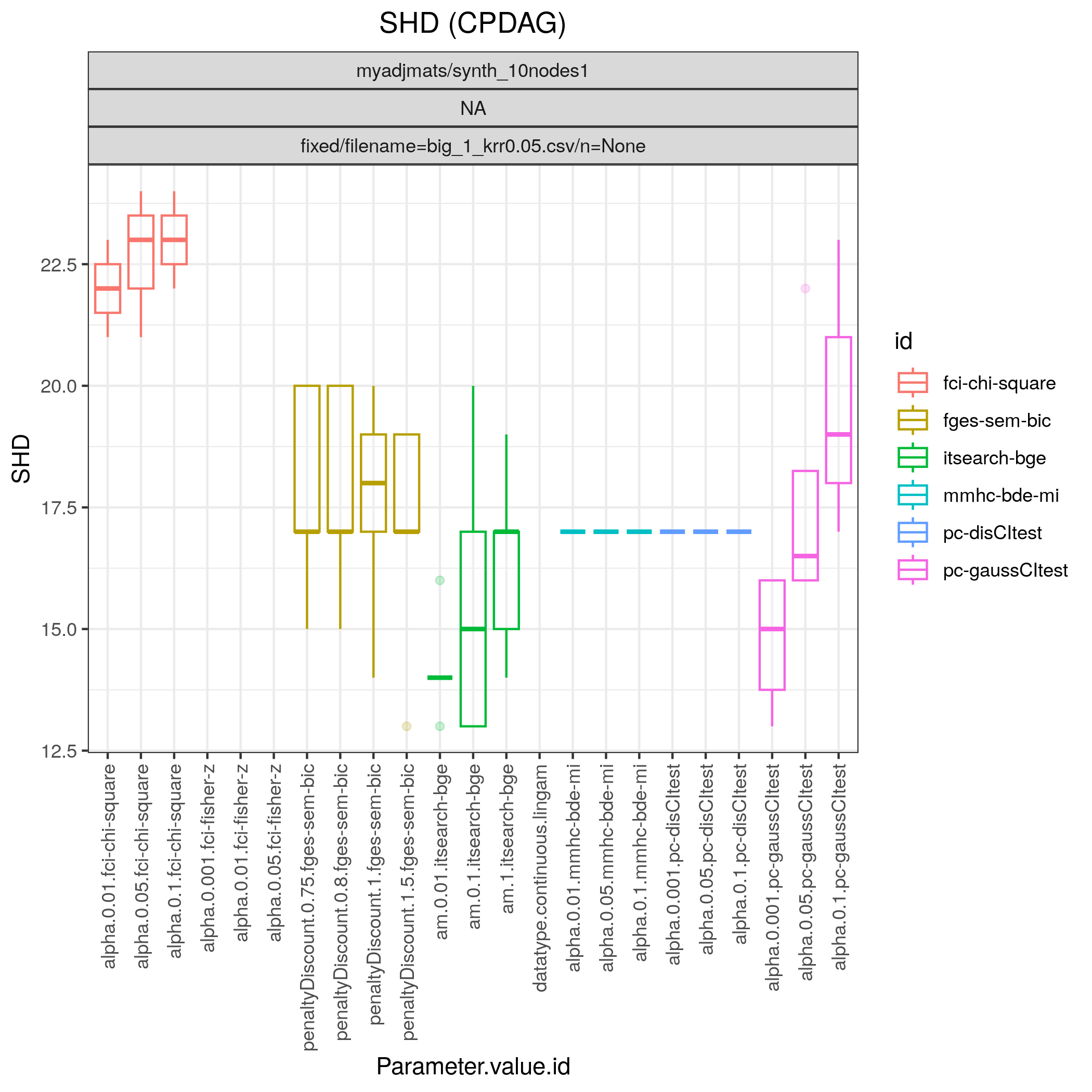}
	\caption{Synthetic 10 nodes data, $k$-RR C-wise mechanism, max probability 0.05.}
\end{minipage}
\begin{minipage}{0.31\linewidth}
\centering
  \includegraphics[scale=0.34]{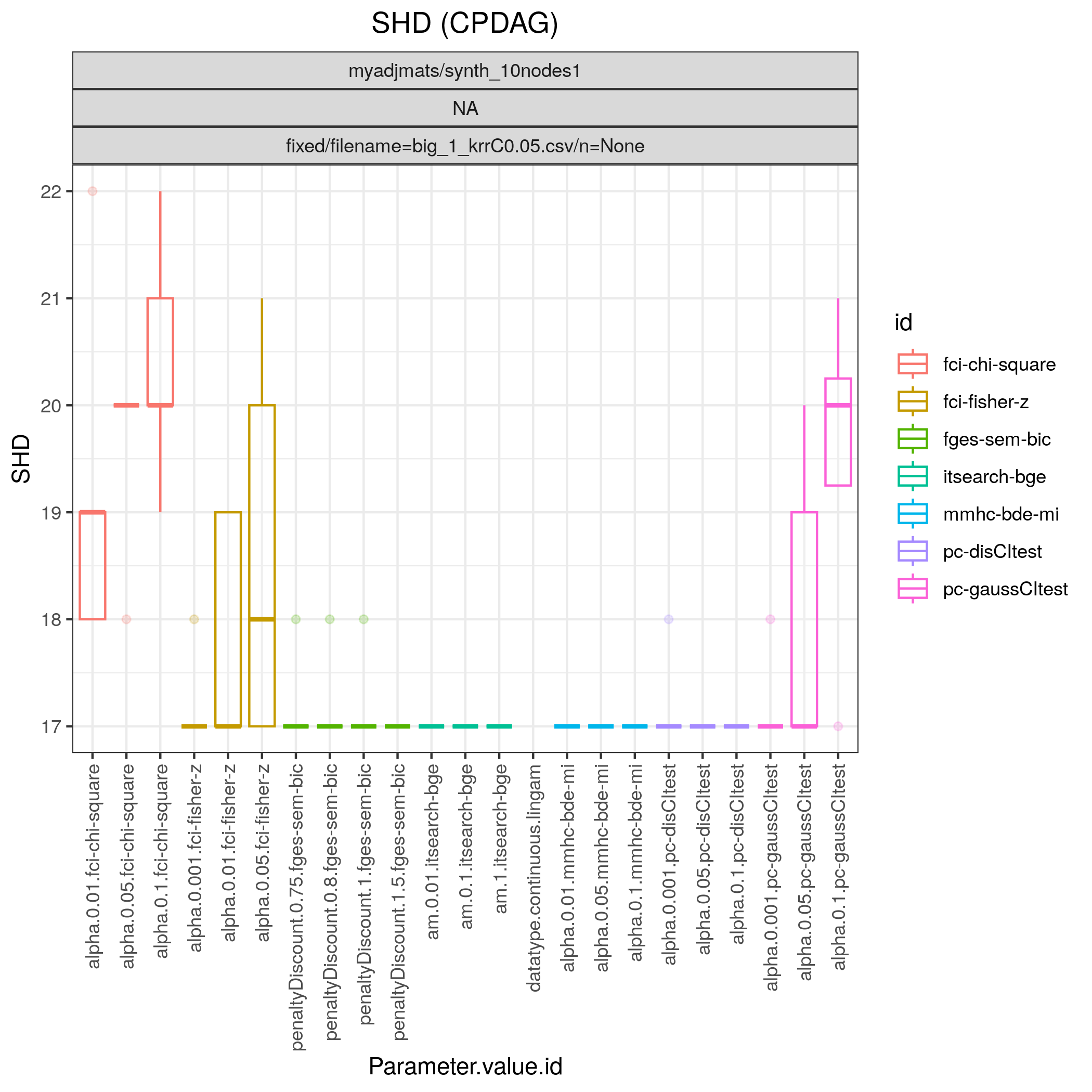}
	\caption{Synthetic 10 nodes data, $k$-RR Comb mechanism, max probability 0.05.}
\end{minipage}
\end{figure}


\noindent
\begin{figure}[H]
\begin{minipage}{0.31\linewidth}
\centering
		\includegraphics[scale=0.34]{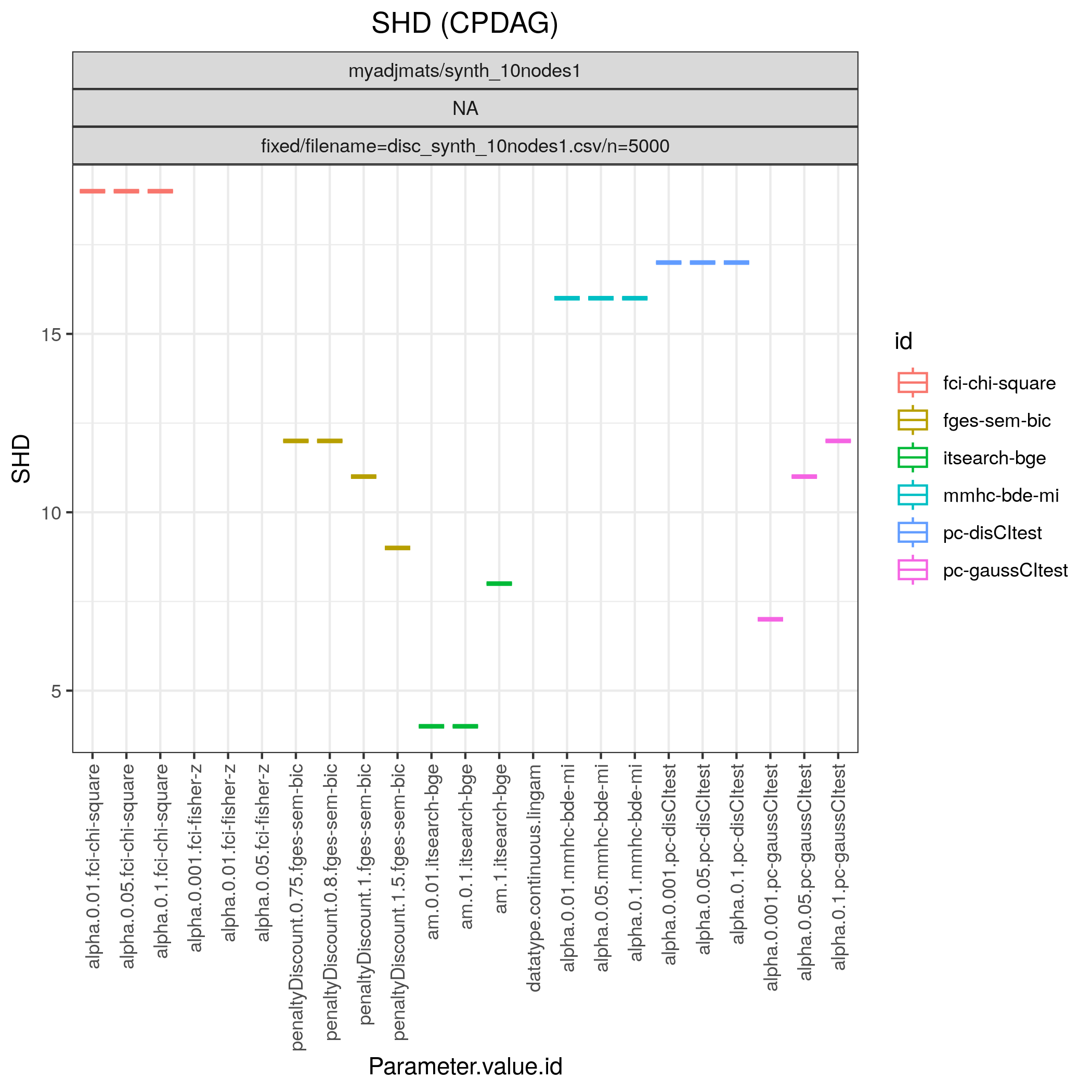}
	\caption{SHD Scores on the Synthetic 10 nodes data set. Discretized, no noise.}
\end{minipage}
\begin{minipage}{0.31\linewidth}
\centering
		\includegraphics[scale=0.34]{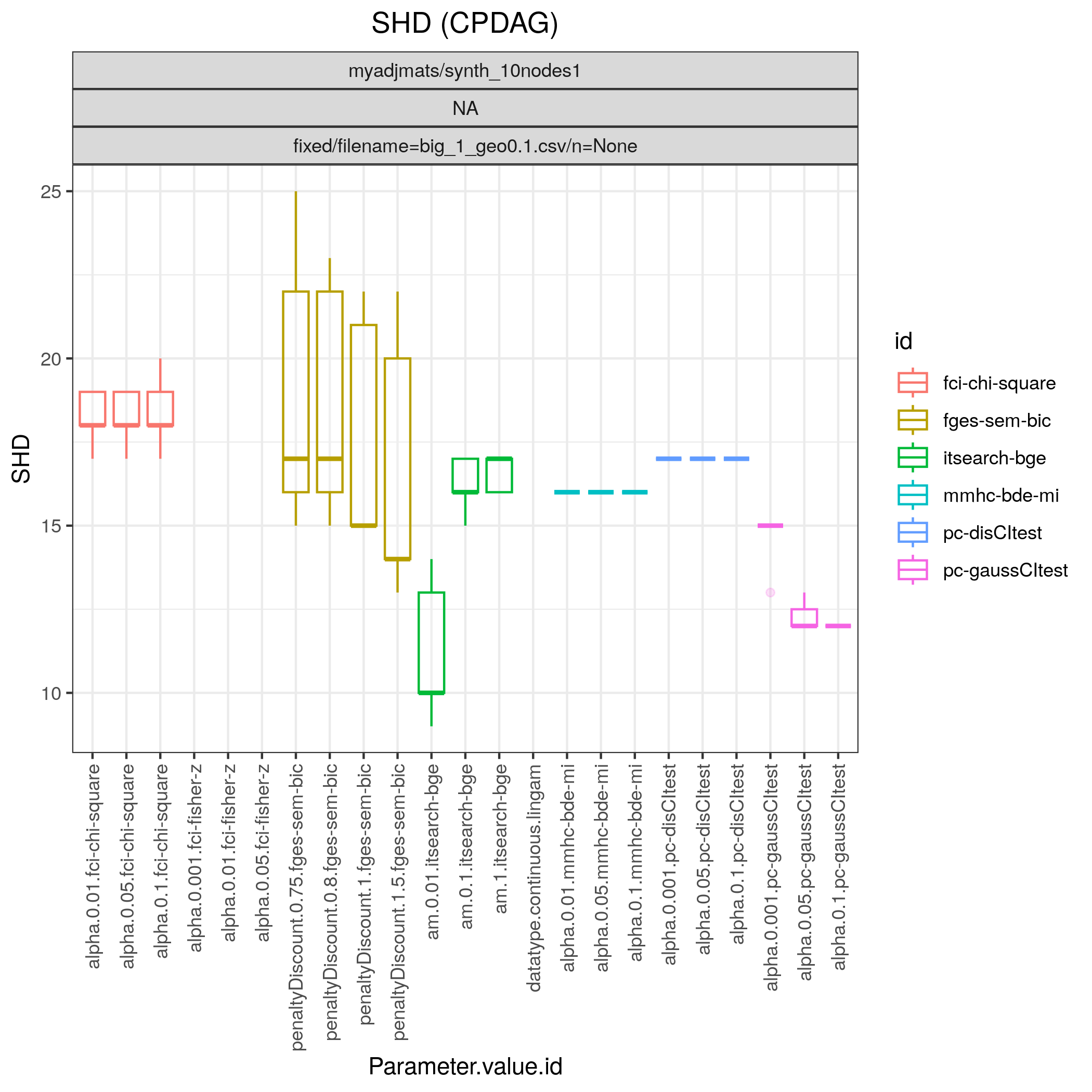}
	\caption{Synthetic 10 nodes data, Geo C-wise mechanism, max probability 0.1.}
\end{minipage}
\begin{minipage}{0.31\linewidth}
\centering
  \includegraphics[scale=0.34]{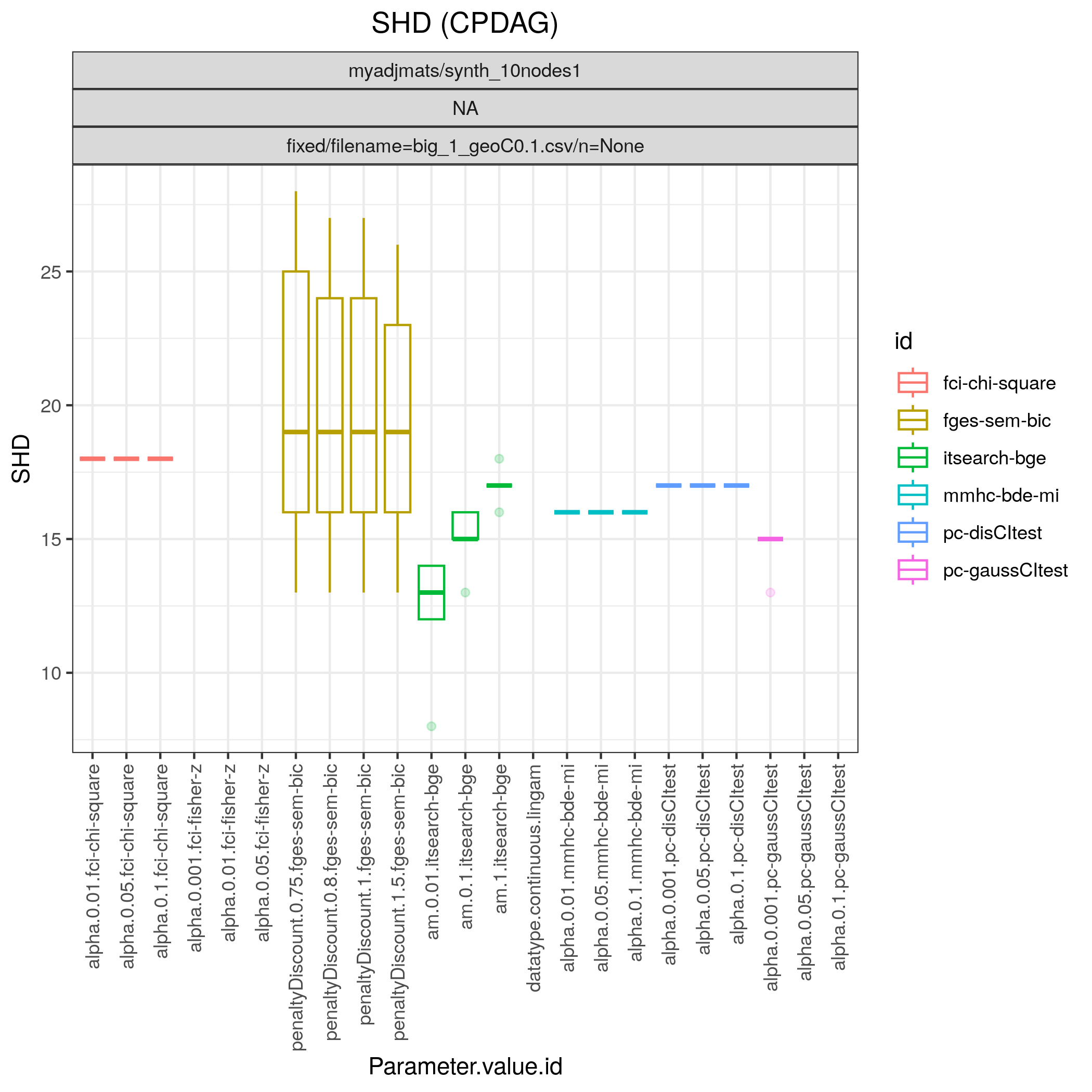}
	\caption{Synthetic 10 nodes data, Geo Comb mechanism, max probability 0.1.}
 \end{minipage}
    \begin{minipage}{0.31\linewidth}
\centering
  \includegraphics[scale=0.34]{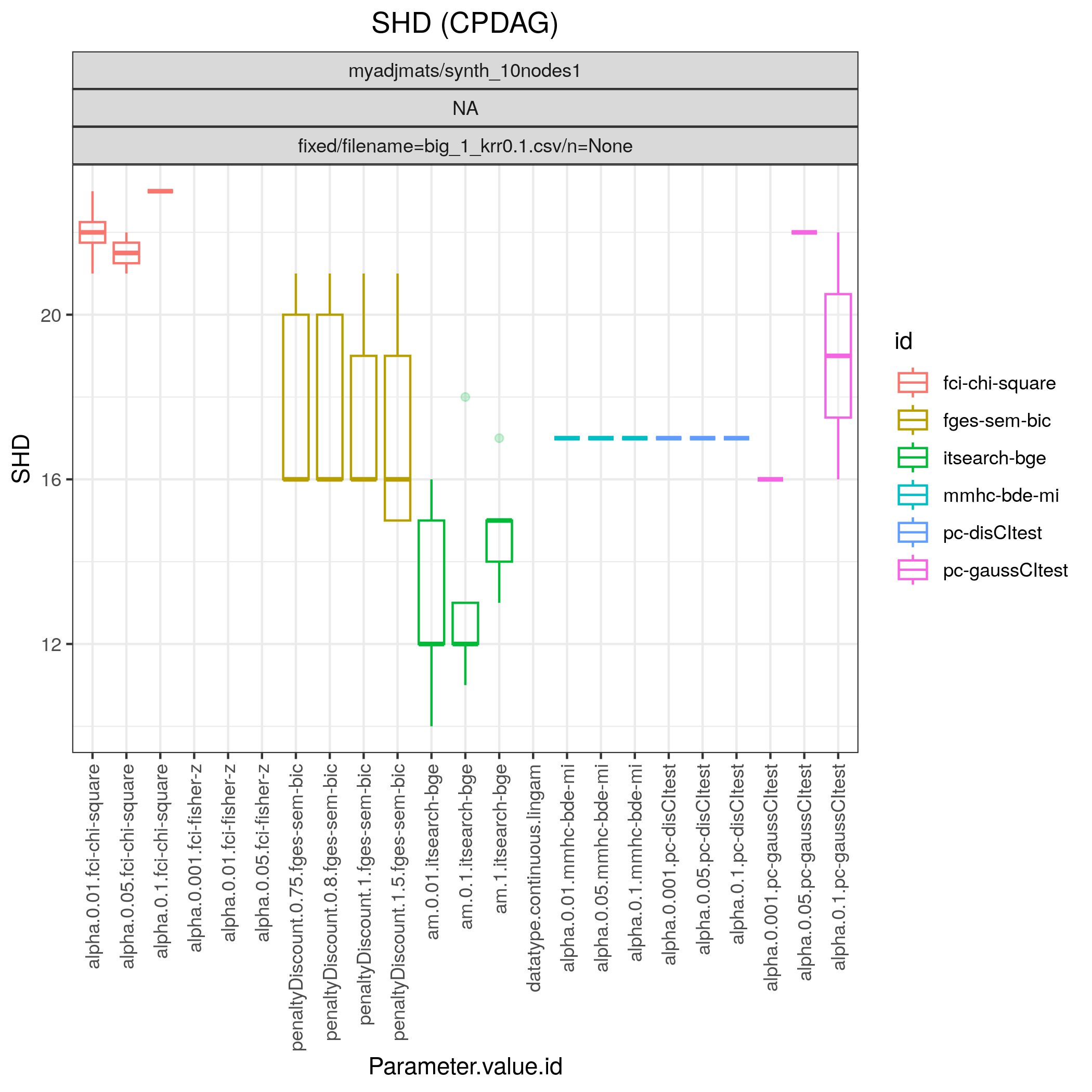}
	\caption{Synthetic 10 nodes data, $k$-RR C-wise mechanism, max probability 0.1.}
\end{minipage}
\begin{minipage}{0.31\linewidth}
\centering
  \includegraphics[scale=0.34]{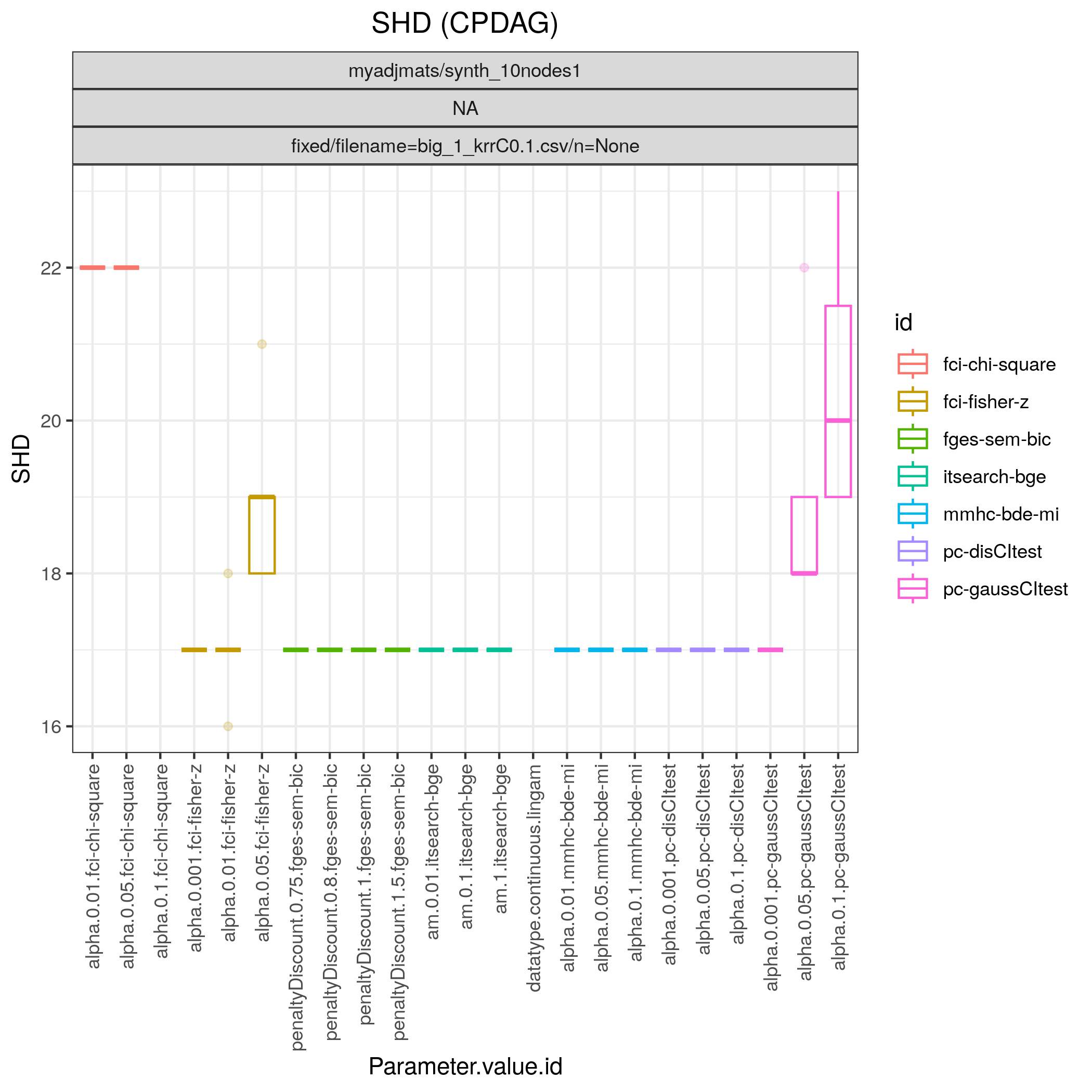}
	\caption{Synthetic 10 nodes data, $k$-RR Comb mechanism, max probability 0.1.}
\end{minipage}
\end{figure}


\noindent
\begin{figure}[H]
\begin{minipage}{0.31\linewidth}
\centering
		\includegraphics[scale=0.34]{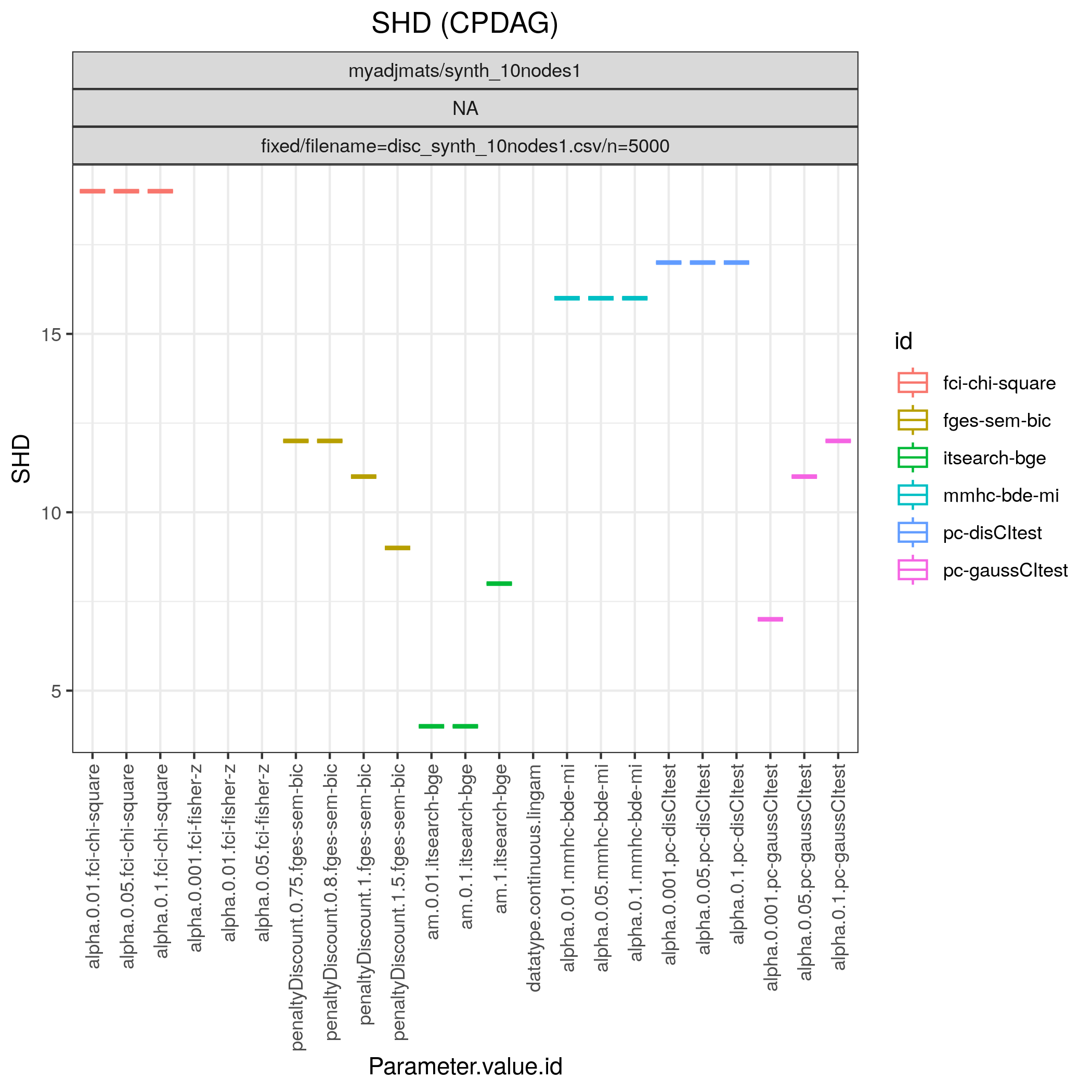}
	\caption{SHD Scores on the Synthetic 10 nodes data set. Discretized, no noise.}
\end{minipage}
\begin{minipage}{0.31\linewidth}
\centering
		\includegraphics[scale=0.34]{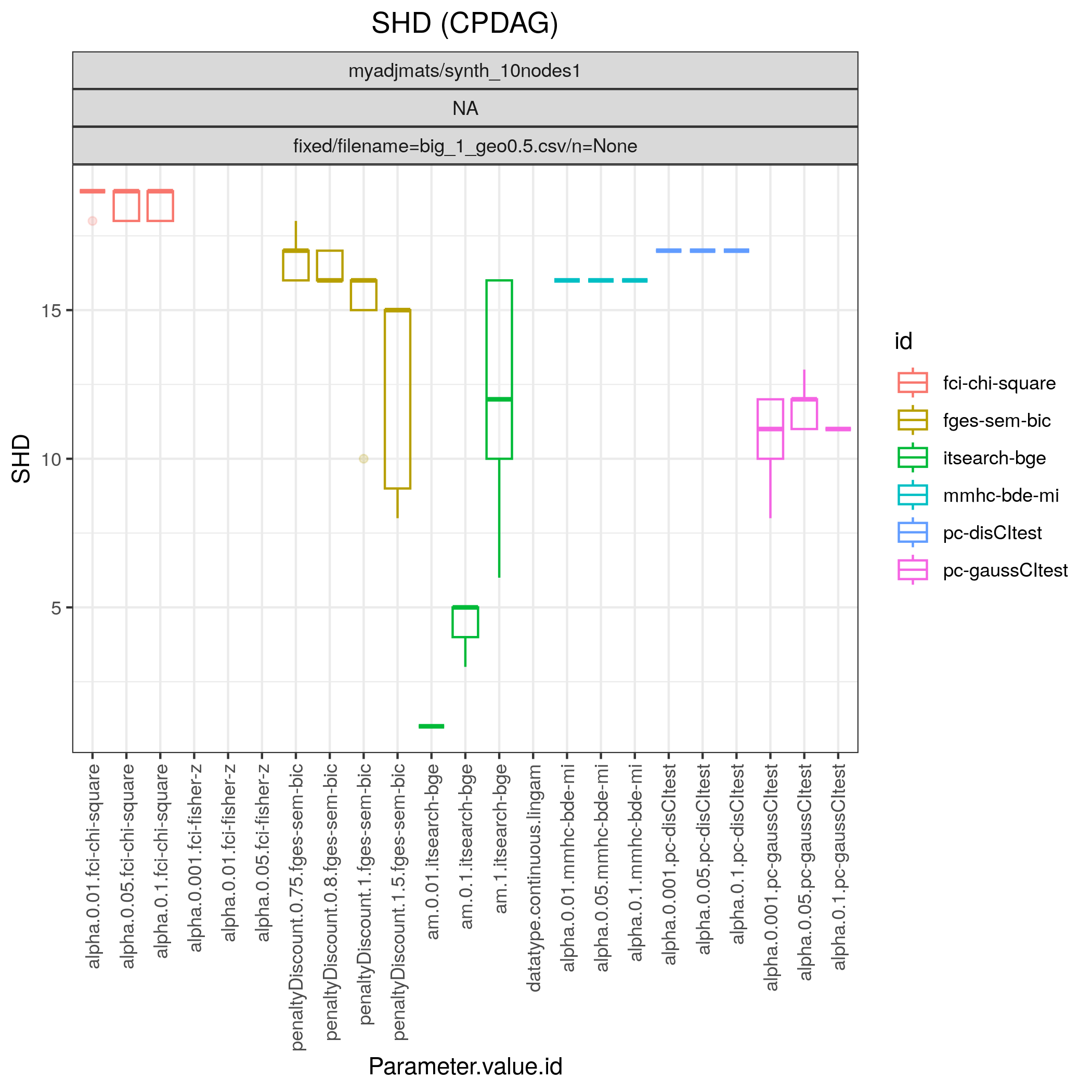}
	\caption{Synthetic 10 nodes data, Geo C-wise mechanism, max probability 0.5.}
\end{minipage}
\begin{minipage}{0.31\linewidth}
\centering
  \includegraphics[scale=0.34]{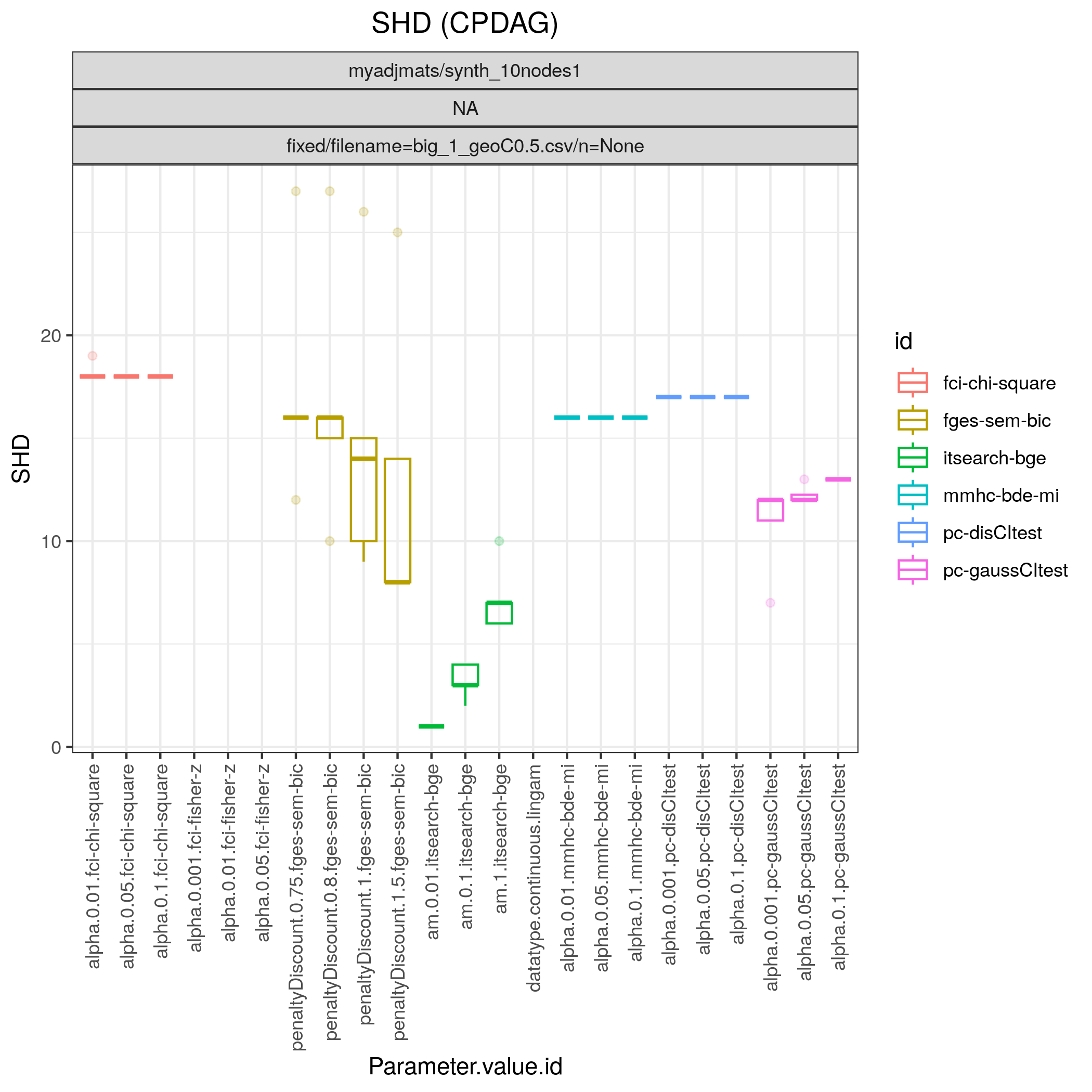}
	\caption{Synthetic 10 nodes data, Geo Comb mechanism, max probability 0.5.}
 \end{minipage}
    \begin{minipage}{0.31\linewidth}
\centering
  \includegraphics[scale=0.34]{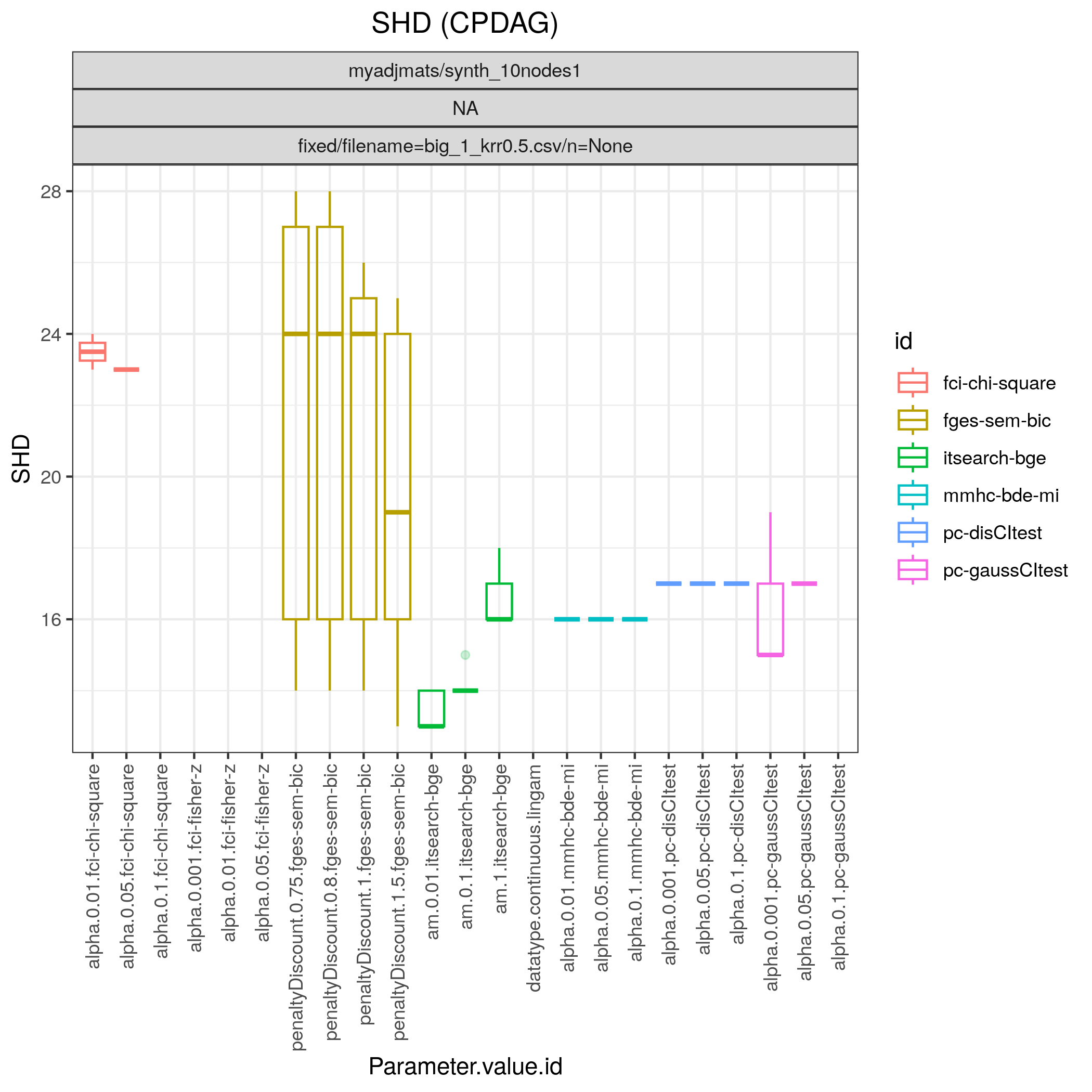}
	\caption{Synthetic 10 nodes data, $k$-RR C-wise mechanism, max probability 0.5.}
\end{minipage}
\begin{minipage}{0.31\linewidth}
\centering
  \includegraphics[scale=0.34]{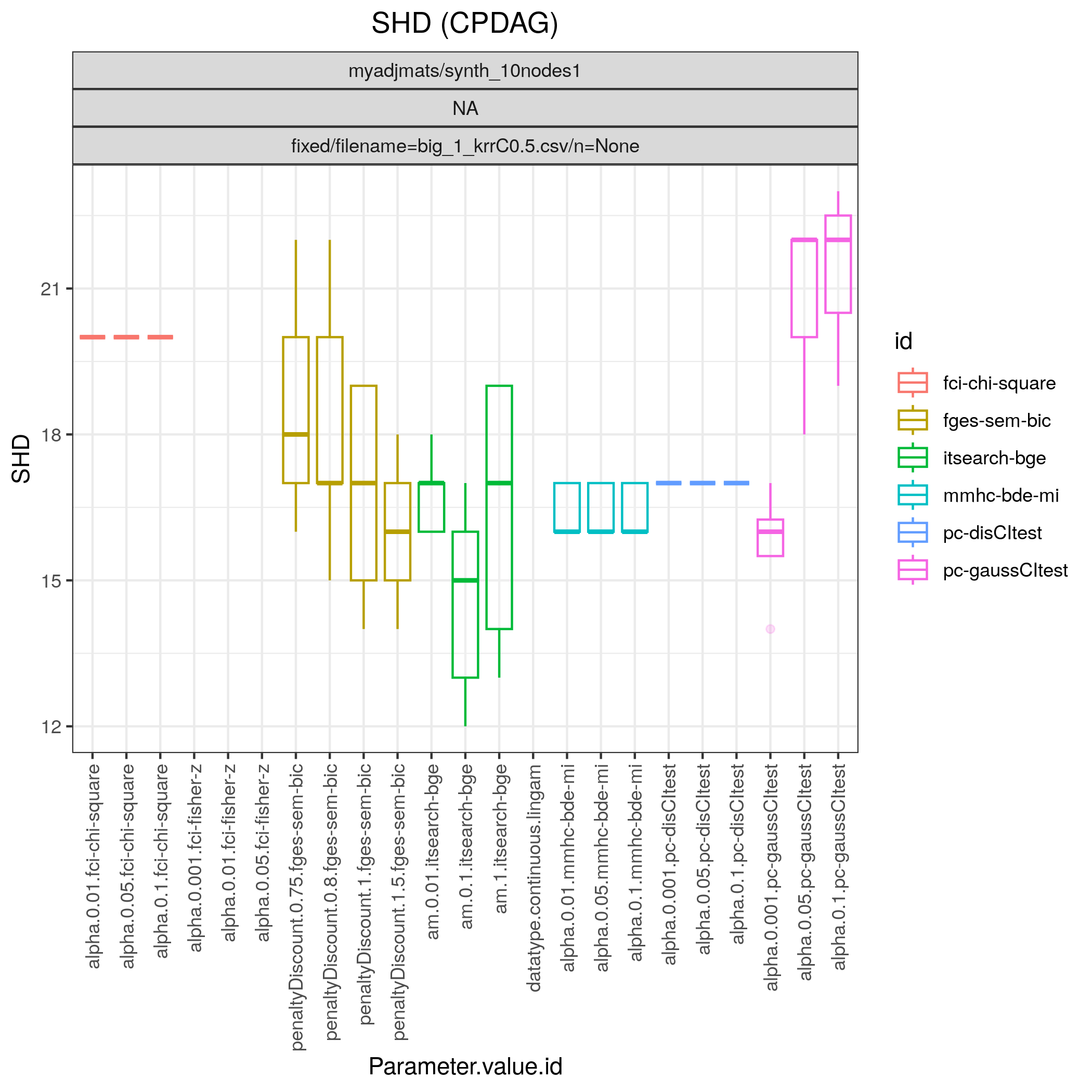}
	\caption{Synthetic 10 nodes data, $k$-RR Comb mechanism, max probability 0.5.}
\end{minipage}
\end{figure}





\end{document}